\documentclass[traditabstract,longauth]{aa}  
\usepackage{graphicx}
\usepackage{natbib}
\usepackage{lscape}
\usepackage{longtable}
\usepackage{soul}
\usepackage{txfonts}

\usepackage{float}

\def\mstar  {$M_{\star}$}
\def\macc   {$\dot{M}_{\rm acc}$}
\def\lacc   {$L_{\rm acc}$}

\def\msun {$M_{\odot}$}
\def\lsun {$L_{\odot}$}
\def\lstar {$L_\star$}
\newcommand{\degree}{\ensuremath{^\circ}}
\def\lline {$L_{\rm line}$}

\newcommand{\teff}{$T_{\rm eff}$}
\newcommand{\logg}{$\log g$}
\newcommand{\vsini}{$v\sin i$}
\def\nodata {...}

\usepackage{xcolor}

\begin{document}

   \title{PENELLOPE: The ESO data legacy program to complement the Hubble UV Legacy Library of Young Stars (ULLYSES)} \subtitle{I. Survey presentation and accretion properties of Orion OB1 and $\sigma$-Orionis\thanks{Based on observations collected at the European Southern Observatory under ESO programme 106.20Z8.} }

\titlerunning{PENELLOPE - I. Survey presentation and Orion}
\authorrunning{Manara et al.}


   \author{C.F. Manara \inst{\ref{instESO}} \and A. Frasca\inst{\ref{instCT}} \and L. Venuti\inst{\ref{instNASA}} \and M. Siwak \inst{\ref{instKO}} \and G.J. Herczeg \inst{\ref{KIAA} }    \and N. Calvet\inst{\ref{instMIC}} \and J. Hernandez\inst{\ref{instUNAM}} \and {\L}. Tychoniec \inst{\ref{instESO}} \and M. Gangi\inst{\ref{instRO}} \and J.M. Alcal\'a \inst{\ref{instNA}} \and H. M. J.  Boffin\inst{\ref{instESO}} \and B. Nisini \inst{\ref{instRO}}  \and M. Robberto\inst{\ref{instSTScI}} \and C. Briceno\inst{\ref{instCeTo}}  \and J. Campbell-White\inst{\ref{UoD}} \and A. Sicilia-Aguilar\inst{\ref{UoD}} \and P. McGinnis\inst{\ref{instDIAS}}\and D. Fedele\inst{\ref{instTO},\ref{instFI}} \and \'A. K\'osp\'al\inst{\ref{instKO},\ref{instHD},\ref{instEL}}  \and 
   P. \'Abrah\'am\inst{\ref{instKO},\ref{instEL}} \and
   J. Alonso-Santiago\inst{\ref{instCT}} \and S. Antoniucci \inst{\ref{instRO}} \and
   N. Arulanantham \inst{\ref{instSTScI}} \and F. Bacciotti \inst{\ref{instFI}}
  \and A. Banzatti \inst{\ref{instTxST}} \and G. Beccari \inst{\ref{instESO}} \and M. Benisty \inst{\ref{instIPAG}} \and K. Biazzo \inst{\ref{instRO}} \and J. Bouvier\inst{\ref{instIPAG}}
  \and S. Cabrit\inst{{\ref{instLERMA}}}
  \and A.Caratti o Garatti\inst{\ref{instDIAS}},\inst{\ref{instUCD}}  \and D. Coffey \inst{\ref{instUCD}},\inst{\ref{instDIAS}} \and E. Covino \inst{\ref{instNA}} \and C. Dougados \inst{\ref{instIPAG}} \and J. Eislöffel \inst{\ref{instTLS}} \and B. Ercolano\inst{\ref{instLMU}} \and C. C. Espaillat\inst{\ref{instBU}} \and J. Erkal \inst{\ref{instUCD},\ref{instESO}} \and S. Facchini \inst{\ref{instESO}} \and M. Fang \inst{\ref{pmo} } \and E. Fiorellino \inst{\ref{instRO},\ref{instKO}} \and W.J. Fischer\inst{\ref{instSTScI}} \and K. France \inst{\ref{LASP} } \and J.F. Gameiro\inst{\ref{instIA}} \and  R. Garcia Lopez\inst{\ref{instUCD}, \ref{instDIAS}} \and T. Giannini \inst{\ref{instRO}} \and C. Ginski\inst{\ref{instAPI}} \and K. Grankin \inst{\ref{instCrAO}} \and H.M. G\"unther\inst{\ref{instMIT}} \and L. Hartmann\inst{\ref{instMIC}} \and L.A. Hillenbrand \inst{\ref{instCaltech}} \and G.A.J. Hussain \inst{\ref{instESTEC}} \and M.M. James \inst{\ref{instIfA}} \and M. Koutoulaki\inst{\ref{instESO}} \and G. Lodato\inst{\ref{instMI}} \and K. Mauc\'o \inst{\ref{instUV},\ref{NPF}}  \and I.  Mendigut\'ia\inst{\ref{instCAB}} \and R. Mentel\inst{\ref{instUCD}, \ref{instDIAS}} \and A. Miotello\inst{\ref{instESO}} \and R.D. Oudmaijer \inst{\ref{leeds}} \and E. Rigliaco\inst{\ref{instPD}} \and G.P. Rosotti \inst{\ref{instLeid},\ref{instLeicester}} \and E. Sanchis\inst{\ref{instESO}} \and
  P.C. Schneider\inst{\ref{instHS}} \and L. Spina\inst{\ref{instPD}} \and 
  B. Stelzer\inst{\ref{instEKUT},\ref{instOAPa}} \and
  L. Testi\inst{\ref{instESO},\ref{instFI}} \and T. Thanathibodee\inst{\ref{instMIC}}  \and J.S. Vink \inst{\ref{aop}} \and F.M. Walter\inst{\ref{instSBU}} \and J. P. Williams \inst{\ref{instIfA}}  \and G. Zsidi \inst{\ref{instKO},\ref{instEL}}     }

   \institute{European Southern Observatory, Karl-Schwarzschild-Strasse 2, 85748 Garching bei M\"unchen, Germany\label{instESO}\\  \email{cmanara@eso.org}
\and 
INAF -- Osservatorio Astrofisico di Catania, via S. Sofia, 78, 95123 Catania, Italy\label{instCT}
\and
NASA Ames Research Center, Moffett Blvd, Mountain View, CA 94035, USA\label{instNASA}
\and
Konkoly Observatory, Research Centre for Astronomy and Earth Sciences, E\"otv\"os Lor\'and Research Network (ELKH), Konkoly-Thege Mikl\'os \'ut 15-17, H-1121 Budapest, Hungary\label{instKO}
\and
Kavli Institute for Astronomy and Astrophysics, Peking University, Yiheyuan 5, Haidian Qu, 100871 Beijing, China
\label{KIAA}
\and
Department of Astronomy, University of Michigan, 1085 S. University Ave., Ann Arbor, MI 48109 USA \label{instMIC}
\and
Instituto de Astronom\'ia, UNAM, Campus Ensenada,  Carretera Tijuana-Ensenada km 103, 22860 Ensenada, B.C. M\'exico \label{instUNAM}
\and
INAF -- Osservatorio Astronomico di Roma, via di Frascati 33, 00078 Monte Porzio Catone, Italy\label{instRO}
\and
INAF -- Osservatorio Astronomico di Capodimonte, via Moiariello 16, 80131 Napoli, Italy\label{instNA}
\and
Space Telescope Science Institute, 3700 San Martin Drive, Baltimore, MD 21218, USA\label{instSTScI}
\and 
Cerro Tololo Inter-American Observatory/NSF's NOIRLab, Casilla 603, La Serena, Chile\label{instCeTo}
\and
SUPA, School of Science and Engineering, University of Dundee, Nethergate, Dundee, DD1 4HN, U.K.\label{UoD}
\and 
School of Cosmic Physics, Dublin Institute for Advanced Studies, 31 Fitzwilliam Place, Dublin 2, Ireland\label{instDIAS}
\and 
INAF -- Osservatorio Astrofisico di Torino, Via Osservatorio 20, I-10025 Pino Torinese, Italy\label{instTO}
\and 
INAF -- Osservatorio Astrofisico di Arcetri, L.go E. Fermi 5, 50125 Firenze, Italy\label{instFI}
\and
Max Planck Institute for Astronomy, K\"onigstuhl 17, D-69117 Heidelberg, Germany\label{instHD}
\and
ELTE E\"otv\"os Lor\'and University, Institute of Physics, P\'azm\'any P\'eter s\'et\'any 1/A, H-1117 Budapest, Hungary\label{instEL}
\and
Department of Physics, Texas State University, 749 N Comanche Street, San Marcos, TX 78666, USA \label{instTxST}
\and 
Univ. Grenoble Alpes, CNRS, IPAG, 38000 Grenoble, France
\label{instIPAG}
\and
Observatoire de Paris, PSL University, Sorbonne University, CNRS, LERMA, 61 Av. de l'Observatoire, 75014 Paris, France\label{instLERMA}
\and
School of Physics, University College Dublin, Belfield, Dublin 4, Ireland\label{instUCD}
\and 
Th\"uringer Landessternwarte, Sternwarte 5, D-07778 Tautenburg, Germany\label{instTLS}
\and 
Universit\"at Sternwarte M\"unchen, Ludwig-Maximillian-Universit\"at, Scheinerstrasse 1, 81679 M\"unchen, Germany \label{instLMU}
\and
Institute for Astrophysical Research, Department of Astronomy, Boston University, 725 Commonwealth Avenue, Boston, MA 02215, USA\label{instBU}
\and
Purple Mountain Observatory, Chinese Academy of Sciences, 10 Yuanhua Road, Nanjing 210023, China \label{pmo}
\and
Laboratory for Atmospheric and Space Physics, University of Colorado.  Boulder, CO 80309, USA \label{LASP}
\and
Instituto de Astrof\'{\i}sica e Ci\^encias do Espa\c co, CAUP \& DFA/FCUP, Universidade do Porto, Rua das Estrelas, 4150-762 Porto, Portugal. \label{instIA}
\and
Sterrenkundig Instituut Anton Pannekoek, Science Park 904, 1098 XH Amsterdam, The Netherlands
\label{instAPI}
\and
Crimean Astrophysical Observatory, 298409 Nauchny, Russia \label{instCrAO}
\and MIT Kavli Institute for Astrophysics and Space Research, 77 Massachusetts Avenue, Cambridge, MA 02139, USA
\label{instMIT}
\and
California Institute of Technology, 1200 East California Blvd, Pasadena, CA 91125, USA\label{instCaltech}
\and 
European Space Agency (ESA), European Space Research and Technology Centre (ESTEC), Keplerlaan 1, 2201 AZ Noordwijk, The Netherlands \label{instESTEC}
\and
Institute for Astronomy, University of Hawaii at Manoa, Honolulu, USA\label{instIfA}
\and
Dipartimento di Fisica, Universit\'a degli Studi di Milano, Via Giovanni Celoria 16, I-20133 Milano, Italy\label{instMI}
\and
Instituto  de  F\'isica  y  Astronom\'ia,  Facultad  de  Ciencias,  Universidad de Valpara\'iso, Av. Gran Breta\~na 1111, 5030 Casilla, Valpara\'iso, Chile \label{instUV}
\and
N\'ucleo Milenio de Formaci\'on Planetaria - NPF, Universidad de Valpara\'iso, Av. Gran Breta\~na 1111, Valpara\'iso, Chile \label{NPF}
\and
Centro de Astrobiolog\'ia (CSIC-INTA), Departamento de Astrof\'isica, ESA-ESAC Campus, PO Box 78, 28691 Villanueva de la Cañada, Madrid, Spain.\label{instCAB}
\and
School of Physics and Astronomy, University of Leeds, Leeds, LS2 9JT, United Kingdom \label{leeds}
\and
INAF -- Osservatorio Astronomico di Padova, Vicolo dell'Osservatorio 5, I-35122, Padova, Italy\label{instPD}
\and
Leiden Observatory, Leiden University, PO Box 9513, 2300 RA Leiden, The Netherlands\label{instLeid}
\and
School of Physics and Astronomy, University of Leicester, Leicester LE1 7RH, UK\label{instLeicester}
\and 
Hamburger Sternwarte, Gojenbergsweg 112, 21029 Hamburg, Germany \label{instHS}
\and 
Eberhard-Karls Universit\"at T\"ubingen, Sand 1, 72076 T\"ubingen, Germany \label{instEKUT}
\and 
INF - Osservatorio Astronomico di Palermo, Piazza del Parlamento 1, 90134 Palermo, Italy \label{instOAPa}
\and
Armagh Observatory and Planetarium, College Hill, Armagh BT61 9DG, Northern Ireland \label{aop}
\and Stony Brook University, Stony Brook, NY 11794, USA
\label{instSBU}
}
             
 \date{Received Feb 23, 2021; accepted Mar 22, 2021}
 
 
  \abstract
  {The evolution of young stars and disks is driven by the interplay of several processes, notably the accretion and ejection of material. These processes, critical to correctly describe the conditions of planet formation, are best probed spectroscopically. Between  2020--2022, about 500 orbits of the Hubble Space Telescope (HST) are being devoted in to the ULLYSES public survey of about 70 low-mass ($M_\star \le 2 M_\odot$) young (age$<$10 Myr) stars at UV wavelengths. Here, we present the PENELLOPE Large Program carried out with the ESO Very Large Telescope (VLT) with the aim of acquiring, contemporaneously to the HST, optical ESPRESSO/UVES high-resolution spectra for the purpose of investigating the kinematics of the emitting gas, along with UV-to-NIR X-Shooter medium-resolution flux-calibrated spectra 
  to provide the fundamental parameters that HST data alone cannot provide, such as extinction and stellar properties. The data obtained by PENELLOPE have no proprietary time and the fully reduced spectra are being made available to the whole community. Here, we describe the data and the first scientific analysis of the accretion properties for the sample of 13 targets located in the Orion OB1 association 
and in the $\sigma$-Orionis cluster, observed in November-December 2020. 
We find that the accretion rates are in line with those observed previously in similarly young star-forming regions, with a variability on a timescale of days  $(\lesssim$3). The comparison of the fits to the continuum excess emission obtained with a slab model on the X-Shooter spectra and the HST/STIS spectra shows a shortcoming in the X-Shooter estimates of $\lesssim$10\%, which is well within the assumed uncertainty. Its origin can be either due to an erroneous UV extinction curve or to the simplicity of the modeling and, thus, this question will form the basis of the investigation undertaken over the course of the PENELLOPE program.
The combined ULLYSES and PENELLOPE data will be key in attaining a better understanding of the accretion and ejection mechanisms in young stars. }


   \maketitle
%

\section{Introduction}
The formation and evolution of stars and planets are processes that are intimately connected and our knowledge of the properties of young stars and their circumstellar disks are key to understanding planet formation \citep[e.g.,][]{morby16}. 
The early evolution of stars and disks is regulated by the interplay of several mechanisms, in particular, accretion onto the star \citep[see reviews by][]{bouvier07,hartmann16} and ejection of matter from the disks through winds and outflows \citep[see reviews by e.g.,][]{frank14,EP17}, as well as processes that are attributed to the effects of the local environment \citep[e.g.,][]{fischer17,winter18}. These processes are driven by the transfer of angular momentum in the disk \citep[e.g.,][]{lyndenbell74,pringle81,bai16,PR19}. Thus,  the study of accretion and ejection it is fundamental
in order to consistently explain how the initial angular momentum is distributed and how disks are dispersed \citep[e.g.,][]{alexander14}. Both of these characteristics are necessary  for the explanation of why planetary systems are formed so differently from each other.

The task, however, is not a simple one. 
It is only by observing large samples of young stars and probing wide ranges of age and mass that we can gain access to a sufficient set of statistics to aid in the understanding of both processes. 
Spectroscopic surveys of young stars in different star-forming regions have shown that the mass accretion rates onto the central star (\macc) slowly decrease with isochronal ages, possibly in accordance with viscous evolution models \citep[e.g.,][]{hartmann98,sicilia-aguilar10,antoniucci14}, although high accretion rates are still observed at ages of $>$5--10 Myr \citep[e.g.,][]{ingleby14,Frasca2015,Rugel2018,venuti19,manara20}. The empirical measurement of the steep dependence of \macc ~ on the stellar mass (\mstar) \citep{hillenbrand92,muzerolle03,calvet04,mohanty05,natta06,manara12,manara16a,manara17a,alcala14,alcala17,venuti14} may be the consequence of a single dominant phenomenon or a mix of many, such as, initial conditions followed by viscous evolution \citep[e.g.,][]{alexander06,dullemond06}, an imprint of internal photoevaporation \citep[e.g.,][]{clarke06,ercolano14}, different accretion regimes at different stellar masses \citep[e.g.,][]{mohanty05,hartmann06}, environmental effects \citep[e.g.,][]{padoan05}, or self-gravity in the disk \citep[e.g.,][]{vorobyov09,desouza16}. Recently, the connection of measurements of \macc~ from spectroscopy and of disk masses with millimeter interferometry, mainly with the Atacama Large Millimeter/submillimeter Array (ALMA), has also revealed the presence of a correlation between these two quantities \citep[e.g.,][]{manara16b,mulders17}, which is possibly in line with expectations drawn from viscous evolution \citep[e.g.,][]{jones12,lodato17,rosotti17}, but this also reveals tensions with respect to this simplified description of disk evolution \citep[e.g.,][]{mulders17,manara20}.   These tensions may be relieved by introducing complications to a simplistic model, including external or internal photoevaporation \citep[e.g.,][]{rosotti17,sellek20a,somigliana20}, or dust evolution \citep{sellek20b}.

Winds from protoplanetary disks affect the disk evolution by carrying away angular momentum and perhaps even driving accretion \citep{ferreira06,bai16}.
Studies of forbidden emission lines emitted from high-velocity jets and outflows have shown that typically the mass loss rate is a factor of $\sim$0.1--0.3 of \macc \citep[e.g.,][]{hartigan95,nisini18}. On the other hand, thanks to the study of large samples of young stellar objects, it is becoming more evident that the low-velocity component of forbidden lines, tracing slow disk winds \citep[e.g.,][]{natta14} is composed of multiple sub-components tracing either internal photoevaporative winds \citep[e.g.,][]{ercolano16,EP17,ballabio20} or, most probably, magneto-hydrodynamical winds \citep[e.g.,][]{rigliaco13,simon16,EP17,mcginnis18,banzatti19,weber20}. The combination of multiple emission lines observed at high spectral resolution ($\Delta v\lesssim$ 5--10 km/s)  allows us to determine the physical conditions in the winds and in the jets \citep[e.g.,][]{giannini19} and to constrain the mass loss rates in these winds \citep[e.g.,][]{fang18}.   

A complete view of the accretion and ejection properties is, nonetheless, achievable solely by combining data at different wavelengths from the ultraviolet (UV) to the optical and to the infrared (IR), since each of these trace different processes and regions of the young stars \citep[e.g.,][]{arulanantham18,banzatti19}.

Finally, this rich phenomenology is characterized by time variability over timescales that can be as short as a few minutes \citep[e.g.,][]{stauffer14, Siwak2018}. The variability of the accretion process has been studied both with spectroscopy \citep[e.g.,][]{basri90,JB95a,JB95b,jay06,biazzo12,costigan14,sousa16} and with photometry \citep[e.g.,][]{cody10,cody14,venuti14}. The variability of emission lines is observed both in their intensity and morphology, which can be understood via adequate modeling \citep[e.g.,][]{muzerolle98,muzerolle01,kurosawa06}.  

About 500 orbits of the Hubble Space Telescope (HST) are being devoted to the study of about 70 young low-mass stars in the Director's Discretionary Time ULLYSES program \citep{ullysesDR1}. This program is aimed at obtaining low- and medium-resolution spectra of young stars covering the wavelength range from $\sim$140 nm to $\sim$1 $\mu$m, hence providing an unprecedented view of accretion and ejection tracers at ultraviolet wavelengths. 
Inspired by this initiative, we proposed (and were granted) a $\sim$250 hours Large Program on the ESO VLT that is aimed at obtaining complementary data to all the targets of the ULLYSES program. This program, named PENELLOPE in order to underline the complementarity to ULLYSES and to match its spelling flaw, is a public (no proprietary time) community-driven effort. PENELLOPE uniquely provides optical high-resolution (even at $R>100,000$) and medium-resolution flux-calibrated optical and infrared spectra (up to 2.5~$\mu$m) at $R>10,000$, providing access to the following information that is otherwise not obtainable with the spectra provided by ULLYSES: main stellar properties (luminosity, temperature, gravity, rotational velocity, mass, age), accurate interstellar extinction, and veiling; kinematics and geometry of the accretion process; kinematics and physical properties of disk winds and jets; properties of the hot molecular content of the disk.
In this way, the marriage of PENELLOPE and ULLYSES 
will provide the whole community with a unique and contemporaneous UV, optical, and near-infrared dataset to study the stellar, accretion, and wind and outflow properties of a significant sample of objects spanning a wide range of disk ages ($\sim$1-10 Myr) and evolutionary stages, including full and transitional disks, and Class~III disk-less targets.

In this work, we present the observing strategy and data reduction and analysis process of the VLT/PENELLOPE program (Sect.~\ref{sect::programme}) and use the sample of targets in the Orion region (see Sect.~\ref{sect::sample}) as a first application. We derive the stellar and accretion properties, as well as the photospheric parameters, the observed variability in key emission lines, and compare the VLT and HST spectra in Sect.~\ref{sect::analysis}. 
A detailed description of the emission line profiles, including the possibility to use this information to study outflows and ejection processes is deferred to future works. 
We then discuss the accretion properties for these targets in the context of current surveys of accretion properties  in Sect.~\ref{sect::discussion}. Finally, we outline the conclusions of this work in Sect.~\ref{sect::conclusions}.

\section{PENELLOPE survey strategy}\label{sect::programme}

\subsection{HST ULLYSES program}
The Hubble Space Telescope (HST) is dedicating 1,000 orbits of Director's Discretionary Time to the ULLYSES program\footnote{\url{https://ullyses.stsci.edu/}} \citep[PI Roman-Duval;][]{ullysesDR1}, which will generate a spectral library of high- and low-mass young stars. In HST cycles 28 and 29 (2020--2022), about half of these orbits are being used to obtain spectra of about 70 low-mass ($\sim$0.1-2 $M_\odot$) young stars with ages from $\sim$1 to 10 Myr. 

Four of the low-mass stars are observed multiple times: twelve times over three consecutive rotation periods and again with the same cadence about a year later. Cosmic Origins Spectrograph (COS) spectra will be obtained for these targets from 140 nm to 180 nm with medium resolution ($R\sim17,000$) and from 250 nm to 310 nm with low resolution ($R\sim3000$). The others, including all the stars discussed in this paper, are observed once, with COS medium-resolution ($R\sim15,000$) spectroscopy in the far-ultraviolet (120 nm $<\lambda<$ 180 nm) and Space Telescope Imaging Spectrograph (STIS) low-resolution ($R<1,000$) spectroscopy in the near-ultraviolet (160 nm $<\lambda<$ 320 nm) and optical ($\lambda<1\mu$m), all obtained within $\sim24$ hours of each other.

The targets are located in nine nearby star-forming regions, namely Chamaeleon~I, Corona Australis, $\epsilon$ Cha, $\eta$ Cha, Lupus, Orion OB1, $\sigma$-Orionis, TW Hydrae, and Taurus.  
All these star-forming regions are accessible by telescopes located in the Southern Hemisphere. This choice was made given that only very few targets in this part of the sky had been observed with COS and STIS to date \citep[e.g.,][]{arulanantham18,arulanantham20} and also for the purpose of enabling synergies between HST and the major ground-based optical to near-infrared (e.g., Very Large Telescope, VLT) and \mbox{(sub-)mm} (e.g., Atacama Large Millimetre Array, ALMA) observatories. 
The HST COS \& STIS observations have been coordinated, when possible, with photometric observations by the NASA Transiting Exoplanet Survey Satellite (TESS, \citealt{Ricker2014}).
In addition, a number of ground-based astronomical observatories are ready to monitor the targets during the HST observations with simultaneous multi-band photometry. These data, which are to be made publicly available as soon as they are taken, will serve as a benchmark for future studies of UV radiation from young stars.

\subsection{Observing strategy at VLT}
The observational strategy of the VLT/PENELLOPE program is such that the HST/ULLYSES targets are observed contemporaneously to HST with two major observing modes, that is, high-resolution spectroscopy (R $>$70,000) with UVES or ESPRESSO (Sect.~\ref{sect::hires_spec}), and medium-resolution (R $\sim$10,000-20,000) spectroscopy with X-Shooter to obtain broad-wavelength, flux-calibrated spectra (Sect.~\ref{sect::midres_spec}). Observations are taken in service mode with tight absolute time constraints of about three days on the X-Shooter observation, and with a time-linked concatenation of observations for the high-resolution data, where the first epochs has a two-day absolute time window, followed by the next with a relative time window of a minimum time separation of one night and a maximum of two nights. The exact intervals of the observations are updated about two weeks before the observations once the HST time windows have been finalized.

\subsubsection{High-resolution spectra}\label{sect::hires_spec}

High-resolution spectroscopy is performed either with the Ultraviolet and Visual Echelle Spectrograph (UVES, \citealt{dekker00})  
or the Echelle SPectrograph for Rocky Exoplanets and Stable Spectroscopic Observations (ESPRESSO, \citealt{pepe20}),
depending on the brightness of the targets. The brighter targets, V $<$ 16.5\,mag, are observed with the instrument ESPRESSO, a fibre-fed (1.0\arcsec~ wide) spectrograph which allows for  $R\sim140,000$ ($\Delta v\sim 2$ km/s) spectra between 380 nm and 788 nm to be obtained. ESPRESSO offers great scheduling flexibility, as it can be operated from any of the four Unit Telescopes (UT) of the VLT. Fainter targets, as well as targets that are to be probed when ESPRESSO is not available (see Sect.~\ref{sect::sample}), are observed with the UVES spectrograph, which provides twice lower $R\sim 70,000$ ($\Delta v\sim 4$ km/s) spectral resolution. 
This instrument, fed through a long slit, 
has two arms, Red ($\lambda\gtrsim 480$ nm) and Blue ($\lambda\lesssim 450$ nm).
The resolution and wavelength range depend on the specific setting. PENELLOPE uses 0.6\arcsec\  wide slits in both arms, leading to a resolving power $R\sim$70,000, centering the dichroics at 580 nm (red arm) and 390 nm (blue arm), respectively. This allows us to simultaneously cover the wavelength ranges $\lambda\sim330-450$ nm, 480--680 nm, with a small gap between 575 nm and 585 nm. 
The exposure times are set so that the signal-to-noise ratio (S/N) at 630 nm is larger than $\sim$50 in each observation. The UVES slits are typically oriented at position angles 0\degree, 120\degree, and 180\degree~ in the three epochs (see later), apart from known cases where the target is a visual binary with separation in the range of $\sim$1\arcsec-8\arcsec. In those cases, the slit is aligned to include both components of the system.

\subsubsection{Medium-resolution flux-calibrated spectra}\label{sect::midres_spec}

The medium-resolution broad-wavelength coverage flux-calibrated spectroscopy is obtained using the X-Shooter instrument \citep{vernet11}. This long-slit (11\arcsec) spectrograph provides simultaneous coverage of the region between $\sim$300 nm and $\sim$2500 nm, divided into three arms, UVB (300 $\lesssim \lambda \lesssim$ 500 nm), VIS (500 $\lesssim \lambda \lesssim$ 1000 nm), and NIR (1000 $\lesssim \lambda \lesssim$ 2500 nm). Each target is observed first using a set of 5.0\arcsec--wide slits in the three arms, leading to a low resolution observation with no slit losses, key to obtain absolute flux calibration of the spectra. Then, using 1.0\arcsec/0.4\arcsec/0.4\arcsec--wide slits for the UVB, VIS, and NIR arms, respectively, high S/N spectra with $R\sim$5400, 18400, and 11600 are obtained in the three arms. The exposure times are set such that the S/N at 400 nm is $\gtrsim 3-5$, resulting in a  S/N in the VIS and NIR arms always $>$100. 
To mitigate the effects of differential atmospheric dispersion, the slits are always oriented at parallactic angle, apart from known cases where the target is a visual binary with separation  $\sim$1\arcsec-8\arcsec. In those cases the slit is also aligned to include both components of the system.

\subsubsection{Sequence of the observations}\label{sect::sequence_obs}

In order to measure all the stellar, accretion, and outflow properties while reducing the uncertainties due to the variability of these processes, the timing and number of observations 
must be closely coordinated with HST. 
Knowing that each HST observation with the COS and STIS instruments spans $\sim$1-2 days, 
per each HST epoch we aim at obtaining:\newline
\indent First, three high-resolution spectra (ESPRESSO or UVES), each one taken on the night before, on the night of the HST observations, and on the following night. These are needed to have a detailed understanding of the line profiles, especially to study accretion and ejection processes, and their short-term variability in order to estimate and correct for any variability of the accretion and ejection processes occurring between the time of the HST and VLT observations.  \newline
\indent Second, a combination of narrow-slit (for the resolution) plus wide-slit (for the flux-calibration) medium-resolution X-Shooter spectra possibly in the same night, in any case, within $\pm$2 days from the HST observation, to provide an accurate measurement across the spectrum of the target brightness at the time of the observation and to study several spectral features from the Balmer continuum to the $K$-band in the infrared.

Any lack of availability of instruments or poor atmospheric conditions may hinder the possibility of executing this strategy. Therefore, these points represent a guideline to be followed as much as possible, but which could be altered if needed. In any case, the goal is to obtain three high-resolution spectra and one X-Shooter spectrum per each HST target and observation.

\subsection{Contemporaneous photometry}
\label{Sec:Photometry}
Photometric data that are close in time to the ULLYSES and PENELLOPE observations presented in this paper were collected from several ground-based observatories with the aim of studying the variability and to provide a reference for absolute flux calibration.  

In this work, we make use of data taken at the {\it M. G. Fracastoro} station (Serra La Nave, Mt. Etna,
1750 m a.s.l.) of the {\it Osservatorio Astrofisico di Catania} (OACT, Italy) from 25 November to 16 December 2020. 
We used the facility imaging camera at the 0.91\,m telescope with a set of broadband Bessel filters ($B$, $V$, $R$, $I$, $Z$) as well as two narrow-band filters, $H\alpha_9$ and $H\alpha_{18}$, centered at 656.8\,nm and at 676.4\,nm, whose full widths at half-maximum are 9 and 18\,nm, respectively. The index $H\alpha_{18}$-$H\alpha_9$ has a weak dependence on the photospheric parameters and is basically a measure of intensity of the H$\alpha$ emission in units of the continuum that can be converted into H$\alpha$ equivalent width \citep{frasca2018}. The $Z$ filter has a transmittance that produces a passband very similar to that of the Sloan $z'$ filter when it is multiplied by the responsive quantum efficiency of the adopted CCD. 
The CCD camera adopts a Kodak KAF1001E\footnote{\scriptsize \tt http://sln.oact.inaf.it/sln\_old/dmdocuments/ccd91rappint2-07.pdf} 1k$\times$1k chip that, with a focal reducer, covers a field of view of about 
11.5$\times$11.5 arcminutes.

The broadband $BVRI$ photometry was calibrated using the following procedure. The stars in the standard areas GD\,71 and SA\,98 \citep{Stetson2000,Landolt2009} were observed on the nights with the best photometric conditions. These observations were used to calculate, for a number of non-variable stars in the same fields of the targets, the zero points and transformation coefficients to the Johnson-Cousins system. The transformations to the Johnson-Cousins system were then used to derive the standard magnitudes of our targets.
The magnitudes of the local comparison stars retrieved from SDSS \citep{SDSS} and Pan-STARRS \citep{Pan-STARRS} survey catalogs were used to get $z'$ magnitudes of our targets. For details about the OACT data and their reduction and the narrow-band H$\alpha$ photometry, we refer to \citet{frasca2018}. 

We also used $BVri$ photometry collected by AAVSOnet\footnote{\scriptsize\tt https://www.aavso.org/aavsonet}, which is a set of robotic telescopes operated by volunteers for the American Association of Variable Star Observers (AAVSO). Stars in the AAVSO Photometric All-Sky Survey (APASS, \citealt{Henden2018}) 
were used to calibrate the AAVSO photometric data for our targets.
The $r,i$ magnitudes were converted to $R_C, I_C$ using the prescription given by Lupton (2005\footnote{\scriptsize\tt http://www.sdss3.org/dr8/algorithms/sdssUBVRITransform.php}). 

Some additional photometry was obtained at the Crimean Astrophysical Observatory (CrAO, Russia) on the AZT-11  1.25 m telescope equipped with a CCD camera (with the ProLine PL23042 detector) and a set of broadband Bessel filters ($B, V, R, I$). For each object, from 5 to 9 photometric points per band were obtained for the time interval between October 16 and December 16, 2020.

\subsection{Spectroscopic data reduction process}\label{sect::reduction}
Spectroscopic data reduction was carried out using the ESO Reflex workflow v2.8.5 \citep{reflex}, specifically, the ESPRESSO v2.2.1 pipeline \citep{pepe20}, the UVES v6.1.3 pipeline \citep{uvespipe}, and the X-Shooter v3.5.0 pipeline \citep{xspipe}. The pipelines carry out the standard steps of flat, bias, and dark correction, wavelength calibration, spectral rectification and extraction of the 1D spectra, and flux calibration using a standard star obtained in the same night. The UVES pipeline also allows extraction of individual spectra when two targets are present in the slit.

Additional steps are then performed as follows. The 1D extraction of the X-Shooter spectra is carried out with IRAF\footnote{{\scshape iraf} is distributed by the National Optical Astronomy Observatories, which are operated by the Association of Universities for Research in Astronomy, Inc., under the cooperative agreement with the National Science Foundation.
NOAO stopped supporting IRAF, see \url{https://iraf-community.github.io/}} from the rectified flux-calibrated 2D spectrum in cases where the S/N of the UVB arm is low, and for resolved binaries. For the latter cases, particular attention must be paid to the selection of the parameters for apertures definition and trace of the spectrum profile to maximize the signal of the extracted flux, avoiding to mix the spectra of two close stars.

Telluric correction is performed using the {\it molecfit} \citep{molecfit1,molecfit2} tool v3.0.3 for the VIS and NIR arms of X-Shooter, and for the high-resolution ESPRESSO spectra. The latter is done by adapting the current molecfit workflow to the ESPRESSO data. The correction is always performed fitting the atmospheric model directly on the science spectra, since the S/N on the continuum is always high enough to ensure a better correction with respect to using the telluric standard star observed in the same night to compute the model. For the UVES spectra, the telluric correction is performed using a standard telluric star spectrum obtained with the same instrument and same configuration. This is possible since only a few O$_2$ telluric lines around the [\ion{O}{i}]$\lambda$6300\AA\ and H$_2$O lines around H$\alpha$ and \ion{Na}{i}\,D$_2$ are present; they can be easily identified and removed thanks to an IDL\footnote{IDL (Interactive Data Language) is a registered trademark of  Harris Corporation.} procedure that allows us to align the telluric features by cross-correlation and rescaling of their intensity \citep[see, e.g.,][]{Frasca2000}.

Finally, the X-Shooter spectra obtained with the narrow slits are scaled to the wide-slit ones to correct for slit losses. This procedure has already been tested in previous works, for example, \citet{mendigutia13}, \citet{alcala17}, \citet{manara16a,manara17b},  \citet{Rugel2018}, and \citet{kospal20}, and leads to a typical absolute flux calibration accuracy of $\sim$10\%.
The reduced, flux-calibrated, and telluric-corrected spectra are available on Zenodo\footnote{For the Orion sample, X-Shooter data: \url{https://zenodo.org/record/4477091#.YBMOdpNKjlx} ;
UVES data: \url{https://zenodo.org/record/4478360#.YBPQRJNKhTY} ;
ESPRESSO data: \url{https://zenodo.org/record/4478376#.YBPQWpNKhTY}} in the ODYSSEUS data community\footnote{\url{https://sites.bu.edu/odysseus/}}, and will also be made available also on the ESO Archive Phase 3 service.

\begin{figure}
\begin{center}
\hspace{-0.7cm}
\includegraphics[width=9.5cm]{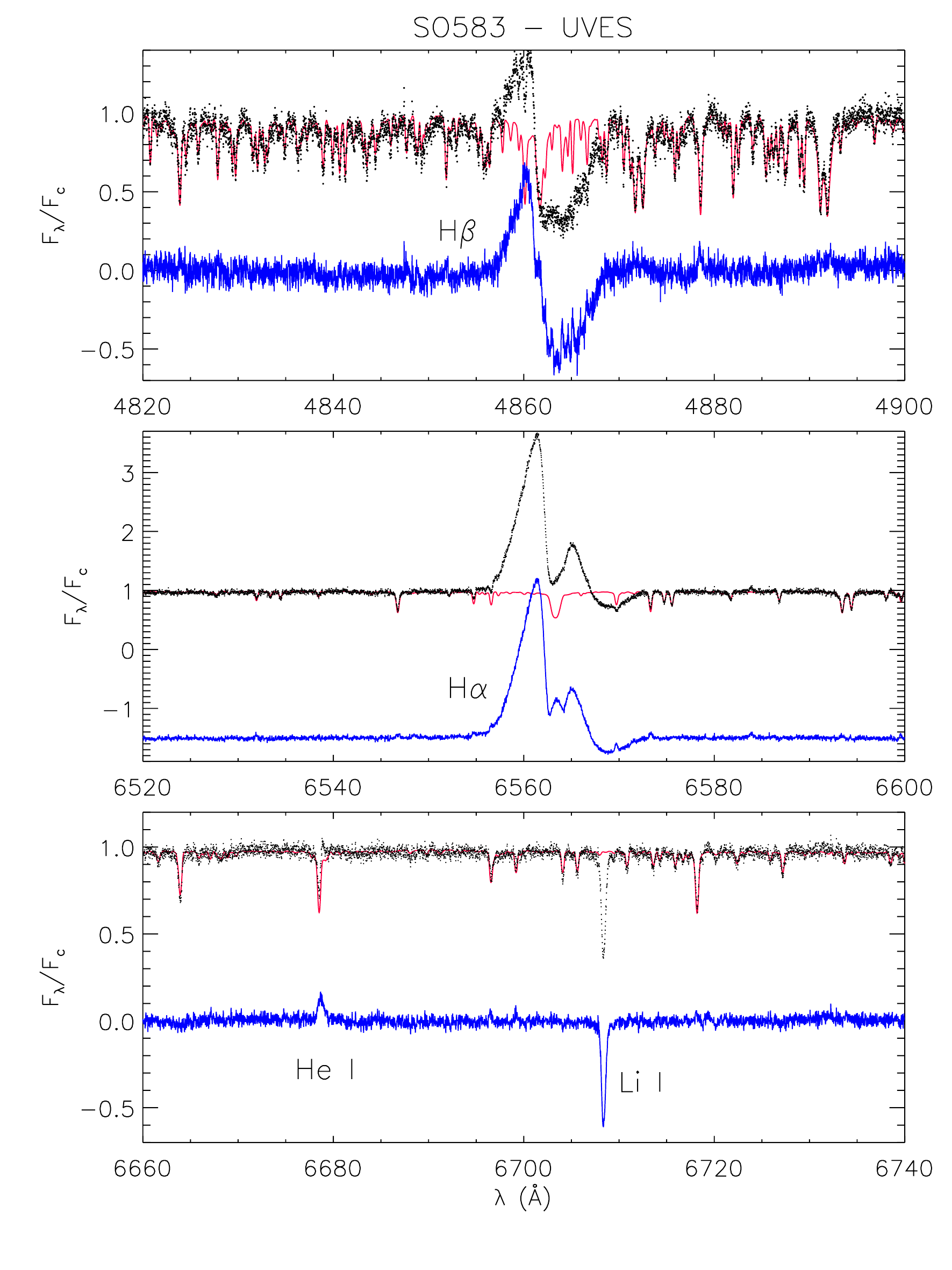}  %
\vspace{-.5cm}
\caption{Subtraction of the non-active, lithium-poor template (red line) from
the spectrum of SO\,583  (black dots), which reveals the complex structure of the H$\beta$ and H$\alpha$ lines (blue
lines in the upper and central panel, respectively) and emphasizes the \ion{Li}{i} $\lambda$6707.8\,\AA\ absorption line,
removing the nearby blended lines (bottom panel). The \ion{He}{i} $\lambda$6678.2\,\AA\ emission line clearly emerges in the subtracted spectrum.
}
\label{fig:subtraction}
\end{center}
\end{figure}

\subsection{Data analysis}\label{sect::methods}
The data are first analyzed with two tools, named here as ``fitter'' and ROTFIT, as described below. The analysis is aimed at deriving the stellar, accretion, and photospheric properties of the targets. 

\subsubsection{Fitter of broadband spectra to obtain stellar and accretion properties}\label{sect::method_fitter}

The analysis of the X-Shooter spectra to derive stellar and accretion properties is carried out with the method originally described in \citet{manara13b} and later applied to a number of X-Shooter studies of young stars \citep[e.g.,][]{alcala14,alcala17,manara16a,manara17a,venuti19}. In short, the observed spectrum is dereddened and fit with the sum of a photospheric template spectrum and a hydrogen slab model with uniform density and temperature gas to reproduce the continuum excess emission due to accretion. The grid of models used to find the best fit comprises Class~III photospheric templates with spectral types (SpT) from G- to late M-type taken from \citet{manara13a,manara17b}, different slab models, and a series of extinction  values ($A_V$) all assuming the reddening law by \citet{cardelli98} and $R_V=3.1$. The integrated flux of the best fit slab models gives an estimate of the excess luminosity due to accretion (\lacc), and the best fit normalization of the Class~III templates gives an estimate of the stellar luminosity (\lstar). By converting the SpT to effective temperature (\teff) using the relation by \citet{luhman03}, together with \citet{kh95}, as described by \citet{manara13b}, we are able to position the targets on the HR diagram. By comparing the position on the HR diagram with an interpolated set of evolutionary models by \citet{B15} or, if \mstar$>$1.4 \msun, by \citet{siess00}, it is possible to infer the stellar mass (\mstar) of the targets. The choice of evolutionary models is done in line with previous works \citep[e.g.,][]{alcala17,manara17a}. Finally, the mass accretion rate (\macc) is obtained from the classic relation \macc = 1.25~$\cdot$ \lacc $R_\star/(G M_\star)$ \citep[e.g.,][]{hartmann98}.  

As several emission lines are present in the X-Shooter spectra, we measure their luminosity (\lline) and convert them to \lacc \ using the relations by \citet{alcala17}. Typically,  
the values of \lacc \ obtained from the fitter described above and the mean value of \lacc \ derived from the emission line fluxes are similar within the uncertainties 
\citep[e.g.,][]{HH08,alcala14,alcala17}. The presence of multiple accretion tracers along the wide wavelength coverage of the X-Shooter spectra provides an additional check on the $A_V$ estimates.

\subsubsection{Photospheric properties from ROTFIT}\label{sect::method_rotfit}
The estimate of the photospheric properties \teff\ and \logg, radial (RV) and projected rotational velocity (\vsini), and veiling is performed on both the medium-resolution and the high-resolution spectra using the ROTFIT code, already tested both on X-Shooter data and on the higher resolution UVES spectra from the Gaia ESO Survey \citep{Frasca2015,frasca2017}. 

ROTFIT uses a grid of template spectra to perform a $\chi^2$ minimization of the difference between the observed and template parameters in selected spectral regions. 
To construct the grid of templates, we collected different photospheric templates with spectral types similar to our targets. We built two grids of templates of high-resolution spectra of real slowly rotating stars with a low activity level: the first one includes spectra retrieved from the ELODIE archive (R$\simeq$42,000, \citealt{Moultaka2004}), which are the same used for the analysis of young stars within the \textit{Gaia}-ESO survey by the OACT node \citep{Frasca2015}; the second is composed of spectra of KM-type stars retrieved from the HARPS archive (R$\simeq$115,000)\footnote{\tt http://archive.eso.org/wdb/wdb/adp/phase3\_main/form}. The main parameters of the HARPS templates are reported in Table~\ref{Tab:HARPS_templates} along with references for the quoted stellar parameters.

To perform the $\chi^2$ minimization, when using the grid of ELODIE templates, the UVES and ESPRESSO spectra are convolved with a Gaussian kernel and resampled to match the resolution of ELODIE ($R=42\,000$). For the analysis with the HARPS grid we have degraded the ESPRESSO spectra to $R=115,000$, while the reverse was done for the UVES ones, that is, the HARPS templates were brought to the UVES resolution ($R=70\,000$) and resampled on the points of the target spectra.  
For the determination of photospheric parameters, \vsini, and veiling from high-resolution spectra, we prefer real  over synthetic spectra,  because generally the former better reproduce the unknown photospheric spectrum. Some photospheric lines may be missing in the synthetic spectra, or the  depths and widths of some of them may be poorly reproduced due to uncertain intensity values, Land\'e factors and broadening effects. 
However, as the non-active templates are mostly main-sequence stars, we need synthetic spectra for a safer determination of \logg. 
To this aim, BT-Settl synthetic spectra \citep{Allard2012} are used to fit specific spectral regions containing gravity-sensitive features. 

For the X-Shooter spectra, which span a much wider wavelength range, we use a grid of BT-Settl synthetic spectra with a solar iron abundance \citep{Allard2012}. The radial velocity is calculated  by means of the cross-correlation of template and target spectrum in specific spectral regions free from broad features and emission lines. The \vsini\ and veiling values are treated as free parameters in the fitting procedure and the best values are found by the $\chi^2$ minimum. 

The photospheric lithium-poor templates with no sign of accretion and absent (or negligible) chromospheric emission in the cores of strong lines fitted by ROTFIT (including rotational broadening and veiling) can be subtracted to remove the photospheric lines. This has the great advantage of emphasizing emission features against underlying photospheric absorption lines, but it is also very important to clean the \ion{Li}{i}$\lambda$6707.8\,\AA\ line from blended nearby lines, which, unlike \ion{Li}{i}$\lambda$6707.8\,\AA, are present in the templates. Finally, this enables us to remove strong photospheric lines around emission lines tracing outflows and winds, such as the [OI]$\lambda$6300\AA~ line (see Fig.\,\ref{fig:subtraction} and Figs.\,\ref{fig:subtraction_suppl},\,\ref{fig:subtraction_suppl2}  for a few examples). The line profiles corrected for telluric absorption and for photospheric absorption lines will be provided to the community through Zenodo and the ESO Phase 3 Archive, and will serve as a great dataset to study the accretion and ejection mechanisms \citep[e.g.,][]{banzatti19,mcginnis18,simon16,rigliaco13}.

\begin{table*}  
\begin{center} 
\footnotesize 
\caption{\label{tab::res} Stellar and accretion properties from this work for the targets in the Orion OB1 Association and $\sigma$-Orionis cluster } 
\begin{tabular}{ll|cc|cc| cccc| cc   } 
\hline \hline 
Target &        Other Name &    RA$_{2000}$ & DEC$_{2000}$ & Parallax & Dist. & SpT & $A_V$ & \lstar & log\lacc &     \mstar & log(\macc)      \\
 &        &     hh:mm:ss.s & dd:mm:ss.s & [mas] & [pc] & & [mag] & [\lsun] & [lsun]  &     [\msun] & [\msun/yr]     \\
\hline
CVSO17  &       ...&    05:23:04.72&    01:37:15.3 &  2.4144$\pm$0.0531 &       414.2$^{+9.3}_{-8.9}$&         M2& 0.0 & 0.30 & -4.62$^*$ &    0.37    & -11.43$^*$\\
CVSO36  &       ...     & 05:25:50.37 &         01:49:37.3& 2.9327 $\pm$ 0.0265 & 335.5$\pm3.0$  & M2& 0.1 & 0.22 & -3.42$^*$ &  0.39 &  -10.31$^*$ \\
CVSO58  &  ...&         05:29:23.26     & -01:25:15.5 & 2.8229 $\pm$  0.0231 &       349.0$\pm2.8$ &         K7& 0.8 & 0.32 & -1.12 &        0.81    & -8.37 \\
CVSO90  &       ...     & 05:31:20.63   & -00:49:19.8   & 2.9107 $\pm$  0.033 &       338.7$^{3.8}_{-3.7}$    & M0.5 & 0.1 & 0.13 & -1.34     & 0.62 &         -8.61 \\
CVSO104 &               Haro 5-64 &     05:32:06.49     & -01:11:00.8   & 2.7296 $\pm$  0.0298 &  360.7$^{+3.9}_{-3.8}$   & M2 & 0.2 & 0.37 & -1.73       & 0.37&   -8.49 \\
CVSO107 &       ...     & 05:32:25.79 &         -00:36:53.4     & 2.9843 $\pm$  0.0226 & 330.4$\pm2.5$   & M0.5  & 0.3 & 0.32 & -1.30 & 0.53&    -7.30 \\
CVSO109&                V462 Ori        & 05:32:32.66 &         -01:13:46.1 & \nodata &     400     & M0.5& 0.1 & 0.92 & -0.77 &    0.46    & -7.49 \\
CVSO146&         V499 Ori&      05:35:46.01 &   -00:57:52.2 &   2.9701 $\pm$  0.0152  & 332.0$\pm1.7$        & K6 &  0.6 & 0.80 & -1.46 &    0.86 &  -8.57    \\
CVSO165 & ...   & 05:39:02.57 &         -01:20:32.3     & \nodata &     400     & K6 & 0.2 & 0.98 & -2.05 & 0.84 &        -9.10 \\
CVSO176&                V609 Ori&       05:40:24.15 &   -00:31:21.3     &         3.2593 $\pm$  0.0314 & 302.4$^{+2.9}_{-2.8}$    & M3.5&  1.0 & 0.34 & -1.27 &       0.25    & -7.84 \\
\hline 
\hline
SO\,518 & V505 Ori & 05:38:27.26 &      -02:45:09.7 & 2.5064 $\pm$  0.025 & 392.3$^{+3.9}_{-3.8}$ & K7 & 1.0 & 0.24 & -1.22 & 0.81 & -8.53 \\ 
SO\,583 & TX Ori & 05:38:33.69 &        -02:44:14.1 & \nodata & 385 & K5 & 0.4 & 3.61 & -0.30 & 1.09 & -7.21 \\
SO\,1153 & V510 Ori & 05:39:39.83 &     -02:31:21.9 & 2.5212 $\pm$  0.0268 & 390.3$^{+4.1}_{-4.0}$ & K7 & 0.1 & 0.17 & -0.88 & 0.76 & -8.24 \\
\hline
\end{tabular} 
\tablefoot{Values obtained fitting the X-Shooter spectra. $^*$These values should be considered as upper limits since the targets are not-accreting.} 
\end{center} 
\end{table*}

\section{First PENELLOPE data: Young stars in Orion}\label{sect::sample}

The first observations of the ULLYSES and PENELLOPE programs have focused on targets located in the Orion region. In particular, ten targets are part of the Orion OB1 association, and three of the $\sigma$-Orionis cluster. In the following, we briefly introduce the two regions, and report the properties from the literature, including the distances from the recent Gaia EDR3 release \citep{gaia,gaiaEDR3}, for the 13 targets analyzed here (see Table~\ref{tab::res}).

\subsection{Information on the Orion sample}

The Orion OB1 association is one of the closest and most populous OB associations,
encompassing subassociations that range in age from $\sim$1 Myr to $\sim$ 10 Myr
\citep{blaauw64}. The
low-mass counterparts to the OBA-type stars characterizing the association
were first identified in the CIDA Variability Survey
\citep[CVSO,][]{briceno01,briceno05,briceno19},
carried out using the
Quest camera in the J. Stock telescope in Venezuela and recently confirmed  spectroscopically
 \citep[cf.][]{briceno19}. 
Circumstellar disks were discovered and characterized with 
Spitzer \citep{hernandez07} and Herschel \citep{mauco18}.
The X-ray luminosities and accretion properties of some of the 
CVSO objects, including HST observations, have been
discussed in 
\citet{ingleby09,ingleby11,ingleby14}
and \citet{thanathibodee18,thanathibodee19}.

The HST/ULLYSES team selected ten targets in the Orion
OB1a and OB1b subassociations. Their  spectral types are between K6 and M3,
and most of them have low interstellar extinction 
\citep[$A_V<0.3$ mag,][]{briceno19}. Eight targets with signatures of accretion are located in Orion OB1b, which has an age of $\sim$ 5 Myr, 
while the two non-accreting targets, CVSO~17 and CVSO~36, 
are located in the 25 Ori group within the Orion OB1a
subassociation, with ages of $\sim$ 10 Myr \citep{briceno19}. 

In addition to distributed populations, the Orion OB1 association includes several stellar clusters such as the $\sigma$ Ori cluster.
Low-mass members of  $\sigma$ Ori were first reported by \citet{walter1997}, who found over 80 X-ray sources and spectroscopically identified more than 100 low-mass, pre-main-sequence (PMS) members lying within 1$^{\circ}$ from the star $\sigma$ Ori, which is 
in fact a massive quintuplet system
of O and early B stars \citep{caballero2014}. 
With an estimated age of $\sim$3-5 Myr 
\citep{caballero2018}, the $\sigma$ Ori cluster is of interest because of its intermediate age
and for being an excellent
laboratory to study protoplanetary disks in dense environments containing OB stars, where far-ultraviolet radiation fields from massive stars can externally illuminate the disks, producing photoevaporating winds. 

Extensive studies of membership to the $\sigma$ Orionis cluster were carried out by 
\citet{hernandez14} and \citet{caballero2019}. Its protoplanetary disks have been followed with Spitzer \citep{hernandez07a}, Herschel PACS \citep{mauco2016}, and ALMA \citep{ansdell17}. Estimates of accretion rates for members of this cluster were obtained with $U$-band photometry by \citet{rigliaco11a}, and, for a small sub-sample, with X-Shooter spectroscopy \citep{rigliaco12}. 
The HST/ULLYSES team selected three targets in $\sigma$ Ori, with low extinction and clear signs of accretion \citep{mauco2016}.
Information on the targets taken from the literature is
presented in Appendix~\ref{app::lit} and summarized in Tables ~\ref{tab::lit1} and \ref{tab::lit2}.

\subsubsection{Distances from Gaia parallaxes}
The Gaia EDR3 astrometric solutions for the ten targets are generally quite good,  showing, for example, low renormalized unit weight errors (RUWEs). Only the astrometric solution for CVSO~17 has RUWE=1.8 and thus slightly exceeds the value of RUWE=1.4, which is considered an appropriate nominal limit for GAIA EDR3. Stars with RUWE$>$1.4 could have an ill-behaved astrometric solution \citep{2020arXiv201201533G}. 
For these, we compare the distance obtained by inverting the parallax (arithmetic distances), after correcting for the zero points estimated from the analytical functions of \citet{lindegren20}, with the geometric Bayesian distance estimated by \citet{bailer-jones2020}. The differences between the arithmetic and Bayesian distance are smaller than 1 pc. Stars with reliable astrometric solutions have uncertainties in parallaxes around 1\%. Thus, we
assumed the arithmetic distances for these targets 
(Table~\ref{tab::res}). \citet{bailer-jones2015} suggests that we cannot apply the inverse relation between parallaxes and distances to stars with fractional parallax errors larger than 20\%, where the estimation of distances becomes an
inference problem in which the use of prior assumptions is necessary.

Different stellar groups in the Orion star-forming complex can be detected as kinematically distinct populations with characteristic proper motions, radial velocities, and distances \citep{kounkel18}. In our case, the distances obtained for the 10 targets with reliable astrometric solutions, together with the radial velocities from \citet{kounkel18} and measured here (Sect.~\ref{sect::results_rotfit}), are consistent with the expected values for the Orion OB1a/OB1b sub association, and the $\sigma$-Ori cluster, respectively.
Even CVSO 17, which has a value of RUWE=1.8, has a distance consistent with the upper boundary of the distance distribution of the 25 Ori cluster \citep{briceno19}, different from the distance obtained for the other star in the cluster (CVSO 36; 345 pc). Regardless of the relative large fractional parallax error of CVSO17 (22\%), the arithmetic distance reported in Table~\ref{tab::res} agrees with the geometric Bayesian distance, 415.3$^{+9.7}_{-10.1}$ pc, estimated by \citet[][]{bailer-jones2020}.
The assumed distance for CVSO 17 has the caveat that the value of RUWE is just above the usual limit.
For CVSO 109 and CVSO 165, for which no reliable astrometric solution is available, we assume the mean distance of the Orion OB1b sub-association, 400 pc \citep{briceno19}, and for SO\,583, we assume the mean distance to the $\sigma$-Orionis cluster, 385 pc \citep{mauco2016}

\subsubsection{Multiplicity of the targets}\label{sect::binaries}
Multiplicity for several of the targets in the Orion OB1 association was recently studied by \citet{tokovinin20}, who reported that the following targets are physical binaries:
CVSO 17 (separation$\sim$8.1\arcsec, same companion present also in Gaia EDR3),
CVSO 36 (sep $\sim$ 3.3\arcsec),
CVSO 109 (sep$\sim$0.7\arcsec).

In two other cases, a visual companion is a background object:
CVSO 104 (sep$\sim$2.35\arcsec, same nearby object present also in Gaia EDR3, with parallax = 1.49 mas, thus background), 
CVSO 165 (sep$\sim$5.4\arcsec, same nearby object present also in Gaia EDR3, also background). For CVSO 165, the observations of the HST/ULLYSES program reveal that the primary component is actually a $\sim$0.3\arcsec~ binary \citep{proffitt21}, which we do not resolve in our VLT observations. The same observations also confirm the binarity of CVSO~109 \citep{proffitt21}.
No binary components were found for CVSO 58, CVSO 90, CVSO 107, and CVSO 146, and no observations for CVSO 176 were carried out by \citet{tokovinin20}.

In our VLT observations, we cannot resolve the close ($<$1\arcsec) binary systems CVSO 109 and CVSO 165, whereas the other binaries are resolved. Moreover, we observe from our multi-epoch high-resolution spectra that the primary component of CVSO 104 is a spectroscopic binary.

\begin{figure}[]
\centering
\includegraphics[width=0.5\textwidth]{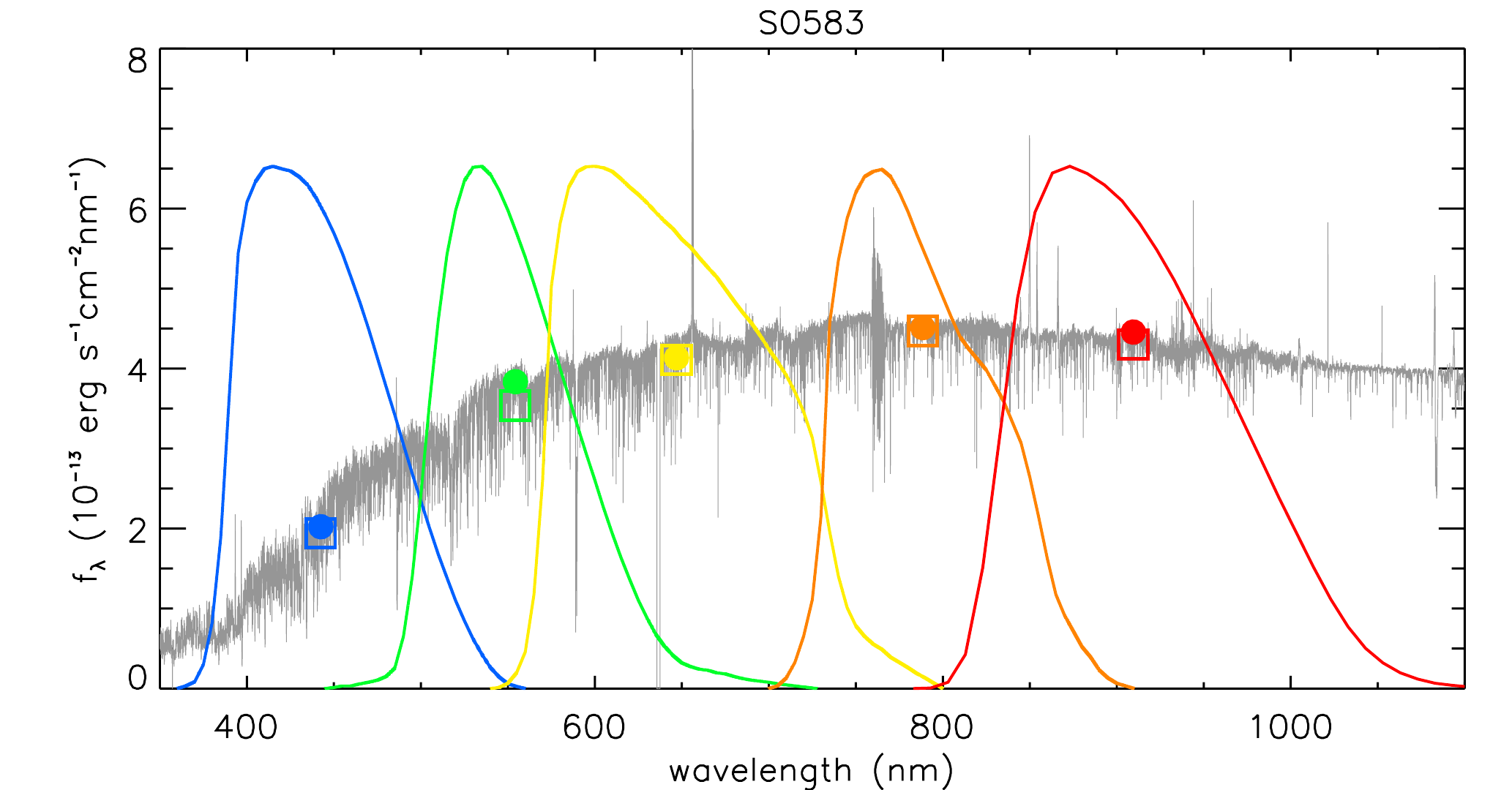}
\includegraphics[width=0.5\textwidth]{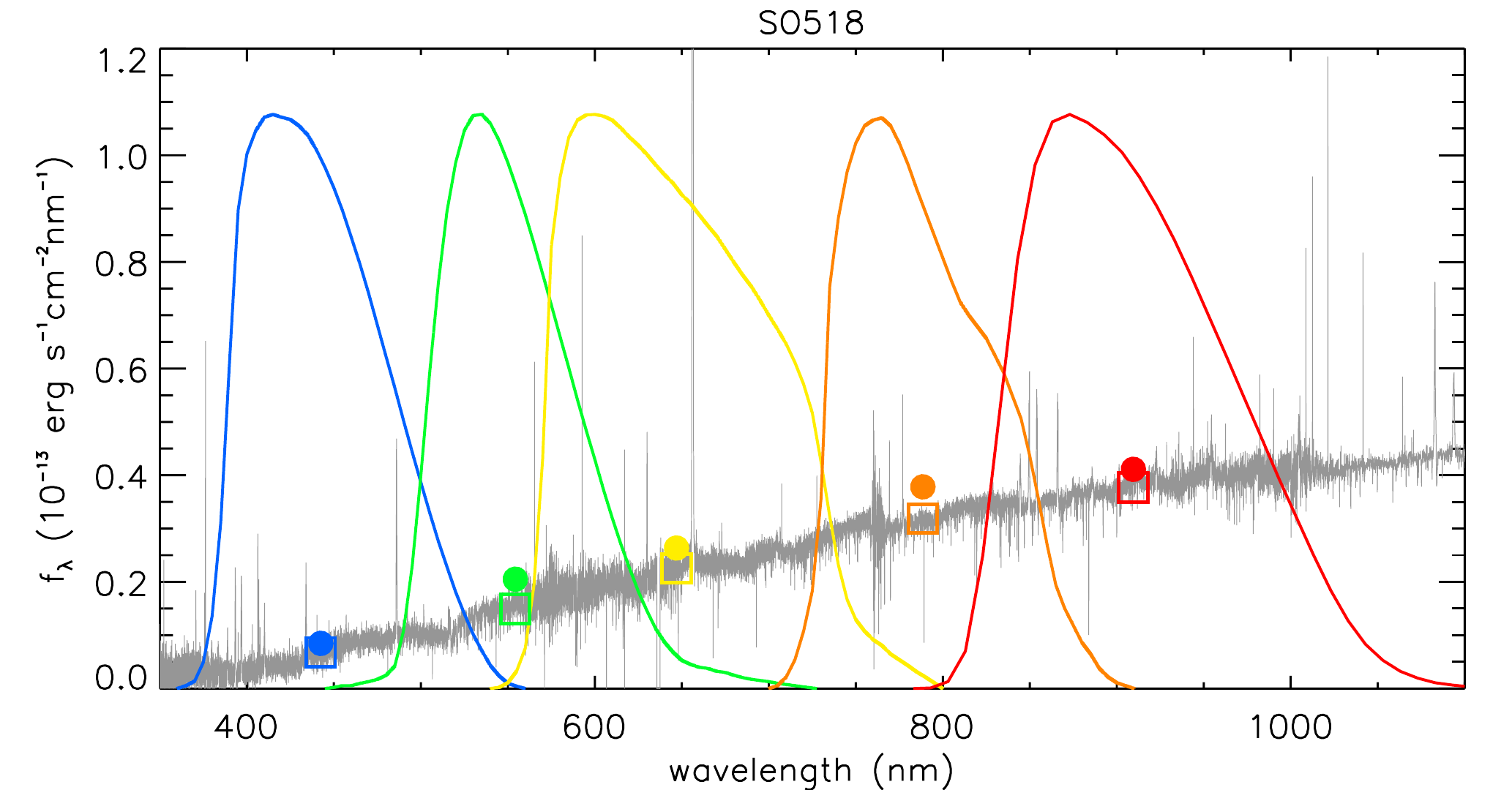}
\includegraphics[width=0.5\textwidth]{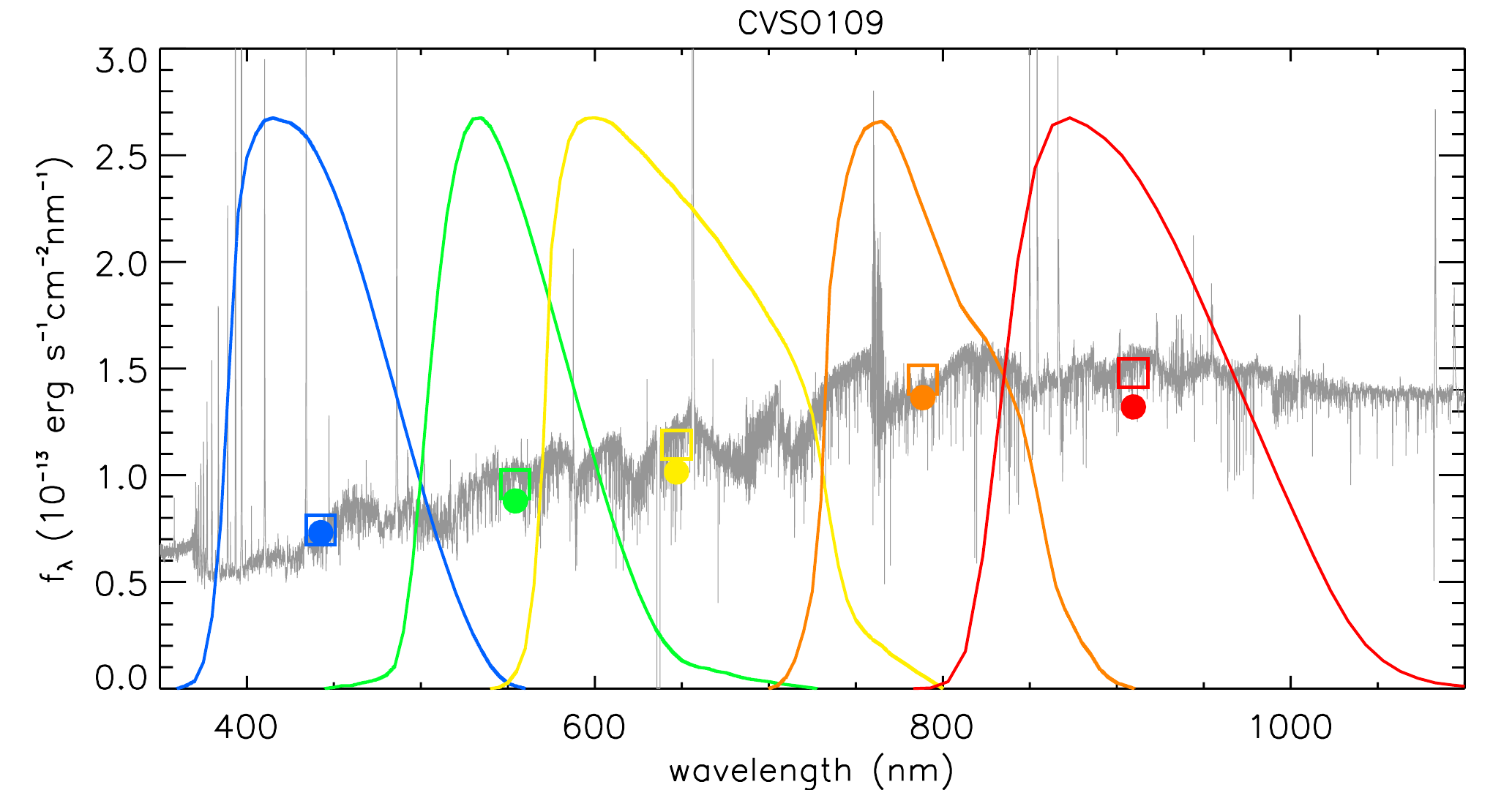}
\caption{X-Shooter flux-calibrated spectrum of three of our targets (grey), as labelled, with quasi-simultaneous photometry for the same target obtained from OACT in the $BVR_{\rm C}I_{\rm C}z'$ bands (dots). The filter bandpasses and the synthetic photometry (open squares) obtained integrating the spectrum over these bandpasses are overlaid.
     \label{fig::synt_pho}}
\end{figure}

\subsection{Observation and data reduction for the first PENELLOPE dataset}
The first observing run of our PENELLOPE Large Program happened soon after VLT was brought back into operations after the shutdown due to the COVID-19 pandemic. During the first two weeks of observations, only X-Shooter and UVES, among the instruments selected for our program, were operational. We thus obtained observations of our bright targets with UVES instead of ESPRESSO in this period. The first observation with UVES was on the target CVSO 104 and happened on November 25, 2020, while the first   X-Shooter observation was performed on November 27, 2020. It was only on December 8, 2020 that ESPRESSO came back into operation and could be used to observe SO\,1153. The last spectra of this run were obtained on December 15, 2020. A technical issue on UT3 made X-Shooter unavailable for three nights, and a technical problem on ESPRESSO made it unavailable for a few days after December 15, 2020. This has caused a larger than desired time separation between the HST and X-Shooter observations for CVSO 176, and resulted in having only one epoch of high-resolution spectra for CVSO 90.

Figure~\ref{fig::ew_ha_var} shows the time of the observations with VLT and HST. 
In summary, for our 13 targets,  we typically obtained  two high-resolution spectra in advance of the HST observations, and one during or after the HST observations. It is only in the case of SO\,1153 that the three spectra were obtained after the HST observations and that is because of the initial unavailability of ESPRESSO. The three epochs with high-resolution spectroscopy were all obtained in three consecutive nights. In most cases the X-Shooter observations were obtained on the night of the HST/STIS observations or the night before, with the exception of CVSO 176, observed with X-Shooter two nights after the HST/STIS observations, and of SO\,518 and SO\,583, both observed one night after HST/STIS due to the unavailability of X-Shooter on the preceding nights. 
With those few exceptions, the plan of our observations (as described in Section~\ref{sect::sequence_obs}) was successfully implemented.
The observations log and additional information is presented in Appendix~\ref{app:log} and Tables~\ref{tab::log_xs}-\ref{tab::log_uves}-\ref{tab::log_espresso}. Observations were carried out under excellent weather conditions and typically resulted in an image quality of $\sim$1\arcsec.
\begin{figure}[]
\centering
\includegraphics[width=0.5\textwidth]{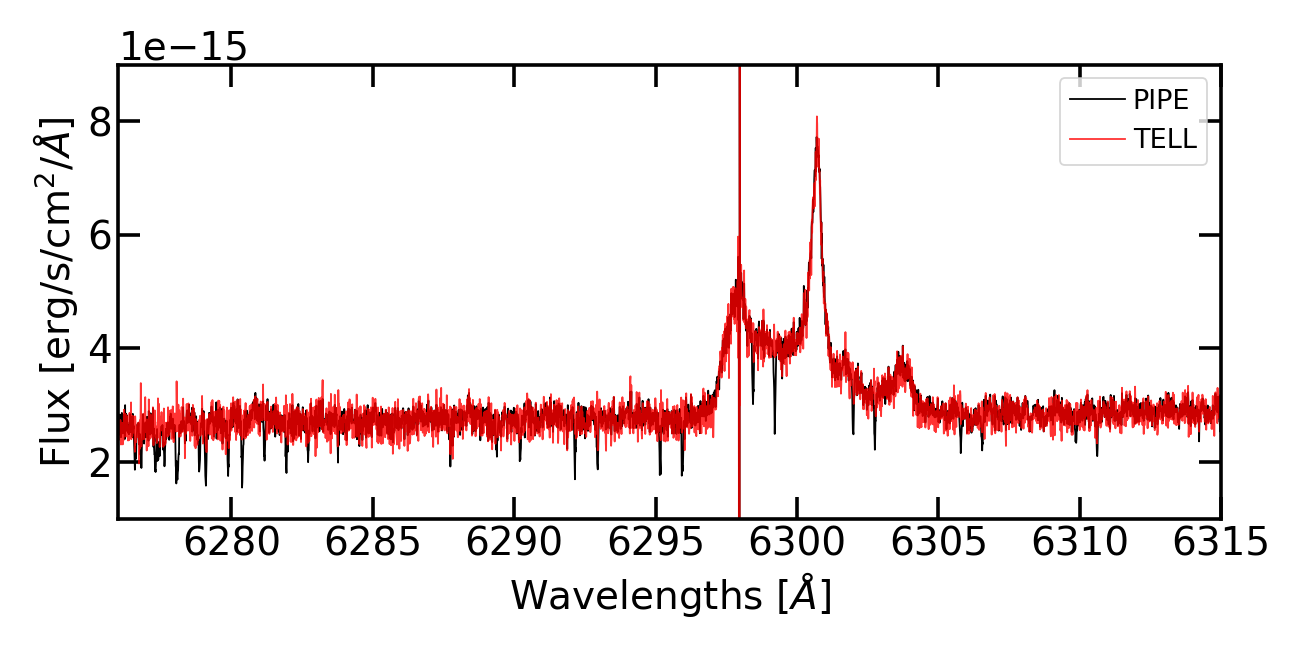}
\includegraphics[width=0.5\textwidth]{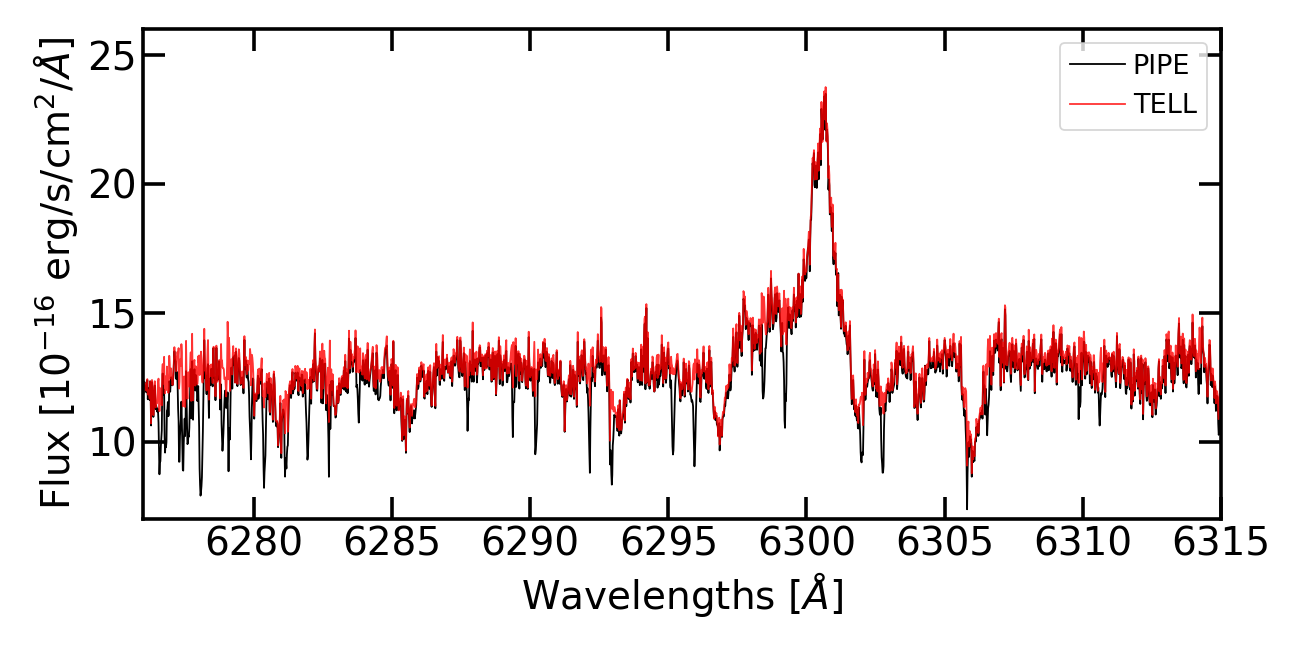}
\caption{Example of telluric correction in the region of the [OI]6300\AA ~ line for the ESPRESSO spectrum of CVSO~90 performed with molecfit (top) and for the UVES spectrum of CVSO~58 carried out with a standard telluric (bottom). The black spectrum is obtained from the pipeline reduction, and the red one shows the telluric corrected spectrum.
     \label{fig::tell_corr}}
\end{figure}

Reduction of the spectroscopic data is carried out as described in Sect.~\ref{sect::reduction}. Here we briefly mention the peculiarities relative to this dataset. 
The 1D spectra have been obtained using the standard extraction provided by the pipelines, with the only exception of the UVB arms of the X-Shooter spectra of SO\,518 and SO\,583, which were extracted with IRAF from the 2D flux calibrated spectra produced by the pipeline. This was needed since the image quality parameter at the time of the observations ($\sim$1.1\arcsec-1.2\arcsec) was much larger than the slit width (0.5\arcsec) used only for these targets in this arm, and the manual extraction with IRAF improved the S/N of the spectra. Also, the X-Shooter spectra of CVSO 104, and of its wide visual companion observed in the slit, were extracted manually with IRAF. The spectrum of the companion (located to the east of the target) is not analyzed here since this is a background star. The UVES spectra of this same target and its visual companion were instead extracted with the pipeline, adapting the extraction window to the position of the two traces in the 2D spectra. 

The rescaling of the narrow slit X-Shooter spectra to those obtained with a wide slit are done using a single scaling factor in the UVB and a linearly wavelength dependent scaling factor in the VIS and NIR arms. An example of the achieved flux calibration of the X-Shooter spectra is shown in Fig.~\ref{fig::synt_pho}. %
The overall agreement between the photometry obtained within less than between two and four hours from the X-Shooter spectrum is typically  within 10\%; in the worst case, which is our assumed (and possibly overestimated) uncertainty on the absolute flux calibration of the X-Shooter spectra. In the case of  SO\,518, the photometric fluxes are systematically higher, at all wavelengths, while for CVSO\,109 they are all lower than those measured on the X-Shooter spectrum. This can be understood by looking at the light curves (Fig.~\ref{fig::lc_phot}). Indeed, OACT photometric data are taken about two hours before the X-Shooter spectrum for SO\,518, whose brightness was dimming, and four hours before the X-Shooter spectrum for CVSO\,109, which was brightening. 

Telluric correction leads to good results in most of the VIS spectra and for the ESPRESSO spectra. Only the regions with very deep O$_2$ telluric absorption lines at around $\sim$760 nm, and sometimes the one around $\sim$690 nm, are poorly corrected. In particular, these absorption lines are saturated in the ESPRESSO data, and are therefore not properly corrected. In the X-Shooter spectra, the correction of the H$_2$O bands at around $\sim$950 nm and $\sim$1130 nm leaves in some cases strong residuals. The correction from telluric lines around the [OI]6300\AA ~ line is always good, as shown in Fig.~\ref{fig::tell_corr}. Similarly, the telluric correction of the UVES spectra using a telluric standard star observed with the same instrument configuration allows us to properly remove the telluric absorption features around the [OI]6300\AA ~ line, as shown in Fig.~\ref{fig::tell_corr}.


\section{Analysis of the Orion data}\label{sect::analysis}
Here, we present the analysis of the X-Shooter, UVES, and ESPRESSO data obtained in the first run of PENELLOPE, focusing on the stellar, photospheric, and accretion properties of the whole sample. Detailed analysis of the individual spectra, focusing on the study of the ejections from the emission line profiles, on the kinematic properties of the accretion process, and of the binarity of the targets and more, are deferred to forthcoming works.

\subsection{Stellar and accretion parameters from the X-Shooter spectra}
 
\begin{figure}[]
\centering
\includegraphics[width=0.45\textwidth]{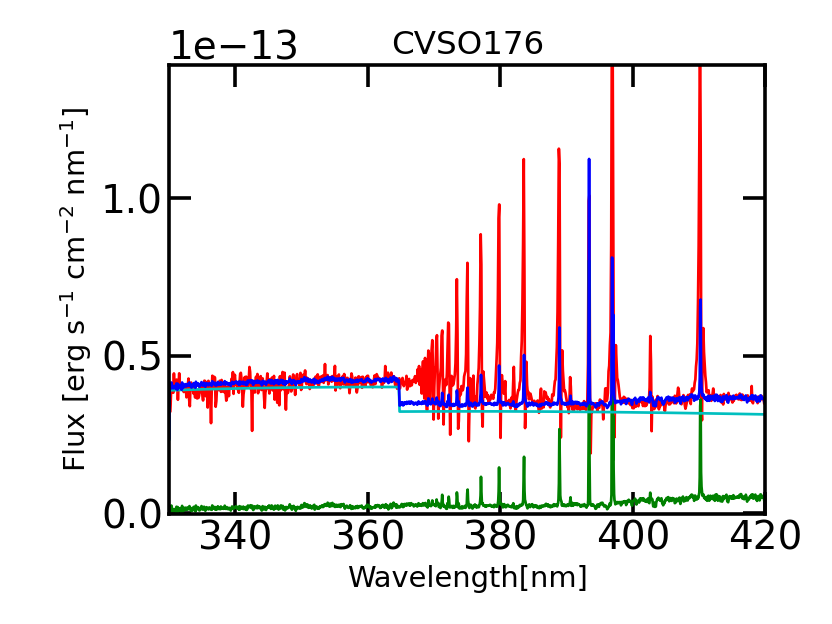}
\includegraphics[width=0.45\textwidth]{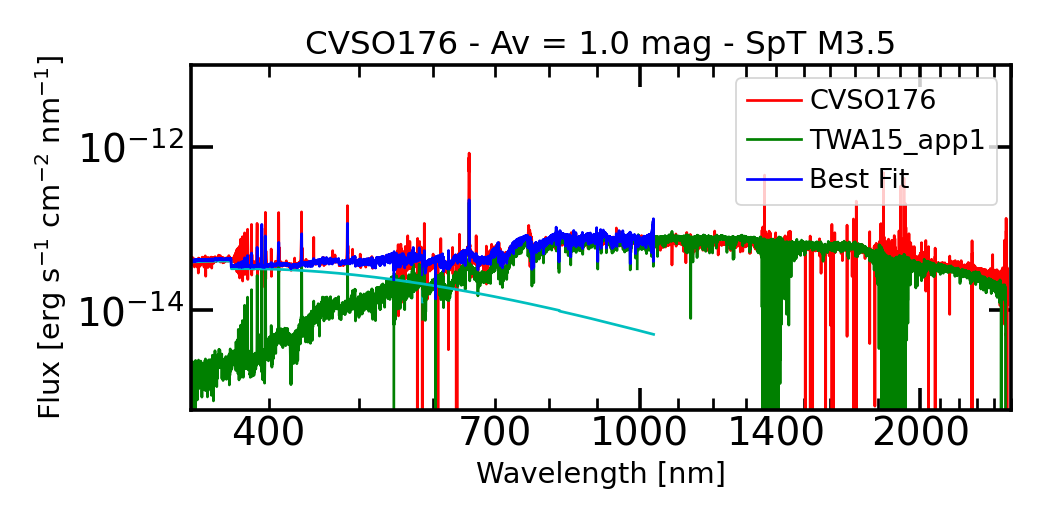}
\caption{Example of a best fit for the X-Shooter spectrum of CVSO176 obtained with the fitter described in Sect.~\ref{sect::method_fitter}. The spectrum of the target is shown with a red line, the best fit photospheric template with a green line, the slab model with a cyan line, and the best-fit sum of the template and slab model with a blue line. The best fit is only shown up to 1000 nm, since the contribution from the disk emission can be significant at longer wavelengths. 
     \label{fig::fit_example}}
\end{figure}

Figure~\ref{fig::fit_example} shows the best fit of the X-Shooter spectrum of the CVSO176 source obtained with the method described in Sect.~\ref{sect::method_fitter}, while best fits for the entire sample are shown in Fig.~\ref{fig::fit_bj_ob1}-\ref{fig::fit_bj_sOri}-\ref{fig::fit_all_ob1}-\ref{fig::fit_all_sOri}. In general, the fits reproduce well the Balmer jump and the continuum of the targets up to  $\sim$1000 nm. In the NIR, a region of the spectrum that is not considered when running the fitter, the best fit is in most cases fainter than the data, which is expected since the flux contribution from the disk emission at NIR wavelengths is not considered in the modeling. The values of spectral type, $A_V$, \lstar, \mstar, \lacc, and \macc~ derived from the fit are listed in Table~\ref{tab::res}. Typical uncertainties on those parameters are 0.1 mag, 0.2 dex, 0.1 dex, 0.25 dex, and 0.45~dex, as described by \citet{alcala14,alcala17} and \citet{manara13a}.

Our analysis of the X-Shooter spectra of CVSO~17 and CVSO~36 shows that these are non-accreting stars. A similar conclusion can be reached in considering the HST spectra (Pittman in prep.). We can consider the estimated values of \lacc \ as upper limits to the accretion rates, which thus fall well below the typically assumed detection limits for accretion rates due to chromospheric activity discussed by \citet{manara13a,manara17b}.

\begin{figure}[]
\centering
\includegraphics[width=0.5\textwidth]{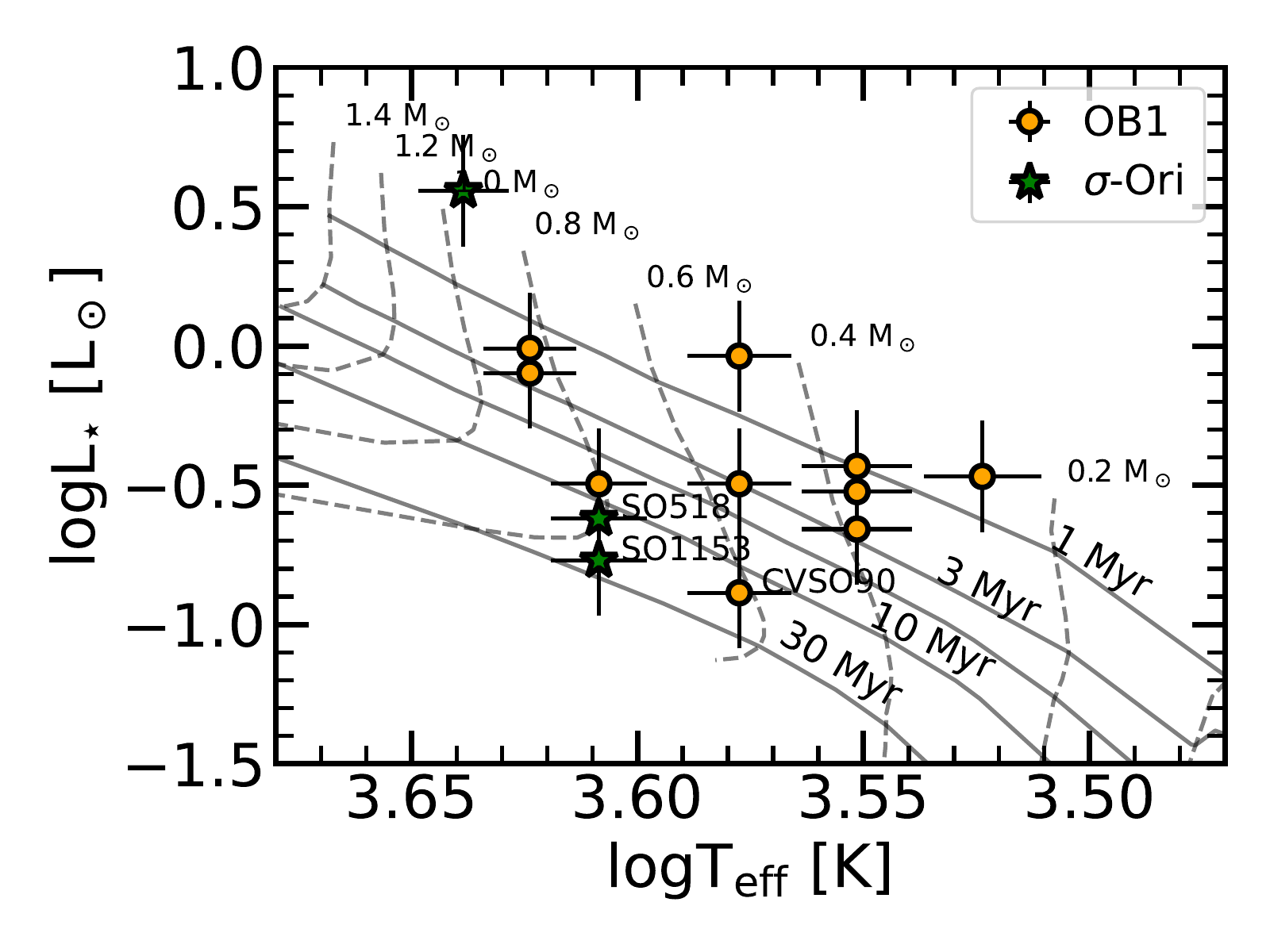}
\caption{Hertzprung-Russel diagram (HRD) of the Orion targets, with evolutionary tracks from \citet{B15}. The isochrones shown here are for ages of 1, 3, 5, 10, and 30 Myr.
     \label{fig::HRD_OB1}}
\end{figure}

As expected, the best fits are generally obtained with low values of $A_V\lesssim$ 0.5 mag. It is only in the case of CVSO176 and SO\,518 that we obtain $A_V$ = 1 mag, along with $A_V=0.8$ mag for CVSO58. These values are in line with previous literature estimates (see Table~\ref{tab::lit1}). They are also consistent with the values that minimize the spread of \lacc~ obtained converting the line luminosity of several lines in the X-Shooter spectrum using the relations by \citet{alcala17}.

The target CVSO104 was reported to be a K7 star in the literature, which may be due to its binarity. Looking at our spectra, the fit improves assuming a later spectral type. A more detailed analysis of this target that  carefully takes its binary status into account will be performed in a forthcoming paper.

The Hertzprung-Russel diagram (HRD) of the data analyzed here is shown in Fig.~\ref{fig::HRD_OB1}. Several targets lie between the 1 and 10 Myr isochrones of \citet{B15}, with only SO\,518, SO\,1153, and CVSO90 lying between the 10 Myr and 30 Myr isochrones.  
\citet{cody10} reported SO\,518 and SO\,1153 to be 
aperiodic type, with typical $\Delta I$ of 0.8 and 0.5 mag, respectively. At the time of our observations, 
SO\,518 and CVSO90 are affected by some significant ($\Delta V>$0.5 mag) photometric variability, in particular by dimming events. We can see from Fig.~\ref{fig::synt_pho} and Fig.~\ref{fig::lc_phot} that SO\,518 was observed during a dimming event of more than 1 mag in the $V$-band. The value of \lstar \ measured on the spectrum is therefore lower than the one we would measure at its peak brightness, as the extinction we calculate here does not account for grey-extinction effects. We note that such a dimming event would affect also the measured accretion luminosity, as described in the extreme case of edge-on disks in \citet{alcala14}.  Another aspect to consider is that a different relation between \teff \ and SpT \citep[e.g.,][]{PM13,HH14} would affect the position of the target on the HRD, moving the star to a colder temperature, and thus an apparently younger isochronal age. This is also the case for SO\,1153.

\begin{figure}[]
\centering
\includegraphics[width=0.4\textwidth]{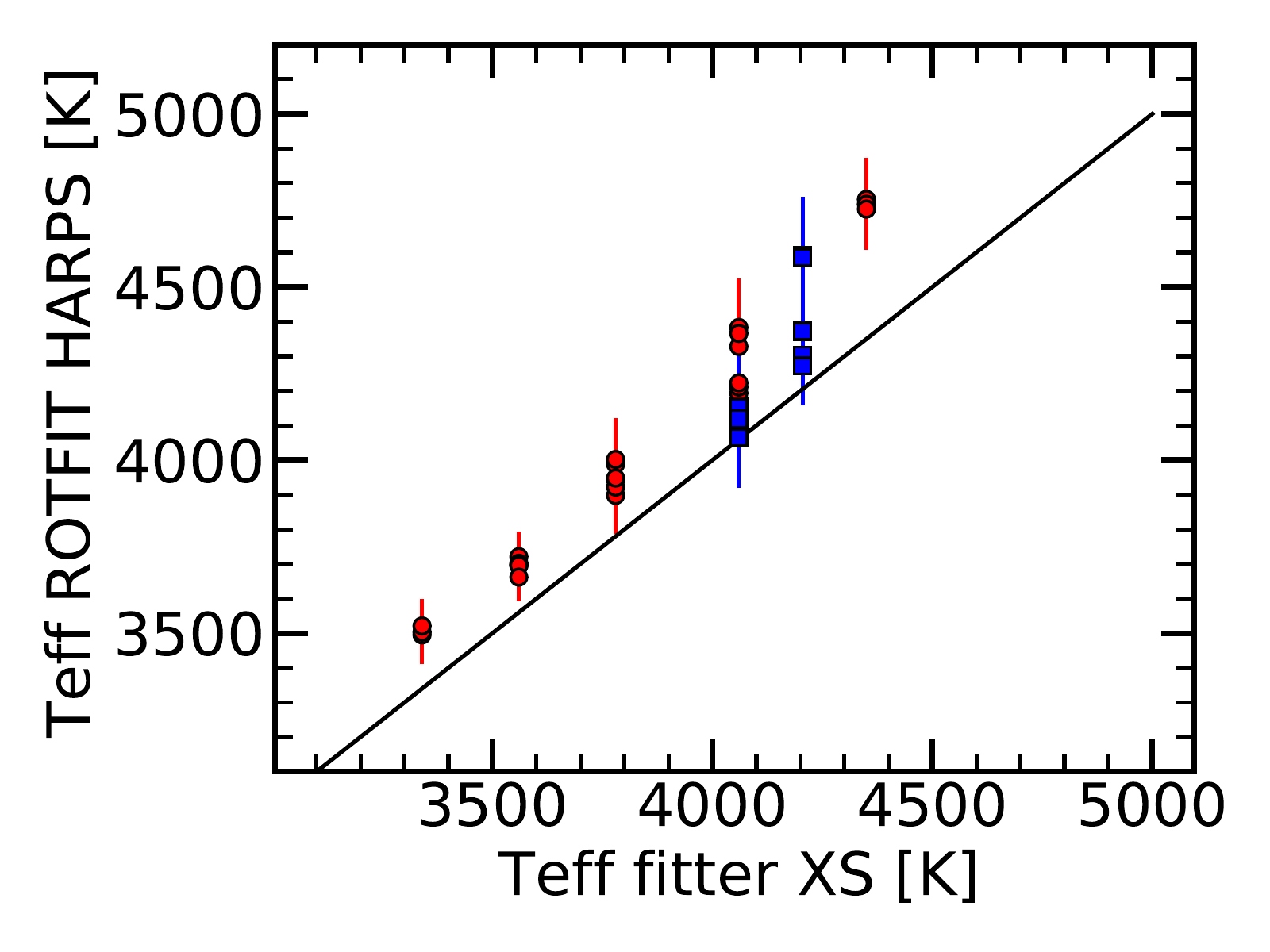}
\includegraphics[width=0.4\textwidth]{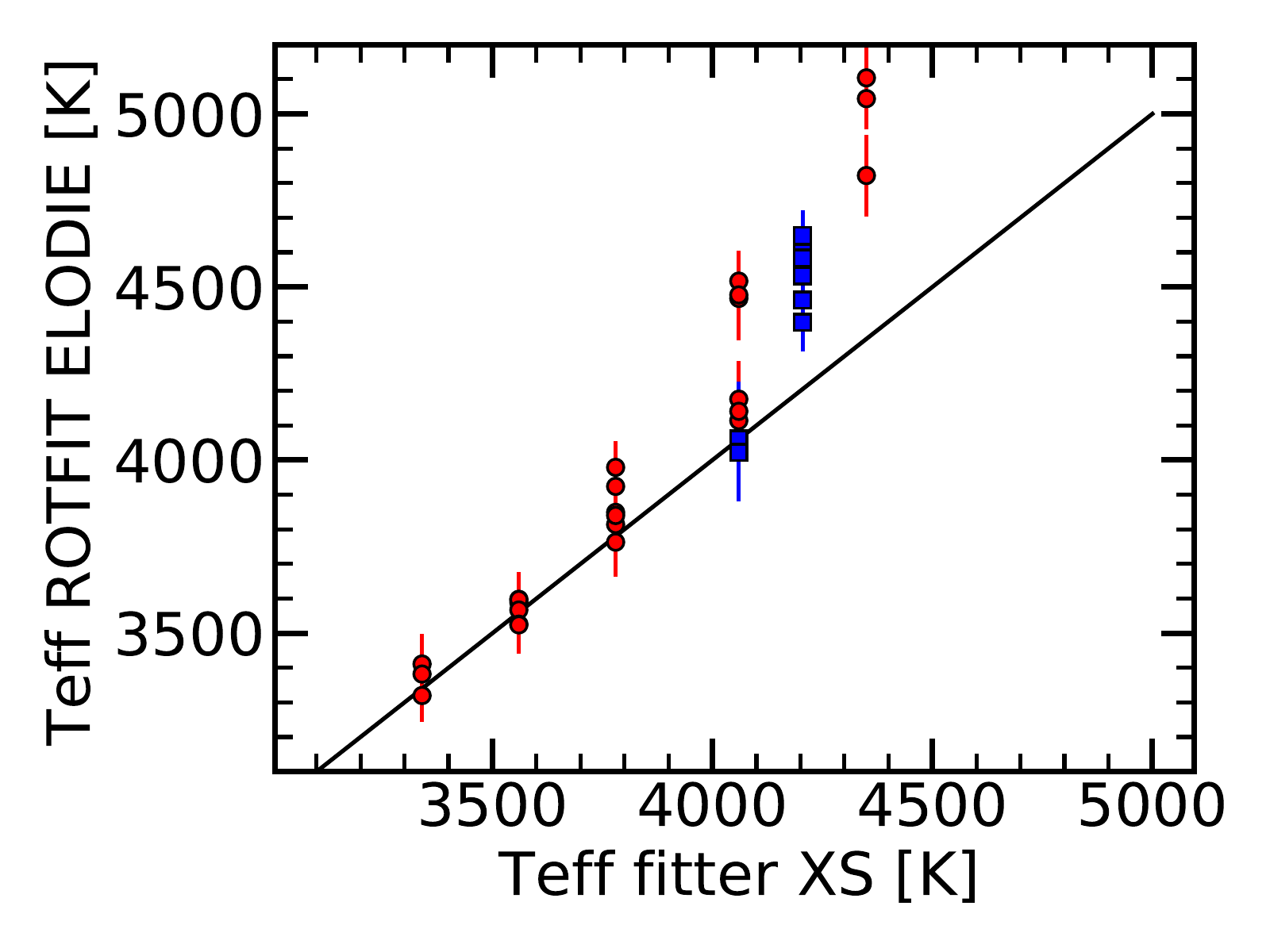}
\caption{Effective temperature (\teff) derived with ROTFIT on the UVES (red) and ESPRESSO (blue) spectra vs the temperature obtained converting the SpT from the fit of the X-Shooter spectra. Top: ROTFIT results using the HARPS templates, bottom: ROTFIT results using the ELODIE templates. 
     \label{fig::rotfit_fit_orion}}
\end{figure}

\begin{figure}[]
\centering
\includegraphics[width=0.4\textwidth]{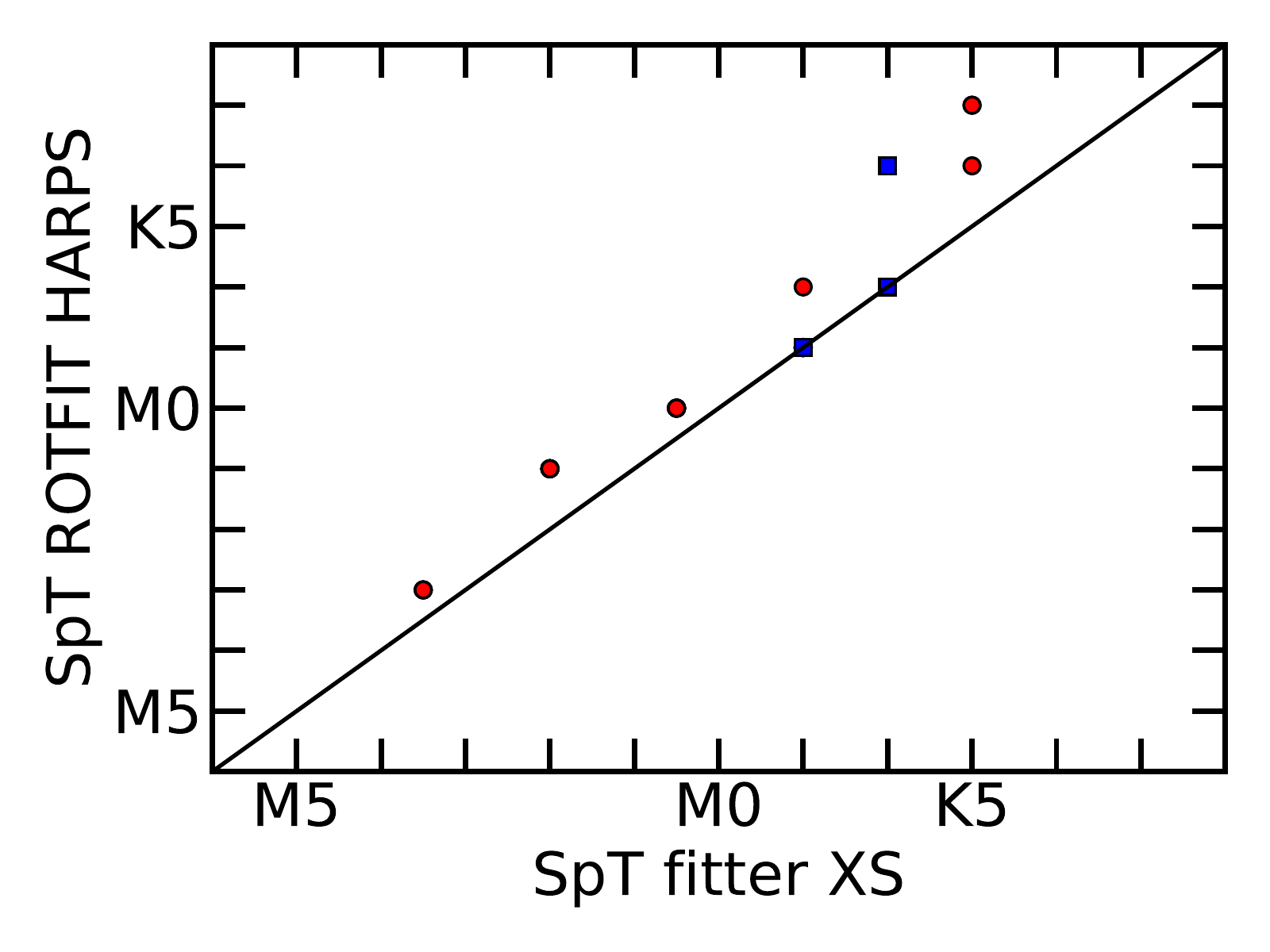}
\includegraphics[width=0.4\textwidth]{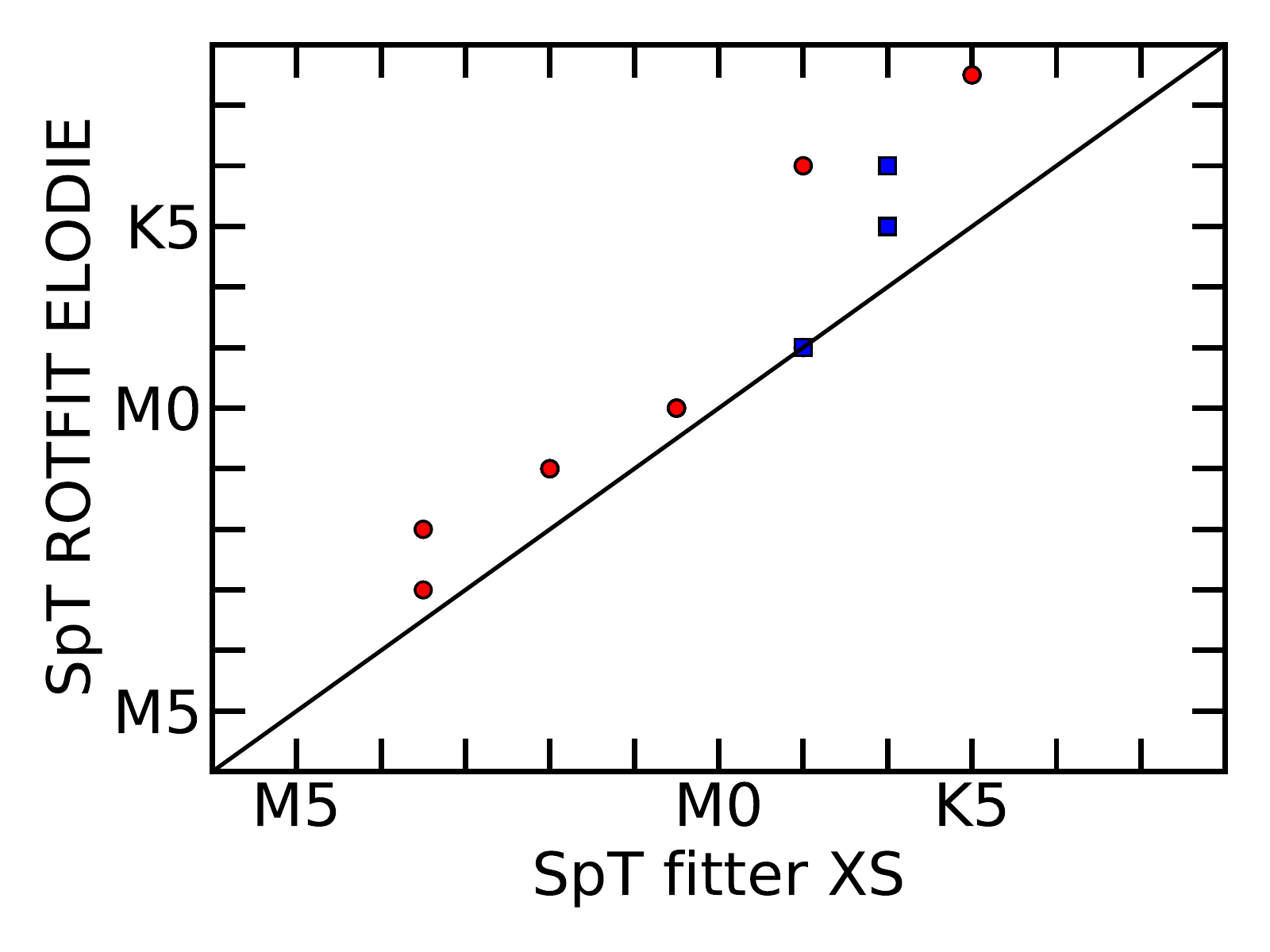}
\caption{Spectral type (SpT) derived with ROTFIT on the UVES (red) and ESPRESSO (blue) spectra vs SpT from the fit of the X-Shooter spectra. Top: ROTFIT results using the HARPS templates, bottom: ROTFIT results using the ELODIE templates. 
     \label{fig::rotfit_spt_fit_orion}}
\end{figure}

\subsection{Photospheric parameters and veiling from high-resolution spectra}\label{sect::results_rotfit}

Both the high-resolution UVES and ESPRESSO spectra and the X-Shooter spectra are analyzed with the ROTFIT tool, described in Sect.~\ref{sect::method_rotfit} and in \citet{Frasca2015,frasca2017}.
The results obtained on the UVES spectra using the HARPS templates are reported in Table~\ref{tab::res_rotfit}, those on the ESPRESSO data in Table~\ref{tab::res_rotfit_esp}, and those on the X-Shooter spectra in Table~\ref{tab::res_rotfit_xs}. Strongly veiled objects or those with too many emission lines in their spectra, in particular CVSO~90 and SO~1153, are challenging for the analysis with ROTFIT. The results for the former are not reported, while for the latter, they are added here but they are more uncertain than for the other targets. For all the accreting objects, the values of \logg \ are also quite uncertain and will be re-examined in a future work.

\begin{figure}[]
\centering
\includegraphics[width=0.4\textwidth]{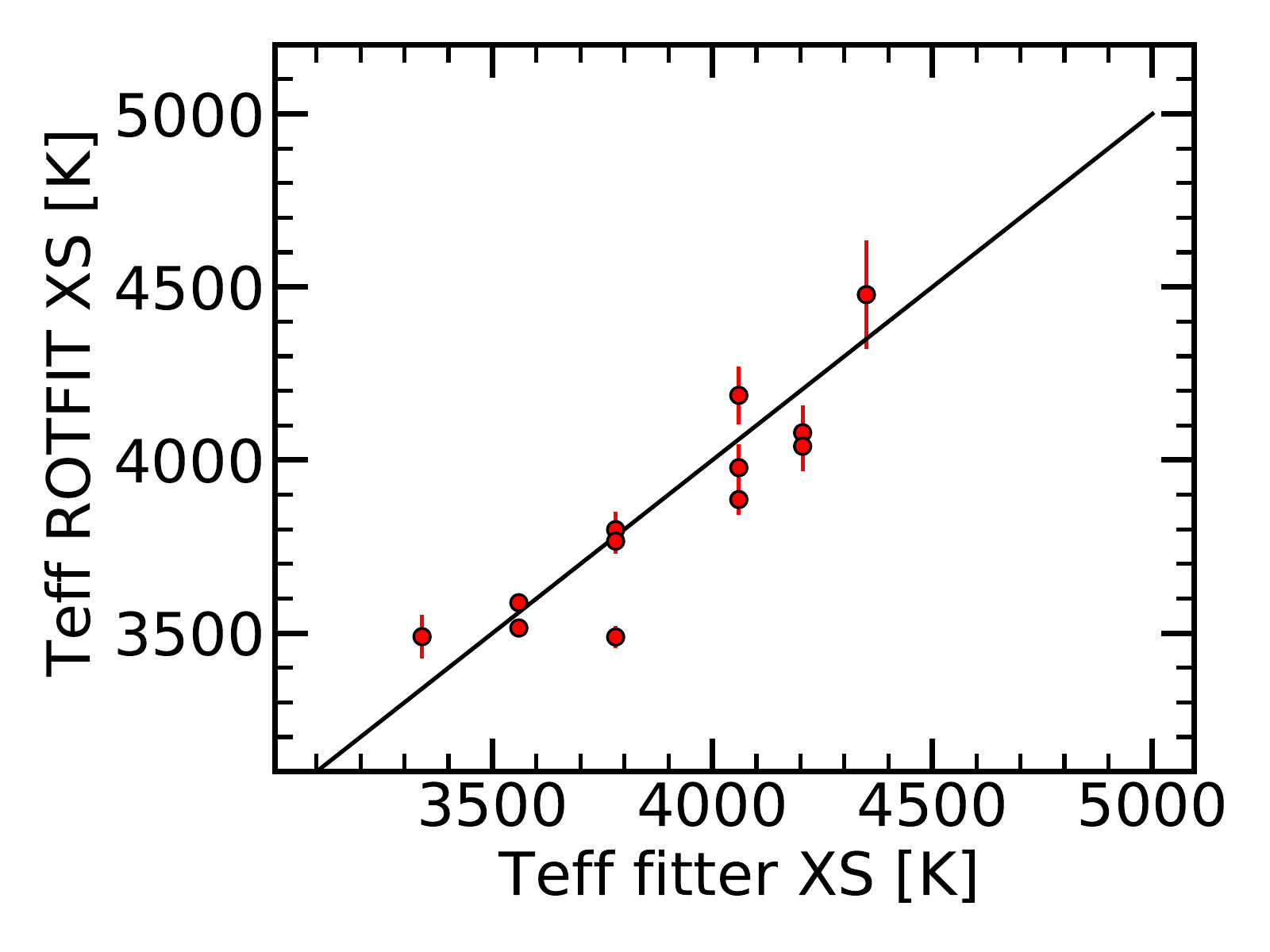}
\caption{Effective temperature (\teff) derived with ROTFIT on the X-Shooter spectra versus the temperature obtained converting the SpT from the fit of the X-Shooter spectra.
     \label{fig::rotfit_fit_orion_xs}}
\end{figure}

The values of \teff \ and the spectral types obtained with ROTFIT are in good agreement with the ones derived by fitting the X-Shooter spectrum with the method by \citet{manara13b}. The comparison is shown in Fig.~\ref{fig::rotfit_fit_orion} and Fig.~\ref{fig::rotfit_spt_fit_orion} for the UVES spectra. The spectral types obtained with the two methods are in good agreement and the differences in \teff \ are due to a different spectral type-\teff \ scale used for the HARPS templates with respect to the one used for the X-Shooter fit. Similarly, the results obtained with ROTFIT on the X-Shooter spectra are in good agreement with the values obtained on the same X-Shooter spectra with the fitter discussed in Sect.~\ref{sect::method_fitter}. This is shown in Fig.~\ref{fig::rotfit_fit_orion_xs}, and is very much in line with previous results \citep[e.g.,][]{manara20,frasca2017}.

\begin{figure}[]
\centering
\includegraphics[width=0.4\textwidth]{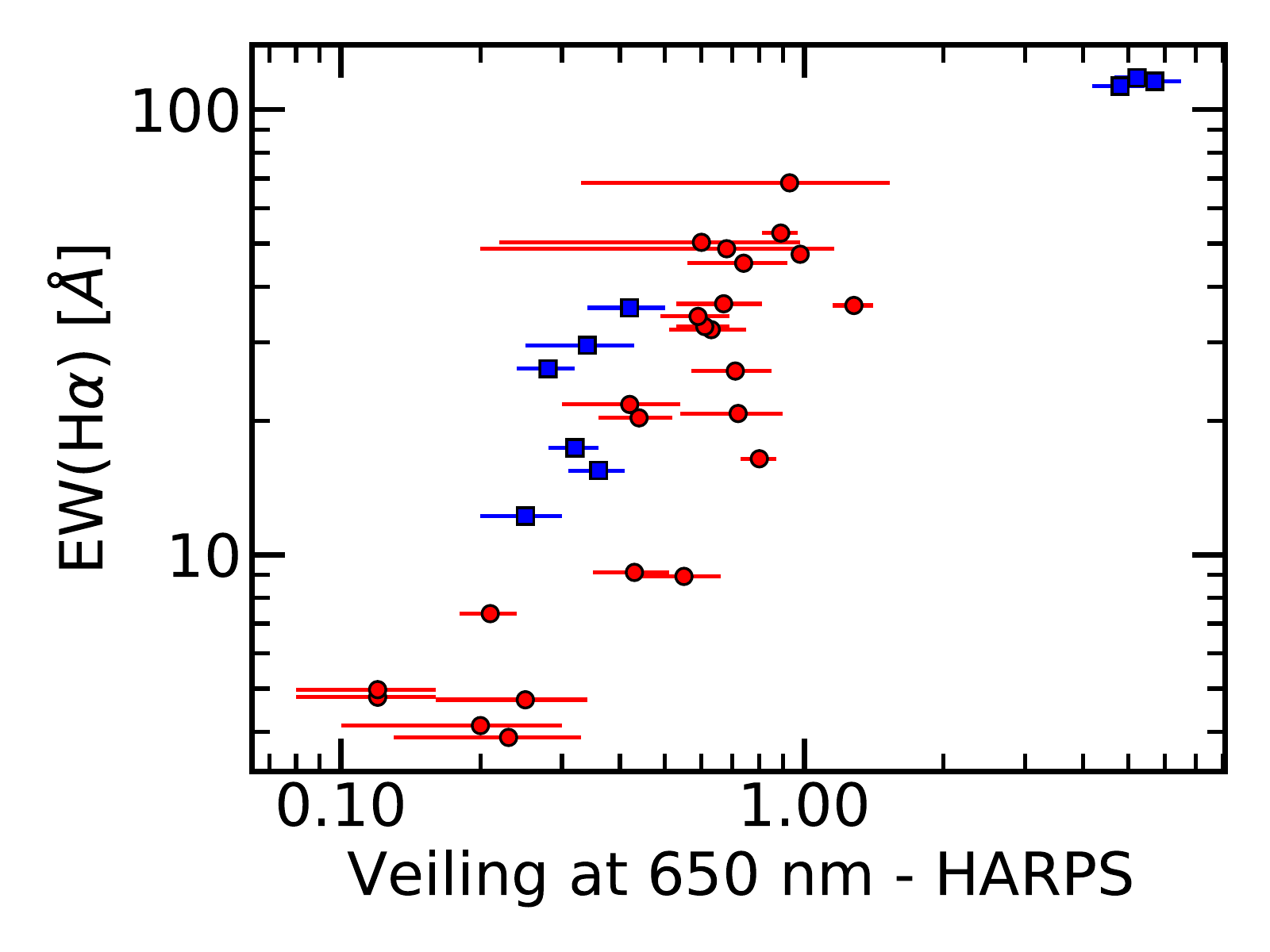}
\caption{Veiling measured on the UVES (red) and ESPRESSO (blue) spectra using the HARPS template versus the EW of the H$\alpha$ line measured on the same spectra. Each target was observed three times. 
     \label{fig::rotfit_ew_veil}}
\end{figure}

A comparison of the veiling values measured with ROTFIT with the equivalent width (EW) of the H$\alpha$ lines shows a general correlated increase of the veiling with the strength of the H$\alpha$ emission line. This is shown in Fig.~\ref{fig::rotfit_ew_veil}. 
The Spearman's rank correlation coefficient for the veiling at 650\,nm, $\rho=0.83$  with a significance $\sigma=2.6\times 10^{-9}$ \citep{Pressetal1992} supports the high degree of correlation.

\begin{figure}[]
\centering
\includegraphics[width=0.4\textwidth]{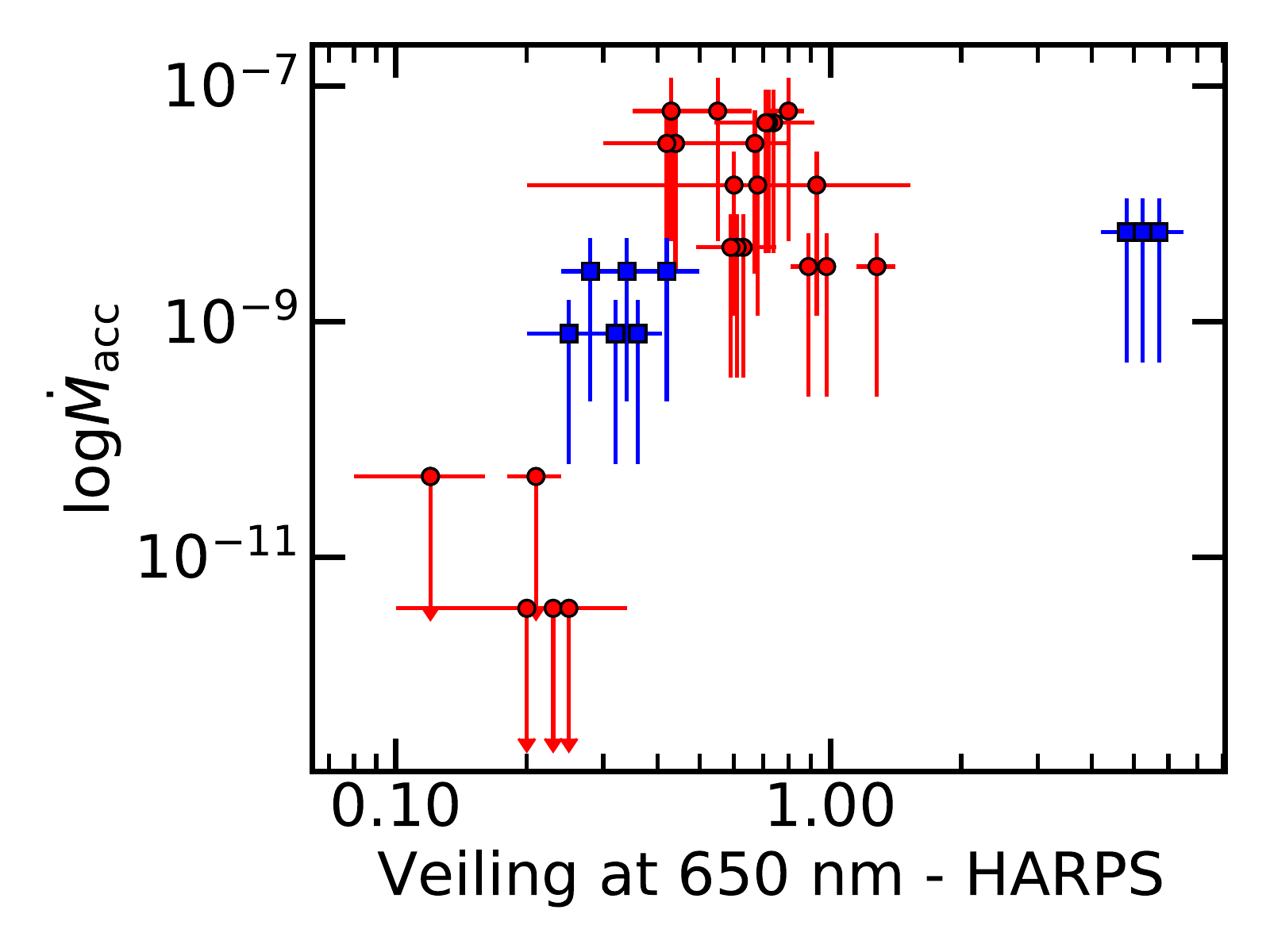}
\caption{Veiling measured on the UVES (red) and ESPRESSO (blue) spectra using the HARPS template versus the \macc ~ measured for each target from the X-Shooter spectra. Each target was observed three times with UVES or ESPRESSO, and only once with X-Shooter. 
     \label{fig::rotfit_macc_veil}}
\end{figure}

\begin{figure}[]
\centering
\includegraphics[width=0.4\textwidth]{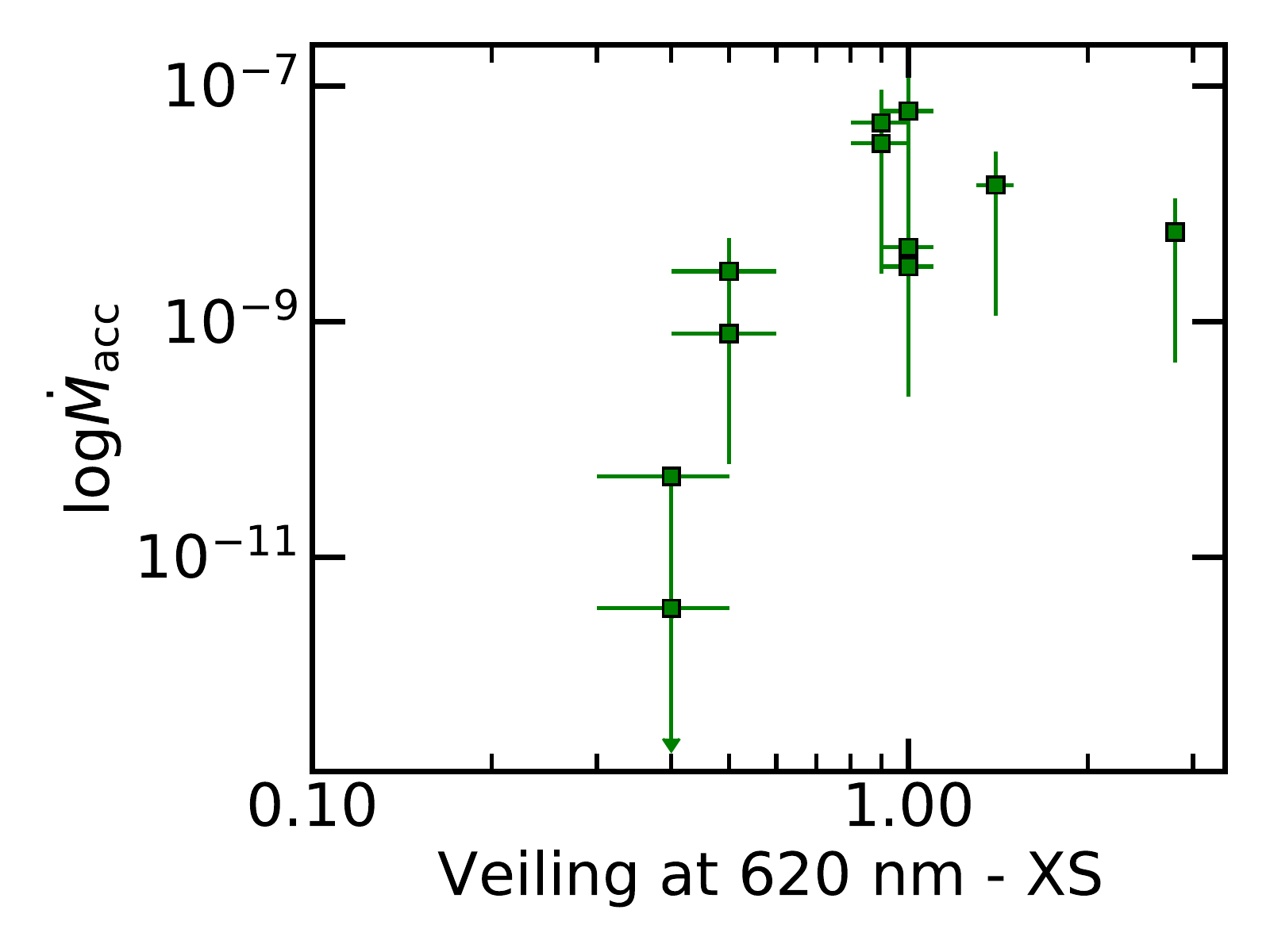}
\caption{Veiling measured on the X-Shooter spectra vs the \macc ~ measured for each target from the same spectra.  
     \label{fig::rotfit_macc_veil_xs}}
\end{figure}

On the other hand, the comparison of the veiling values measured with ROTFIT on the UVES or ESPRESSO spectra with \macc ~ derived from  X-Shooter data, which are not exactly simultaneous, shows in general a larger value of \macc ~ at higher values of veiling, but with no clear correlation ($\rho=0.62$, $\sigma=1.09\times 10^{-4}$), possibly due to the narrow range of (high) \macc ~ covered by the observations. This is shown in Fig.~\ref{fig::rotfit_macc_veil}. 
A similar degree of correlation ($\rho=0.45$, $\sigma=0.27$) is found when comparing the veiling measurements from the X-Shooter spectra to \macc~ from the same spectra, as shown in Fig.~\ref{fig::rotfit_macc_veil_xs}. A further investigation of these correlation\textbf{s} should be carried out with larger statistical sample during the course of the PENELLOPE program.

Among the values that can be derived with ROTFIT on these spectra, we have a first estimate of metallicity, which is in the range [Fe/H]=[$-0.20,+0.05$] with average values of $-0.07$ and $-0.01$ for Ori~OB1 and $\sigma$~Ori, respectively. A more detailed analysis of the spectra to determine the values of metallicity  and photospheric element abundances is deferred to a future work.

\subsection{Variability from high-resolution spectra}

Multiple emission lines are present in our high-resolution spectra (see as an example Figures~\ref{fig::lines_CVSO17}-\ref{fig::lines_SO1153}). These lines trace several processes, from accretion (e.g., H$\alpha$, H$\beta$, CaIRT) to winds and outflows (e.g., [OI]$\lambda$6300\AA). Here, we focus only on the accretion aspect and, in particular, we consider only the H$\alpha$ line for this analysis. Analyses of the other emission lines are  underway and will be presented in companion papers. In addition, we only use the EW of the H$\alpha$ line in the analysis, as this is sufficient for our purposes, but we note that future works may make use of the wealth of simultaneous photometry to convert these values into line luminosity.  

The EW of the H$\alpha$ line, measured in both the X-Shooter and the high-resolution spectra, is shown as a function of time of observations in Figures~\ref{fig::ew_ha_var}-\ref{fig::ew_ha_var_sori}. The EW values for the two non-accreting targets, CVSO17 and CVSO36, are always smaller than 10 \AA, and constant within $\sim$1-2 \AA\ during our observations. Their emission is entirely ascribable to chromospheric emission \citep[e.g.,][]{manara13a,WB03}. For the other targets, variability of up to a factor $\lesssim$2 in the EW of the H$\alpha$ line is observed within the three to four days of observations. Although the EW cannot always be linked to a line flux \citep[e.g.,][]{mendigutia13}, the relatively small variations in the photometry observed for our targets during the observations imply that the variability in the EW can be related to variations of \lacc, and thus \macc, by less than a factor of $\sim$3. This is in line with the typical variability observed on timescales of $\sim$a week, or less, by other works  \citep[e.g.,][]{costigan12,costigan14,venuti14}. 

\begin{figure}[]
\centering
\includegraphics[width=0.4\textwidth]{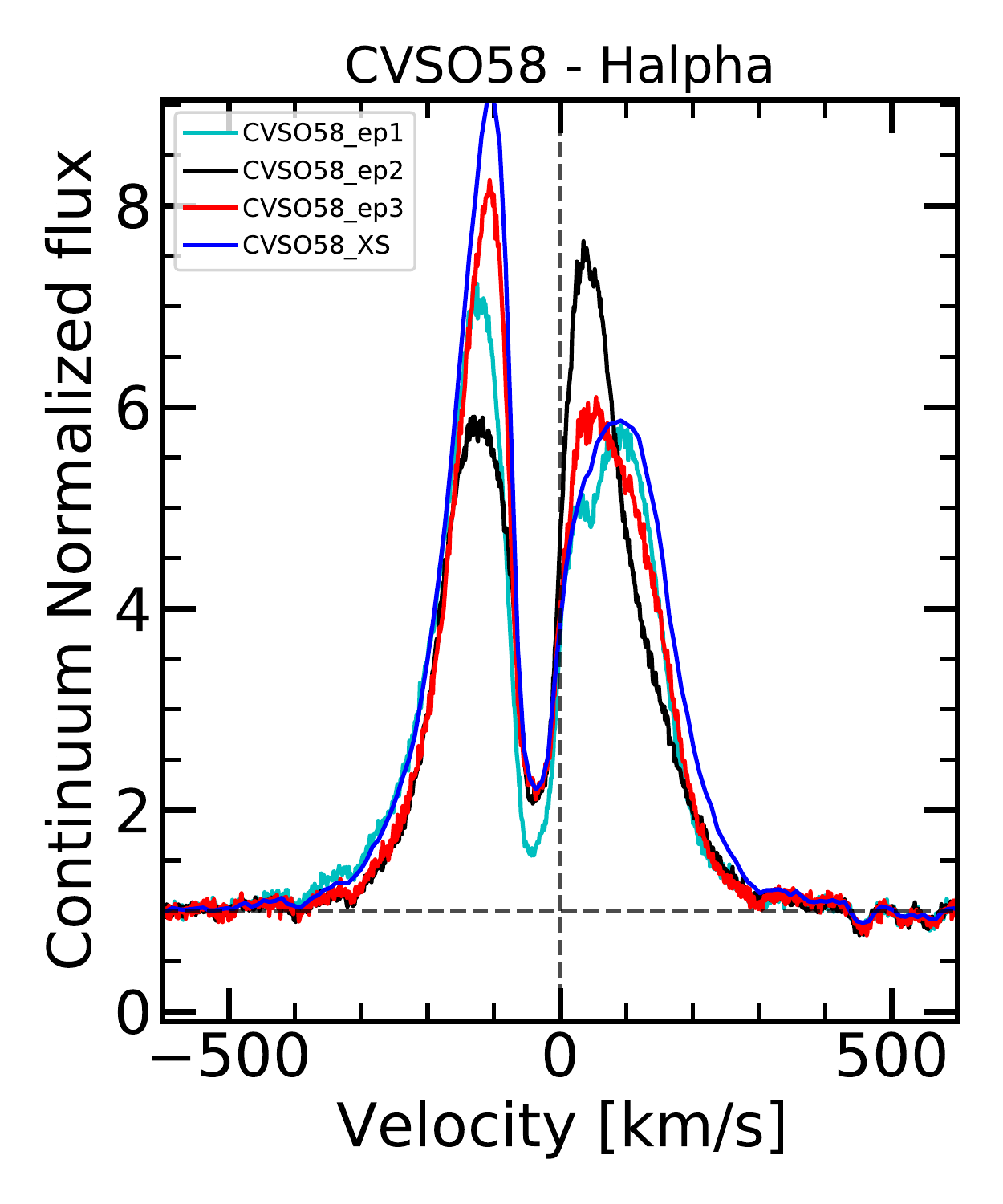}
\includegraphics[width=0.4\textwidth]{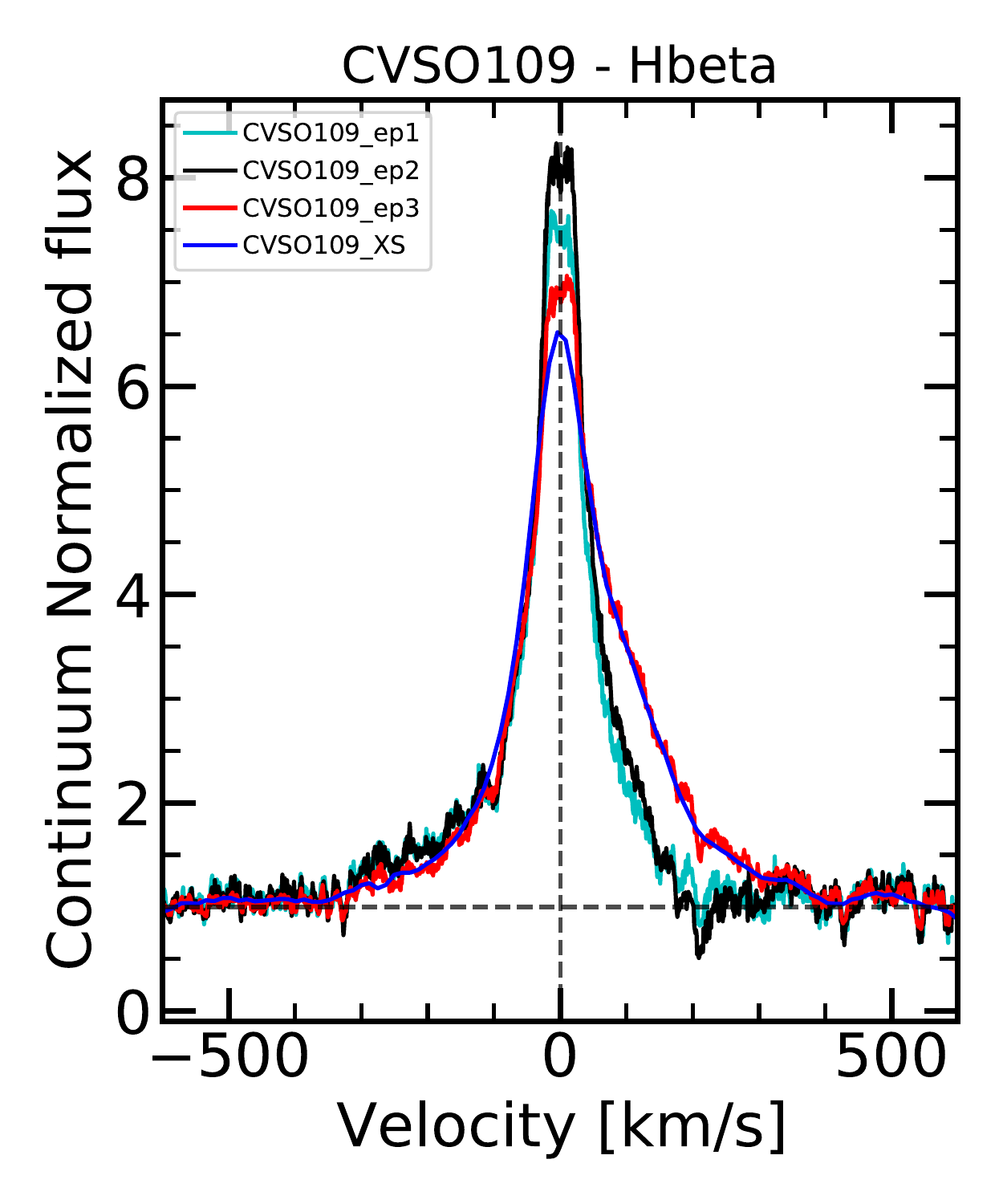}
\caption{Example of emission line profiles variability observed with UVES and X-Shooter for two targets, as labelled. The time difference between the UVES epochs is always one day, whereas the X-Shooter spectrum is obtained on the same night as the last UVES observation.
     \label{fig::lines_var_example}}
\end{figure}

The line profiles, however, can vary substantially, as we can see from the example shown in Fig.~\ref{fig::lines_var_example}, and for all the targets in  Fig.~\ref{fig::lines_CVSO58}-\ref{fig::lines_SO1153}. Some of this variability in the line profile is similar to what has been observed by other authors \citep[e.g.,][]{jay06,biazzo12,Fang2013,sousa16,bonito2020}. 
A detailed analysis of the kinematic variations traced by the shape of the emission lines, as well as by the profiles of the forbidden lines, will be subject of future works. Hereafter, we only highlight a few features.
The variations observed in the line profiles of CVSO17 and CVSO36 (Fig.~\ref{fig::lines_CVSO17}-\ref{fig::lines_CVSO36}) are modest and consistent with our previous assessment that these lines are almost totally of chromospheric origin. 

For CVSO~176, a different line profile shape is present in the X-Shooter spectrum, taken two days after the end of the UVES and HST observations, with respect to the profiles of the emission lines in the UVES spectra. In the X-Shooter spectra, most lines show a strong red-shifted absorption feature, possibly due to an enhanced accretion event or to a more favorable viewing geometry of the accretion column(s).

For CVSO109, we observe clear variability in the spectral features across the four epochs. Out of 20 emission lines identified above a 3$\sigma$ detection threshold and common to the X-Shooter and UVES coverage, we detect a defined redshifted emission wing in nine H lines and in the Ca II K line. The feature is also prominent in the Ca II H line, although unresolved due to proximity with the adjacent H7 line, and in the Ca II IR triplet, although the latter is not covered by UVES. The redshifted component, detected in the X-Shooter and final UVES epoch taken with $\sim$0.5 hr difference, is highly variable, being absent in the first two UVES epochs taken one and two days earlier. 
Such a temporal feature is indicative of a non-axisymmetric structure in the inner disk, similar to that identified from the spectra of EX Lupi \citep{2012A&A...544A..93S,2015A&A...580A..82S}. Preliminary analysis of the H lines reveal that the central velocities of the wings are correlated with the energy of the upper level (Ek) and the transition probability (Aki) of the corresponding emission lines. This, together with the range in velocities, suggests that we are tracing rotating material at  different temperatures and densities with varying radii. A more detailed investigation of this target is being carried out using the STAR-MELT analysis package, and will be presented in Campbell-White, et al. (in prep.). 

CVSO104 exhibits a peculiar HeI profile with central absorption, as well as unusual variability in the H$\alpha$ and H$\beta$ profiles, where the blueshifted emission wings fully disappear at times. This is most probably due to the presence of two unresolved components of a spectroscopic binary, but could also be related to intermittent ejection of material in either or both of the components.

Finally, variations in the (continuum normalized) line intensity and profile of the [OI]$\lambda$6300\AA ~ and in other forbidden lines is observed in a couple of targets, such as CVSO107, CVSO109, CVSO176, SO\,518, and SO\,1153. All these targets have been observed with UVES using multiple position angle to orient the slit (see Sect.~\ref{sect::programme}). It might be possible that the variations are related to the presence of (micro-)jets, or to an intrinsic variation of the ejection process. Future detailed analysis of these line profiles is needed, and will be the subject of future works.

\subsection{Variability from photometry}

As described in Sect.~\ref{Sec:Photometry}, a ground-based multiband photometric monitoring of our targets was performed from different sites contemporaneously with our observations. As an example, we show in Fig.~\ref{fig::lc_phot}
the light curves from OACT, CrAO and AAVSO from mid-November to mid-December for CVSO\,109, SO\,518, and SO\,583.  
The synthetic photometry made on the flux-calibrated X-Shooter spectra is overlaid with purple asterisks. The index  $H\alpha_{18}$--$H\alpha_{9}$, which measures the intensity of the H$\alpha$ emission, has been derived from OACT narrow-band photometry as well as from UVES and X-Shooter spectra and is displayed for the two sources in $\sigma$~Ori. The agreement between observed and synthetic photometry is apparent. 

Erratic or quasi-periodic variations with an amplitude decreasing with the increasing central wavelength of the band are visible. The largest variation amplitudes ($\approx$\,1.5--2 mag) were observed for CVSO\,109 and SO\,518 in the $B$ band. Smaller variations are observed for SO\,583. We also note, for all these sources, the nearly anticorrelated behaviour of the H$\alpha$ intensity, which is stronger when the star is fainter and redder, maybe in relation to the location of the accretion funnel with respect to the line of sight. 
The variability of accretion from photometry and line fluxes will be the subject of a future work.

%
\section{Discussion}\label{sect::discussion}

\subsection{Comparison of X-Shooter and HST Balmer continua}

One of the main goals of our program is to compare the overall flux calibration and shape of the spectra obtained with VLT/X-Shooter and HST/STIS. Figures~\ref{fig::hst_vs_xs}-\ref{fig::hst_vs_xs2} show the spectra obtained with the two instruments. 
Overall, a good agreement is observed, especially in the Balmer continuum slope. 
The COS and STIS observations of SO\,1153 were repeated on 12 February 2021, as the first observations suffered from difficulties in acquiring guide stars and therefore have issues with their wavelength and flux calibrations.

The observations with the smallest time difference between HST/STIS and VLT/X-Shooter are for CVSO58 ($\Delta t\sim$6 hours) and CVSO109 ($\Delta t\sim$1.5 hours). 
In these cases, indeed, the difference between the observed flux in the HST/STIS and VLT/X-Shooter spectra in the Balmer continuum region is only a factor $\sim$1.3 and 1, respectively. However, the slopes of the HST and X-Shooter spectra are different at wavelengths longer than 500 nm for CVSO109, suggesting either a slightly different extinction toward the object, although probably too large for  such a small time difference, or a significant contamination of the X-Shooter spectrum by the redder unresolved companion. Despite the larger time difference, other targets also show a good agreement between the X-Shooter and HST/STIS spectra, in particular CVSO107, CVSO146, CVSO165, and SO\,518.

\begin{figure}[]
\centering
\includegraphics[width=0.4\textwidth]{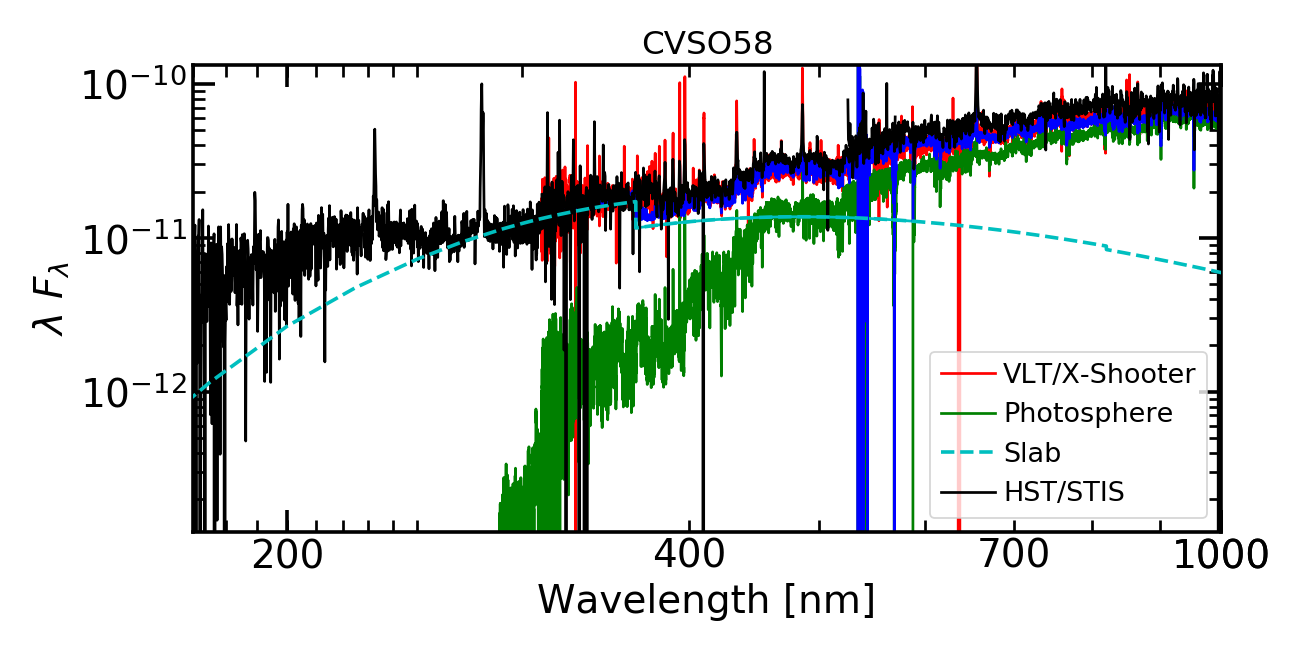}
\includegraphics[width=0.4\textwidth]{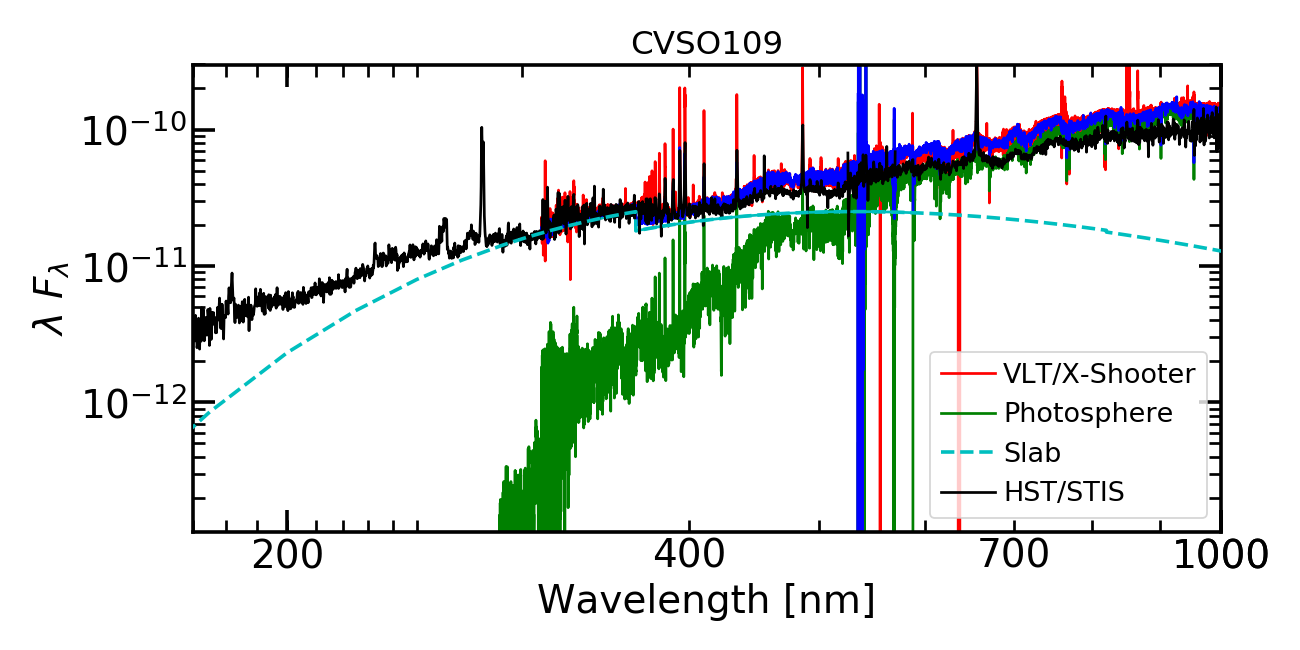}
\includegraphics[width=0.4\textwidth]{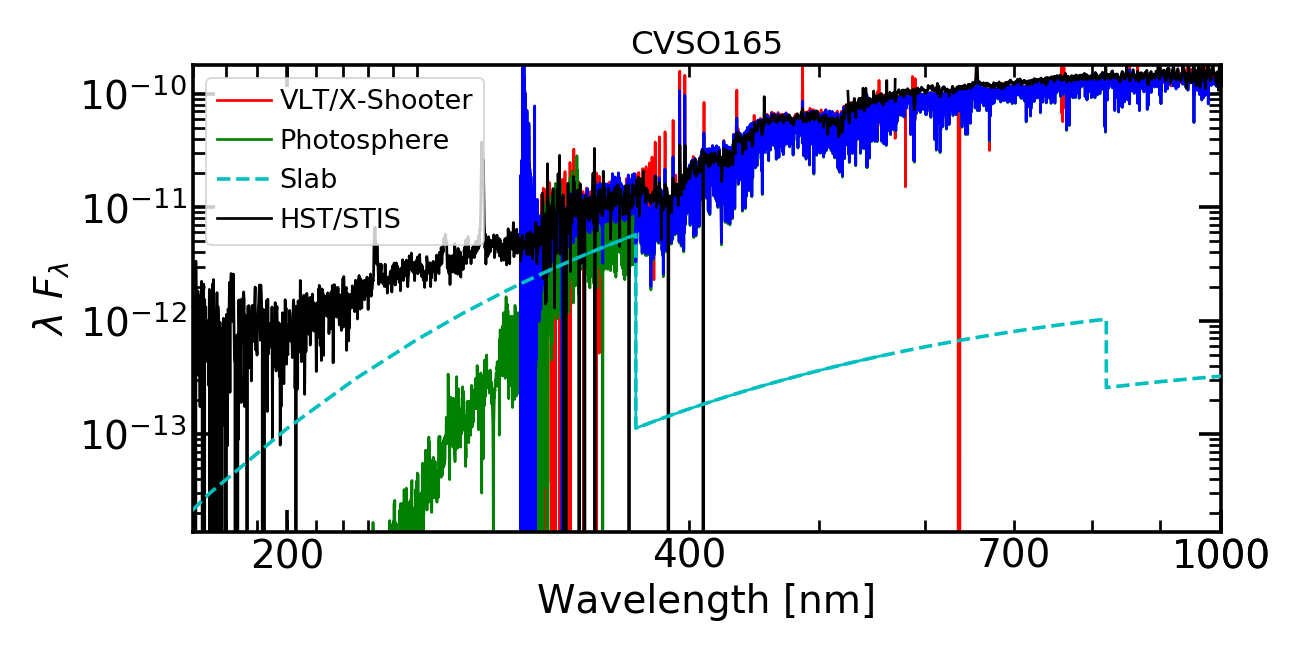}
\caption{Comparison of the extrapolation of the results obtained fitting the VLT/X-Shooter spectrum (best fit in blue) with the HST spectrum in the UV.
     \label{fig::slab_test_hst}}
\end{figure}

We focus, in particular, on three targets, CVSO58,  CVSO109, and CVSO165, for which the VLT/X-Shooter and HST/STIS spectra are observed close in time and have similar shapes, in order to test whether our slab model leading to the best fit of the X-Shooter spectra (see Sect.~\ref{sect::method_fitter}) reproduces the emission from HST at wavelengths shorter than 300 nm. The comparison, performed by scaling the de-reddened HST/STIS spectrum to that of the VLT/X-Shooter, is shown in Fig.~\ref{fig::slab_test_hst}. In general, the HST spectra at $\lambda<$ 300 nm are brighter than the extrapolation of the slab model in this wavelength range. 

To quantify this discrepancy, we compute the excess emission in the HST/STIS spectra relative to the total flux of the slab model, the latter being proportional to \lacc. To extract the excess we subtract the photosphere from the X-Shooter spectra extended to shorter wavelengths using a BT-Settl synthetic spectrum \citep{Allard2012} of the same temperature. We also remove the flux of bright emission lines in the HST spectrum. 
In the cases of CVSO58 and CVSO109, the excess emission in the HST spectra with respect to the slab model is $\sim$20\% and 10\% of the total slab model flux, respectively. In both cases, the contribution of the photospheric emission subtracted from the total flux is negligible. This means that the accretion luminosity derived from the slab for these two objects could be underestimated by up to a factor 20\%, which is well within the uncertainty of the \lacc~ values from the fit of the X-Shooter spectra (typically 0.2 dex). For CVSO165, instead, a larger factor $\sim$0.5 is obtained. This target is the one with the lowest value of \lacc/\lstar ~ among the accreting targets in our sample. Similarly to \citet{ingleby11} and \citet{alcala19}, we see that the excess emission in the UV range could be underestimated more substantially for objects accreting at lower rates and with spectral type earlier than about K3. The latter is a consequence of a lower contrast between the continuum  excess  emission  and  the  photospheric+chromospheric emission. However, we ought to consider that this target is actually a binary that is unresolved in the VLT observations, but resolved by the HST.

A comparison with the other spectra leads to a relative additional flux in the HST spectra of typically less than 10\% to $\sim$20\%, with a maximum of $\sim$25\%. With this small sample statistics it is not possible to constrain the origin of this discrepancy. Several factors are at play, including the assumed reddening law \citep[][$R_V$=3.1]{cardelli98} and a single temperature to describe the excess emission due to accretion with a slab model. The former affects the shape of the slab model in the UV range more than at optical wavelengths; thus,   future works should investigate the kind of reddening law that could reproduce the observed emission at all wavelengths \citep[see also discussion in][]{alcala19}. The single temperature and density assumption is also a simplification of the physical conditions in the accretion region \citep[e.g.,][]{hartmann16} and multiple components should be used to better reproduce the shape of the emission at all wavelengths \citep[e.g.,][]{robinson19}. However, the small underestimate in the UV of the slab model shown here also indicates that the value of \lacc \ derived with this procedure is a good estimate of the accretion luminosity of the target. 

This kind of analysis, carried out on a larger sample provided by the spectra obtained by the PENELLOPE and ULLYSES programs, will allow us to firmly quantify whether a different reddening law is needed to describe the extinction in the UV part of the spectrum or if the single temperature slab models systemically underestimate the UV flux produced by the accretion process and by how much.

\subsection{Accretion properties of young low-mass stars}

The first set of data obtained by the PENELLOPE program related to young stars in the Orion star-forming region has allowed us to derive stellar and accretion parameters for the 13 targets, two of which were found to be non-accreting. For the eleven accretors, we show that the short-time ($<$1 week) variability of the accretion rate at the time of the observation is small, accounting for a factor $\lesssim3$ in \lacc. This factor is obtained by analyzing the variability of the equivalent width of the H$\alpha$ line, and is well in line with previous results \citep[e.g.,][]{biazzo12,costigan14,venuti14}. However, the variability in the kinematics of the accretion-related emission lines is substantial in multiple objects, suggesting that the structure of the accretion flows onto the star is complex and variable, but the overall energy released by the process is relatively constant, at least on these short timescales probed by our observations. 

Based on this result, we are confident that the accretion rates measured with our fitting procedure from the broadband X-Shooter spectra are a good estimate of the typical accretion rate of the targets at around the time of the observations. Moreover, we have shown that the fit of the X-Shooter spectra with a slab model falls short in reproducing the excess emission seen in the UV spectra by $\sim$10\% in most cases. All in all, the typically assumed uncertainties on the values of \lacc ~ of 0.2 dex and on \macc ~ of 0.4 dex from our fitter are in line with both the observed short-term variability and the missing UV flux by the model.
\begin{figure}[]
\centering
\includegraphics[width=0.4\textwidth]{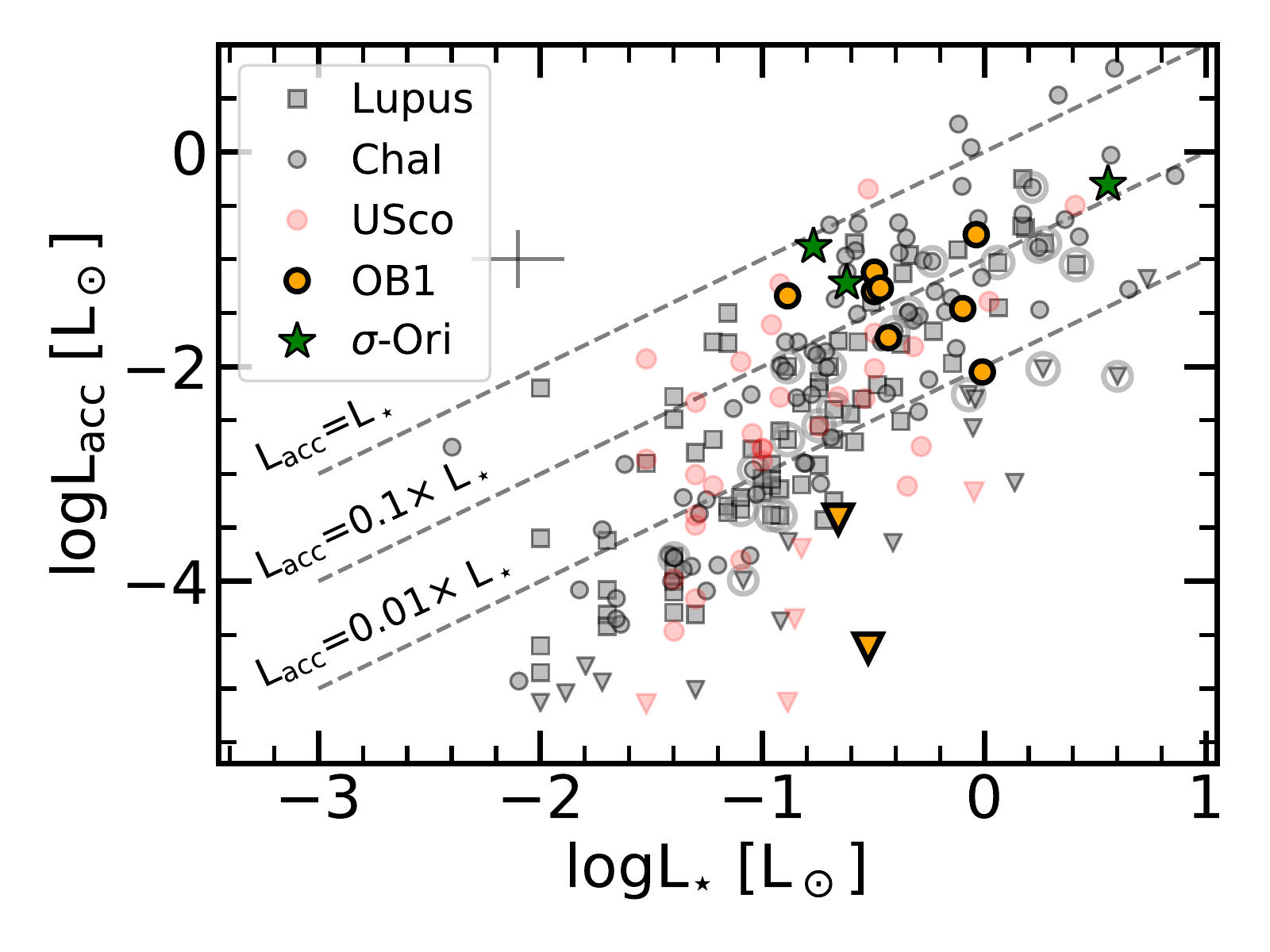}
\includegraphics[width=0.4\textwidth]{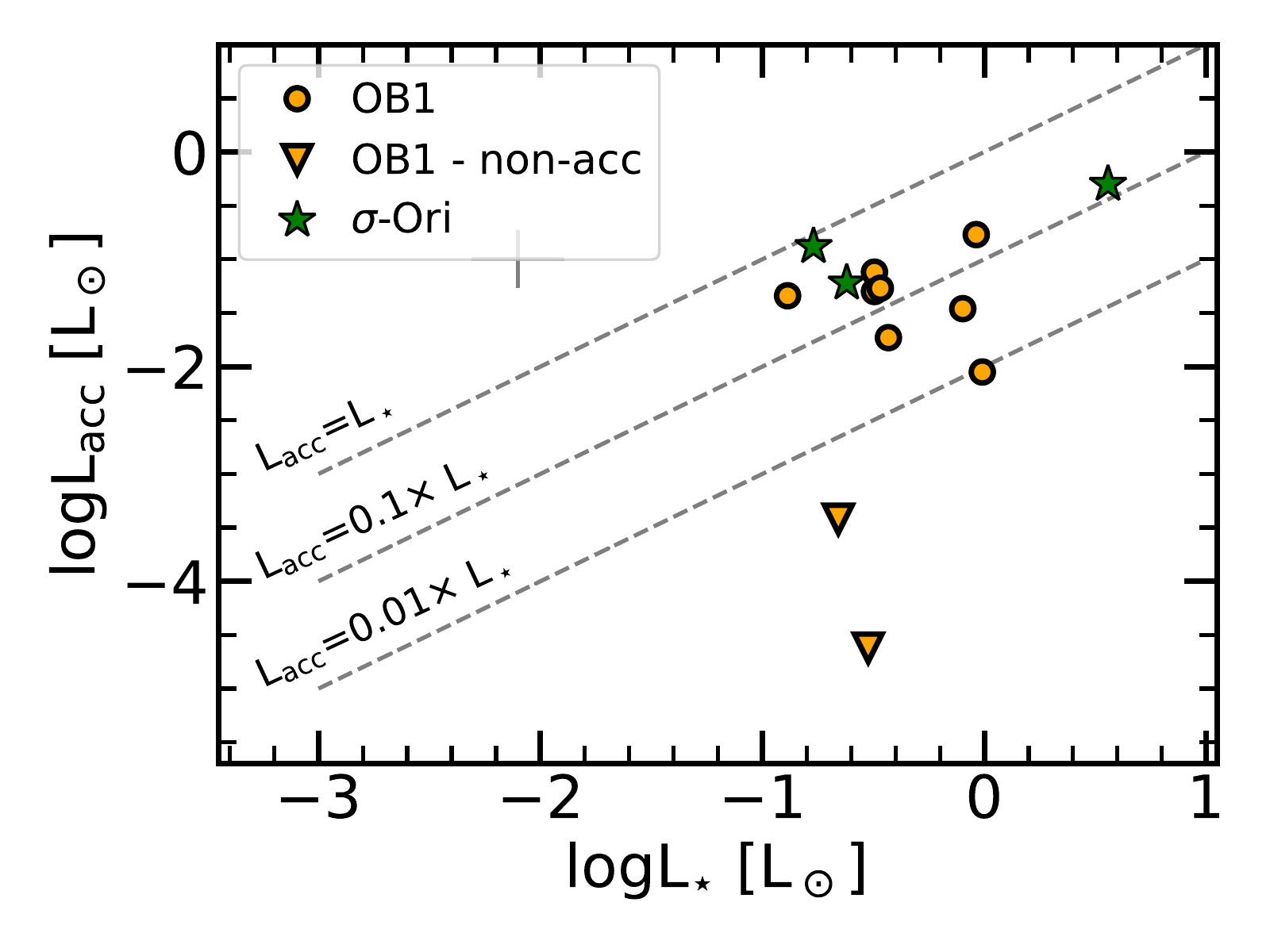}
\caption{Relation between the accretion luminosity and stellar luminosity obtained from the X-Shooter for the PENELLOPE targets in  the Orion region. The top panel shows the comparison with data obtained with the same technique in the literature, while the bottom panel highlights solely the new results from this work.
     \label{fig::lacc_lstar}}
\end{figure}

With the values of \lacc ~ and \macc ~ derived here, we can compare the accretion rates of this population of young stars with previous estimates of accretion rates with the same method in other young star forming regions, in particular, Lupus \citep[$\sim 1-3$ Myr old,][]{alcala14,alcala17} rescaled to the Gaia DR2 distances in \citet{alcala19}, Chamaeleon~I \citep[$\sim 1-3$ Myr old,][]{manara16a,manara17b} rescaled to the Gaia DR2 distances in \citet{manara19b}, and Upper Scorpius \citep[$\sim 5-10$ Myr old,][]{manara20}. 
The relation between \lacc~ and \lstar ~ for the targets in these regions is shown in Fig.~\ref{fig::lacc_lstar}. In line with previous results, the data of accreting targets, including those presented here, follow the same \lacc-\lstar~ relation as target in other star-forming regions with younger and older ages. No dependence on age or location is observed. The upper limits on \lacc ~ derived for the two non-accreting targets are found at values of \lacc/\lstar$<0.01$, in line with other non-accreting systems observed in young regions. 

\begin{figure}[]
\centering
\includegraphics[width=0.4\textwidth]{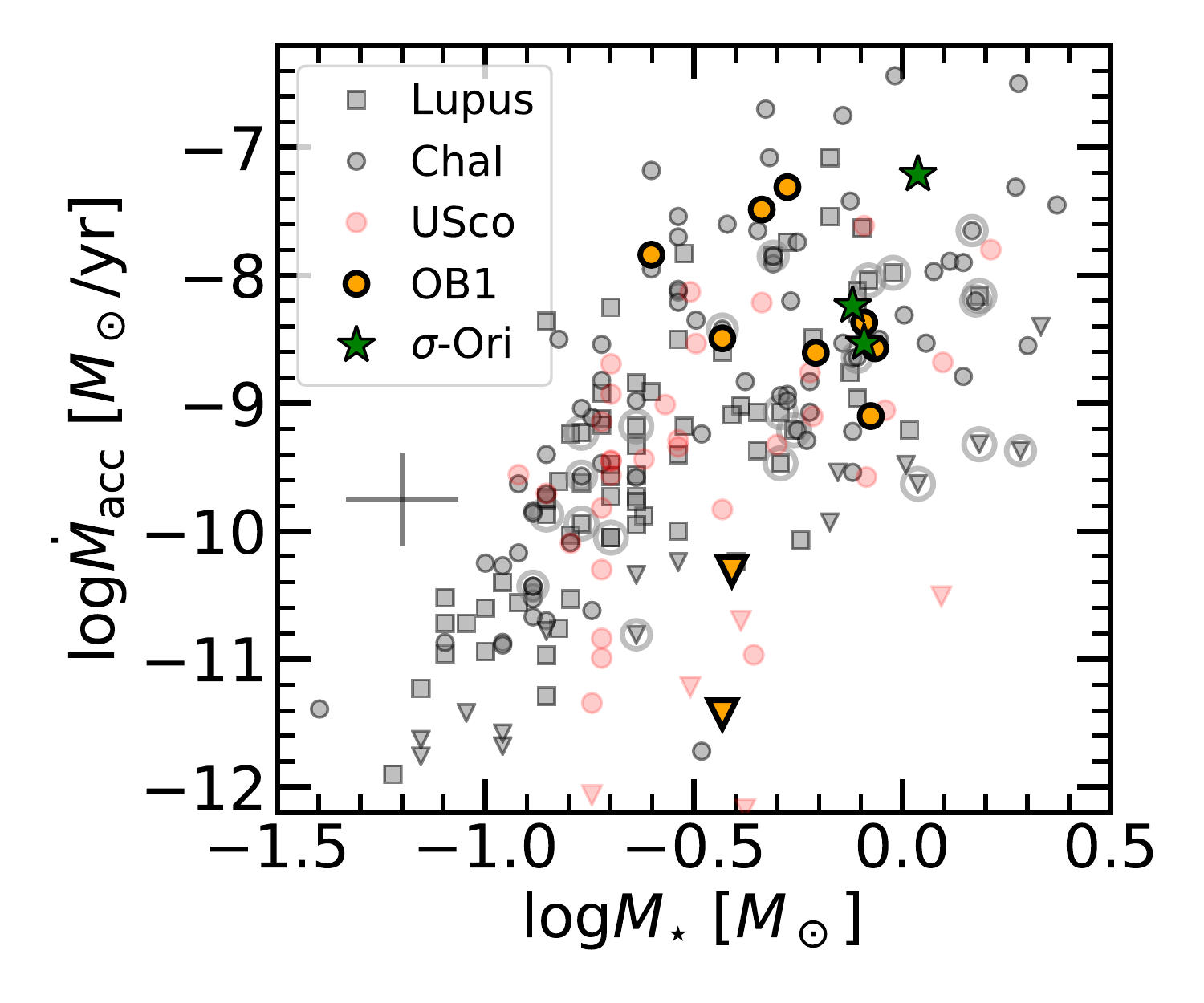}
\includegraphics[width=0.4\textwidth]{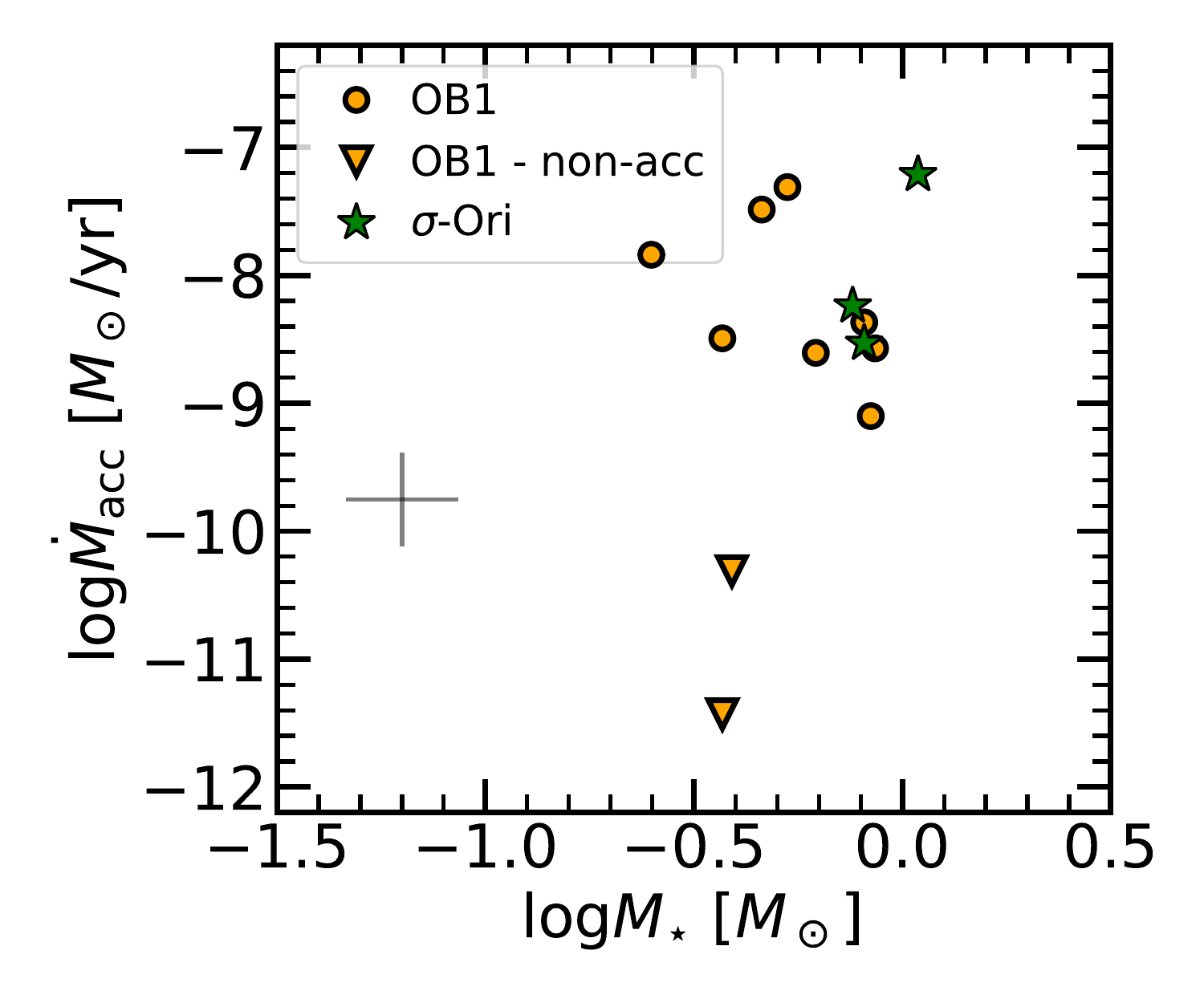}
\caption{Relation between the mass accretion rates and stellar mass obtained from the X-Shooter for the PENELLOPE targets in  the Orion region. The top panel shows the comparison with data obtained with the same technique in the literature, while the bottom panel highlights solely the new results from this work.
     \label{fig::macc_mstar}}
\end{figure}

The values of \macc ~ vs \mstar ~ for the targets in Orion and in the other young regions is shown in Fig.~\ref{fig::macc_mstar}. Also in this case, the data presented here on the Orion targets populate the same part of the parameter space as targets observed in other star-forming regions. The observed scatter of \macc ~ at a given value of \mstar ~ is large even in this small sample, $\sim$2 dex when considering only the accreting objects. This scatter is much larger than the variability we observe on timescales of a few days, which is thought to be the maximum observed variability on timescales of years \citep{costigan12,costigan14}. The origin of the scatter should be then ascribed to real differences in the accretion properties of targets in a single star-forming region, possibly due to different disk evolutionary stages \citep[e.g.,][]{najita07,najita15,manara14} or properties of the magnetic field \citep[e.g.,][]{gregory06} or of the details of the star-disk interaction, or even of the surroundings, such as the presence of companions \citep[e.g.,][]{rosotti19}. A detailed analysis of the observed emission lines for these targets will help to understand the origin of these differences. The non-accreting targets observed here have upper limits on \macc ~ much smaller than the typical \macc ~ observed for stars of the same stellar mass. This is in line with what is observed in other star-forming regions and may also indicate rapid dispersal of the (inner) disk as processes such as photoevaporation take over \citep[e.g.,][]{EP17}.

Overall, we do not see noticeable differences between the accretion rates measured in the 13 targets in the Orion OB1 and $\sigma$-Ori regions with respect to what is observed in other young star-forming regions with similar techniques. This result is in agreement with the findings of \citet{manara20}, \citet{venuti19}, \citet{Rugel2018}, and \citet{ingleby14}, among others, who found that the accretion rates of still accreting objects at ages of $>3-5$ Myr are still comparable to the accretion rates of younger stars. However, this result is still puzzling as these high accretion rates are difficult to reconcile with the current framework to describe disk evolution \citep[e.g.,][]{ingleby14,manara20}.  Larger samples can help to statistically study the differences between the populations of various regions and provide the necessary data for constraining models based on observations.

%

\section{Conclusions}\label{sect::conclusions}

In this paper, we present the survey strategy and a set of initial results from the PENELLOPE Large Program aimed at obtaining observations with VLT/X-Shooter, and VLT/ESPRESSO or VLT/UVES, contemporaneously with the HST STIS and COS observations of the public ULLYSES survey. Complemented with contemporaneous photometry, these programs will provide an unprecedented view of the accretion and ejection processes in young stars. 

We showcase the observing strategy, data reduction procedure, and data analysis tools on a sample of ten targets in the Orion OB1 association and further three targets in the $\sigma$-Orionis region, observed with HST and VLT in November and December 2020. We have presented the contemporaneous multi-band photometry obtained at OACT and show the good agreement between the absolute flux-calibrated X-Shooter spectra and this photometric data.

We show that the agreement between the temperature and spectral type estimates obtained with the two methods used here on the X-Shooter spectra and on the high-resolution spectra is consistent within a sub-class. We then found that the veiling measured on the high-resolution spectra with ROTFIT correlates with the EW of the H$\alpha$ line measured on the same spectra, whereas no strong correlation is found between the veiling and \macc. 

We were able to constrain the short-term ($<$1 week) variability of the accretion luminosity for these targets to be a factor $\lesssim$3, in line with previous works \citep[e.g..][]{cody14,cody17,venuti14,costigan12,costigan14}. Future works on the dataset from this program will also allow for the the long-term (a few years) variability of these targets to be constrained thanks to the availability of previous spectra obtained with the same or with different instruments. 
On the other hand, the profiles of accretion and ejection tracing lines vary substantially between the various epochs, suggesting that the structure of the accretion flow is complex and variabile, whereas the energy released is relatively constant. The data from this program will allow for detailed analyses of the structure of the accretion and ejection emitting regions for all the targets to further understand the accretion and ejection processes as well as their interplay. 

The first comparison of the HST/STIS and X-Shooter spectra on a limited number of targets suggests that the slab models used to fit the X-Shooter spectra underestimate the UV emission seen in the HST spectra by $\sim 10$\%, and usually less than $\sim 20$\%, in the targets observed here. Several causes can be ascribed for this discrepancy, such as an inappropriate extinction curve in the UV part of the spectrum, or a real shortcoming of a single temperature model to reproduce the excess emission in the UV. These hypotheses will be investigated in future works with a larger sample of contemporaneous HST/STIS and X-Shooter spectra. 
Nevertheless, these findings demonstrate the general validity of the accretion rates measured with X-Shooter on these spectra, and that the reported uncertainties of $\sim$0.2 dex on \lacc ~ and $\sim$0.4 dex on \macc ~ are reasonable.
Finally, we compared the relation between the accretion rate and stellar mass, showing similar values of \macc ~ and similar spread of values at any \mstar ~ as in previous surveys. 

The VLT/PENELLOPE program will continue to obtain data alongside the HST/ULLYSES program for the next $\lesssim$2 years, providing a public dataset that will be key to creating a coherent and comprehensive view of the accretion process and its impact on the evolution of disks in nearby star-forming regions. 
This dataset will also be key to constraining other processes not showcased in this work, such as the ejection of matter in young stellar objects, probed by the high-resolution multi-epoch spectra available from this program. 
We invite the community to make use of these public datasets, acknowledging the effort of the group of people who initiated this effort, and with an open desire to enhance our knowledge. 

\begin{acknowledgements}
We thank the referee for the constructive report. 
We thank the ESO support staff for the help in the preparation of the observations, in the scheduling, and for carrying out the observations at Paranal. We also thank the ESO staff, in particular John Pritchard and Lodovico Coccato, for implementing the Molecfit workflow for ESPRESSO and making it available to us. We thank Antonella Natta for insightful discussions.
Based on observations obtained with the NASA/ESA Hubble Space Telescope, retrieved from the Mikulski Archive for Space Telescopes (MAST) at the Space Telescope Science Institute (STScI). STScI is operated by the Association of Universities for Research in Astronomy, Inc. under NASA contract NAS 5-26555.
This work is supported in part by the ODYSSEUS team (HST AR-16129), \url{https://sites.bu.edu/odysseus/}.
This project has received funding from the European Union's Horizon 2020 research and innovation programme under the Marie Sklodowska-Curie grant agreement No 823823 (DUSTBUSTERS).
This research received financial support from the project PRIN-INAF-MAINSTREAM "Protoplanetary disks seen through the eyes of new-generation instruments" (CUPC54I19000600005) and PRIN-INAF 2019  ``Spectroscopically Tracing the Disk Dispersal Evolution".
This work was partly supported by the Deutsche Forschungs-Gemeinschaft (DFG, German Research Foundation) - Ref no. FOR 2634/1 TE 1024/1-1. 
This work has made use of data from the European Space Agency (ESA) mission
{\it Gaia} (\url{https://www.cosmos.esa.int/gaia}), processed by the {\it Gaia}
Data Processing and Analysis Consortium (DPAC,
\url{https://www.cosmos.esa.int/web/gaia/dpac/consortium}). Funding for the DPAC
has been provided by national institutions, in particular the institutions
participating in the {\it Gaia} Multilateral Agreement.

JCW and ASA are supported by the  STFC grant number ST/S000399/1 ("The Planet-Disk Connection: Accretion, Disk Structure, and Planet Formation").
ACG and PMG acknowledge funding from the European Research Council under Advanced Grant No. 743029, Ejection, Accretion Structures in YSOs (EASY).
BE acknowledges support from the DFG Research Unit ''Transition Disks" (FOR 2634/1, ER 685/8-1) and from the DFG cluster of excellence “Origin and Structure of the
Universe” (http://www.universe-cluster.de/).
R.G.L acknowledges support by  Science  Foundation  Ireland under Grant No. 18/SIRG/5597
This project has received funding from the European Research Council (ERC) under the European Union's Horizon 2020 research and innovation programme under grant agreement No 716155 (SACCRED). 
L.V. acknowledges support by an appointment to the NASA Postdoctoral Program at the NASA Ames Research Center, administered by Universities Space Research Association under contract with NASA. G.R. acknowledges support from the Netherlands Organisation for Scientific Research (NWO, program number 016.Veni.192.233) and from an STFC Ernest Rutherford Fellowship (grant number ST/T003855/1). J.H. acknowledges support from the National Research Council of M\'exico (CONACyT) project No. 86372 and the PAPIIT UNAM project  IA102921.

\end{acknowledgements}

\bibliography{bibliography.bib}

\begin{thebibliography}{186}
\expandafter\ifx\csname natexlab\endcsname\relax\def\natexlab#1{#1}\fi

\bibitem[{{Alam} {et~al.}(2015){Alam}, {Albareti}, {Allende Prieto}, {Anders},
  {Anderson}, {Anderton}, {Andrews}, {Armengaud}, {Aubourg}, {Bailey}, {Basu},
  {Bautista}, {Beaton}, {Beers}, {Bender}, {Berlind}, {Beutler}, {Bhardwaj},
  {Bird}, {Bizyaev}, {Blake}, {Blanton}, {Blomqvist}, {Bochanski}, {Bolton},
  {Bovy}, {Shelden Bradley}, {Brandt}, {Brauer}, {Brinkmann}, {Brown},
  {Brownstein}, {Burden}, {Burtin}, {Busca}, {Cai}, {Capozzi}, {Carnero
  Rosell}, {Carr}, {Carrera}, {Chambers}, {Chaplin}, {Chen}, {Chiappini},
  {Chojnowski}, {Chuang}, {Clerc}, {Comparat}, {Covey}, {Croft}, {Cuesta},
  {Cunha}, {da Costa}, {Da Rio}, {Davenport}, {Dawson}, {De Lee}, {Delubac},
  {Deshpande}, {Dhital}, {Dutra-Ferreira}, {Dwelly}, {Ealet}, {Ebelke},
  {Edmondson}, {Eisenstein}, {Ellsworth}, {Elsworth}, {Epstein}, {Eracleous},
  {Escoffier}, {Esposito}, {Evans}, {Fan}, {Fern{\'a}ndez-Alvar}, {Feuillet},
  {Filiz Ak}, {Finley}, {Finoguenov}, {Flaherty}, {Fleming}, {Font-Ribera},
  {Foster}, {Frinchaboy}, {Galbraith-Frew}, {Garc{\'\i}a},
  {Garc{\'\i}a-Hern{\'a}ndez}, {Garc{\'\i}a P{\'e}rez}, {Gaulme}, {Ge},
  {G{\'e}nova-Santos}, {Georgakakis}, {Ghezzi}, {Gillespie}, {Girardi},
  {Goddard}, {Gontcho}, {Gonz{\'a}lez Hern{\'a}ndez}, {Grebel}, {Green},
  {Grieb}, {Grieves}, {Gunn}, {Guo}, {Harding}, {Hasselquist}, {Hawley},
  {Hayden}, {Hearty}, {Hekker}, {Ho}, {Hogg}, {Holley-Bockelmann}, {Holtzman},
  {Honscheid}, {Huber}, {Huehnerhoff}, {Ivans}, {Jiang}, {Johnson},
  {Kinemuchi}, {Kirkby}, {Kitaura}, {Klaene}, {Knapp}, {Kneib}, {Koenig},
  {Lam}, {Lan}, {Lang}, {Laurent}, {Le Goff}, {Leauthaud}, {Lee}, {Lee},
  {Licquia}, {Liu}, {Long}, {L{\'o}pez-Corredoira}, {Lorenzo-Oliveira},
  {Lucatello}, {Lundgren}, {Lupton}, {Mack}, {Mahadevan}, {Maia}, {Majewski},
  {Malanushenko}, {Malanushenko}, {Manchado}, {Manera}, {Mao}, {Maraston},
  {Marchwinski}, {Margala}, {Martell}, {Martig}, {Masters}, {Mathur},
  {McBride}, {McGehee}, {McGreer}, {McMahon}, {M{\'e}nard}, {Menzel},
  {Merloni}, {M{\'e}sz{\'a}ros}, {Miller}, {Miralda-Escud{\'e}}, {Miyatake},
  {Montero-Dorta}, {More}, {Morganson}, {Morice-Atkinson}, {Morrison},
  {Mosser}, {Muna}, {Myers}, {Nandra}, {Newman}, {Neyrinck}, {Nguyen},
  {Nichol}, {Nidever}, {Noterdaeme}, {Nuza}, {O'Connell}, {O'Connell},
  {O'Connell}, {Ogando}, {Olmstead}, {Oravetz}, {Oravetz}, {Osumi}, {Owen},
  {Padgett}, {Padmanabhan}, {Paegert}, {Palanque-Delabrouille}, {Pan},
  {Parejko}, {P{\^a}ris}, {Park}, {Pattarakijwanich}, {Pellejero-Ibanez},
  {Pepper}, {Percival}, {P{\'e}rez-Fournon}, {\u1e54rez-Ra`fols}, {Petitjean},
  {Pieri}, {Pinsonneault}, {Porto de Mello}, {Prada}, {Prakash},
  {Price-Whelan}, {Protopapas}, {Raddick}, {Rahman}, {Reid}, {Rich}, {Rix},
  {Robin}, {Rockosi}, {Rodrigues}, {Rodr{\'\i}guez-Torres}, {Roe}, {Ross},
  {Ross}, {Rossi}, {Ruan}, {Rubi{\~n}o-Mart{\'\i}n}, {Rykoff},
  {Salazar-Albornoz}, {Salvato}, {Samushia}, {S{\'a}nchez}, {Santiago},
  {Sayres}, {Schiavon}, {Schlegel}, {Schmidt}, {Schneider}, {Schultheis},
  {Schwope}, {Sc{\'o}ccola}, {Scott}, {Sellgren}, {Seo}, {Serenelli}, {Shane},
  {Shen}, {Shetrone}, {Shu}, {Silva Aguirre}, {Sivarani}, {Skrutskie},
  {Slosar}, {Smith}, {Sobreira}, {Souto}, {Stassun}, {Steinmetz}, {Stello},
  {Strauss}, {Streblyanska}, {Suzuki}, {Swanson}, {Tan}, {Tayar}, {Terrien},
  {Thakar}, {Thomas}, {Thomas}, {Thompson}, {Tinker}, {Tojeiro}, {Troup},
  {Vargas-Maga{\~n}a}, {Vazquez}, {Verde}, {Viel}, {Vogt}, {Wake}, {Wang},
  {Weaver}, {Weinberg}, {Weiner}, {White}, {Wilson}, {Wisniewski},
  {Wood-Vasey}, {Ye`che}, {York}, {Zakamska}, {Zamora}, {Zasowski}, {Zehavi},
  {Zhao}, {Zheng}, {Zhou}, {Zhou}, {Zou}, \& {Zhu}}]{SDSS}
{Alam}, S., {Albareti}, F.~D., {Allende Prieto}, C., {et~al.} 2015, \apjs, 219,
  12

\bibitem[{{Alcal{\'a}} {et~al.}(2019){Alcal{\'a}}, {Manara}, {France},
  {Schneider}, {Arulanantham}, {Miotello}, {G{\"u}nther}, \&
  {Brown}}]{alcala19}
{Alcal{\'a}}, J.~M., {Manara}, C.~F., {France}, K., {et~al.} 2019, \aap, 629,
  A108

\bibitem[{{Alcal{\'a}} {et~al.}(2017){Alcal{\'a}}, {Manara}, {Natta}, {Frasca},
  {Testi}, {Nisini}, {Stelzer}, {Williams}, {Antoniucci}, {Biazzo}, {Covino},
  {Esposito}, {Getman}, \& {Rigliaco}}]{alcala17}
{Alcal{\'a}}, J.~M., {Manara}, C.~F., {Natta}, A., {et~al.} 2017, \aap, 600,
  A20

\bibitem[{{Alcal{\'a}} {et~al.}(2014){Alcal{\'a}}, {Natta}, {Manara}, {Spezzi},
  {Stelzer}, {Frasca}, {Biazzo}, {Covino}, {Randich}, {Rigliaco}, {Testi},
  {Comer{\'o}n}, {Cupani}, \& {D'Elia}}]{alcala14}
{Alcal{\'a}}, J.~M., {Natta}, A., {Manara}, C.~F., {et~al.} 2014, \aap, 561, A2

\bibitem[{{Alexander} {et~al.}(2014){Alexander}, {Pascucci}, {Andrews},
  {Armitage}, \& {Cieza}}]{alexander14}
{Alexander}, R., {Pascucci}, I., {Andrews}, S., {Armitage}, P., \& {Cieza}, L.
  2014, Protostars and Planets VI, 475

\bibitem[{{Alexander} {et~al.}(2006){Alexander}, {Clarke}, \&
  {Pringle}}]{alexander06}
{Alexander}, R.~D., {Clarke}, C.~J., \& {Pringle}, J.~E. 2006, \mnras, 369, 229

\bibitem[{{Allard} {et~al.}(2012){Allard}, {Homeier}, \&
  {Freytag}}]{Allard2012}
{Allard}, F., {Homeier}, D., \& {Freytag}, B. 2012, Philosophical Transactions
  of the Royal Society of London Series A, 370, 2765

\bibitem[{{Ansdell} {et~al.}(2017){Ansdell}, {Williams}, {Manara}, {Miotello},
  {Facchini}, {van der Marel}, {Testi}, \& {van Dishoeck}}]{ansdell17}
{Ansdell}, M., {Williams}, J.~P., {Manara}, C.~F., {et~al.} 2017, \aj, 153, 240

\bibitem[{{Antoniucci} {et~al.}(2014){Antoniucci}, {Garc{\'{\i}}a L{\'o}pez},
  {Nisini}, {Caratti o Garatti}, {Giannini}, \& {Lorenzetti}}]{antoniucci14}
{Antoniucci}, S., {Garc{\'{\i}}a L{\'o}pez}, R., {Nisini}, B., {et~al.} 2014,
  \aap, 572, A62

\bibitem[{{Arulanantham} {et~al.}(2020){Arulanantham}, {France}, {Cazzoletti},
  {Miotello}, {Manara}, {Schneider}, {Hoadley}, {van Dishoeck}, \&
  {G{\"u}nther}}]{arulanantham20}
{Arulanantham}, N., {France}, K., {Cazzoletti}, P., {et~al.} 2020, \aj, 159,
  168

\bibitem[{{Arulanantham} {et~al.}(2018){Arulanantham}, {France}, {Hoadley},
  {Manara}, {Schneider}, {Alcal{\'a}}, {Banzatti}, {G{\"u}nther}, {Miotello},
  {van der Marel}, {van Dishoeck}, {Walsh}, \& {Williams}}]{arulanantham18}
{Arulanantham}, N., {France}, K., {Hoadley}, K., {et~al.} 2018, \apj, 855, 98

\bibitem[{{Bai}(2016)}]{bai16}
{Bai}, X.-N. 2016, \apj, 821, 80

\bibitem[{Bailer-Jones(2015)}]{bailer-jones2015}
Bailer-Jones, C. A.~L. 2015, Publications of the Astronomical Society of the
  Pacific, 127, 994

\bibitem[{{Bailer-Jones} {et~al.}(2021){Bailer-Jones}, {Rybizki}, {Fouesneau},
  {Demleitner}, \& {Andrae}}]{bailer-jones2020}
{Bailer-Jones}, C.~A.~L., {Rybizki}, J., {Fouesneau}, M., {Demleitner}, M., \&
  {Andrae}, R. 2021, \aj, 161, 147

\bibitem[{{Ballabio} {et~al.}(2020){Ballabio}, {Alexander}, \&
  {Clarke}}]{ballabio20}
{Ballabio}, G., {Alexander}, R.~D., \& {Clarke}, C.~J. 2020, \mnras, 496, 2932

\bibitem[{{Ballester} {et~al.}(2000){Ballester}, {Modigliani}, {Boitquin},
  {Cristiani}, {Hanuschik}, {Kaufer}, \& {Wolf}}]{uvespipe}
{Ballester}, P., {Modigliani}, A., {Boitquin}, O., {et~al.} 2000, The
  Messenger, 101, 31

\bibitem[{{Banzatti} {et~al.}(2019){Banzatti}, {Pascucci}, {Edwards}, {Fang},
  {Gorti}, \& {Flock}}]{banzatti19}
{Banzatti}, A., {Pascucci}, I., {Edwards}, S., {et~al.} 2019, \apj, 870, 76

\bibitem[{{Baraffe} {et~al.}(2015){Baraffe}, {Homeier}, {Allard}, \&
  {Chabrier}}]{B15}
{Baraffe}, I., {Homeier}, D., {Allard}, F., \& {Chabrier}, G. 2015, \aap, 577,
  A42

\bibitem[{{Basri} \& {Batalha}(1990)}]{basri90}
{Basri}, G. \& {Batalha}, C. 1990, \apj, 363, 654

\bibitem[{{Biazzo} {et~al.}(2012){Biazzo}, {Alcal{\'a}}, {Covino}, {Frasca},
  {Getman}, \& {Spezzi}}]{biazzo12}
{Biazzo}, K., {Alcal{\'a}}, J.~M., {Covino}, E., {et~al.} 2012, \aap, 547, A104

\bibitem[{{Blaauw}(1964)}]{blaauw64}
{Blaauw}, A. 1964, \araa, 2, 213

\bibitem[{{Bonito} {et~al.}(2020){Bonito}, {Prisinzano}, {Venuti}, {Damiani},
  {Micela}, {Sacco}, {Traven}, {Biazzo}, {Sbordone}, {Masseron}, {Zwitter},
  {Gonneau}, {Bayo}, {Roccatagliata}, {Randich}, {Vink}, {Jofre}, {Flaccomio},
  {Magrini}, {Carraro}, {Morbidelli}, {Frasca}, {Monaco}, {Rigliaco}, {Worley},
  {Hourihane}, {Gilmore}, {Franciosini}, {Lewis}, \& {Koposov}}]{bonito2020}
{Bonito}, R., {Prisinzano}, L., {Venuti}, L., {et~al.} 2020, \aap, 642, A56

\bibitem[{{Bouvier} {et~al.}(2007){Bouvier}, {Alencar}, {Harries},
  {Johns-Krull}, \& {Romanova}}]{bouvier07}
{Bouvier}, J., {Alencar}, S.~H.~P., {Harries}, T.~J., {Johns-Krull}, C.~M., \&
  {Romanova}, M.~M. 2007, in Protostars and Planets V, ed. B.~{Reipurth},
  D.~{Jewitt}, \& K.~{Keil}, 479

\bibitem[{{Brewer} {et~al.}(2016){Brewer}, {Fischer}, {Valenti}, \&
  {Piskunov}}]{Brewer2016}
{Brewer}, J.~M., {Fischer}, D.~A., {Valenti}, J.~A., \& {Piskunov}, N. 2016,
  \apjs, 225, 32

\bibitem[{{Brice{\~n}o} {et~al.}(2005){Brice{\~n}o}, {Calvet}, {Hern{\'a}ndez},
  {Vivas}, {Hartmann}, {Downes}, \& {Berlind}}]{briceno05}
{Brice{\~n}o}, C., {Calvet}, N., {Hern{\'a}ndez}, J., {et~al.} 2005, \aj, 129,
  907

\bibitem[{{Brice{\~n}o} {et~al.}(2019){Brice{\~n}o}, {Calvet}, {Hern{\'a}ndez},
  {Vivas}, {Mateu}, {Downes}, {Loerincs}, {P{\'e}rez-Blanco}, {Berlind},
  {Espaillat}, {Allen}, {Hartmann}, {Mateo}, \& {Bailey}}]{briceno19}
{Brice{\~n}o}, C., {Calvet}, N., {Hern{\'a}ndez}, J., {et~al.} 2019, \aj, 157,
  85

\bibitem[{{Brice{\~n}o} {et~al.}(2001){Brice{\~n}o}, Vivas, Calvet, Hartmann,
  Pacheco, Herrera, Romero, Berlind, S{\'a}nchez, Snyder, \&
  Andrews}]{briceno01}
{Brice{\~n}o}, C., Vivas, A.~K., Calvet, N., {et~al.} 2001, Science, 291, 93

\bibitem[{{Caballero}(2014)}]{caballero2014}
{Caballero}, J.~A. 2014, The Observatory, 134, 273

\bibitem[{{Caballero}(2018)}]{caballero2018}
{Caballero}, J.~A. 2018, Research Notes of the American Astronomical Society,
  2, 25

\bibitem[{{Caballero} {et~al.}(2019){Caballero}, {de Burgos},
  {Alonso-Floriano}, {Cabrera-Lavers}, {Garc{\'\i}a-{\'A}lvarez}, \&
  {Montes}}]{caballero2019}
{Caballero}, J.~A., {de Burgos}, A., {Alonso-Floriano}, F.~J., {et~al.} 2019,
  \aap, 629, A114

\bibitem[{{Calvet} {et~al.}(2004){Calvet}, {Muzerolle}, {Brice{\~n}o},
  {Hern{\'a}ndez}, {Hartmann}, {Saucedo}, \& {Gordon}}]{calvet04}
{Calvet}, N., {Muzerolle}, J., {Brice{\~n}o}, C., {et~al.} 2004, \aj, 128, 1294

\bibitem[{{Cardelli} {et~al.}(1989){Cardelli}, {Clayton}, \&
  {Mathis}}]{cardelli98}
{Cardelli}, J.~A., {Clayton}, G.~C., \& {Mathis}, J.~S. 1989, \apj, 345, 245

\bibitem[{{Chambers} {et~al.}(2016){Chambers}, {Magnier}, {Metcalfe},
  {Flewelling}, {Huber}, {Waters}, {Denneau}, {Draper}, {Farrow}, {Finkbeiner},
  {Holmberg}, {Koppenhoefer}, {Price}, {Rest}, {Saglia}, {Schlafly}, {Smartt},
  {Sweeney}, {Wainscoat}, {Burgett}, {Chastel}, {Grav}, {Heasley}, {Hodapp},
  {Jedicke}, {Kaiser}, {Kudritzki}, {Luppino}, {Lupton}, {Monet}, {Morgan},
  {Onaka}, {Shiao}, {Stubbs}, {Tonry}, {White}, {Ba{\~n}ados}, {Bell},
  {Bender}, {Bernard}, {Boegner}, {Boffi}, {Botticella}, {Calamida},
  {Casertano}, {Chen}, {Chen}, {Cole}, {Deacon}, {Frenk}, {Fitzsimmons},
  {Gezari}, {Gibbs}, {Goessl}, {Goggia}, {Gourgue}, {Goldman}, {Grant},
  {Grebel}, {Hambly}, {Hasinger}, {Heavens}, {Heckman}, {Henderson}, {Henning},
  {Holman}, {Hopp}, {Ip}, {Isani}, {Jackson}, {Keyes}, {Koekemoer}, {Kotak},
  {Le}, {Liska}, {Long}, {Lucey}, {Liu}, {Martin}, {Masci}, {McLean}, {Mindel},
  {Misra}, {Morganson}, {Murphy}, {Obaika}, {Narayan}, {Nieto-Santisteban},
  {Norberg}, {Peacock}, {Pier}, {Postman}, {Primak}, {Rae}, {Rai}, {Riess},
  {Riffeser}, {Rix}, {R{\"o}ser}, {Russel}, {Rutz}, {Schilbach}, {Schultz},
  {Scolnic}, {Strolger}, {Szalay}, {Seitz}, {Small}, {Smith}, {Soderblom},
  {Taylor}, {Thomson}, {Taylor}, {Thakar}, {Thiel}, {Thilker}, {Unger},
  {Urata}, {Valenti}, {Wagner}, {Walder}, {Walter}, {Watters}, {Werner},
  {Wood-Vasey}, \& {Wyse}}]{Pan-STARRS}
{Chambers}, K.~C., {Magnier}, E.~A., {Metcalfe}, N., {et~al.} 2016, arXiv
  e-prints, arXiv:1612.05560

\bibitem[{{Clarke} \& {Pringle}(2006)}]{clarke06}
{Clarke}, C.~J. \& {Pringle}, J.~E. 2006, \mnras, 370, L10

\bibitem[{{Cody} \& {Hillenbrand}(2010)}]{cody10}
{Cody}, A.~M. \& {Hillenbrand}, L.~A. 2010, \apjs, 191, 389

\bibitem[{{Cody} {et~al.}(2017){Cody}, {Hillenbrand}, {David}, {Carpenter},
  {Everett}, \& {Howell}}]{cody17}
{Cody}, A.~M., {Hillenbrand}, L.~A., {David}, T.~J., {et~al.} 2017, \apj, 836,
  41

\bibitem[{{Cody} {et~al.}(2014){Cody}, {Stauffer}, {Baglin}, {Micela},
  {Rebull}, {Flaccomio}, {Morales-Calder{\'o}n}, {Aigrain}, {Bouvier},
  {Hillenbrand}, {Gutermuth}, {Song}, {Turner}, {Alencar}, {Zwintz},
  {Plavchan}, {Carpenter}, {Findeisen}, {Carey}, {Terebey}, {Hartmann},
  {Calvet}, {Teixeira}, {Vrba}, {Wolk}, {Covey}, {Poppenhaeger}, {G{\"u}nther},
  {Forbrich}, {Whitney}, {Affer}, {Herbst}, {Hora}, {Barrado}, {Holtzman},
  {Marchis}, {Wood}, {Medeiros Guimar{\~a}es}, {Lillo Box}, {Gillen},
  {McQuillan}, {Espaillat}, {Allen}, {D'Alessio}, \& {Favata}}]{cody14}
{Cody}, A.~M., {Stauffer}, J., {Baglin}, A., {et~al.} 2014, \aj, 147, 82

\bibitem[{{Costigan} {et~al.}(2012){Costigan}, {Scholz}, {Stelzer}, {Ray},
  {Vink}, \& {Mohanty}}]{costigan12}
{Costigan}, G., {Scholz}, A., {Stelzer}, B., {et~al.} 2012, \mnras, 427, 1344

\bibitem[{{Costigan} {et~al.}(2014){Costigan}, {Vink}, {Scholz}, {Ray}, \&
  {Testi}}]{costigan14}
{Costigan}, G., {Vink}, J.~S., {Scholz}, A., {Ray}, T., \& {Testi}, L. 2014,
  \mnras, 440, 3444

\bibitem[{{Dekker} {et~al.}(2000){Dekker}, {D'Odorico}, {Kaufer}, {Delabre}, \&
  {Kotzlowski}}]{dekker00}
{Dekker}, H., {D'Odorico}, S., {Kaufer}, A., {Delabre}, B., \& {Kotzlowski}, H.
  2000, in Society of Photo-Optical Instrumentation Engineers (SPIE) Conference
  Series, Vol. 4008, Optical and IR Telescope Instrumentation and Detectors,
  ed. M.~{Iye} \& A.~F. {Moorwood}, 534--545

\bibitem[{{DeSouza} \& {Basu}(2017)}]{desouza16}
{DeSouza}, A.~L. \& {Basu}, S. 2017, \na, 51, 113

\bibitem[{{Dullemond} {et~al.}(2006){Dullemond}, {Natta}, \&
  {Testi}}]{dullemond06}
{Dullemond}, C.~P., {Natta}, A., \& {Testi}, L. 2006, \apjl, 645, L69

\bibitem[{{Ercolano} {et~al.}(2014){Ercolano}, {Mayr}, {Owen}, {Rosotti}, \&
  {Manara}}]{ercolano14}
{Ercolano}, B., {Mayr}, D., {Owen}, J.~E., {Rosotti}, G., \& {Manara}, C.~F.
  2014, \mnras, 439, 256

\bibitem[{{Ercolano} \& {Owen}(2016)}]{ercolano16}
{Ercolano}, B. \& {Owen}, J.~E. 2016, \mnras, 460, 3472

\bibitem[{{Ercolano} \& {Pascucci}(2017)}]{EP17}
{Ercolano}, B. \& {Pascucci}, I. 2017, Royal Society Open Science, 4, 170114

\bibitem[{{Fang} {et~al.}(2013){Fang}, {Kim}, {van Boekel}, {Sicilia-Aguilar},
  {Henning}, \& {Flaherty}}]{Fang2013}
{Fang}, M., {Kim}, J.~S., {van Boekel}, R., {et~al.} 2013, \apjs, 207, 5

\bibitem[{{Fang} {et~al.}(2018){Fang}, {Pascucci}, {Edwards}, {Gorti},
  {Banzatti}, {Flock}, {Hartigan}, {Herczeg}, \& {Dupree}}]{fang18}
{Fang}, M., {Pascucci}, I., {Edwards}, S., {et~al.} 2018, \apj, 868, 28

\bibitem[{{Ferreira} {et~al.}(2006){Ferreira}, {Dougados}, \&
  {Cabrit}}]{ferreira06}
{Ferreira}, J., {Dougados}, C., \& {Cabrit}, S. 2006, \aap, 453, 785

\bibitem[{{Fischer} {et~al.}(2017){Fischer}, {Megeath}, {Furlan}, {Ali},
  {Stutz}, {Tobin}, {Osorio}, {Stanke}, {Manoj}, {Poteet}, {Booker},
  {Hartmann}, {Wilson}, {Myers}, \& {Watson}}]{fischer17}
{Fischer}, W.~J., {Megeath}, S.~T., {Furlan}, E., {et~al.} 2017, \apj, 840, 69

\bibitem[{{Frank} {et~al.}(2014){Frank}, {Ray}, {Cabrit}, {Hartigan}, {Arce},
  {Bacciotti}, {Bally}, {Benisty}, {Eisl{\"o}ffel}, {G{\"u}del}, {Lebedev},
  {Nisini}, \& {Raga}}]{frank14}
{Frank}, A., {Ray}, T.~P., {Cabrit}, S., {et~al.} 2014, Protostars and Planets
  VI, 451

\bibitem[{{Frasca} {et~al.}(2017){Frasca}, {Biazzo}, {Alcal{\'a}}, {Manara},
  {Stelzer}, {Covino}, \& {Antoniucci}}]{frasca2017}
{Frasca}, A., {Biazzo}, K., {Alcal{\'a}}, J.~M., {et~al.} 2017, \aap, 602, A33

\bibitem[{{Frasca} {et~al.}(2015){Frasca}, {Biazzo}, {Lanzafame}, {Alcal{\'a}},
  {Brugaletta}, {Klutsch}, {Stelzer}, {Sacco}, {Spina}, {Jeffries}, {Montes},
  {Alfaro}, {Barentsen}, {Bonito}, {Gameiro}, {L{\'o}pez-Santiago}, {Pace},
  {Pasquini}, {Prisinzano}, {Sousa}, {Gilmore}, {Randich}, {Micela},
  {Bragaglia}, {Flaccomio}, {Bayo}, {Costado}, {Franciosini}, {Hill},
  {Hourihane}, {Jofr{\'e}}, {Lardo}, {Maiorca}, {Masseron}, {Morbidelli}, \&
  {Worley}}]{Frasca2015}
{Frasca}, A., {Biazzo}, K., {Lanzafame}, A.~C., {et~al.} 2015, \aap, 575, A4

\bibitem[{{Frasca} {et~al.}(2000){Frasca}, {Freire Ferrero}, {Marilli}, \&
  {Catalano}}]{Frasca2000}
{Frasca}, A., {Freire Ferrero}, R., {Marilli}, E., \& {Catalano}, S. 2000,
  \aap, 364, 179

\bibitem[{{Frasca} {et~al.}(2018){Frasca}, {Montes}, {Alcal{\`a}}, {Klutsch},
  \& {Guillout}}]{frasca2018}
{Frasca}, A., {Montes}, D., {Alcal{\`a}}, J.~M., {Klutsch}, A., \& {Guillout},
  P. 2018, \actaa, 68, 403

\bibitem[{{Freudling} {et~al.}(2013){Freudling}, {Romaniello}, {Bramich},
  {Ballester}, {Forchi}, {Garc{\'{\i}}a-Dabl{\'o}}, {Moehler}, \&
  {Neeser}}]{reflex}
{Freudling}, W., {Romaniello}, M., {Bramich}, D.~M., {et~al.} 2013, \aap, 559,
  A96

\bibitem[{{Gagn{\'e}} {et~al.}(2016){Gagn{\'e}}, {Plavchan}, {Gao},
  {Anglada-Escude}, {Furlan}, {Davison}, {Tanner}, {Henry}, {Riedel},
  {Brinkworth}, {Latham}, {Bottom}, {White}, {Mills}, {Beichman}, {Johnson},
  {Ciardi}, {Wallace}, {Mennesson}, {von Braun}, {Vasisht}, {Prato}, {Kane},
  {Mamajek}, {Walp}, {Crawford}, {Rougeot}, {Geneser}, \&
  {Catanzarite}}]{Gagne2016}
{Gagn{\'e}}, J., {Plavchan}, P., {Gao}, P., {et~al.} 2016, \apj, 822, 40

\bibitem[{{Gaia Collaboration} {et~al.}(2020{\natexlab{a}}){Gaia
  Collaboration}, {Brown}, {Vallenari}, {Prusti}, {de Bruijne}, {Babusiaux}, \&
  {Biermann}}]{gaiaEDR3}
{Gaia Collaboration}, {Brown}, A.~G.~A., {Vallenari}, A., {et~al.}
  2020{\natexlab{a}}, arXiv e-prints, arXiv:2012.01533

\bibitem[{{Gaia Collaboration} {et~al.}(2020{\natexlab{b}}){Gaia
  Collaboration}, {Brown}, {Vallenari}, {Prusti}, {de Bruijne}, {Babusiaux}, \&
  {Biermann}}]{2020arXiv201201533G}
{Gaia Collaboration}, {Brown}, A.~G.~A., {Vallenari}, A., {et~al.}
  2020{\natexlab{b}}, arXiv e-prints, arXiv:2012.01533

\bibitem[{{Gaia Collaboration} {et~al.}(2016){Gaia Collaboration}, {Prusti},
  {de Bruijne}, {Brown}, {Vallenari}, {Babusiaux}, {Bailer-Jones}, {Bastian},
  {Biermann}, {Evans}, {Eyer}, {Jansen}, {Jordi}, {Klioner}, {Lammers},
  {Lindegren}, {Luri}, {Mignard}, {Milligan}, {Panem}, {Poinsignon},
  {Pourbaix}, {Randich}, {Sarri}, {Sartoretti}, {Siddiqui}, {Soubiran},
  {Valette}, {van Leeuwen}, {Walton}, {Aerts}, {Arenou}, {Cropper}, {Drimmel},
  {H{\o}g}, {Katz}, {Lattanzi}, {O'Mullane}, {Grebel}, {Holland}, {Huc},
  {Passot}, {Bramante}, {Cacciari}, {Casta{\~n}eda}, {Chaoul}, {Cheek}, {De
  Angeli}, {Fabricius}, {Guerra}, {Hern{\'a}ndez}, {Jean-Antoine-Piccolo},
  {Masana}, {Messineo}, {Mowlavi}, {Nienartowicz}, {Ord{\'o}{\~n}ez-Blanco},
  {Panuzzo}, {Portell}, {Richards}, {Riello}, {Seabroke}, {Tanga},
  {Th{\'e}venin}, {Torra}, {Els}, {Gracia-Abril}, {Comoretto},
  {Garcia-Reinaldos}, {Lock}, {Mercier}, {Altmann}, {Andrae}, {Astraatmadja},
  {Bellas-Velidis}, {Benson}, {Berthier}, {Blomme}, {Busso}, {Carry},
  {Cellino}, {Clementini}, {Cowell}, {Creevey}, {Cuypers}, {Davidson}, {De
  Ridder}, {de Torres}, {Delchambre}, {Dell'Oro}, {Ducourant}, {Fr{\'e}mat},
  {Garc{\'\i}a-Torres}, {Gosset}, {Halbwachs}, {Hambly}, {Harrison}, {Hauser},
  {Hestroffer}, {Hodgkin}, {Huckle}, {Hutton}, {Jasniewicz}, {Jordan},
  {Kontizas}, {Korn}, {Lanzafame}, {Manteiga}, {Moitinho}, {Muinonen},
  {Osinde}, {Pancino}, {Pauwels}, {Petit}, {Recio-Blanco}, {Robin}, {Sarro},
  {Siopis}, {Smith}, {Smith}, {Sozzetti}, {Thuillot}, {van Reeven}, {Viala},
  {Abbas}, {Abreu Aramburu}, {Accart}, {Aguado}, {Allan}, {Allasia},
  {Altavilla}, {{\'A}lvarez}, {Alves}, {Anderson}, {Andrei}, {Anglada Varela},
  {Antiche}, {Antoja}, {Ant{\'o}n}, {Arcay}, {Atzei}, {Ayache}, {Bach},
  {Baker}, {Balaguer-N{\'u}{\~n}ez}, {Barache}, {Barata}, {Barbier}, {Barblan},
  {Baroni}, {Barrado y Navascu{\'e}s}, {Barros}, {Barstow}, {Becciani},
  {Bellazzini}, {Bellei}, {Bello Garc{\'\i}a}, {Belokurov}, {Bendjoya},
  {Berihuete}, {Bianchi}, {Bienaym{\'e}}, {Billebaud}, {Blagorodnova},
  {Blanco-Cuaresma}, {Boch}, {Bombrun}, {Borrachero}, {Bouquillon}, {Bourda},
  {Bouy}, {Bragaglia}, {Breddels}, {Brouillet}, {Br{\"u}semeister},
  {Bucciarelli}, {Budnik}, {Burgess}, {Burgon}, {Burlacu}, {Busonero}, {Buzzi},
  {Caffau}, {Cambras}, {Campbell}, {Cancelliere}, {Cantat-Gaudin}, {Carlucci},
  {Carrasco}, {Castellani}, {Charlot}, {Charnas}, {Charvet}, {Chassat},
  {Chiavassa}, {Clotet}, {Cocozza}, {Collins}, {Collins}, {Costigan}, {Crifo},
  {Cross}, {Crosta}, {Crowley}, {Dafonte}, {Damerdji}, {Dapergolas}, {David},
  {David}, {De Cat}, {de Felice}, {de Laverny}, {De Luise}, {De March}, {de
  Martino}, {de Souza}, {Debosscher}, {del Pozo}, {Delbo}, {Delgado},
  {Delgado}, {di Marco}, {Di Matteo}, {Diakite}, {Distefano}, {Dolding}, {Dos
  Anjos}, {Drazinos}, {Dur{\'a}n}, {Dzigan}, {Ecale}, {Edvardsson}, {Enke},
  {Erdmann}, {Escolar}, {Espina}, {Evans}, {Eynard Bontemps}, {Fabre},
  {Fabrizio}, {Faigler}, {Falc{\~a}o}, {Farr{\`a}s Casas}, {Faye}, {Federici},
  {Fedorets}, {Fern{\'a}ndez-Hern{\'a}ndez}, {Fernique}, {Fienga}, {Figueras},
  {Filippi}, {Findeisen}, {Fonti}, {Fouesneau}, {Fraile}, {Fraser}, {Fuchs},
  {Furnell}, {Gai}, {Galleti}, {Galluccio}, {Garabato}, {Garc{\'\i}a-Sedano},
  {Gar{\'e}}, {Garofalo}, {Garralda}, {Gavras}, {Gerssen}, {Geyer}, {Gilmore},
  {Girona}, {Giuffrida}, {Gomes}, {Gonz{\'a}lez-Marcos},
  {Gonz{\'a}lez-N{\'u}{\~n}ez}, {Gonz{\'a}lez-Vidal}, {Granvik}, {Guerrier},
  {Guillout}, {Guiraud}, {G{\'u}rpide}, {Guti{\'e}rrez-S{\'a}nchez}, {Guy},
  {Haigron}, {Hatzidimitriou}, {Haywood}, {Heiter}, {Helmi}, {Hobbs},
  {Hofmann}, {Holl}, {Holland}, {Hunt}, {Hypki}, {Icardi}, {Irwin}, {Jevardat
  de Fombelle}, {Jofr{\'e}}, {Jonker}, {Jorissen}, {Julbe}, {Karampelas},
  {Kochoska}, {Kohley}, {Kolenberg}, {Kontizas}, {Koposov}, {Kordopatis},
  {Koubsky}, {Kowalczyk}, {Krone-Martins}, {Kudryashova}, {Kull}, {Bachchan},
  {Lacoste-Seris}, {Lanza}, {Lavigne}, {Le Poncin-Lafitte}, {Lebreton},
  {Lebzelter}, {Leccia}, {Leclerc}, {Lecoeur-Taibi}, {Lemaitre}, {Lenhardt},
  {Leroux}, {Liao}, {Licata}, {Lindstr{\o}m}, {Lister}, {Livanou}, {Lobel},
  {L{\"o}ffler}, {L{\'o}pez}, {Lopez-Lozano}, {Lorenz}, {Loureiro},
  {MacDonald}, {Magalh{\~a}es Fernandes}, {Managau}, {Mann}, {Mantelet},
  {Marchal}, {Marchant}, {Marconi}, {Marie}, {Marinoni}, {Marrese},
  {Marschalk{\'o}}, {Marshall}, {Mart{\'\i}n-Fleitas}, {Martino}, {Mary},
  {Matijevi{\v{c}}}, {Mazeh}, {McMillan}, {Messina}, {Mestre}, {Michalik},
  {Millar}, {Miranda}, {Molina}, {Molinaro}, {Molinaro}, {Moln{\'a}r},
  {Moniez}, {Montegriffo}, {Monteiro}, {Mor}, {Mora}, {Morbidelli}, {Morel},
  {Morgenthaler}, {Morley}, {Morris}, {Mulone}, {Muraveva}, {Musella},
  {Narbonne}, {Nelemans}, {Nicastro}, {Noval}, {Ord{\'e}novic},
  {Ordieres-Mer{\'e}}, {Osborne}, {Pagani}, {Pagano}, {Pailler}, {Palacin},
  {Palaversa}, {Parsons}, {Paulsen}, {Pecoraro}, {Pedrosa}, {Pentik{\"a}inen},
  {Pereira}, {Pichon}, {Piersimoni}, {Pineau}, {Plachy}, {Plum}, {Poujoulet},
  {Pr{\v{s}}a}, {Pulone}, {Ragaini}, {Rago}, {Rambaux}, {Ramos-Lerate},
  {Ranalli}, {Rauw}, {Read}, {Regibo}, {Renk}, {Reyl{\'e}}, {Ribeiro},
  {Rimoldini}, {Ripepi}, {Riva}, {Rixon}, {Roelens}, {Romero-G{\'o}mez},
  {Rowell}, {Royer}, {Rudolph}, {Ruiz-Dern}, {Sadowski}, {Sagrist{\`a}
  Sell{\'e}s}, {Sahlmann}, {Salgado}, {Salguero}, {Sarasso}, {Savietto},
  {Schnorhk}, {Schultheis}, {Sciacca}, {Segol}, {Segovia}, {Segransan},
  {Serpell}, {Shih}, {Smareglia}, {Smart}, {Smith}, {Solano}, {Solitro},
  {Sordo}, {Soria Nieto}, {Souchay}, {Spagna}, {Spoto}, {Stampa}, {Steele},
  {Steidelm{\"u}ller}, {Stephenson}, {Stoev}, {Suess}, {S{\"u}veges}, {Surdej},
  {Szabados}, {Szegedi-Elek}, {Tapiador}, {Taris}, {Tauran}, {Taylor},
  {Teixeira}, {Terrett}, {Tingley}, {Trager}, {Turon}, {Ulla}, {Utrilla},
  {Valentini}, {van Elteren}, {Van Hemelryck}, {van Leeuwen}, {Varadi},
  {Vecchiato}, {Veljanoski}, {Via}, {Vicente}, {Vogt}, {Voss}, {Votruba},
  {Voutsinas}, {Walmsley}, {Weiler}, {Weingrill}, {Werner}, {Wevers},
  {Whitehead}, {Wyrzykowski}, {Yoldas}, {{\v{Z}}erjal}, {Zucker}, {Zurbach},
  {Zwitter}, {Alecu}, {Allen}, {Allende Prieto}, {Amorim},
  {Anglada-Escud{\'e}}, {Arsenijevic}, {Azaz}, {Balm}, {Beck}, {Bernstein},
  {Bigot}, {Bijaoui}, {Blasco}, {Bonfigli}, {Bono}, {Boudreault}, {Bressan},
  {Brown}, {Brunet}, {Bunclark}, {Buonanno}, {Butkevich}, {Carret}, {Carrion},
  {Chemin}, {Ch{\'e}reau}, {Corcione}, {Darmigny}, {de Boer}, {de Teodoro}, {de
  Zeeuw}, {Delle Luche}, {Domingues}, {Dubath}, {Fodor}, {Fr{\'e}zouls},
  {Fries}, {Fustes}, {Fyfe}, {Gallardo}, {Gallegos}, {Gardiol}, {Gebran},
  {Gomboc}, {G{\'o}mez}, {Grux}, {Gueguen}, {Heyrovsky}, {Hoar}, {Iannicola},
  {Isasi Parache}, {Janotto}, {Joliet}, {Jonckheere}, {Keil}, {Kim},
  {Klagyivik}, {Klar}, {Knude}, {Kochukhov}, {Kolka}, {Kos}, {Kutka}, {Lainey},
  {LeBouquin}, {Liu}, {Loreggia}, {Makarov}, {Marseille}, {Martayan},
  {Martinez-Rubi}, {Massart}, {Meynadier}, {Mignot}, {Munari}, {Nguyen},
  {Nordlander}, {Ocvirk}, {O'Flaherty}, {Olias Sanz}, {Ortiz}, {Osorio},
  {Oszkiewicz}, {Ouzounis}, {Palmer}, {Park}, {Pasquato}, {Peltzer}, {Peralta},
  {P{\'e}turaud}, {Pieniluoma}, {Pigozzi}, {Poels}, {Prat}, {Prod'homme},
  {Raison}, {Rebordao}, {Risquez}, {Rocca-Volmerange}, {Rosen}, {Ruiz-Fuertes},
  {Russo}, {Sembay}, {Serraller Vizcaino}, {Short}, {Siebert}, {Silva},
  {Sinachopoulos}, {Slezak}, {Soffel}, {Sosnowska}, {Strai{\v{z}}ys}, {ter
  Linden}, {Terrell}, {Theil}, {Tiede}, {Troisi}, {Tsalmantza}, {Tur},
  {Vaccari}, {Vachier}, {Valles}, {Van Hamme}, {Veltz}, {Virtanen}, {Wallut},
  {Wichmann}, {Wilkinson}, {Ziaeepour}, \& {Zschocke}}]{gaia}
{Gaia Collaboration}, {Prusti}, T., {de Bruijne}, J.~H.~J., {et~al.} 2016,
  \aap, 595, A1

\bibitem[{{Garc{\'\i}a P{\'e}rez} {et~al.}(2016){Garc{\'\i}a P{\'e}rez},
  {Allende Prieto}, {Holtzman}, {Shetrone}, {M{\'e}sz{\'a}ros}, {Bizyaev},
  {Carrera}, {Cunha}, {Garc{\'\i}a-Hern{\'a}ndez}, {Johnson}, {Majewski},
  {Nidever}, {Schiavon}, {Shane}, {Smith}, {Sobeck}, {Troup}, {Zamora},
  {Weinberg}, {Bovy}, {Eisenstein}, {Feuillet}, {Frinchaboy}, {Hayden},
  {Hearty}, {Nguyen}, {O'Connell}, {Pinsonneault}, {Wilson}, \&
  {Zasowski}}]{Garcia2016}
{Garc{\'\i}a P{\'e}rez}, A.~E., {Allende Prieto}, C., {Holtzman}, J.~A.,
  {et~al.} 2016, \aj, 151, 144

\bibitem[{{Giannini} {et~al.}(2019){Giannini}, {Nisini}, {Antoniucci},
  {Biazzo}, {Alcal{\'a}}, {Bacciotti}, {Fedele}, {Frasca}, {Harutyunyan},
  {Munari}, {Rigliaco}, \& {Vitali}}]{giannini19}
{Giannini}, T., {Nisini}, B., {Antoniucci}, S., {et~al.} 2019, \aap, 631, A44

\bibitem[{{Glebocki} \& {Gnacinski}(2005)}]{Glebocki2005}
{Glebocki}, R. \& {Gnacinski}, P. 2005, VizieR Online Data Catalog, III/244

\bibitem[{{Grandjean} {et~al.}(2020){Grandjean}, {Lagrange}, {Keppler},
  {Meunier}, {Mignon}, {Borgniet}, {Chauvin}, {Desidera}, {Galland}, {Messina},
  {Sterzik}, {Pantoja}, {Rodet}, \& {Zicher}}]{harps_templates}
{Grandjean}, A., {Lagrange}, A.~M., {Keppler}, M., {et~al.} 2020, \aap, 633,
  A44

\bibitem[{{Gregory} {et~al.}(2006){Gregory}, {Jardine}, {Simpson}, \&
  {Donati}}]{gregory06}
{Gregory}, S.~G., {Jardine}, M., {Simpson}, I., \& {Donati}, J.~F. 2006,
  \mnras, 371, 999

\bibitem[{{Hartigan} {et~al.}(1995){Hartigan}, {Edwards}, \&
  {Ghandour}}]{hartigan95}
{Hartigan}, P., {Edwards}, S., \& {Ghandour}, L. 1995, \apj, 452, 736

\bibitem[{{Hartmann} {et~al.}(1998){Hartmann}, {Calvet}, {Gullbring}, \&
  {D'Alessio}}]{hartmann98}
{Hartmann}, L., {Calvet}, N., {Gullbring}, E., \& {D'Alessio}, P. 1998, \apj,
  495, 385

\bibitem[{{Hartmann} {et~al.}(2006){Hartmann}, {D'Alessio}, {Calvet}, \&
  {Muzerolle}}]{hartmann06}
{Hartmann}, L., {D'Alessio}, P., {Calvet}, N., \& {Muzerolle}, J. 2006, \apj,
  648, 484

\bibitem[{{Hartmann} {et~al.}(2016){Hartmann}, {Herczeg}, \&
  {Calvet}}]{hartmann16}
{Hartmann}, L., {Herczeg}, G., \& {Calvet}, N. 2016, \araa, 54, 135

\bibitem[{{Henden} {et~al.}(2018){Henden}, {Levine}, {Terrell}, {Welch},
  {Munari}, \& {Kloppenborg}}]{Henden2018}
{Henden}, A.~A., {Levine}, S., {Terrell}, D., {et~al.} 2018, in American
  Astronomical Society Meeting Abstracts, Vol. 232, American Astronomical
  Society Meeting Abstracts \#232, 223.06

\bibitem[{{Herczeg} \& {Hillenbrand}(2008)}]{HH08}
{Herczeg}, G.~J. \& {Hillenbrand}, L.~A. 2008, \apj, 681, 594

\bibitem[{{Herczeg} \& {Hillenbrand}(2014)}]{HH14}
{Herczeg}, G.~J. \& {Hillenbrand}, L.~A. 2014, \apj, 786, 97

\bibitem[{Hern\'andez {et~al.}(2007)Hern\'andez, Calvet, Brice\~no, Hartmann,
  Vivas, Muzerolle, Downes, Allen, \& Gutermuth}]{hernandez07a}
Hern\'andez, J., Calvet, N., Brice\~no, C., {et~al.} 2007, The Astrophysical
  Journal, 671, 1784

\bibitem[{{Hern{\'a}ndez} {et~al.}(2014{\natexlab{a}}){Hern{\'a}ndez},
  {Calvet}, {Perez}, {Brice{\~n}o}, {Olguin}, {Contreras}, {Hartmann}, {Allen},
  {Espaillat}, \& {Hernan}}]{Hernandez2014}
{Hern{\'a}ndez}, J., {Calvet}, N., {Perez}, A., {et~al.} 2014{\natexlab{a}},
  \apj, 794, 36

\bibitem[{{Hern{\'a}ndez} {et~al.}(2014{\natexlab{b}}){Hern{\'a}ndez},
  {Calvet}, {Perez}, {Brice{\~n}o}, {Olguin}, {Contreras}, {Hartmann}, {Allen},
  {Espaillat}, \& {Hernan}}]{hernandez14}
{Hern{\'a}ndez}, J., {Calvet}, N., {Perez}, A., {et~al.} 2014{\natexlab{b}},
  \apj, 794, 36

\bibitem[{{Hern{\'a}ndez} {et~al.}(2007){Hern{\'a}ndez}, {Hartmann}, {Megeath},
  {Gutermuth}, {Muzerolle}, {Calvet}, {Vivas}, {Brice{\~n}o}, {Allen},
  {Stauffer}, {Young}, \& {Fazio}}]{hernandez07}
{Hern{\'a}ndez}, J., {Hartmann}, L., {Megeath}, T., {et~al.} 2007, \apj, 662,
  1067

\bibitem[{{Hillenbrand} {et~al.}(1992){Hillenbrand}, {Strom}, {Vrba}, \&
  {Keene}}]{hillenbrand92}
{Hillenbrand}, L.~A., {Strom}, S.~E., {Vrba}, F.~J., \& {Keene}, J. 1992, \apj,
  397, 613

\bibitem[{{Hojjatpanah} {et~al.}(2019){Hojjatpanah}, {Figueira}, {Santos},
  {Adibekyan}, {Sousa}, {Delgado-Mena}, {Alibert}, {Cristiani}, {Gonz{\'a}lez
  Hern{\'a}ndez}, {Lanza}, {Di Marcantonio}, {Martins}, {Micela}, {Molaro},
  {Neves}, {Oshagh}, {Pepe}, {Poretti}, {Rojas-Ayala}, {Rebolo}, {Su{\'a}rez
  Mascare{\~n}o}, \& {Zapatero Osorio}}]{Hojjatpanah2019}
{Hojjatpanah}, S., {Figueira}, P., {Santos}, N.~C., {et~al.} 2019, \aap, 629,
  A80

\bibitem[{{Houdebine} {et~al.}(2016){Houdebine}, {Mullan}, {Paletou}, \&
  {Gebran}}]{Houdebine2016}
{Houdebine}, E.~R., {Mullan}, D.~J., {Paletou}, F., \& {Gebran}, M. 2016, \apj,
  822, 97

\bibitem[{{Ingleby} {et~al.}(2009){Ingleby}, {Calvet}, {Bergin}, {Yerasi},
  {Espaillat}, {Herczeg}, {Roueff}, {Abgrall}, {Hern{\'a}ndez}, {Brice{\~n}o},
  {Pascucci}, {Miller}, {Fogel}, {Hartmann}, {Meyer}, {Carpenter}, {Crockett},
  \& {McClure}}]{ingleby09}
{Ingleby}, L., {Calvet}, N., {Bergin}, E., {et~al.} 2009, \apjl, 703, L137

\bibitem[{{Ingleby} {et~al.}(2011){Ingleby}, {Calvet}, {Hern{\'a}ndez},
  {Brice{\~n}o}, {Espaillat}, {Miller}, {Bergin}, \& {Hartmann}}]{ingleby11}
{Ingleby}, L., {Calvet}, N., {Hern{\'a}ndez}, J., {et~al.} 2011, \aj, 141, 127

\bibitem[{{Ingleby} {et~al.}(2014){Ingleby}, {Calvet}, {Hern{\'a}ndez},
  {Hartmann}, {Briceno}, {Miller}, {Espaillat}, \& {McClure}}]{ingleby14}
{Ingleby}, L., {Calvet}, N., {Hern{\'a}ndez}, J., {et~al.} 2014, \apj, 790, 47

\bibitem[{{Jayawardhana} {et~al.}(2006){Jayawardhana}, {Coffey}, {Scholz},
  {Brandeker}, \& {van Kerkwijk}}]{jay06}
{Jayawardhana}, R., {Coffey}, J., {Scholz}, A., {Brandeker}, A., \& {van
  Kerkwijk}, M.~H. 2006, \apj, 648, 1206

\bibitem[{{Jeffers} {et~al.}(2018){Jeffers}, {Sch{\"o}fer}, {Lamert},
  {Reiners}, {Montes}, {Caballero}, {Cort{\'e}s-Contreras}, {Marvin},
  {Passegger}, {Zechmeister}, {Quirrenbach}, {Alonso-Floriano}, {Amado},
  {Bauer}, {Casal}, {Diez Alonso}, {Herrero}, {Morales}, {Mundt}, {Ribas}, \&
  {Sarmiento}}]{Jeffers2018}
{Jeffers}, S.~V., {Sch{\"o}fer}, P., {Lamert}, A., {et~al.} 2018, \aap, 614,
  A76

\bibitem[{{Jeffries} {et~al.}(2006){Jeffries}, {Maxted}, {Oliveira}, \&
  {Naylor}}]{Jeffries2006}
{Jeffries}, R.~D., {Maxted}, P.~F.~L., {Oliveira}, J.~M., \& {Naylor}, T. 2006,
  \mnras, 371, L6

\bibitem[{{Johns} \& {Basri}(1995{\natexlab{a}})}]{JB95a}
{Johns}, C.~M. \& {Basri}, G. 1995{\natexlab{a}}, \aj, 109, 2800

\bibitem[{{Johns} \& {Basri}(1995{\natexlab{b}})}]{JB95b}
{Johns}, C.~M. \& {Basri}, G. 1995{\natexlab{b}}, \apj, 449, 341

\bibitem[{{Jones} {et~al.}(2012){Jones}, {Pringle}, \& {Alexander}}]{jones12}
{Jones}, M.~G., {Pringle}, J.~E., \& {Alexander}, R.~D. 2012, \mnras, 419, 925

\bibitem[{{Kausch} {et~al.}(2015){Kausch}, {Noll}, {Smette}, {Kimeswenger},
  {Barden}, {Szyszka}, {Jones}, {Sana}, {Horst}, \& {Kerber}}]{molecfit2}
{Kausch}, W., {Noll}, S., {Smette}, A., {et~al.} 2015, \aap, 576, A78

\bibitem[{{Kenyon} \& {Hartmann}(1995{\natexlab{a}})}]{kh95}
{Kenyon}, S.~J. \& {Hartmann}, L. 1995{\natexlab{a}}, \apjs, 101, 117

\bibitem[{{Kenyon} \& {Hartmann}(1995{\natexlab{b}})}]{kenyon95}
{Kenyon}, S.~J. \& {Hartmann}, L. 1995{\natexlab{b}}, \apjs, 101, 117

\bibitem[{{K{\'o}sp{\'a}l} {et~al.}(2020){K{\'o}sp{\'a}l}, {Szab{\'o}},
  {{\'A}brah{\'a}m}, {Kraus}, {Takami}, {Lucas}, {Contreras Pe{\~n}a}, \&
  {Udalski}}]{kospal20}
{K{\'o}sp{\'a}l}, {\'A}., {Szab{\'o}}, Z.~M., {{\'A}brah{\'a}m}, P., {et~al.}
  2020, \apj, 889, 148

\bibitem[{{Kounkel} {et~al.}(2018){Kounkel}, {Covey}, {Su{\'a}rez},
  {Rom{\'a}n-Z{\'u}{\~n}iga}, {Hernandez}, {Stassun}, {Jaehnig}, {Feigelson},
  {Pe{\~n}a Ram{\'\i}rez}, {Roman-Lopes}, {Da Rio}, {Stringfellow}, {Kim},
  {Borissova}, {Fern{\'a}ndez-Trincado}, {Burgasser},
  {Garc{\'\i}a-Hern{\'a}ndez}, {Zamora}, {Pan}, \& {Nitschelm}}]{kounkel18}
{Kounkel}, M., {Covey}, K., {Su{\'a}rez}, G., {et~al.} 2018, \aj, 156, 84

\bibitem[{{Kurosawa} {et~al.}(2006){Kurosawa}, {Harries}, \&
  {Symington}}]{kurosawa06}
{Kurosawa}, R., {Harries}, T.~J., \& {Symington}, N.~H. 2006, \mnras, 370, 580

\bibitem[{{Landolt}(2009)}]{Landolt2009}
{Landolt}, A.~U. 2009, \aj, 137, 4186

\bibitem[{{Lindegren} {et~al.}(2020){Lindegren}, {Bastian}, {Biermann},
  {Bombrun}, {de Torres}, {Gerlach}, {Geyer}, {Hern{\'a}ndez}, {Hilger},
  {Hobbs}, {Klioner}, {Lammers}, {McMillan}, {Ramos-Lerate},
  {Steidelm{\"u}ller}, {Stephenson}, \& {van Leeuwen}}]{lindegren20}
{Lindegren}, L., {Bastian}, U., {Biermann}, M., {et~al.} 2020, arXiv e-prints,
  arXiv:2012.01742

\bibitem[{{Lindgren} \& {Heiter}(2017)}]{Lindgren2017}
{Lindgren}, S. \& {Heiter}, U. 2017, \aap, 604, A97

\bibitem[{{Lodato} {et~al.}(2017){Lodato}, {Scardoni}, {Manara}, \&
  {Testi}}]{lodato17}
{Lodato}, G., {Scardoni}, C.~E., {Manara}, C.~F., \& {Testi}, L. 2017, \mnras,
  472, 4700

\bibitem[{{Luck}(2017)}]{Luck2017}
{Luck}, R.~E. 2017, \aj, 153, 21

\bibitem[{{Luck}(2018)}]{Luck2018}
{Luck}, R.~E. 2018, \aj, 155, 111

\bibitem[{{Luhman} {et~al.}(2003){Luhman}, {Stauffer}, {Muench}, {Rieke},
  {Lada}, {Bouvier}, \& {Lada}}]{luhman03}
{Luhman}, K.~L., {Stauffer}, J.~R., {Muench}, A.~A., {et~al.} 2003, \apj, 593,
  1093

\bibitem[{{Lynden-Bell} \& {Pringle}(1974)}]{lyndenbell74}
{Lynden-Bell}, D. \& {Pringle}, J.~E. 1974, \mnras, 168, 603

\bibitem[{{Manara} {et~al.}(2013{\natexlab{a}}){Manara}, {Beccari}, {Da Rio},
  {De Marchi}, {Natta}, {Ricci}, {Robberto}, \& {Testi}}]{manara13b}
{Manara}, C.~F., {Beccari}, G., {Da Rio}, N., {et~al.} 2013{\natexlab{a}},
  \aap, 558, A114

\bibitem[{{Manara} {et~al.}(2016{\natexlab{a}}){Manara}, {Fedele}, {Herczeg},
  \& {Teixeira}}]{manara16a}
{Manara}, C.~F., {Fedele}, D., {Herczeg}, G.~J., \& {Teixeira}, P.~S.
  2016{\natexlab{a}}, \aap, 585, A136

\bibitem[{{Manara} {et~al.}(2017{\natexlab{a}}){Manara}, {Frasca},
  {Alcal{\'a}}, {Natta}, {Stelzer}, \& {Testi}}]{manara17b}
{Manara}, C.~F., {Frasca}, A., {Alcal{\'a}}, J.~M., {et~al.}
  2017{\natexlab{a}}, \aap, 605, A86

\bibitem[{{Manara} {et~al.}(2019){Manara}, {Mordasini}, {Testi}, {Williams},
  {Miotello}, {Lodato}, \& {Emsenhuber}}]{manara19b}
{Manara}, C.~F., {Mordasini}, C., {Testi}, L., {et~al.} 2019, \aap, 631, L2

\bibitem[{{Manara} {et~al.}(2020){Manara}, {Natta}, {Rosotti}, {Alcal{\'a}},
  {Nisini}, {Lodato}, {Testi}, {Pascucci}, {Hillenbrand}, {Carpenter},
  {Scholz}, {Fedele}, {Frasca}, {Mulders}, {Rigliaco}, {Scardoni}, \&
  {Zari}}]{manara20}
{Manara}, C.~F., {Natta}, A., {Rosotti}, G.~P., {et~al.} 2020, \aap, 639, A58

\bibitem[{{Manara} {et~al.}(2012){Manara}, {Robberto}, {Da Rio}, {Lodato},
  {Hillenbrand}, {Stassun}, \& {Soderblom}}]{manara12}
{Manara}, C.~F., {Robberto}, M., {Da Rio}, N., {et~al.} 2012, \apj, 755, 154

\bibitem[{{Manara} {et~al.}(2016{\natexlab{b}}){Manara}, {Rosotti}, {Testi},
  {Natta}, {Alcal{\'a}}, {Williams}, {Ansdell}, {Miotello}, {van der Marel},
  {Tazzari}, {Carpenter}, {Guidi}, {Mathews}, {Oliveira}, {Prusti}, \& {van
  Dishoeck}}]{manara16b}
{Manara}, C.~F., {Rosotti}, G., {Testi}, L., {et~al.} 2016{\natexlab{b}}, \aap,
  591, L3

\bibitem[{{Manara} {et~al.}(2017{\natexlab{b}}){Manara}, {Testi}, {Herczeg},
  {Pascucci}, {Alcal{\'a}}, {Natta}, {Antoniucci}, {Fedele}, {Mulders},
  {Henning}, {Mohanty}, {Prusti}, \& {Rigliaco}}]{manara17a}
{Manara}, C.~F., {Testi}, L., {Herczeg}, G.~J., {et~al.} 2017{\natexlab{b}},
  \aap, 604, A127

\bibitem[{{Manara} {et~al.}(2014){Manara}, {Testi}, {Natta}, {Rosotti},
  {Benisty}, {Ercolano}, \& {Ricci}}]{manara14}
{Manara}, C.~F., {Testi}, L., {Natta}, A., {et~al.} 2014, \aap, 568, A18

\bibitem[{{Manara} {et~al.}(2013{\natexlab{b}}){Manara}, {Testi}, {Rigliaco},
  {Alcal{\'a}}, {Natta}, {Stelzer}, {Biazzo}, {Covino}, {Covino}, {Cupani},
  {D'Elia}, \& {Randich}}]{manara13a}
{Manara}, C.~F., {Testi}, L., {Rigliaco}, E., {et~al.} 2013{\natexlab{b}},
  \aap, 551, A107

\bibitem[{{Mauc{\'o}} {et~al.}(2018){Mauc{\'o}}, {Brice{\~n}o}, {Calvet},
  {Hern{\'a}ndez}, {Ballesteros-Paredes}, {Gonz{\'a}lez}, {Espaillat}, {Li},
  {Telesco}, {Downes}, {Mac{\'\i}as}, {Qi}, {Michel}, {D'Alessio}, \&
  {Ali}}]{mauco18}
{Mauc{\'o}}, K., {Brice{\~n}o}, C., {Calvet}, N., {et~al.} 2018, \apj, 859, 1

\bibitem[{{Mauc{\'o}} {et~al.}(2016){Mauc{\'o}}, {Hern{\'a}ndez}, {Calvet},
  {Ballesteros-Paredes}, {Brice{\~n}o}, {McClure}, {D'Alessio}, {Anderson}, \&
  {Ali}}]{mauco2016}
{Mauc{\'o}}, K., {Hern{\'a}ndez}, J., {Calvet}, N., {et~al.} 2016, \apj, 829,
  38

\bibitem[{{McDonald} {et~al.}(2017){McDonald}, {Zijlstra}, \&
  {Watson}}]{McDonald2017}
{McDonald}, I., {Zijlstra}, A.~A., \& {Watson}, R.~A. 2017, \mnras, 471, 770

\bibitem[{{McGinnis} {et~al.}(2018){McGinnis}, {Dougados}, {Alencar},
  {Bouvier}, \& {Cabrit}}]{mcginnis18}
{McGinnis}, P., {Dougados}, C., {Alencar}, S.~H.~P., {Bouvier}, J., \&
  {Cabrit}, S. 2018, \aap, 620, A87

\bibitem[{{Mendigut{\'\i}a} {et~al.}(2013){Mendigut{\'\i}a}, {Brittain},
  {Eiroa}, {Meeus}, {Montesinos}, {Mora}, {Muzerolle}, {Oudmaijer}, \&
  {Rigliaco}}]{mendigutia13}
{Mendigut{\'\i}a}, I., {Brittain}, S., {Eiroa}, C., {et~al.} 2013, \apj, 776,
  44

\bibitem[{{Mishenina} {et~al.}(2012){Mishenina}, {Soubiran}, {Kovtyukh},
  {Katsova}, \& {Livshits}}]{Mishenina2012}
{Mishenina}, T.~V., {Soubiran}, C., {Kovtyukh}, V.~V., {Katsova}, M.~M., \&
  {Livshits}, M.~A. 2012, \aap, 547, A106

\bibitem[{{Modigliani} {et~al.}(2010){Modigliani}, {Goldoni}, {Royer},
  {Haigron}, {Guglielmi}, {Fran{\c{c}}ois}, {Horrobin}, {Bristow}, {Vernet},
  {Moehler}, {Kerber}, {Ballester}, {Mason}, \& {Christensen}}]{xspipe}
{Modigliani}, A., {Goldoni}, P., {Royer}, F., {et~al.} 2010, in Society of
  Photo-Optical Instrumentation Engineers (SPIE) Conference Series, Vol. 7737,
  Observatory Operations: Strategies, Processes, and Systems III, ed. D.~R.
  {Silva}, A.~B. {Peck}, \& B.~T. {Soifer}, 773728

\bibitem[{{Mohanty} {et~al.}(2005){Mohanty}, {Jayawardhana}, \&
  {Basri}}]{mohanty05}
{Mohanty}, S., {Jayawardhana}, R., \& {Basri}, G. 2005, \apj, 626, 498

\bibitem[{{Montes} {et~al.}(2018){Montes}, {Gonz{\'a}lez-Peinado}, {Tabernero},
  {Caballero}, {Marfil}, {Alonso-Floriano}, {Cort{\'e}s-Contreras},
  {Gonz{\'a}lez Hern{\'a}ndez}, {Klutsch}, \& {Moreno-J{\'o}dar}}]{Montes2018}
{Montes}, D., {Gonz{\'a}lez-Peinado}, R., {Tabernero}, H.~M., {et~al.} 2018,
  \mnras, 479, 1332

\bibitem[{{Morbidelli} \& {Raymond}(2016)}]{morby16}
{Morbidelli}, A. \& {Raymond}, S.~N. 2016, JGR, 121, 1962

\bibitem[{{Moultaka} {et~al.}(2004){Moultaka}, {Ilovaisky}, {Prugniel}, \&
  {Soubiran}}]{Moultaka2004}
{Moultaka}, J., {Ilovaisky}, S.~A., {Prugniel}, P., \& {Soubiran}, C. 2004,
  \pasp, 116, 693

\bibitem[{{Mulders} {et~al.}(2017){Mulders}, {Pascucci}, {Manara}, {Testi},
  {Herczeg}, {Henning}, {Mohanty}, \& {Lodato}}]{mulders17}
{Mulders}, G.~D., {Pascucci}, I., {Manara}, C.~F., {et~al.} 2017, \apj, 847, 31

\bibitem[{{Muzerolle} {et~al.}(1998){Muzerolle}, {Calvet}, \&
  {Hartmann}}]{muzerolle98}
{Muzerolle}, J., {Calvet}, N., \& {Hartmann}, L. 1998, \apj, 492, 743

\bibitem[{{Muzerolle} {et~al.}(2001){Muzerolle}, {Calvet}, \&
  {Hartmann}}]{muzerolle01}
{Muzerolle}, J., {Calvet}, N., \& {Hartmann}, L. 2001, \apj, 550, 944

\bibitem[{{Muzerolle} {et~al.}(2003){Muzerolle}, {Hillenbrand}, {Calvet},
  {Brice{\~n}o}, \& {Hartmann}}]{muzerolle03}
{Muzerolle}, J., {Hillenbrand}, L., {Calvet}, N., {Brice{\~n}o}, C., \&
  {Hartmann}, L. 2003, \apj, 592, 266

\bibitem[{{Najita} {et~al.}(2015){Najita}, {Andrews}, \&
  {Muzerolle}}]{najita15}
{Najita}, J.~R., {Andrews}, S.~M., \& {Muzerolle}, J. 2015, \mnras, 450, 3559

\bibitem[{{Najita} {et~al.}(2007){Najita}, {Strom}, \& {Muzerolle}}]{najita07}
{Najita}, J.~R., {Strom}, S.~E., \& {Muzerolle}, J. 2007, \mnras, 378, 369

\bibitem[{{Natta} {et~al.}(2014){Natta}, {Testi}, {Alcal{\'a}}, {Rigliaco},
  {Covino}, {Stelzer}, \& {D'Elia}}]{natta14}
{Natta}, A., {Testi}, L., {Alcal{\'a}}, J.~M., {et~al.} 2014, \aap, 569, A5

\bibitem[{{Natta} {et~al.}(2006){Natta}, {Testi}, \& {Randich}}]{natta06}
{Natta}, A., {Testi}, L., \& {Randich}, S. 2006, \aap, 452, 245

\bibitem[{{Nisini} {et~al.}(2018){Nisini}, {Antoniucci}, {Alcal{\'a}},
  {Giannini}, {Manara}, {Natta}, {Fedele}, \& {Biazzo}}]{nisini18}
{Nisini}, B., {Antoniucci}, S., {Alcal{\'a}}, J.~M., {et~al.} 2018, \aap, 609,
  A87

\bibitem[{{Padoan} {et~al.}(2005){Padoan}, {Kritsuk}, {Norman}, \&
  {Nordlund}}]{padoan05}
{Padoan}, P., {Kritsuk}, A., {Norman}, M.~L., \& {Nordlund}, {\r{A}}. 2005,
  \apjl, 622, L61

\bibitem[{{Passegger} {et~al.}(2018){Passegger}, {Reiners}, {Jeffers},
  {Wende-von Berg}, {Sch{\"o}fer}, {Caballero}, {Schweitzer}, {Amado},
  {B{\'e}jar}, {Cort{\'e}s-Contreras}, {Hatzes}, {K{\"u}rster}, {Montes},
  {Pedraz}, {Quirrenbach}, {Ribas}, \& {Seifert}}]{Passegger2018}
{Passegger}, V.~M., {Reiners}, A., {Jeffers}, S.~V., {et~al.} 2018, \aap, 615,
  A6

\bibitem[{{Passegger} {et~al.}(2019){Passegger}, {Schweitzer}, {Shulyak},
  {Nagel}, {Hauschildt}, {Reiners}, {Amado}, {Caballero},
  {Cort{\'e}s-Contreras}, {Dom{\'\i}nguez-Fern{\'a}ndez}, {Quirrenbach},
  {Ribas}, {Azzaro}, {Anglada-Escud{\'e}}, {Bauer}, {B{\'e}jar}, {Dreizler},
  {Guenther}, {Henning}, {Jeffers}, {Kaminski}, {K{\"u}rster}, {Lafarga},
  {Mart{\'\i}n}, {Montes}, {Morales}, {Schmitt}, \&
  {Zechmeister}}]{Passegger2019}
{Passegger}, V.~M., {Schweitzer}, A., {Shulyak}, D., {et~al.} 2019, \aap, 627,
  A161

\bibitem[{{Passegger} {et~al.}(2016){Passegger}, {Wende-von Berg}, \&
  {Reiners}}]{Passegger2016}
{Passegger}, V.~M., {Wende-von Berg}, S., \& {Reiners}, A. 2016, \aap, 587, A19

\bibitem[{{Pecaut} \& {Mamajek}(2013)}]{PM13}
{Pecaut}, M.~J. \& {Mamajek}, E.~E. 2013, \apjs, 208, 9

\bibitem[{{Pepe} {et~al.}(2021){Pepe}, {Cristiani}, {Rebolo}, {Santos},
  {Dekker}, {Cabral}, {Di Marcantonio}, {Figueira}, {Lo Curto}, {Lovis},
  {Mayor}, {M{\'e}gevand}, {Molaro}, {Riva}, {Zapatero Osorio}, {Amate},
  {Manescau}, {Pasquini}, {Zerbi}, {Adibekyan}, {Abreu}, {Affolter}, {Alibert},
  {Aliverti}, {Allart}, {Allende Prieto}, {{\'A}lvarez}, {Alves}, {Avila},
  {Baldini}, {Bandy}, {Barros}, {Benz}, {Bianco}, {Borsa}, {Bourrier},
  {Bouchy}, {Broeg}, {Calderone}, {Cirami}, {Coelho}, {Conconi}, {Coretti},
  {Cumani}, {Cupani}, {D'Odorico}, {Damasso}, {Deiries}, {Delabre},
  {Demangeon}, {Dumusque}, {Ehrenreich}, {Faria}, {Fragoso}, {Genolet},
  {Genoni}, {G{\'e}nova Santos}, {Gonz{\'a}lez Hern{\'a}ndez}, {Hughes},
  {Iwert}, {Kerber}, {Knudstrup}, {Landoni}, {Lavie}, {Lillo-Box}, {Lizon},
  {Maire}, {Martins}, {Mehner}, {Micela}, {Modigliani}, {Monteiro}, {Monteiro},
  {Moschetti}, {Murphy}, {Nunes}, {Oggioni}, {Oliveira}, {Oshagh}, {Pall{\'e}},
  {Pariani}, {Poretti}, {Rasilla}, {Rebord{\~a}o}, {Redaelli}, {Santana
  Tschudi}, {Santin}, {Santos}, {S{\'e}gransan}, {Schmidt}, {Segovia},
  {Sosnowska}, {Sozzetti}, {Sousa}, {Span{\`o}}, {Su{\'a}rez Mascare{\~n}o},
  {Tabernero}, {Tenegi}, {Udry}, \& {Zanutta}}]{pepe20}
{Pepe}, F., {Cristiani}, S., {Rebolo}, R., {et~al.} 2021, \aap, 645, A96

\bibitem[{{P{\'e}rez-Blanco} {et~al.}(2018){P{\'e}rez-Blanco}, {Mauc{\'o}},
  {Hern{\'a}ndez}, {Calvet}, {Espaillat}, {McClure}, {Brice{\~n}o}, {Robinson},
  {Feldman}, {Villarreal}, \& {D'Alessio}}]{Perez2018}
{P{\'e}rez-Blanco}, A., {Mauc{\'o}}, K., {Hern{\'a}ndez}, J., {et~al.} 2018,
  \apj, 867, 116

\bibitem[{{Press} {et~al.}(1992){Press}, {Teukolsky}, {Vetterling}, \&
  {Flannery}}]{Pressetal1992}
{Press}, W.~H., {Teukolsky}, S.~A., {Vetterling}, W.~T., \& {Flannery}, B.~P.
  1992, {Numerical recipes in FORTRAN. The art of scientific computing}

\bibitem[{{Pringle}(1981)}]{pringle81}
{Pringle}, J.~E. 1981, \araa, 19, 137

\bibitem[{{Proffitt} {et~al.}(2021){Proffitt}, {Roman-Duval}, {Taylor},
  {Monroe}, {Fischer}, {Fischer}, {Fullerton}, {Aloisi}, {Britt}, {Busko},
  {Carlberg}, {De Rosa}, {Frazer}, {Hernandez}, {Hirschauer}, {James},
  {Jedrzejewski}, {Lockwood}, {Oliveira}, {Plesha}, {Riedel}, {Riley},
  {Sahnow}, {Sankrit}, {Shaw}, {Smith}, {Sohn}, {Som}, {Ubeda}, \&
  {Welty}}]{proffitt21}
{Proffitt}, C.~R., {Roman-Duval}, J., {Taylor}, J.~M., {et~al.} 2021, Research
  Notes of the American Astronomical Society, 5, 36

\bibitem[{{Pudritz} \& {Ray}(2019)}]{PR19}
{Pudritz}, R.~E. \& {Ray}, T.~P. 2019, Frontiers in Astronomy and Space
  Sciences, 6, 54

\bibitem[{{Rajpurohit} {et~al.}(2018){Rajpurohit}, {Allard}, {Teixeira},
  {Homeier}, {Rajpurohit}, \& {Mousis}}]{Rajpurohit2018}
{Rajpurohit}, A.~S., {Allard}, F., {Teixeira}, G.~D.~C., {et~al.} 2018, \aap,
  610, A19

\bibitem[{{Reiners} {et~al.}(2012){Reiners}, {Joshi}, \&
  {Goldman}}]{Reiners2012}
{Reiners}, A., {Joshi}, N., \& {Goldman}, B. 2012, \aj, 143, 93

\bibitem[{{Reiners} {et~al.}(2018){Reiners}, {Zechmeister}, {Caballero},
  {Ribas}, {Morales}, {Jeffers}, {Sch{\"o}fer}, {Tal-Or}, {Quirrenbach},
  {Amado}, {Kaminski}, {Seifert}, {Abril}, {Aceituno}, {Alonso-Floriano},
  {Ammler-von Eiff}, {Antona}, {Anglada-Escud{\'e}}, {Anwand-Heerwart},
  {Arroyo-Torres}, {Azzaro}, {Baroch}, {Barrado}, {Bauer}, {Becerril},
  {B{\'e}jar}, {Ben{\'\i}tez}, {Berdinas\u0303}, {Bergond}, {Bl{\"u}mcke},
  {Brinkm{\"o}ller}, {del Burgo}, {Cano}, {C{\'a}rdenas V{\'a}zquez}, {Casal},
  {Cifuentes}, {Claret}, {Colom{\'e}}, {Cort{\'e}s-Contreras}, {Czesla},
  {D{\'\i}ez-Alonso}, {Dreizler}, {Feiz}, {Fern{\'a}ndez}, {Ferro},
  {Fuhrmeister}, {Galad{\'\i}-Enr{\'\i}quez}, {Garcia-Piquer}, {Garc{\'\i}a
  Vargas}, {Gesa}, {G{\'o}mez Galera}, {Gonz{\'a}lez Hern{\'a}ndez},
  {Gonz{\'a}lez-Peinado}, {Gr{\"o}zinger}, {Grohnert}, {Gu{\`a}rdia},
  {Guenther}, {Guijarro}, {de Guindos}, {Guti{\'e}rrez-Soto}, {Hagen},
  {Hatzes}, {Hauschildt}, {Hedrosa}, {Helmling}, {Henning}, {Hermelo},
  {Hern{\'a}ndez Arab{\'\i}}, {Hern{\'a}ndez Casta{\~n}o}, {Hern{\'a}ndez
  Hernando}, {Herrero}, {Huber}, {Huke}, {Johnson}, {de Juan}, {Kim}, {Klein},
  {Kl{\"u}ter}, {Klutsch}, {K{\"u}rster}, {Lafarga}, {Lamert}, {Lamp{\'o}n},
  {Lara}, {Laun}, {Lemke}, {Lenzen}, {Launhardt}, {L{\'o}pez del Fresno},
  {L{\'o}pez-Gonz{\'a}lez}, {L{\'o}pez-Puertas}, {L{\'o}pez Salas},
  {L{\'o}pez-Santiago}, {Luque}, {Mag{\'a}n Madinabeitia}, {Mall}, {Mancini},
  {Mandel}, {Marfil}, {Mar{\'\i}n Molina}, {Maroto Fern{\'a}ndez},
  {Mart{\'\i}n}, {Mart{\'\i}n-Ruiz}, {Marvin}, {Mathar}, {Mirabet}, {Montes},
  {Moreno-Raya}, {Moya}, {Mundt}, {Nagel}, {Naranjo}, {Nortmann}, {Nowak},
  {Ofir}, {Oreiro}, {Pall{\'e}}, {Panduro}, {Pascual}, {Passegger}, {Pavlov},
  {Pedraz}, {P{\'e}rez-Calpena}, {P{\'e}rez Medialdea}, {Perger}, {Perryman},
  {Pluto}, {Rabaza}, {Ram{\'o}n}, {Rebolo}, {Redondo}, {Reffert}, {Reinhart},
  {Rhode}, {Rix}, {Rodler}, {Rodr{\'\i}guez}, {Rodr{\'\i}guez-L{\'o}pez},
  {Rodr{\'\i}guez Trinidad}, {Rohloff}, {Rosich}, {Sadegi},
  {S{\'a}nchez-Blanco}, {S{\'a}nchez Carrasco}, {S{\'a}nchez-L{\'o}pez},
  {Sanz-Forcada}, {Sarkis}, {Sarmiento}, {Sch{\"a}fer}, {Schmitt}, {Schiller},
  {Schweitzer}, {Solano}, {Stahl}, {Strachan}, {St{\"u}rmer}, {Su{\'a}rez},
  {Tabernero}, {Tala}, {Trifonov}, {Tulloch}, {Ulbrich}, {Veredas}, {Vico
  Linares}, {Vilardell}, {Wagner}, {Winkler}, {Wolthoff}, {Xu}, {Yan}, \&
  {Zapatero Osorio}}]{Reiners2018}
{Reiners}, A., {Zechmeister}, M., {Caballero}, J.~A., {et~al.} 2018, \aap, 612,
  A49

\bibitem[{{Rich} {et~al.}(2017){Rich}, {Wisniewski}, {McElwain}, {Hashimoto},
  {Kudo}, {Kusakabe}, {Okamoto}, {Abe}, {Akiyama}, {Brandner}, {Brandt},
  {Cargile}, {Carson}, {Currie}, {Egner}, {Feldt}, {Fukagawa}, {Goto}, {Grady},
  {Guyon}, {Hayano}, {Hayashi}, {Hayashi}, {Hebb}, {He{\l}miniak}, {Henning},
  {Hodapp}, {Ishii}, {Iye}, {Janson}, {Kandori}, {Knapp}, {Kuzuhara}, {Kwon},
  {Matsuo}, {Mayama}, {Miyama}, {Momose}, {Morino}, {Moro-Martin}, {Nakagawa},
  {Nishimura}, {Oh}, {Pyo}, {Schlieder}, {Serabyn}, {Sitko}, {Suenaga}, {Suto},
  {Suzuki}, {Takahashi}, {Takami}, {Takato}, {Terada}, {Thalmann}, {Tomono},
  {Turner}, {Watanabe}, {Yamada}, {Takami}, {Usuda}, \& {Tamura}}]{Rich2017}
{Rich}, E.~A., {Wisniewski}, J.~P., {McElwain}, M.~W., {et~al.} 2017, \mnras,
  472, 1736

\bibitem[{{Ricker} {et~al.}(2014){Ricker}, {Winn}, {Vanderspeck}, {Latham},
  {Bakos}, {Bean}, {Berta-Thompson}, {Brown}, \& {et al.}}]{Ricker2014}
{Ricker}, G.~R., {Winn}, J.~N., {Vanderspeck}, R., {et~al.} 2014, Journal of
  Astronomical Telescopes, Instruments, and Systems, 014003

\bibitem[{{Rigliaco} {et~al.}(2011){Rigliaco}, {Natta}, {Randich}, {Testi}, \&
  {Biazzo}}]{rigliaco11a}
{Rigliaco}, E., {Natta}, A., {Randich}, S., {Testi}, L., \& {Biazzo}, K. 2011,
  \aap, 525, A47

\bibitem[{{Rigliaco} {et~al.}(2012){Rigliaco}, {Natta}, {Testi}, {Randich},
  {Alcal{\`a}}, {Covino}, \& {Stelzer}}]{rigliaco12}
{Rigliaco}, E., {Natta}, A., {Testi}, L., {et~al.} 2012, \aap, 548, A56

\bibitem[{{Rigliaco} {et~al.}(2013){Rigliaco}, {Pascucci}, {Gorti}, {Edwards},
  \& {Hollenbach}}]{rigliaco13}
{Rigliaco}, E., {Pascucci}, I., {Gorti}, U., {Edwards}, S., \& {Hollenbach}, D.
  2013, \apj, 772, 60

\bibitem[{{Robinson} \& {Espaillat}(2019)}]{robinson19}
{Robinson}, C.~E. \& {Espaillat}, C.~C. 2019, \apj, 874, 129

\bibitem[{{Rojas-Ayala} {et~al.}(2012){Rojas-Ayala}, {Covey}, {Muirhead}, \&
  {Lloyd}}]{Rojas-Ayala2012}
{Rojas-Ayala}, B., {Covey}, K.~R., {Muirhead}, P.~S., \& {Lloyd}, J.~P. 2012,
  \apj, 748, 93

\bibitem[{{Roman-Duval} {et~al.}(2020){Roman-Duval}, {Proffitt}, {Taylor},
  {Monroe}, {Fischer}, {Fischer}, {Fullerton}, {Aloisi}, {Britt}, {Busko},
  {Carlberg}, {De Rosa}, {Jedrzejewski}, {Lockwood}, {Frazer}, {Hernandez},
  {James}, {Oliveira}, {Plesha}, {Riedel}, {Riley}, {Sahnow}, {Sankrit},
  {Shaw}, {Smith}, {Sohn}, {Som}, {Ubeda}, \& {Welty}}]{ullysesDR1}
{Roman-Duval}, J., {Proffitt}, C.~R., {Taylor}, J.~M., {et~al.} 2020, Research
  Notes of the American Astronomical Society, 4, 205

\bibitem[{{Rosotti} \& {Clarke}(2018)}]{rosotti19}
{Rosotti}, G.~P. \& {Clarke}, C.~J. 2018, \mnras, 473, 5630

\bibitem[{{Rosotti} {et~al.}(2017){Rosotti}, {Clarke}, {Manara}, \&
  {Facchini}}]{rosotti17}
{Rosotti}, G.~P., {Clarke}, C.~J., {Manara}, C.~F., \& {Facchini}, S. 2017,
  \mnras, 468, 1631

\bibitem[{{Rugel} {et~al.}(2018){Rugel}, {Fedele}, \& {Herczeg}}]{Rugel2018}
{Rugel}, M., {Fedele}, D., \& {Herczeg}, G. 2018, \aap, 609, A70

\bibitem[{{Schweitzer} {et~al.}(2019){Schweitzer}, {Passegger}, {Cifuentes},
  {B{\'e}jar}, {Cort{\'e}s-Contreras}, {Caballero}, {del Burgo}, {Czesla},
  {K{\"u}rster}, {Montes}, {Zapatero Osorio}, {Ribas}, {Reiners},
  {Quirrenbach}, {Amado}, {Aceituno}, {Anglada-Escud{\'e}}, {Bauer},
  {Dreizler}, {Jeffers}, {Guenther}, {Henning}, {Kaminski}, {Lafarga},
  {Marfil}, {Morales}, {Schmitt}, {Seifert}, {Solano}, {Tabernero}, \&
  {Zechmeister}}]{Schweitzer2019}
{Schweitzer}, A., {Passegger}, V.~M., {Cifuentes}, C., {et~al.} 2019, \aap,
  625, A68

\bibitem[{{Sellek} {et~al.}(2020{\natexlab{a}}){Sellek}, {Booth}, \&
  {Clarke}}]{sellek20b}
{Sellek}, A.~D., {Booth}, R.~A., \& {Clarke}, C.~J. 2020{\natexlab{a}}, \mnras,
  498, 2845

\bibitem[{{Sellek} {et~al.}(2020{\natexlab{b}}){Sellek}, {Booth}, \&
  {Clarke}}]{sellek20a}
{Sellek}, A.~D., {Booth}, R.~A., \& {Clarke}, C.~J. 2020{\natexlab{b}}, \mnras,
  492, 1279

\bibitem[{{Sicilia-Aguilar} {et~al.}(2015){Sicilia-Aguilar}, {Fang},
  {Roccatagliata}, {Collier Cameron}, {K{\'o}sp{\'a}l}, {Henning},
  {{\'A}brah{\'a}m}, \& {Sipos}}]{2015A&A...580A..82S}
{Sicilia-Aguilar}, A., {Fang}, M., {Roccatagliata}, V., {et~al.} 2015, \aap,
  580, A82

\bibitem[{{Sicilia-Aguilar} {et~al.}(2010){Sicilia-Aguilar}, {Henning}, \&
  {Hartmann}}]{sicilia-aguilar10}
{Sicilia-Aguilar}, A., {Henning}, T., \& {Hartmann}, L.~W. 2010, \apj, 710, 597

\bibitem[{{Sicilia-Aguilar} {et~al.}(2012){Sicilia-Aguilar}, {K{\'o}sp{\'a}l},
  {Setiawan}, {{\'A}brah{\'a}m}, {Dullemond}, {Eiroa}, {Goto}, {Henning}, \&
  {Juh{\'a}sz}}]{2012A&A...544A..93S}
{Sicilia-Aguilar}, A., {K{\'o}sp{\'a}l}, {\'A}., {Setiawan}, J., {et~al.} 2012,
  \aap, 544, A93

\bibitem[{{Siess} {et~al.}(2000){Siess}, {Dufour}, \& {Forestini}}]{siess00}
{Siess}, L., {Dufour}, E., \& {Forestini}, M. 2000, \aap, 358, 593

\bibitem[{{Simon} {et~al.}(2016){Simon}, {Pascucci}, {Edwards}, {Feng},
  {Gorti}, {Hollenbach}, {Rigliaco}, \& {Keane}}]{simon16}
{Simon}, M.~N., {Pascucci}, I., {Edwards}, S., {et~al.} 2016, \apj, 831, 169

\bibitem[{{Siwak} {et~al.}(2018){Siwak}, {Ogloza}, {Moffat}, {Matthews},
  {Rucinski}, {Kallinger}, {Kuschnig}, {Cameron}, {Weiss}, {Rowe}, {Guenther},
  \& {Sasselov}}]{Siwak2018}
{Siwak}, M., {Ogloza}, W., {Moffat}, A. F.~J., {et~al.} 2018, \mnras, 478, 758

\bibitem[{{Smette} {et~al.}(2015){Smette}, {Sana}, {Noll}, {Horst}, {Kausch},
  {Kimeswenger}, {Barden}, {Szyszka}, {Jones}, {Gallenne}, {Vinther},
  {Ballester}, \& {Taylor}}]{molecfit1}
{Smette}, A., {Sana}, H., {Noll}, S., {et~al.} 2015, \aap, 576, A77

\bibitem[{{Somigliana} {et~al.}(2020){Somigliana}, {Toci}, {Lodato}, {Rosotti},
  \& {Manara}}]{somigliana20}
{Somigliana}, A., {Toci}, C., {Lodato}, G., {Rosotti}, G., \& {Manara}, C.~F.
  2020, \mnras, 492, 1120

\bibitem[{{Sousa} {et~al.}(2016){Sousa}, {Alencar}, {Bouvier}, {Stauffer},
  {Venuti}, {Hillenbrand}, {Cody}, {Teixeira}, {Guimar\~aes}, {McGinnis},
  {Flaccomio}, {F\H{u}r\'esz}, {Micela}, \& {Gameiro}}]{sousa16}
{Sousa}, A.~P., {Alencar}, S.~H.~P., {Bouvier}, J., {et~al.} 2016, \aap, 586,
  A47

\bibitem[{{Stauffer} {et~al.}(2014){Stauffer}, {Cody}, {Baglin}, {Alencar},
  {Rebull}, {Hillenbrand}, {Venuti}, {Turner}, {Carpenter}, {Plavchan},
  {Findeisen}, {Carey}, {Terebey}, {Morales-Calder{\'o}n}, {Bouvier}, {Micela},
  {Flaccomio}, {Song}, {Gutermuth}, {Hartmann}, {Calvet}, {Whitney}, {Barrado},
  {Vrba}, {Covey}, {Herbst}, {Furesz}, {Aigrain}, \& {Favata}}]{stauffer14}
{Stauffer}, J., {Cody}, A.~M., {Baglin}, A., {et~al.} 2014, \aj, 147, 83

\bibitem[{{Stetson}(2000)}]{Stetson2000}
{Stetson}, P.~B. 2000, \pasp, 112, 925

\bibitem[{{Thanathibodee} {et~al.}(2018){Thanathibodee}, {Calvet}, {Herczeg},
  {Brice{\~n}o}, {Clark}, {Reiter}, {Ingleby}, {McClure}, {Mauc{\'o}}, \&
  {Hern{\'a}ndez}}]{thanathibodee18}
{Thanathibodee}, T., {Calvet}, N., {Herczeg}, G., {et~al.} 2018, \apj, 861, 73

\bibitem[{{Thanathibodee} {et~al.}(2019){Thanathibodee}, {Calvet}, {Muzerolle},
  {Brice{\~n}o}, {Hern{\'a}ndez}, \& {Mauc{\'o}}}]{thanathibodee19}
{Thanathibodee}, T., {Calvet}, N., {Muzerolle}, J., {et~al.} 2019, \apj, 884,
  86

\bibitem[{{Tokovinin} {et~al.}(2020){Tokovinin}, {Petr-Gotzens}, \&
  {Brice{\~n}o}}]{tokovinin20}
{Tokovinin}, A., {Petr-Gotzens}, M.~G., \& {Brice{\~n}o}, C. 2020, \aj, 160,
  268

\bibitem[{{Torres} {et~al.}(2006){Torres}, {Quast}, {da Silva}, {de La Reza},
  {Melo}, \& {Sterzik}}]{Torres2006}
{Torres}, C.~A.~O., {Quast}, G.~R., {da Silva}, L., {et~al.} 2006, \aap, 460,
  695

\bibitem[{{Valenti} \& {Fischer}(2005)}]{Valenti2005}
{Valenti}, J.~A. \& {Fischer}, D.~A. 2005, \apjs, 159, 141

\bibitem[{{Venuti} {et~al.}(2014){Venuti}, {Bouvier}, {Flaccomio}, {Alencar},
  {Irwin}, {Stauffer}, {Cody}, {Teixeira}, {Sousa}, {Micela}, {Cuillandre}, \&
  {Peres}}]{venuti14}
{Venuti}, L., {Bouvier}, J., {Flaccomio}, E., {et~al.} 2014, \aap, 570, A82

\bibitem[{{Venuti} {et~al.}(2019){Venuti}, {Stelzer}, {Alcal{\'a}}, {Manara},
  {Frasca}, {Jayawardhana}, {Antoniucci}, {Argiroffi}, {Natta}, {Nisini},
  {Randich}, \& {Scholz}}]{venuti19}
{Venuti}, L., {Stelzer}, B., {Alcal{\'a}}, J.~M., {et~al.} 2019, \aap, 632, A46

\bibitem[{{Vernet} {et~al.}(2011){Vernet}, {Dekker}, {D'Odorico}, {Kaper},
  {Kjaergaard}, {Hammer}, {Randich}, {Zerbi}, {Groot}, {Hjorth}, {Guinouard},
  {Navarro}, {Adolfse}, {Albers}, {Amans}, {Andersen}, {Andersen}, {Binetruy},
  {Bristow}, {Castillo}, {Chemla}, {Christensen}, {Conconi}, {Conzelmann},
  {Dam}, {de Caprio}, {de Ugarte Postigo}, {Delabre}, {di Marcantonio},
  {Downing}, {Elswijk}, {Finger}, {Fischer}, {Flores}, {Fran{\c{c}}ois},
  {Goldoni}, {Guglielmi}, {Haigron}, {Hanenburg}, {Hendriks}, {Horrobin},
  {Horville}, {Jessen}, {Kerber}, {Kern}, {Kiekebusch}, {Kleszcz}, {Klougart},
  {Kragt}, {Larsen}, {Lizon}, {Lucuix}, {Mainieri}, {Manuputy}, {Martayan},
  {Mason}, {Mazzoleni}, {Michaelsen}, {Modigliani}, {Moehler}, {M{\o}ller},
  {Norup S{\o}rensen}, {N{\o}rregaard}, {P{\'e}roux}, {Patat}, {Pena}, {Pragt},
  {Reinero}, {Rigal}, {Riva}, {Roelfsema}, {Royer}, {Sacco}, {Santin},
  {Schoenmaker}, {Spano}, {Sweers}, {Ter Horst}, {Tintori}, {Tromp}, {van
  Dael}, {van der Vliet}, {Venema}, {Vidali}, {Vinther}, {Vola}, {Winters},
  {Wistisen}, {Wulterkens}, \& {Zacchei}}]{vernet11}
{Vernet}, J., {Dekker}, H., {D'Odorico}, S., {et~al.} 2011, \aap, 536, A105

\bibitem[{{Vorobyov} \& {Basu}(2009)}]{vorobyov09}
{Vorobyov}, E.~I. \& {Basu}, S. 2009, \apj, 703, 922

\bibitem[{{Walter} {et~al.}(1997){Walter}, {Wolk}, {Freyberg}, \&
  {Schmitt}}]{walter1997}
{Walter}, F.~M., {Wolk}, S.~J., {Freyberg}, M., \& {Schmitt}, J.~H.~M.~M. 1997,
  \memsai, 68, 1081

\bibitem[{{Weber} {et~al.}(2020){Weber}, {Ercolano}, {Picogna}, {Hartmann}, \&
  {Rodenkirch}}]{weber20}
{Weber}, M.~L., {Ercolano}, B., {Picogna}, G., {Hartmann}, L., \& {Rodenkirch},
  P.~J. 2020, \mnras, 496, 223

\bibitem[{{White} \& {Basri}(2003)}]{WB03}
{White}, R.~J. \& {Basri}, G. 2003, \apj, 582, 1109

\bibitem[{{Winter} {et~al.}(2018){Winter}, {Clarke}, {Rosotti}, {Ih},
  {Facchini}, \& {Haworth}}]{winter18}
{Winter}, A.~J., {Clarke}, C.~J., {Rosotti}, G., {et~al.} 2018, \mnras, 478,
  2700

\bibitem[{{Woolf} \& {Wallerstein}(2005)}]{Woolf2005}
{Woolf}, V.~M. \& {Wallerstein}, G. 2005, \mnras, 356, 963

\bibitem[{{Yee} {et~al.}(2017){Yee}, {Petigura}, \& {von Braun}}]{Yee2017}
{Yee}, S.~W., {Petigura}, E.~A., \& {von Braun}, K. 2017, \apj, 836, 77

\bibitem[{{Zboril} \& {Byrne}(1998)}]{Zboril1998}
{Zboril}, M. \& {Byrne}, P.~B. 1998, \mnras, 299, 753

\end{thebibliography}



\setlength{\tabcolsep}{4pt}

\begin{table*}  
\begin{center} 
\footnotesize 
\caption{\label{tab::res_rotfit} Photospheric properties from the UVES spectra fit with ROTFIT for the targets in the Orion OB1 Association and $\sigma$-Orionis cluster } 
\begin{tabular}{l|cc | lccrc | cccc   } 
\hline \hline 
     Name & epoch &    HJD &   $~~~~T_{\rm eff}$ &    log$g$  &   SpT &  $v$sin$i$~~~  & RV &  r500 &   r550  &   r600  &   r650  \\
        &    &  2459100+  & ~~~~[K] &  & & [km/s]~~ & [km/s] &  &    \\
\hline
     CVSO17  & ep1  &  81.63530  &  3721 $\pm$ 72  & 4.84 $\pm$  0.11 &  M1V  & 7.2 $\pm$ 0.8 & 25.89\,$\pm$\,0.55 & 0.00 $\pm$ 0.00 & 0.01 $\pm$ 0.02 & 0.24 $\pm$ 0.14 & 0.20 $\pm$ 0.10 \\
     CVSO17  & ep2  &  82.62335  &  3697 $\pm$ 88  & 4.85 $\pm$  0.11 &  M1V  & 7.4 $\pm$ 0.7 & 25.40\,$\pm$\,0.74 & 0.02 $\pm$ 0.04 & 0.06 $\pm$ 0.06 & 0.18 $\pm$ 0.14 & 0.23 $\pm$ 0.10 \\
     CVSO17  & ep3  &  83.66124  &  3695 $\pm$ 88  & 4.85 $\pm$  0.11 &  M1V  & 7.7 $\pm$ 0.6 & 25.31\,$\pm$\,0.74 & 0.01 $\pm$ 0.03 & 0.04 $\pm$ 0.05 & 0.12 $\pm$ 0.20 & 0.25 $\pm$ 0.09 \\
\hline
     CVSO36  & ep1  &  85.68189  &  3702 $\pm$ 88  & 4.84 $\pm$  0.11 &  M1V  & 3.0 $\pm$ 0.5 & 19.05\,$\pm$\,0.47 & 0.00 $\pm$ 0.00 & 0.00 $\pm$ 0.00 & 0.18 $\pm$ 0.08 & 0.21 $\pm$ 0.03 \\
     CVSO36  & ep2  &  86.62187  &  3696 $\pm$ 91  & 4.84 $\pm$  0.11 &  M1V  & 2.8 $\pm$ 0.7 & 18.95\,$\pm$\,0.47 & 0.00 $\pm$ 0.00 & 0.00 $\pm$ 0.00 & 0.17 $\pm$ 0.06 & 0.12 $\pm$ 0.04 \\
     CVSO36  & ep3  &  87.63843  &  3662 $\pm$ 69  & 4.84 $\pm$  0.11 &  M1V  & 3.0 $\pm$ 0.6 & 19.24\,$\pm$\,0.52 &  0.00 $\pm$ 0.00 & 0.00 $\pm$ 0.00 & 0.11 $\pm$ 0.06 & 0.12 $\pm$ 0.04 \\
\hline
     CVSO58  & ep1  &  83.61246  &  4193 $\pm$  103  & 4.67 $\pm$  0.10 &  K7V & 17.4 $\pm$ 1.1 & 22.29\,$\pm$\,0.63 & 0.81 $\pm$ 0.26 & 0.85 $\pm$ 0.11 & 0.54 $\pm$ 0.05 & 0.63 $\pm$ 0.12 \\
     CVSO58  & ep2  &  84.61185  &  4211 $\pm$  110  & 4.66 $\pm$  0.10 &  K7V & 17.1 $\pm$ 1.6 & 19.87\,$\pm$\,0.59 & 0.76 $\pm$ 0.26 & 0.87 $\pm$ 0.14 & 0.55 $\pm$ 0.05 & 0.61 $\pm$ 0.08 \\
     CVSO58  & ep3  &  85.60270  &  4223 $\pm$  105  & 4.66 $\pm$  0.11 &  K7V & 17.9 $\pm$ 1.3 & 23.65\,$\pm$\,0.67 & 0.81 $\pm$ 0.22 & 0.81 $\pm$ 0.04 & 0.59 $\pm$ 0.12 & 0.59 $\pm$ 0.10 \\
\hline
    CVSO107  & ep1  &  86.66418  &  3988 $\pm$  118  & 4.68 $\pm$  0.10 &  K8V  & 7.7 $\pm$ 0.9 & 17.29\,$\pm$\,0.52 & 0.68 $\pm$ 0.24 & 0.86 $\pm$ 0.16 & 0.67 $\pm$ 0.12 & 0.74 $\pm$ 0.18 \\
    CVSO107  & ep2  &  87.60273  &  3943  $\pm$  93  & 4.69 $\pm$  0.10 &  M0V  & 6.9 $\pm$ 0.9 & 15.33\,$\pm$\,0.48 & 0.67 $\pm$ 0.24 & 0.85 $\pm$ 0.22 & 0.63 $\pm$ 0.08 & 0.72 $\pm$ 0.18 \\
    CVSO107  & ep3  &  88.65286  &  4002 $\pm$  119  & 4.68 $\pm$  0.10 &  M0V  & 5.9 $\pm$ 0.9 & 14.73\,$\pm$\,0.44 & 0.71 $\pm$ 0.27 & 0.98 $\pm$ 0.11 & 0.70 $\pm$ 0.00 & 0.71 $\pm$ 0.14 \\
\hline
    CVSO109  & ep1  &  79.65665  &  3898 $\pm$  112  & 4.70 $\pm$  0.11 &  M0V  & 3.5 $\pm$ 1.1 & 16.99\,$\pm$\,0.40 & 0.25 $\pm$ 0.11 & 0.55 $\pm$ 0.10 & 0.43 $\pm$ 0.05 & 0.44 $\pm$ 0.08 \\
    CVSO109  & ep2  &  80.63793  &  3922 $\pm$  106  & 4.69 $\pm$  0.11 &  M0V  & 3.3 $\pm$ 1.0 & 16.72\,$\pm$\,0.41 & 0.26 $\pm$ 0.14 & 0.45 $\pm$ 0.10 & 0.39 $\pm$ 0.06 & 0.42 $\pm$ 0.12 \\
    CVSO109  & ep3  &  81.67743  &  3948  $\pm$  91  & 4.69 $\pm$  0.11 &  M0V  & 3.2 $\pm$ 0.9 & 15.98\,$\pm$\,0.38 &  0.71 $\pm$ 0.23 & 0.90 $\pm$ 0.08 & 0.69 $\pm$ 0.05 & 0.67 $\pm$ 0.14 \\
\hline
    CVSO176  & ep1  &  87.68387  &  3495  $\pm$  85  & 4.89 $\pm$  0.11 &  M3V & 19.0 $\pm$ 1.8 & ~7.42\,$\pm$\,1.34 & 0.66 $\pm$ 0.15 & 0.72 $\pm$ 0.28 & 0.90 $\pm$ 0.55 & 0.93 $\pm$ 0.60 \\
    CVSO176  & ep2  &  88.61053  &  3503  $\pm$  82  & 4.86 $\pm$  0.10 &  M3V & 18.4 $\pm$ 1.0 & 10.04\,$\pm$\,1.25 & 0.39 $\pm$ 0.10 & 0.33 $\pm$ 0.18 & 0.57 $\pm$ 0.40 & 0.60 $\pm$ 0.38 \\
    CVSO176  & ep3  &  89.59468  &  3521  $\pm$  77  & 4.86 $\pm$  0.10 &  M3V & 18.4 $\pm$ 1.2 & 10.65\,$\pm$\,1.27 & 0.25 $\pm$ 0.05 & 0.34 $\pm$ 0.16 & 0.63 $\pm$ 0.40 & 0.68 $\pm$ 0.48 \\
\hline \hline
       SO\,518 & ep1  & 82.66445  &  4328 $\pm$  168  & 4.66 $\pm$ 0.11 &  K7V & 12.6 $\pm$ 0.7 & 30.98\,$\pm$\,0.43 & 2.27 $\pm$ 0.14 & 2.12 $\pm$ 0.16 & 1.46 $\pm$ 0.11 & 1.28 $\pm$ 0.13 \\
       SO\,518 & ep2  & 83.70244  &  4383 $\pm$  141  & 4.63 $\pm$ 0.10 &  K7V & 13.3 $\pm$ 0.9 & 30.42\,$\pm$\,0.50 & 1.44 $\pm$ 0.18 & 1.44 $\pm$ 0.09 & 1.04 $\pm$ 0.09 & 0.98 $\pm$ 0.04 \\
       SO\,518 & ep3  & 84.65642  &  4366 $\pm$  150  & 4.63 $\pm$ 0.10 &  K6V & 14.0 $\pm$ 1.1 & 32.83\,$\pm$\,0.61 & 1.38 $\pm$ 0.17 & 1.23 $\pm$ 0.12 & 0.93 $\pm$ 0.08 & 0.89 $\pm$ 0.08 \\
\hline
       SO\,583 & ep1  & 82.68932  &  4753 $\pm$  119  & 4.65 $\pm$  0.11 & K3V  & 8.7 $\pm$ 0.9 & 30.97\,$\pm$\,0.28 & 0.24 $\pm$ 0.12 & 0.40 $\pm$ 0.00 & 0.40 $\pm$ 0.17 & 0.43 $\pm$ 0.08 \\
       SO\,583 & ep2  & 83.64056  &  4739 $\pm$  118  & 4.58 $\pm$  0.10 & K3V & 10.0 $\pm$ 1.0 & 30.95\,$\pm$\,0.30  & 0.32 $\pm$ 0.11 & 0.53 $\pm$ 0.14 & 0.52 $\pm$ 0.11 & 0.55 $\pm$ 0.11 \\
       SO\,583 & ep3  & 84.63973  &  4725 $\pm$  117  & 4.59 $\pm$  0.10 & K4V & 10.8 $\pm$ 1.5 & 29.61\,$\pm$\,0.34  & 0.49 $\pm$ 0.21 & 0.95 $\pm$ 0.07 & 0.84 $\pm$ 0.05 & 0.80 $\pm$ 0.07 \\

\hline
\end{tabular} 
\tablefoot{Values obtained fitting the UVES spectra with the HARPS templates. The measured veiling at different wavelengths is reported in the column labelled 'r' followed by the wavelength in nm.} 
\end{center} 
\end{table*}  

\begin{table*}  
\begin{center} 
\footnotesize 
\caption{\label{tab::res_rotfit_esp} Photospheric properties from the ESPRESSO spectra fit with ROTFIT for the targets in the Orion OB1 Association and $\sigma$-Orionis cluster } 
\begin{tabular}{l|cc | lccrc | cccc   } 
\hline \hline 
     Name & epoch &    HJD & ~~~~$T_{\rm eff}$ &    log$g$  &   SpT &  $v$sin$i$~~~ &  RV  &  r500 &   r550  &   r600  &   r650  \\
        &    &  2459100+  & ~~~~[K] &  & & [km/s]~~ & [km/s] &  &    \\
\hline
    CVSO146  & ep1  & 92.62465 &  4303\,$\pm$\,97 &  4.66\,$\pm$\,0.11 &  K6V & 4.4\,$\pm$\,0.8 &  20.70\,$\pm$\,0.29 & 0.29\,$\pm$\,0.11 &  0.38\,$\pm$\,0.07 &  0.25\,$\pm$\,0.11 & 0.28\,$\pm$\,0.04 \\
    CVSO146  & ep2  & 93.60337 &  4372\,$\pm$\,101 &  4.63\,$\pm$\,0.10 &  K6V & 5.0\,$\pm$\,0.8 & 20.40\,$\pm$\,0.30 & 0.40\,$\pm$\,0.10 & 0.44\,$\pm$\,0.10 & 0.31\,$\pm$\,0.10 & 0.34\,$\pm$\,0.09 \\
    CVSO146  & ep3  & 94.59294 & 4272\,$\pm$\,113 & 4.70\,$\pm$\,0.10 &  K6V & 6.4\,$\pm$\,0.9 & 19.82\,$\pm$\,0.31 & 0.56\,$\pm$\,0.08 & 0.60\,$\pm$\,0.02 & 0.53\,$\pm$\,0.10 & 0.42\,$\pm$\,0.08 \\
\hline
    CVSO165 & ep1 & 96.57902 & 4591\,$\pm$\,167 & 4.60\,$\pm$\,0.10 & K4V & 15.5\,$\pm$\,0.9 & 27.94\,$\pm$\,0.59 & 0.10\,$\pm$\,0.07 &  0.28\,$\pm$\,0.04 &  0.24\,$\pm$\,0.12 & 0.25\,$\pm$\,0.05 \\
    CVSO165 & ep2  & 97.60896  &  4591\,$\pm$\,169 &  4.60\,$\pm$\,0.10 & K4V & 15.4\,$\pm$\,0.9 & 28.00\,$\pm$\,0.38 & 0.13\,$\pm$\,0.05 & 0.32\,$\pm$\,0.04 & 0.28\,$\pm$\,0.07 &  0.32\,$\pm$\,0.04 \\
    CVSO165 & ep3  & 98.61706  &  4585\,$\pm$\,167 &  4.60\,$\pm$\,0.10 & K4V & 15.2\,$\pm$\,0.8 & 27.77\,$\pm$\,0.39 &  0.21\,$\pm$\,0.05 & 0.36\,$\pm$\,0.05 & 0.29\,$\pm$\,0.11 &  0.36\,$\pm$\,0.05 \\
\hline \hline
      SO\,1153  &   ep1  & 91.83202  &  4152\,$\pm$\,158 &  4.67\,$\pm$\,0.10 &  K7V &  9.9\,$\pm$\,1.6 & 34.43\,$\pm$\,0.57 & 5.52\,$\pm$\,0.54 &  5.54\,$\pm$\,0.85 &  4.08\,$\pm$\,0.66 &   4.81\,$\pm$\,0.62 \\
      SO\,1153  &   ep2  & 92.59172  &  4119\,$\pm$\,181 &  4.68\,$\pm$\,0.11 &  K7V &  9.1\,$\pm$\,1.2 & 36.82\,$\pm$\,0.48 & 6.28\,$\pm$\,0.71 &  6.13\,$\pm$\,0.78 &  5.08\,$\pm$\,0.71 &   5.23\,$\pm$\,0.56 \\
      SO\,1153  &   ep3  & 93.63656  &  4065\,$\pm$\,146 &  4.69\,$\pm$\,0.11 &  K7V & 11.6\,$\pm$\,1.9 & 34.66\,$\pm$\,0.57 & 6.52\,$\pm$\,0.44 &  6.83\,$\pm$\,0.85 &  5.22\,$\pm$\,0.65 &   5.71\,$\pm$\,0.80 \\

\hline
\end{tabular} 
\tablefoot{Values obtained fitting the ESPRESSO spectra with the HARPS templates. The measured veiling at different wavelengths is reported in the column labelled 'r' followed by the wavelength in nm.} 
\end{center} 
\end{table*}

\setlength{\tabcolsep}{3pt}

\begin{table*}  
\begin{center} 
\footnotesize 
\caption{\label{tab::res_rotfit_xs} Photospheric properties from the X-Shooter spectra fit with ROTFIT for the targets in the Orion OB1 Association and $\sigma$-Orionis cluster } 
\begin{tabular}{lc|cc | lcccc   } 
\hline \hline 
        Name    & HJD      & $T_{\rm eff}$ & $\log g$  &  $v$sin$i$     & RV      & r620  & r710 & r970 \\
                & 2459100+ &  [K]          &           &    [km/s]  &  [km/s] &       &      &      \\ 
        \hline
          CVSO17 & 88.62128  & 3588 $\pm$ 22  &  4.78 $\pm$ 0.24 &   $<$6.0   & 27.4 $\pm$ 1.9  & 0.4 & 0.2 & 0.1  \\
          CVSO36 & 86.62703  & 3515 $\pm$ 24  &  4.58 $\pm$ 0.14 &   $<$6.0   & 18.6 $\pm$ 1.7   & 0.4 & 0.4 & 0.1 \\
          CVSO58 & 85.76194  & 3886 $\pm$ 45  &  4.81 $\pm$ 0.21 &  19.3 $\pm$ 1.0  & 23.2 $\pm$ 2.2 & 1.0 & 0.2 & 0.5 \\
          CVSO90 & 98.56264  & 3489 $\pm$ 32  &  4.26 $\pm$ 0.30 &  13.0 $\pm$ 16.0  & 23.8 $\pm$ 2.7 & \dots & 1.8 & 1.4  \\
          CVSO107 & 87.58635 & 3766 $\pm$ 37  &  4.50 $\pm$ 0.11 &  9.4 $\pm$ 2.0  & 16.1 $\pm$ 2.0 & 0.9 & 0.2 & 0.3  \\
          CVSO109 & 81.65721 & 3799 $\pm$ 53  &  4.42 $\pm$ 0.12 &  $<$6.0 & 17.8 $\pm$ 1.7  & 0.9 & 0.3 & 0.3  \\
          CVSO146 & 92.58387 & 4079 $\pm$ 80  &  4.70 $\pm$ 0.33 &  $<$6.0  & 16.9 $\pm$ 1.7 & 0.5 & 0.7 & 0.2   \\
          CVSO165 & 97.58340 & 4040 $\pm$ 73  &  4.09 $\pm$ 0.24 & 12.4 $\pm$ 1.0 & 26.5 $\pm$ 1.8  & 0.5 & 0.6 & 0.3 \\
          CVSO176 & 85.68282 & 3490 $\pm$ 62  &  4.21 $\pm$ 0.22 & 17.0 $\pm$ 7.0 & 11.1 $\pm$ 1.8  & 1.4 & 0.5 & 0.6  \\
\hline \hline
          SO\,518 & 85.64623 & 3978 $\pm$ 68  & 4.42 $\pm$ 0.46 &  7.0 $\pm$ 12.0  & 32.3 $\pm$ 2.4 & 1.0 & 0.9 & 1.3  \\
          SO\,583 & 85.66222 & 4478 $\pm$ 157 & 4.26 $\pm$ 0.33 &  7.0 $\pm$ 8.0  & 30.9 $\pm$ 1.8  & 1.0 & 1.3 & 1.3  \\
          SO\,1153 & 90.70681 & 4086 $\pm$ 63  & 3.92 $\pm$ 1.02 & 13.0 $\pm$ 6.0  & 35.0 $\pm$ 2.2 & 4.2 & 2.4 & 2.1   \\
\hline
\end{tabular} 
\tablefoot{The measured veiling at different wavelengths is reported in the column labelled 'r' followed by the wavelength in nm. We estimate uncertainties of 0.1--0.2 for r$<1.5$ and 0.3--0.4 for larger values. } 
\end{center} 
\end{table*}  

\clearpage

\appendix
\section{Information on the targets from the literature}\label{app::lit}

Membership and location of the targets within the Orion OB1 association have
been thoroughly discussed in
\citet{briceno19}. The subassociation to which each target
belongs, as well as spectral types and extinctions
from \citet{briceno19} are given in
Table \ref{tab::lit1}.
The extinctions were determined from the V-I$_c$ color,
using standard colors from \citet{kenyon95}.
Masses and luminosities of the star and accretion luminosities
and mass accretion rates have been determined for a subset of the targets.
\citet{ingleby14} used a combination of low and medium resolution
spectrographs (MagE and MIKE on the Magellan observatory)
to obtain spectra from 3400 {\AA} to 9000 {\AA} 
for a sample of CVSO stars, which included
CVSO 58, CVSO 90, CVSO 107 and CVSO 109. They
determined spectral types, veiling and extinction
following procedures similar to \citet{manara13a}. They
fitted the excess flux above the photosphere with
multicolumn accretion shock models, and obtained
accretion values given in Table \ref{tab::lit1}.
Another subset of the targets,
CVSO 104, CVSO 107, and CVSO 109,  was included
in \citet{mauco18}, who studied the disks of
CVSO stars detected by the PACS instrument on board Herschel.
\citet{mauco18} used spectral types and extinctions from
\citet{briceno19} and determined accretion luminosities and
mass accretion rates (Table \ref{tab::lit1})
from the H$\alpha$ luminosity, estimated from the equivalent width
of the line and the R$_c$ magnitude in \citet{briceno19}.
A comparison of tables \ref{tab::res}  and \ref{tab::lit1} indicates
general agreement between the results of this study and
especially those of \citet{ingleby14}, who used 
similar methods. The differences may be due to intrinsic
variability in CVSO 58, and maybe mismatch of spectral types in
CVSO 107.

Young stars located in the field of view of the $\sigma$ Ori cluster belong to one of the two stellar groups kinematically separated in radial velocities \citep{Jeffries2006}. 
One group (RV $\sim$ 27-37 km/s) is associated with the $\sigma$ Ori cluster, and the other group (RV $\sim$ 20-27 km/s) includes stars that belong to a sparse stellar population located in front of the $\sigma$ Orionis cluster \citep[e.g.,][]{Perez2018}.
Based on radial velocity measurements, the kinematic membership of the stars SO\,518 and SO\,583 was confirmed by \citet{Hernandez2014}. 
The radial velocity of 33.4 km/s obtained from the Sloan Digital Sky Survey Sky Archive Server (SDDS-SAS \footnote{https://dr16.sdss.org/infrared/spectrum/search}), suggests that the star SO1183 is also a bona fide member of the $\sigma$ Ori cluster. This radial velocity was measured from high-resolution H-band spectra using the APOGEE Stellar Parameter and Chemical Abundance Pipeline  \citep[ASPCAP;][]{Garcia2016}. The radial velocities obtained in this work (Tables \ref{tab::res_rotfit}, \ref{tab::res_rotfit_esp}, \& \ref{tab::res_rotfit_xs})
are in agreement with these previous estimates.

The youth of SO\,518, SO\,583, and SO\,1153  is also confirmed by 
lithium in absorption, infrared excesses, and strong H$\alpha$ in emission \citep{Hernandez2014,mauco2016}. In addition, they have been detected by Spitzer
\citep{hernandez07}, Hershell \citep{mauco2016}, and ALMA \citep{ansdell17}.
Stellar and accretion properties from the literature are shown in Table
\ref{tab::lit2}. 
For stars SO\,518 and SO\,583, spectral types from
\citet{hernandez14} agree within the uncertainties with the values used in this work; however, for SO\,1153 there is a difference of 1.5 sub-types. 
The spectral energy distribution (SED) of this star suggest that it is a Class I object \citep{hernandez07a}, with variability amplitude of 0.53 magnitudes \citep{cody10} that could be
due to disk obscuration. These conditions could make it  difficult to obtain more precise stellar parameters. Extinction values and stellar parameters were estimated by \citet{mauco2016}, as well as mass accretion rates, from the luminosity of the H$\alpha$ line. 
Significant differences exist between 
estimates of $A_V$, stellar luminosities, and accretion rates from the literature,
Table \ref{tab::lit2}, and those in this work,
Table \ref{tab::res}. They are mostly due to including veiling estimates in the determinations, variability, as well as to the different methods for obtaining the extinction.

\begin{table*}
\begin{center}
\footnotesize
\caption{\label{tab::lit1} Stellar and accretion properties from the literature for the targets in the Orion OB1a and OB1b Associations }
\begin{tabular}{ l c c c c c c c c c c c }
\hline \hline
Target &        2MASS & Dist. & SpT$^a$ & SpT$^b$ & A$_V^a$ & A$_V^b$ & \mstar$^b$ & \lstar$^b$ & log(\macc)$^b$& log(\macc)$^c$ & Loc \\
\hline
CVSO-17 & J05230470+0137148     &       330     & M3 & --   & 0.0 & --- & --- & --- & ----- & ----- & 1a \\
CVSO-36 & J05255035+0149370     &       330     & M3 & --   & 0.0 & --- & --- & --- & ----- & ----- & 1a \\
CVSO-58 & J05292326-0125153     &       440     & K7 & K7.5 & 1.2 & 0.8 & 0.8 & 0.8 & -7.80 & ----- & 1b \\
CVSO-90 & J05312062-0049197     &       440     & K7 & M0.5 & 0.0 & 0.0 & 0.4 & 0.3 & -8.00 & ----- & 1b \\
CVSO-104& J05320638-0111000     &       440     & K7 & --   & 0.3 & --- & --- & --- & ----- & -8.25 & 1b \\
CVSO-107& J05322578-0036533     &       440     & K7 & K7.5 & 1.4 & 0.7 & 0.8 & 0.9 & -8.60 & -8.53 & 1b \\
CVSO-109& J05323265-0113461     &       440     & M0 & K7.5 & 0.4 & 0.8 & 0.6 & 1.5 & -7.52 & -8.17 & 1b \\
CVSO-146& J05354600-0057522     &       440     & K6 & --   & 0.2 & --- & --- & --- & ----- & ----- & 1b \\
CVSO-165& J05390257-0120323     &       440     & K6 & --   & 0.6 & --- & --- & --- & ----- & ----- & 1b \\
CVSO-176& J05402414-0031213     &       440     & M3 & --   & 0.8 & --- & --- & --- & ----- & ----- & 1b \\
\hline
\end{tabular}
\tablefoot{
$^a$ \citet{briceno19},
$^b$ \citet{ingleby14},
$^c$ \citet{mauco18}.
Distances are reported in parsec.}
\end{center}
\end{table*}

\begin{table*}
\begin{center}
\footnotesize
\caption{\label{tab::lit2} Stellar and accretion properties from the literature for the targets in the $\sigma$ Ori cluster}
\begin{tabular}{ l c c c c c c c c c }
\hline \hline
Target & 2MASS & RV$^a$ & RV$^b$ & SpT$^a$ & A$_V^a$ & \mstar$^c$ & $L_*^c$ & Disk Type$^c$ & log(\macc)$^c$\\
\hline
SO\,518 & J05382725-0245096    & 28.8$\pm$0.6 & 30.3 & K6.0$\pm$1.0  & 0.0 & 0.75 & 0.44 & II &  -8.54 \\
SO\,583 & J05383368-0244141 & 29.5$\pm$0.5 & 31.5  & K4.5$\pm$1.5   & 0.0 & 1.09 & 2.57 & II & -8.13 \\
SO\,1153 & J05393982-0231217 & \nodata & 33.4 & K5.5$\pm$1.0 & 0.15 & 0.88 & 0.52  & I & -8.38  \\
\hline
\end{tabular}
\tablefoot{
$^a$ \citet{Hernandez2014},
$^b$ SDSS-SAS,
$^c$ \citet{mauco18}.
Radial Velocities are reported in km/s, stellar masses are reported in solar masses, and ages are reported in Myr.}
\end{center}
\end{table*}

\section{Observation log}\label{app:log}
The weather conditions at the time of the observations were always extremely good in this campaign, with clear (CLR) or photometric (PHO) sky conditions.  There was no major issue observed in the data, apart from the saturation of the H$\alpha$ line in the third epoch of the UVES observations of CVSO109. Typically, image quality - seeing corrected for airmass of the observations - was about 1\arcsec, as reported in Table~\ref{tab::log_xs}-\ref{tab::log_espresso}.

\begin{table*}  
\begin{center}  
\footnotesize  
\caption{\label{tab::log_xs} X-Shooter observations log}  
\begin{tabular}{l|c|c| ccc | c }    
\hline \hline  
Name               &        Date of observation [UT]    & Exp. Time  & \multicolumn{3}{c}{Slit width [\arcsec]} &      I.Q. \\
& & [Nexp x (s)] & UVB & VIS & NIR & [\arcsec] \\
\hline 
CVSO 17   &	2020-12-05T02:39:59.716	&	2x900  &   	1.0  &	0.4  &  	0.4   &	1.06 \\
CVSO 36   &	2020-12-03T02:48:38.197	&	2x900  &	1.0  &	0.4  &  	0.4  &  	0.95 \\
CVSO 58  &	2020-12-02T06:06:48.094	&	2x460  & 	1.0  &  	0.4   &    	0.4  &  	0.91 \\
CVSO 90  &	2020-12-15T01:19:38.340	&	2x460   &	1.0  &   	0.4  &     	0.4 &  	1.18 \\
CVSO 104 & 2020-11-27T03:06:13.092 & 2x370 & 1.0 & 0.4 & 0.4 & 1.15 \\
CVSO 107   & 2020-12-04T01:54:01.513	&       2x460 &   	1.0 &    	0.4 &	0.4& 	0.99 \\
CVSO 109  &      	2020-11-28T03:36:03.569	&2x	470  & 	1.0&	0.4 &	0.4&	0.89 \\
CVSO 146 &  	2020-12-09T01:50:50.818	&	2x390   &	1.0  & 	0.4 &   	0.4 & 	1.05 \\
CVSO 165  &	2020-12-14T01:50:09.831	&2x	390  & 	1.0 & 	0.4 &  	0.4 &      	1.16 \\
CVSO 176 &    	2020-12-02T04:09:15.119  &	2x900  &   	1.0 &  	0.4  &      	0.4   &    	1.11 \\
\hline
\hline
SO 518   &	2020-12-02T03:21:40.090	&	2x300  &  	0.5  &    	0.4  &      	0.4  & 	1.21 \\
SO 583  &   	2020-12-02T03:44:41.430	&   2x	300  &  	0.5  &       	0.4 &  	0.4  &    	1.11 \\
SO 1153   & 	2020-12-07T04:47:11.315	&	2x490  &   	0.5  &  	0.4  &  	0.4  &	0.79 \\
\hline 
\end{tabular} 
\tablefoot{Typical resolutions are:  in the UVB arm $R\sim$9700 and $R\sim$5400 for the 0.5\arcsec \ and 1.0\arcsec \ wide slit, respectively; in the VIS arm $R\sim$18400 and $R\sim$8900 for the 0.4\arcsec \ and 0.9\arcsec \ wide slit, respectively; in the NIR arm $R\sim$11600 and $R\sim$5600 for the 0.4\arcsec \ and 0.9\arcsec \ wide slit, respectively. Exposure times are reported in the UVB arm, similar to the NIR arms within $\sim$5-10 s, whereas the corresponding VIS exposures are $\sim$60-90 s shorter.  I.Q. is the airmass corrected seeing.}
\end{center} 
\end{table*}  

\begin{table}  
\begin{center}  
\footnotesize  
\caption{\label{tab::log_uves} UVES observations log}  
\begin{tabular}{l|c|c|c }    
\hline \hline  
Name               &        Date of observation [UT]    & Exp. Time  &       I.Q. \\
& & [Nexp x (s)] & [\arcsec] \\
\hline 

CVSO 17  & 2020-12-04T04:02:13.793	& 2x 1800 &	0.81       \\
CVSO 17  & 2020-12-05T02:16:35.606	& 2x 1800 &	0.97       \\
CVSO 17  & 2020-12-06T01:53:46.058	& 2x 1800 &	1.15       \\
\hline
CVSO 36  & 2020-12-02T03:59:25.264	& 2x 1800 &	1.04       \\
CVSO 36  & 2020-12-03T02:32:58.168	& 2x 1800 &	0.88       \\
CVSO 36  & 2020-12-04T02:56:47.849	& 2x 1800 &	0.99       \\
\hline
CVSO 58  & 2020-11-30T02:23:51.801	& 2x 1300 &	1.24       \\
CVSO 58  & 2020-12-01T02:22:57.299	& 2x 1300 &	1.04       \\
CVSO 58  & 2020-12-02T02:09:45.709	& 2x 1300 &	1.23       \\
\hline
CVSO 104 & 2020-11-25T03:09:25.349	& 2x 1100 &	1.12       \\
CVSO 104 & 2020-11-26T02:48:15.196	& 2x 1100 &	1.13       \\
CVSO 104 & 2020-11-27T02:22:14.815	& 2x 1100 &	1.10     \\
\hline
CVSO 107 & 2020-12-03T03:38:39.578	& 2x 1250 &	1.31       \\
CVSO 107 & 2020-12-04T02:10:09.193	& 2x 1250 &	1.00     \\
CVSO 107 & 2020-12-05T03:22:18.754	& 2x 1250 &	1.23       \\
\hline
CVSO 109 & 2020-11-26T03:29:45.199	& 2x 1050 &	0.91       \\
CVSO 109 & 2020-11-27T03:02:45.220	& 2x 1050 &	0.98       \\
CVSO 109 & 2020-11-28T03:59:35.437	& 2x 1050 &	0.88       \\
\hline
CVSO 176 & 2020-11-28T02:52:44.394	& 2x 1800 &	0.98       \\
CVSO 176 & 2020-11-29T02:35:28.950	& 2x 1800 &	0.88       \\
CVSO 176 & 2020-11-30T03:30:00.408	& 2x 1800 &	1.15       \\
\hline
\hline
SO\,518    & 2020-11-29T03:40:36.060	& 2x 1100 &	0.86       \\
SO\,518    & 2020-11-30T04:35:15.871	& 2x 1100 &	1.07       \\
SO\,518    & 2020-12-01T03:28:58.125	& 2x 1100 &	0.91       \\
\hline
SO\,583    & 2020-11-29T04:22:04.942	& 2x 420 &	0.69       \\
SO\,583    & 2020-11-30T03:11:49.882	& 2x 420 &	1.06       \\
SO\,583    & 2020-12-01T03:10:36.338	& 2x 420 &	0.84       \\

\hline 
\end{tabular} 
\tablefoot{I.Q. is the airmass corrected seeing.}
\end{center} 
\end{table}

\begin{table}  
\begin{center}  
\footnotesize  
\caption{\label{tab::log_espresso} ESPRESSO observations log}  
\begin{tabular}{l|c|c|c }    
\hline \hline  
Name               &        Date of observation [UT]    & Exp. Time  &       I.Q. \\
& & [Nexp x (s)] & [\arcsec] \\
\hline 
CVSO 146 & 2020-12-09T02:42:01.462	& 2x 1205  &	1.26   \\
CVSO 146 & 2020-12-10T02:09:50.465	& 2x 1205  &	1.17   \\
CVSO 146 & 2020-12-11T01:55:43.819	& 2x 1205  &	1.21   \\
\hline
CVSO 165 & 2020-12-13T01:35:50.344	& 2x 1080  &	       \\
CVSO 165 & 2020-12-14T02:18:48.659	& 2x 1080  &	       \\
CVSO 165 & 2020-12-15T02:30:53.897	& 2x 1080  &	       \\
\hline
CVSO 90  & 2020-12-15T03:11:36.839	& 2x 1670  &	       \\
\hline
\hline
SO\,1153   & 2020-12-08T07:38:54.786	& 2x 1330  &	       \\
SO\,1153   & 2020-12-09T01:52:07.779	& 2x 1330  &	0.98   \\
SO\,1153   & 2020-12-10T02:55:38.503	& 2x 1330  &	0.89   \\
\hline 
\end{tabular} 
\tablefoot{I.Q. is the airmass corrected seeing.}
\end{center} 
\end{table}

\begin{figure*}[]
\centering
\includegraphics[width=0.4\textwidth]{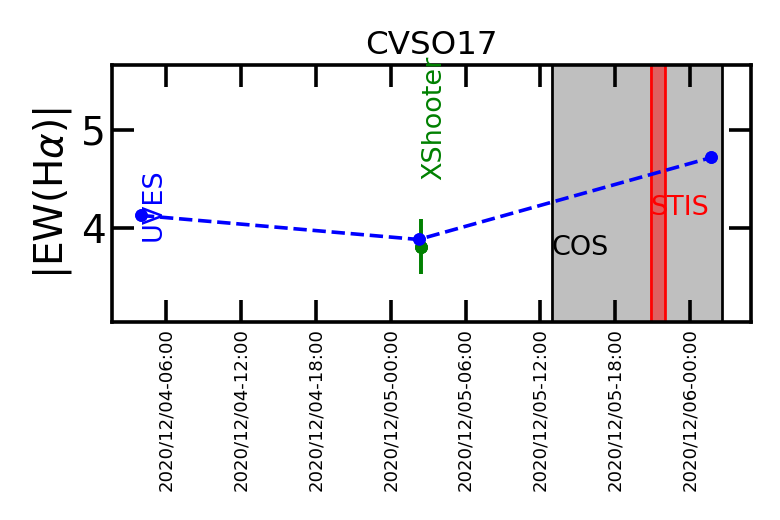}
\includegraphics[width=0.4\textwidth]{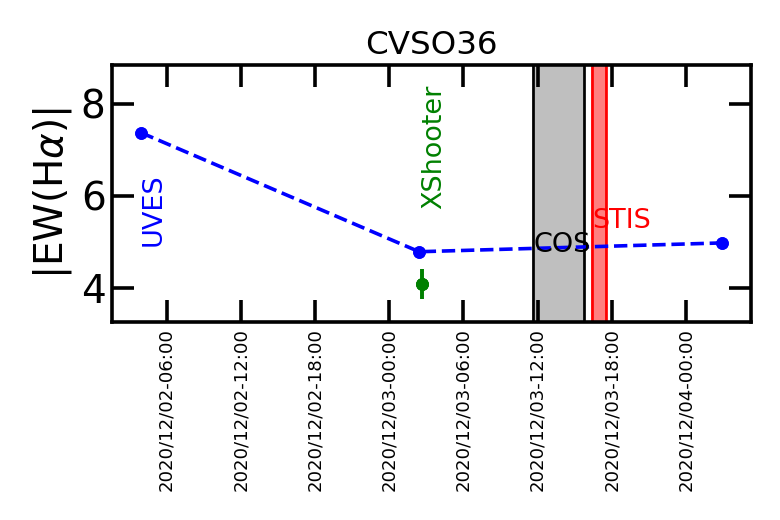}
\includegraphics[width=0.4\textwidth]{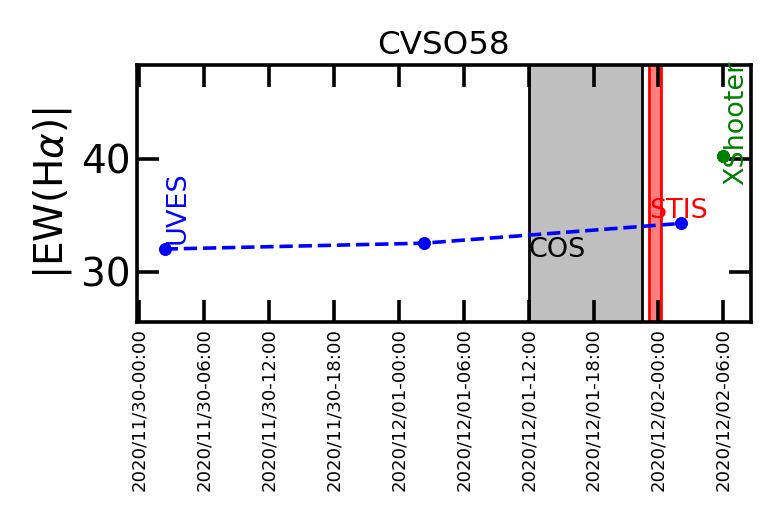}
\includegraphics[width=0.4\textwidth]{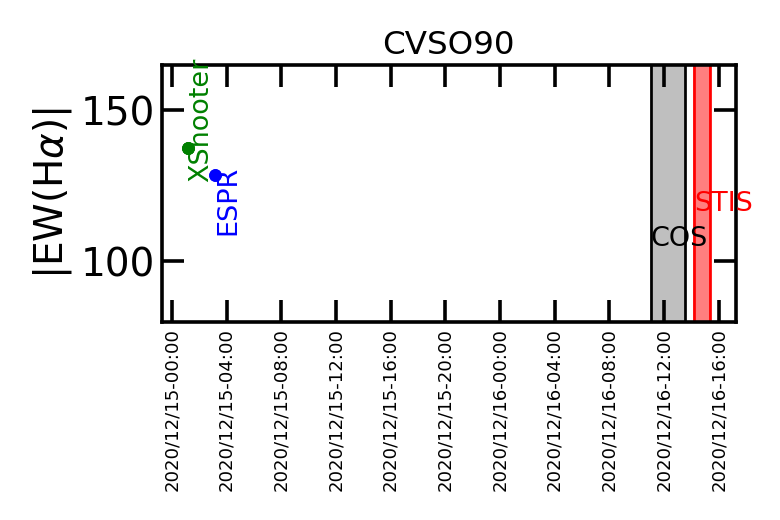}
\includegraphics[width=0.4\textwidth]{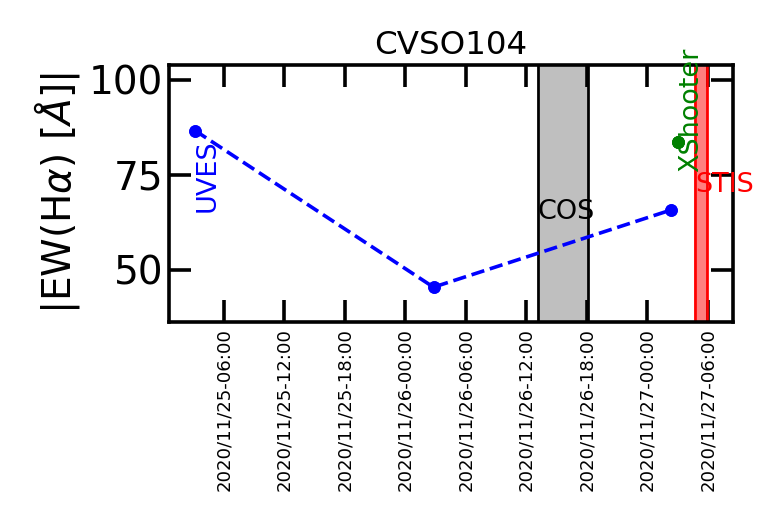}
\includegraphics[width=0.4\textwidth]{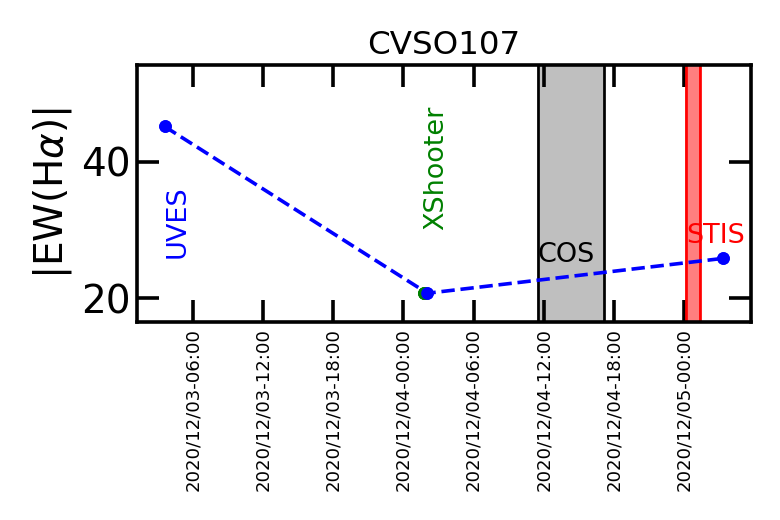}
\includegraphics[width=0.4\textwidth]{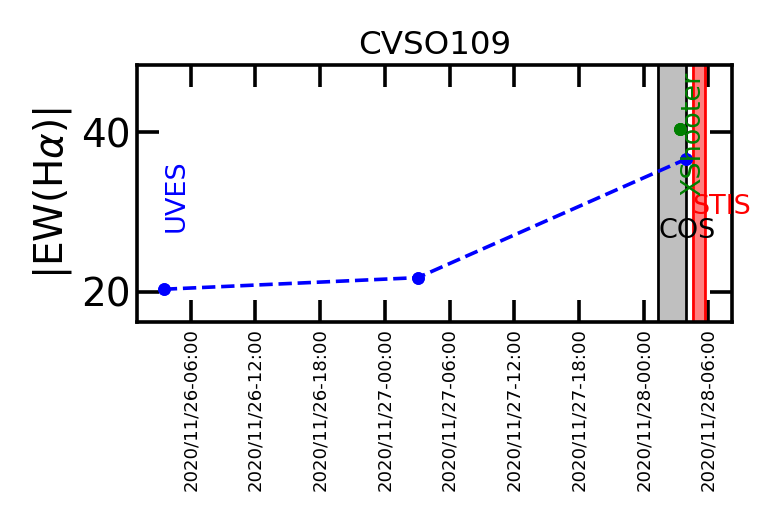}
\includegraphics[width=0.4\textwidth]{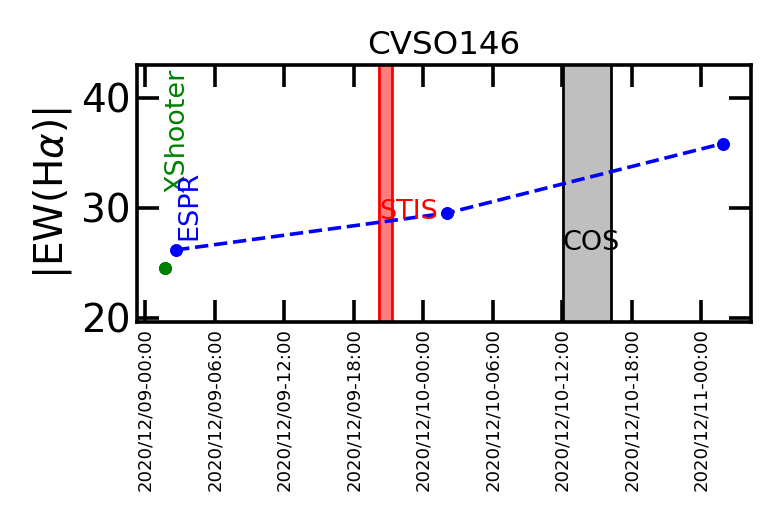}
\includegraphics[width=0.4\textwidth]{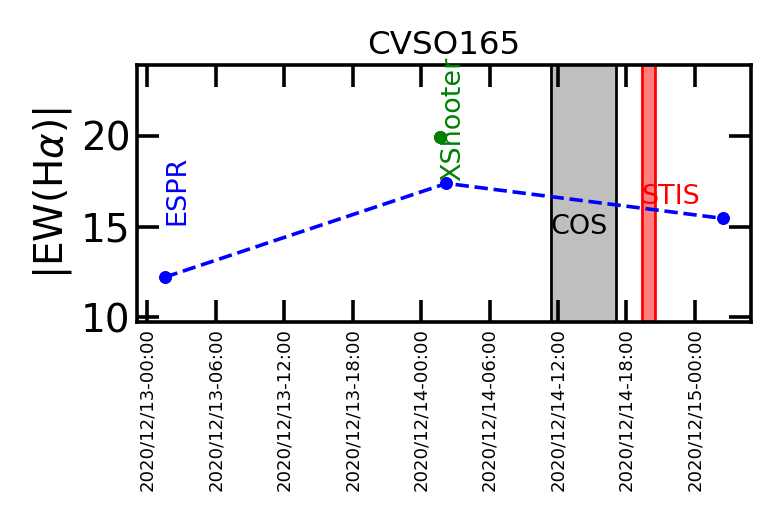}
\includegraphics[width=0.4\textwidth]{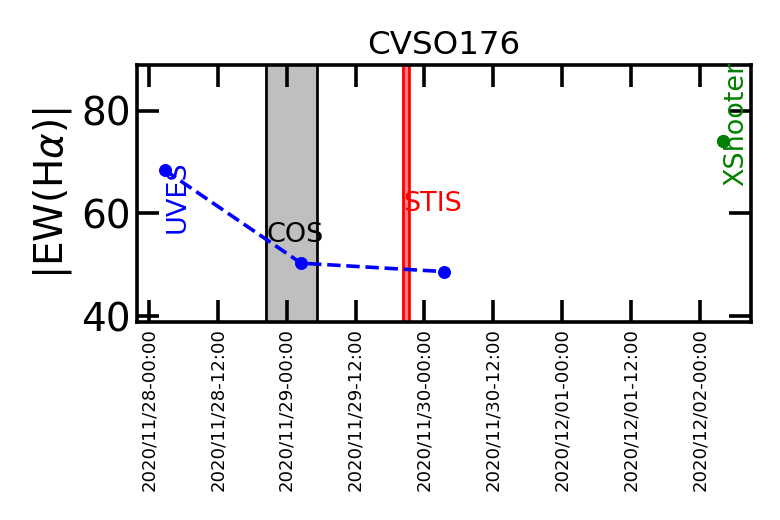}
\caption{Variability of the H$\alpha$ equivalent width for the Orion OB1 targets as measured by EPRESSO or UVES (blue) and X-Shooter (green), together with the timining of the COS (grey) and STIS (red) HST observations.
     \label{fig::ew_ha_var}}
\end{figure*}

\begin{figure*}[]
\centering
\includegraphics[width=0.4\textwidth]{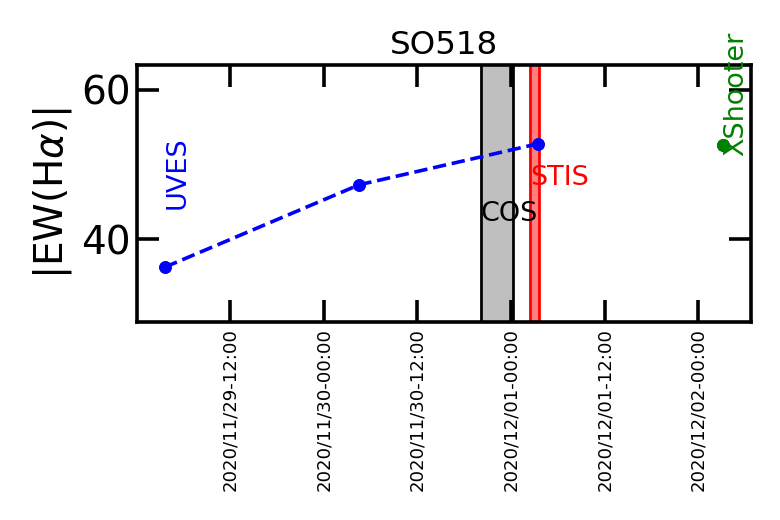}
\includegraphics[width=0.4\textwidth]{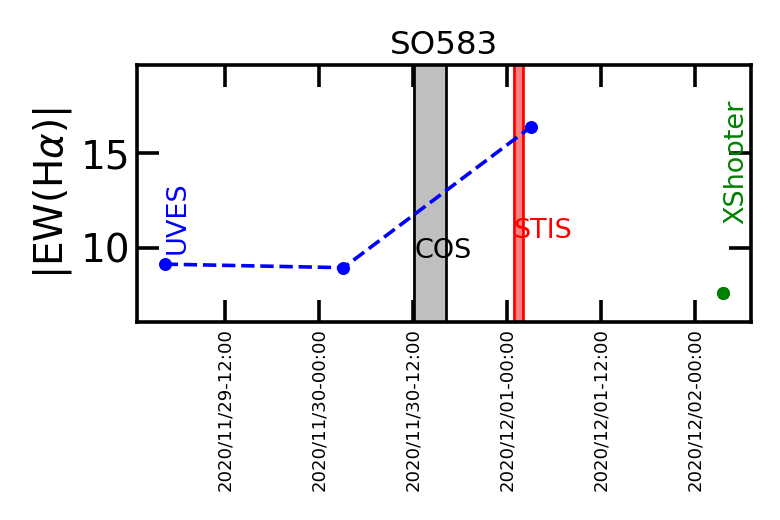}
\includegraphics[width=0.4\textwidth]{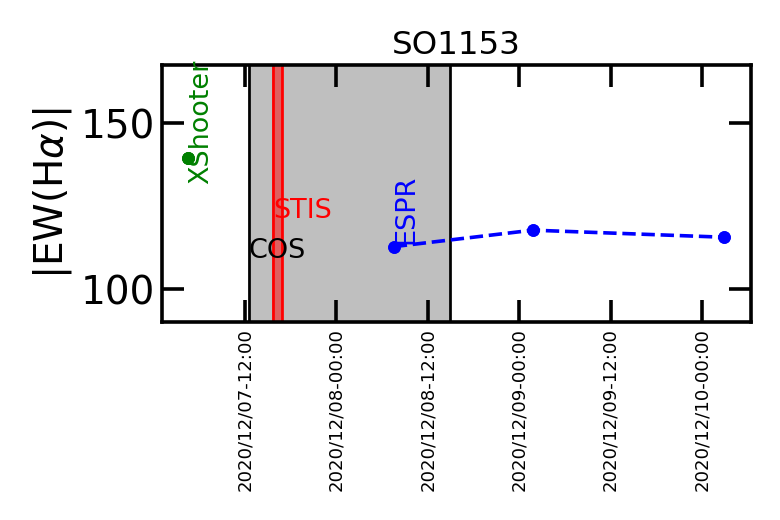}
\caption{Variability of the H$\alpha$ equivalent width for the $\sigma$-Orionis targets as measured by EPRESSO or UVES (blue) and X-Shooter (green), together with the timining of the COS (grey) and STIS (red) HST observations.
     \label{fig::ew_ha_var_sori}}
\end{figure*}

Figures~\ref{fig::ew_ha_var}-\ref{fig::ew_ha_var_sori} show the time at which the targets have been observed, together with highlights of the time of start and end of the HST observations.

\section{Templates used for the ROTFIT analysis}\label{sect::templ_rotfit}
As described in Sect.~\ref{sect::method_rotfit}, a number of main sequence stars observed with the HARPS spectrograph are used to fit the high-resolution spectra of our targets. Table~\ref{Tab:HARPS_templates} reports the information about these targets, collected from \citet{harps_templates}.

\begin{table*}[htb]
\footnotesize
\begin{center}
\caption{HARPS templates.}
\begin{tabular}{llccrrrl}
\hline
\hline
\noalign{\smallskip}
 Name       & SpT    & \teff  & \logg &  [Fe/H] & RV & \vsini   & References \\  
            &         &  [K]   & \multicolumn{2}{c}{[dex]}  & \multicolumn{2}{c}{[km/s]} &      \\  
\hline
\noalign{\smallskip}
 HD4628     & K2V    &  5035  & 4.60  & $-0.27$  & $ -10.165$ & $ 0.7 $  &  Lu2018, Br2016    \\ 
 HD16160    & K3V    &  4858  & 4.89  & $-0.18$  & $  25.874$ & $ 0.9 $  &  Ri2017, Mi2012    \\ 
 HD32147    & K3V    &  4780  & 4.56  & $ 0.37$  & $  21.624$ & $ 1.7 $  &  Lu2018, Va2005    \\ 
 HD24916    & K4V    &  4696  & 4.50  & $-0.12$  & $  3.64  $ & $ 3.5 $  &  Mo2018, Gl2005    \\ 
 HD35171    & K4V    &  4576  & 4.70  & $ 0.02$  & $  38.267$ & $ 3.4 $  &  Mi2012, Mi2012    \\    
 HD154363   & K5V    &  4373  & 4.66  & $-0.32$  & $  34.146$ & $ 2.5 $  &  Lu2017             \\ 
 HD218511   & K5.5V  &  4361  & 4.61  & $-0.09$  & $   4.654$ & $ 1.0 $  &  Lu2018             \\ 
 HD200779   & K6V    &  4406  & 4.62  & $ 0.08$  & $ -66.85 $ & $ 2.0 $  &  Lu2018, Hou2016   \\   
 HD35650    & K6V    &  4269  & 4.65  & $ 0.06$  & $  32.264$ & $ 4.0 $  &  Lu2018, To2006    \\                       
 HIP116384  & K7V    &  4180  & 4.70  & $-0.10$  & $ -10.3  $ & $<2.0 $  &  Zb1998, Hoj2019   \\       
 HIP17157   & K7V    &  4128  & 4.28  & $ 0.08$  & $  23.90 $ & $ 5.0 $  &  Mc2017, Gr2020    \\               
 BD-08\,2582& K8V    &  4044  & 4.66  & $ 0.13$  & $  37.035$ & $<2.0 $  &  Li2017, Hoj2019   \\               
 GJ488      & M0V    &  3989  & 4.66  & $ 0.24$  & $   5.041$ & $ 2.7 $  &  Ye2017, Hoj2019   \\               
 HD209290   & M0.5V  &  3914  & 4.69  & $ 0.05$  & $  18.363$ & $<2.0 $  &  Pa2018, Je2018    \\                      
 HD42581    & M1V    &  3822  & 4.71  & $ 0.06$  & $   4.734$ & $ 2.6 $  &  Pa2018, Hou2016   \\ 
 GJ514      & M1.0V  &  3727  & 4.78  & $ 0.07$  & $  14.606$ & $ 2.1 $  &  Li2017, Hou2016   \\                     
 HD165222   & M1.5V  &  3664  & 4.87  & $-0.21$  & $  33.045$ & $ 1.6 $  &  Pa2018, Hou2016   \\    
 HD119850   & M1.5V  &  3677  & 4.79  & $-0.04$  & $  15.570$ & $ 1.0 $  &  Pa2018, Hou2016   \\   
 HD217987   & M2V    &  3680  & 4.88  & $-0.22$  & $   8.809$ & $ 1.0 $  &  Wo2005, To2006    \\                    
 HIP51317   & M2V    &  3575  & 4.89  & $-0.09$  & $   8.35 $ & $ 0.1 $  &  Pa2018, Ga2016    \\                       
 GJ2066     & M2V    &  3589  & 4.86  & $-0.06$  & $  62.215$ & $<2.0 $  &  Pa2018, Je2018    \\  
 GJ250B     & M2.5V  &  3569  & 4.84  & $ 0.01$  & $  -5.40 $ & $<2.5 $  &  Ro2012, Re2012    \\                       
 GJ752A     & M3V    &  3557  & 4.86  & $ 0.00$  & $  35.737$ & $ 1.2 $  &  Pa2018, Hou2016   \\               
 GJ581      & M3.5V  &  3441  & 4.98  & $-0.08$  & $  -9.662$ & $<2.0 $  &  Sc2019, Re2018    \\                       
 GJ105B     & M4V    &  3392  & 4.81  & $ 0.05$  & $  26.815$ & $ 2.4 $  &  Pa2019, Hou2016   \\                     
 GJ699      & M4.5V  &  3278  & 5.10  & $-0.12$  & $-110.506$ & $ 2.0 $  &  Pa2018, Hou2016   \\                     
 GJ447      & M4.5V  &  3251  & 5.10  & $-0.04$  & $ -31.087$ & $ 2.2 $  &  Pa2018, Hou2016   \\                     
 GJ83.1     & M5.0V  &  3185  & 5.15  & $-0.18$  & $ -28.832$ & $<2.0 $  &  Pa2018, Je2018    \\                      
 GJ1002     & M5.5V  &  3038  & 5.04  & $-0.10$  & $ -40.058$ & $<2.0 $  &  Pa2019, Je2018    \\                      
 GJ551      & M5.5V  &  2927  & 5.02  & $-0.07$  & $ -20.471$ & $ 2.7 $  &  Pa2016, To2016    \\                    
 Wolf359    & M6.5V  &  2900  & 5.40  & $ 0.18$  & $  19.413$ & $<2.0 $  &  Ra2018, Je2018    \\                       
\hline
\noalign{\smallskip}
\end{tabular}
\end{center}
Br2016 = \citet{Brewer2016}; Ga2016 = \citet{Gagne2016}; Gl2005 = \citet{Glebocki2005}; Gr2020 = \citet{harps_templates}; 
Hoj2019 = \citet{Hojjatpanah2019}; Hou2016 = \citet{Houdebine2016}; Je2018 = \citet{Jeffers2018}; Li2017 = \citet{Lindgren2017}; Lu2017 = \citet{Luck2017}; Lu2018 = \citet{Luck2018};
Mc2017 = \citet{McDonald2017}; Mi2012 = \citet{Mishenina2012}; Mo2018 = \citet{Montes2018}; Pa2016 = \citet{Passegger2016}; Pa2018 = \citet{Passegger2018};
Pa2019 = \citet{Passegger2019}; Ra2018 = \citet{Rajpurohit2018}; Re2012 = \citet{Reiners2012}; Re2018 = \citet{Reiners2018}; Ri2017 = \citep{Rich2017}; 
Ro2012 = \citet{Rojas-Ayala2012}; Sc2019 = \citet{Schweitzer2019}; To2006 = \citet{Torres2006}; Va2005 = \citet{Valenti2005}; Wo2005 = \citet{Woolf2005};
 Ye2017 = \citet{Yee2017}; Zb1998 = \citet{Zboril1998}.
\label{Tab:HARPS_templates}
\end{table*}

\section{Light curves from multi-band photometry}\label{sect::lc_OACT}
\begin{figure*}[]
\centering
\hspace{-1cm}
\includegraphics[width=8.7cm]{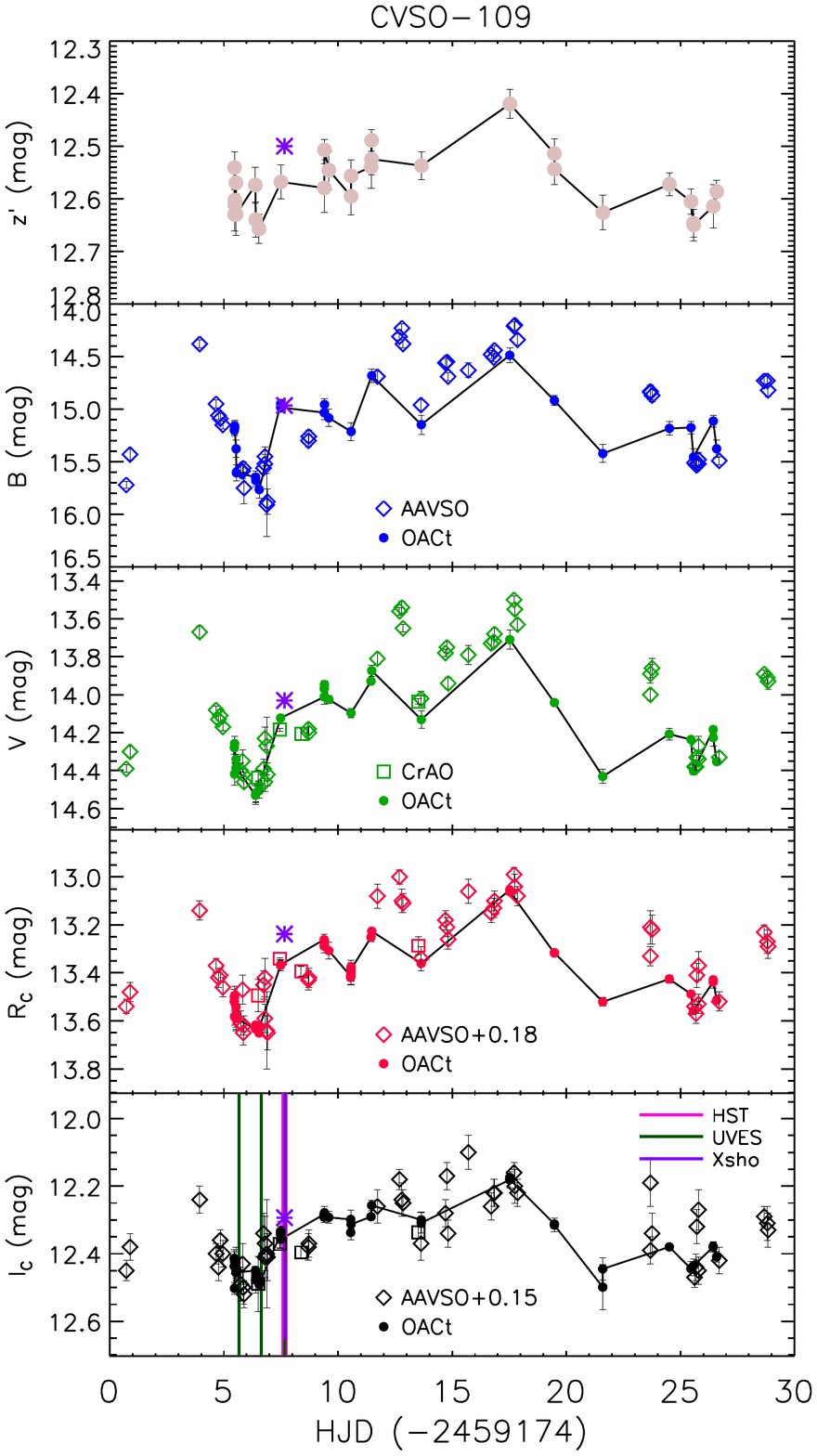}
\hspace{-3cm}
\includegraphics[width=6.2cm]{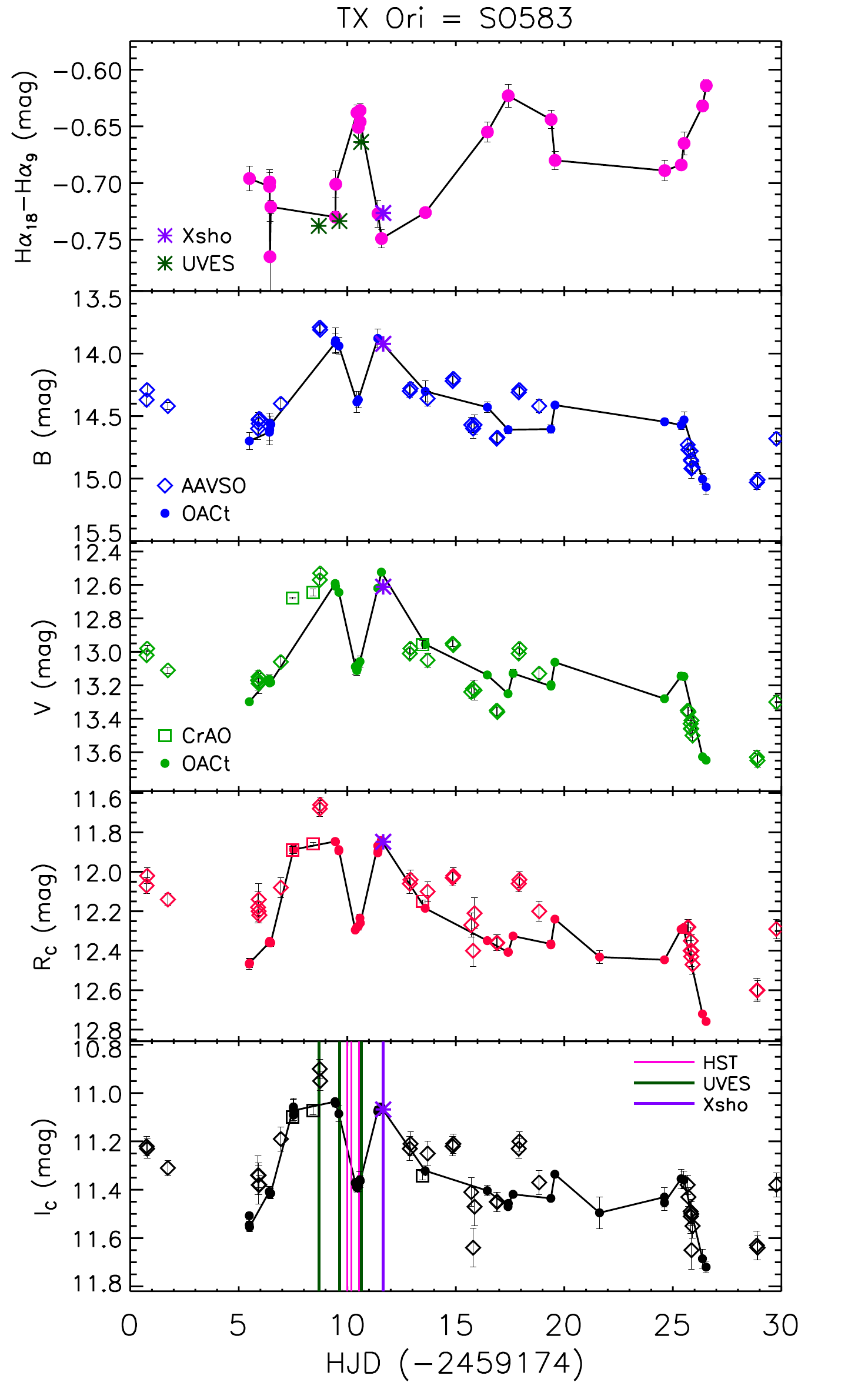}
\hspace{-.7cm}
\includegraphics[width=6.2cm]{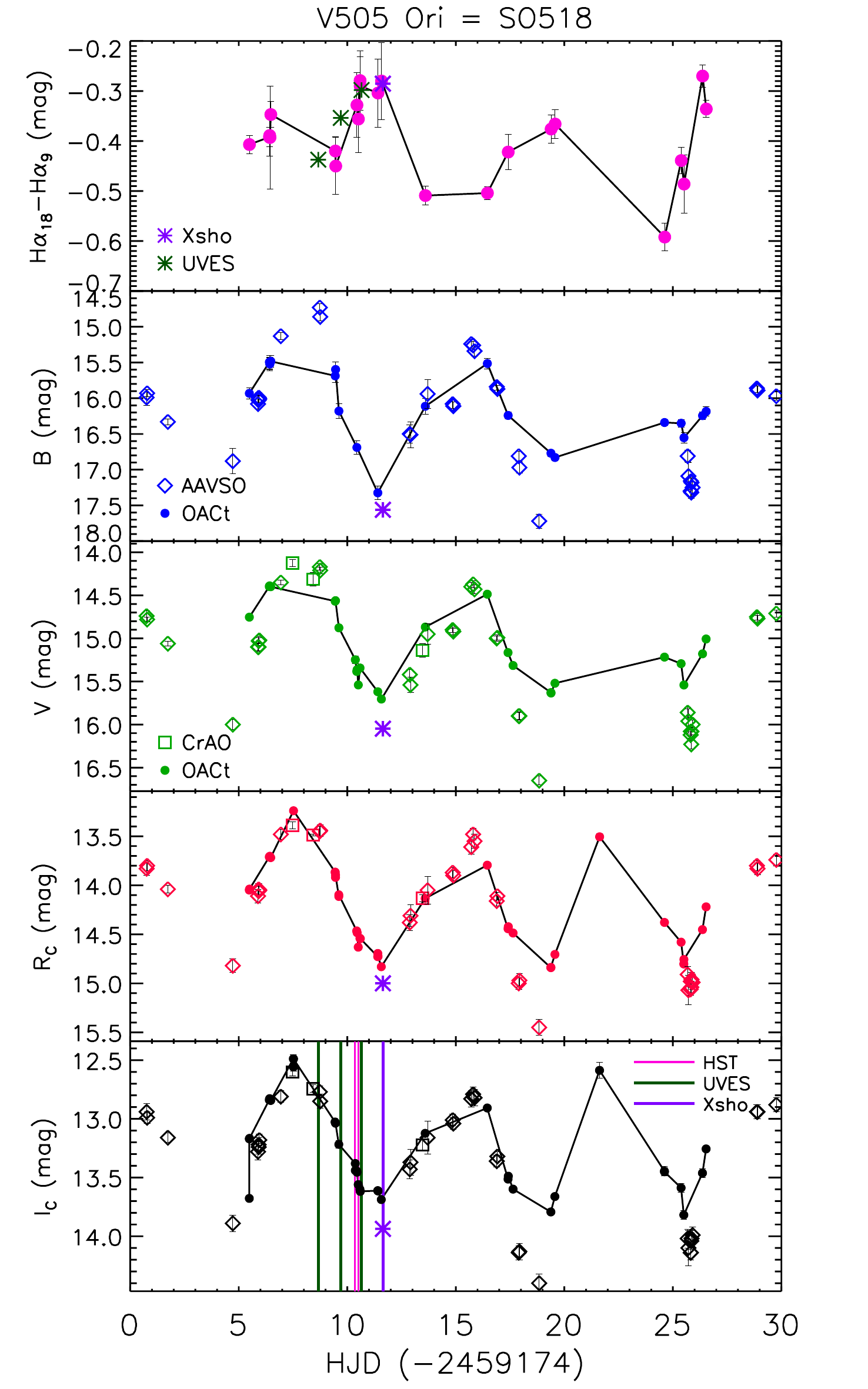}
\caption{Light curves for our targets around the time of our VLT observations. Synthetic photometry on the X-Shooter spectra is shown with purple asterisks.
     \label{fig::lc_phot}}
\end{figure*}

We show in Fig.~\ref{fig::lc_phot}
the plots with the light curves from OACT, CrAO and AAVSO from mid-November to mid-December.  
The epochs of HST and VLT observations are also marked as vertical lines in the boxes with $I_C$ light curves.
The synthetic photometry made on the flux-calibrated X-Shooter spectra is overlaid with purple asterisks. For SO\,583 and SO\,518, we display in the upper panels the index $H\alpha_{18}$--$H\alpha_{9}$, which measures the intensity of the H$\alpha$ emission, from OACT narrow-band photometry and from UVES (green asterisks) and X-Shooter (purple asterisks) spectra. 

\section{Plots of the Balmer jump fits}\label{app::plot_bj}

\begin{figure*}[]
\centering
\includegraphics[width=0.4\textwidth]{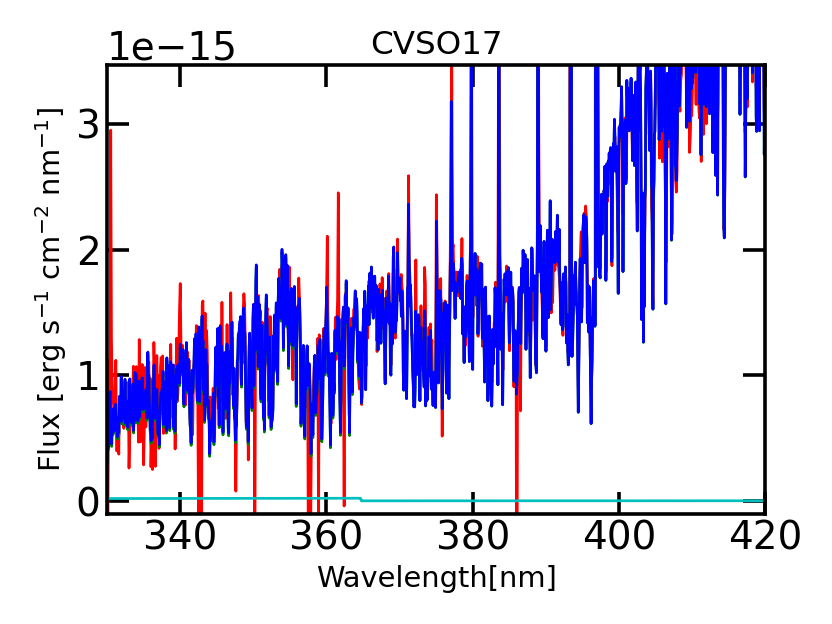}
\includegraphics[width=0.4\textwidth]{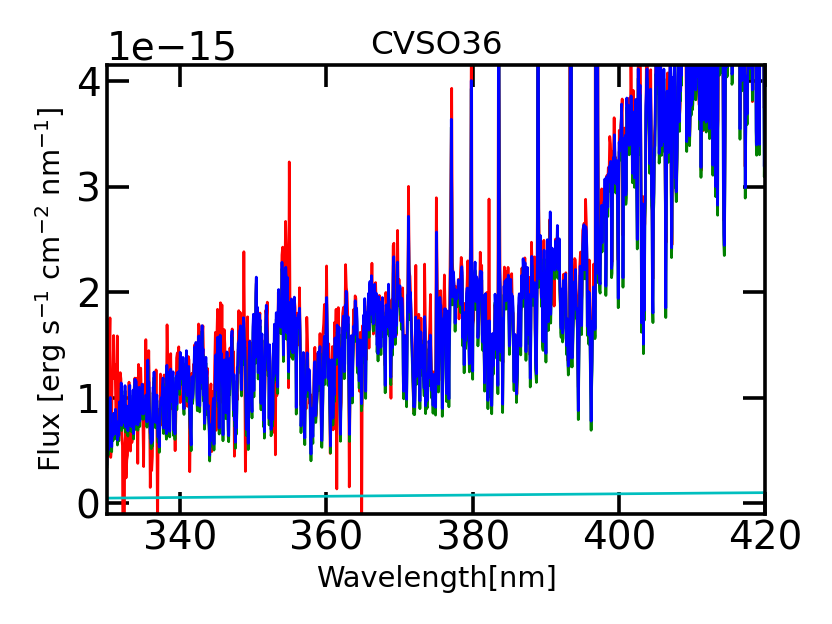}
\includegraphics[width=0.4\textwidth]{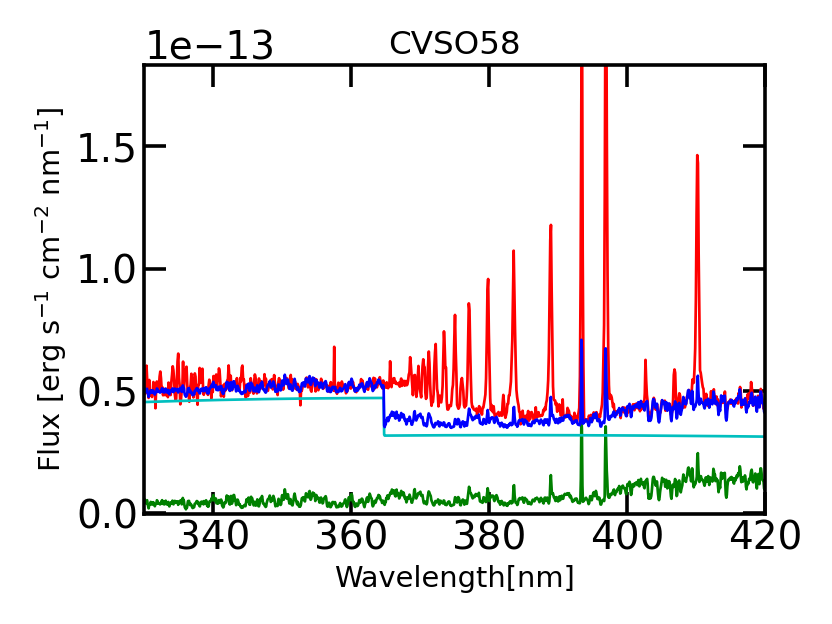}
\includegraphics[width=0.4\textwidth]{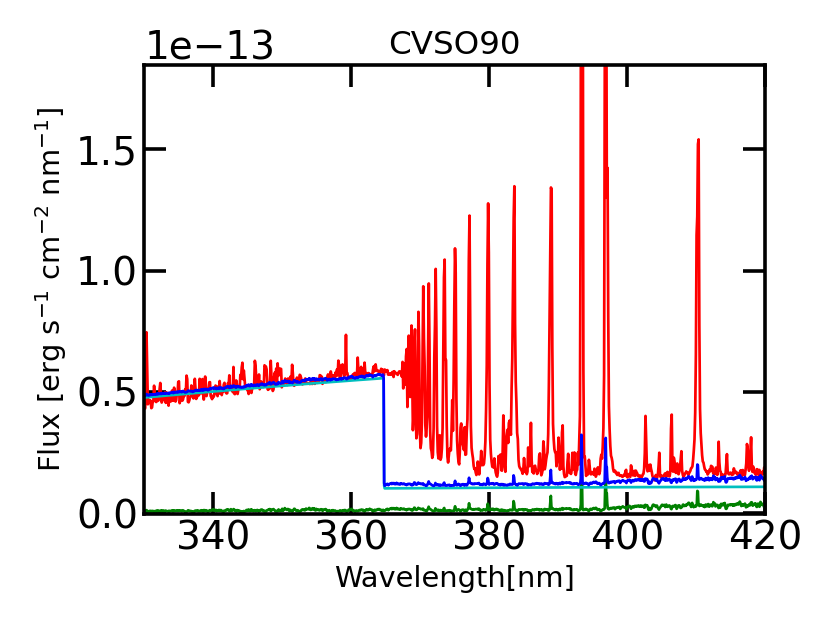}
\includegraphics[width=0.4\textwidth]{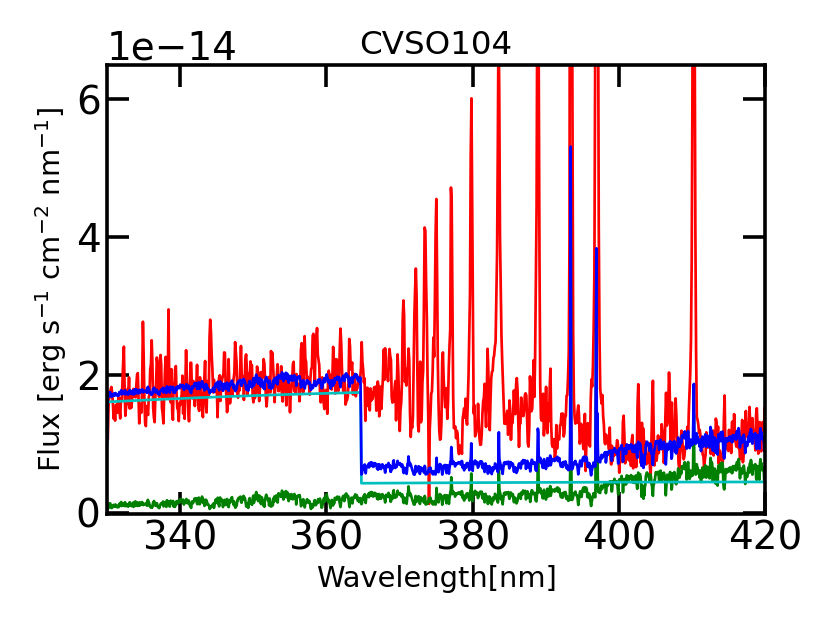}
\includegraphics[width=0.4\textwidth]{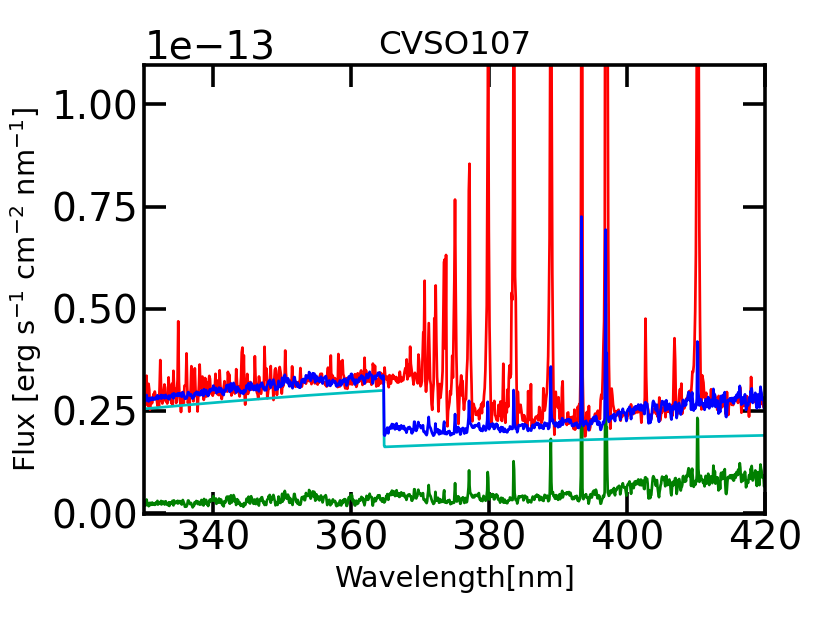}
\includegraphics[width=0.4\textwidth]{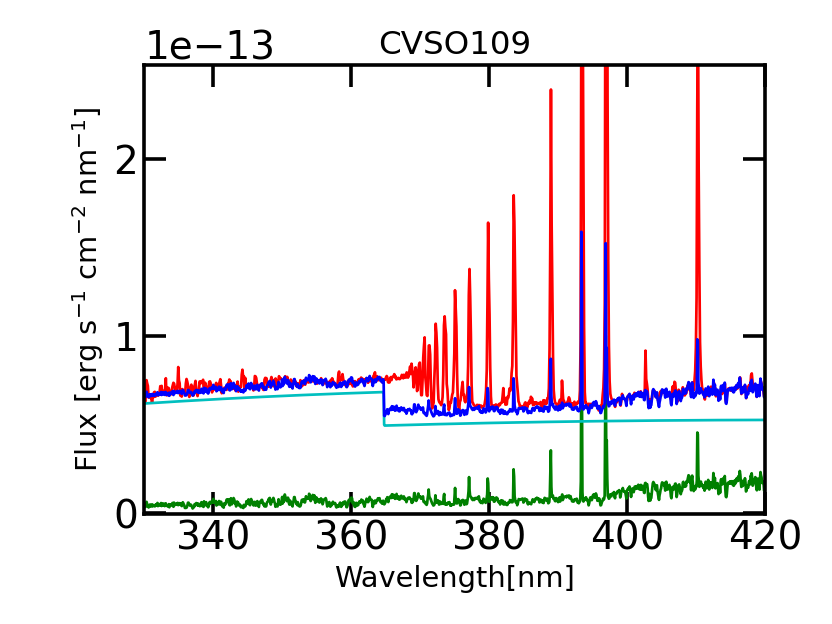}
\includegraphics[width=0.4\textwidth]{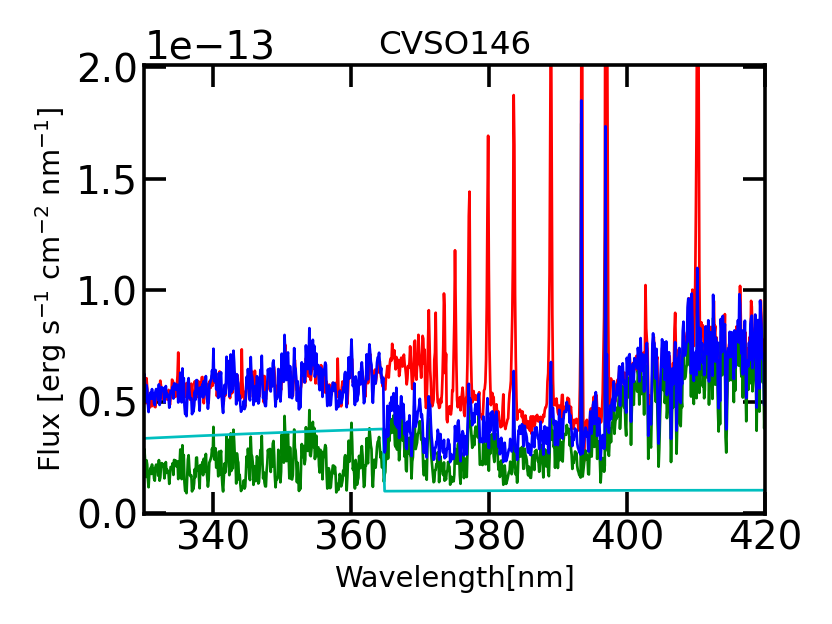}
\caption{Best fit for the Balmer continuum region for the targets in the OB1 association.
     \label{fig::fit_bj_ob1}}
\end{figure*}

\begin{figure*}[]
\centering
\includegraphics[width=0.4\textwidth]{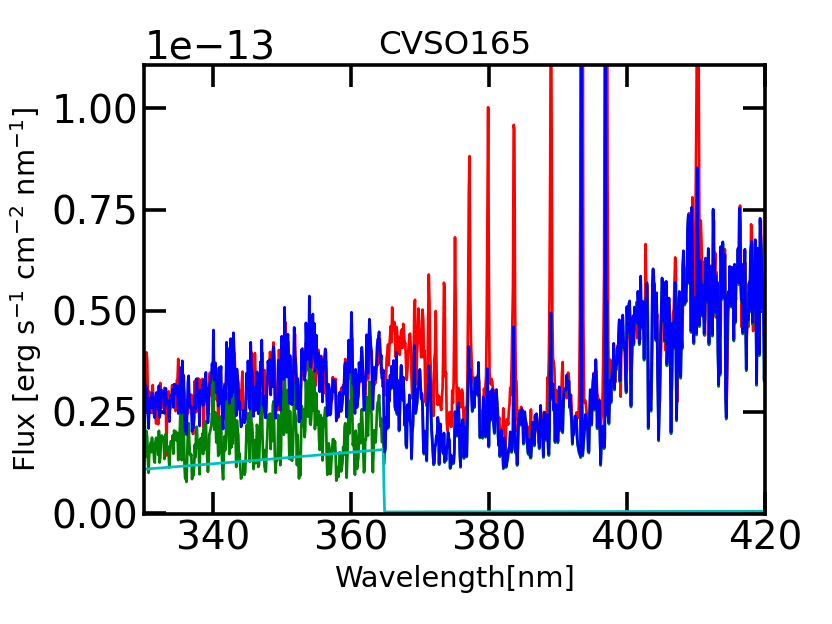}
\includegraphics[width=0.4\textwidth]{CVSO176_BJ_rebin.png}
\includegraphics[width=0.4\textwidth]{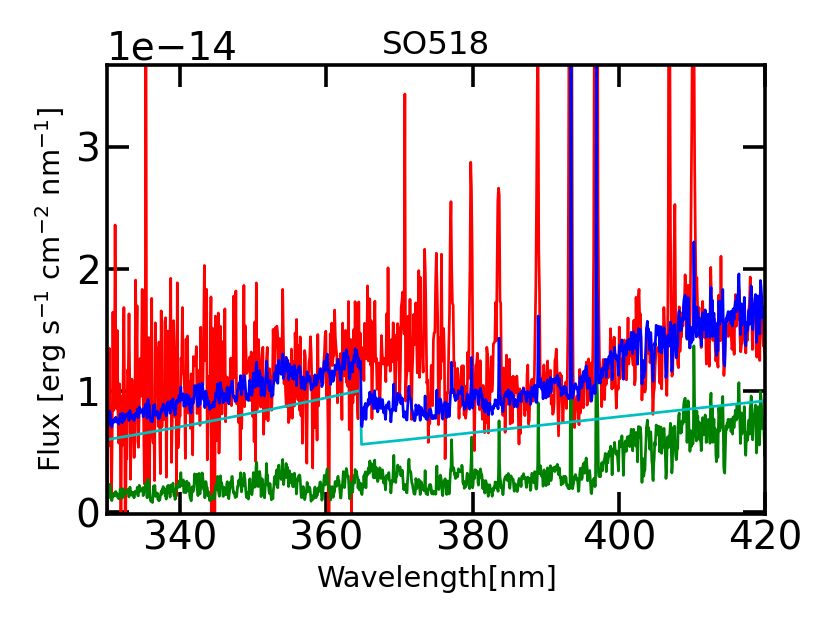}
\includegraphics[width=0.4\textwidth]{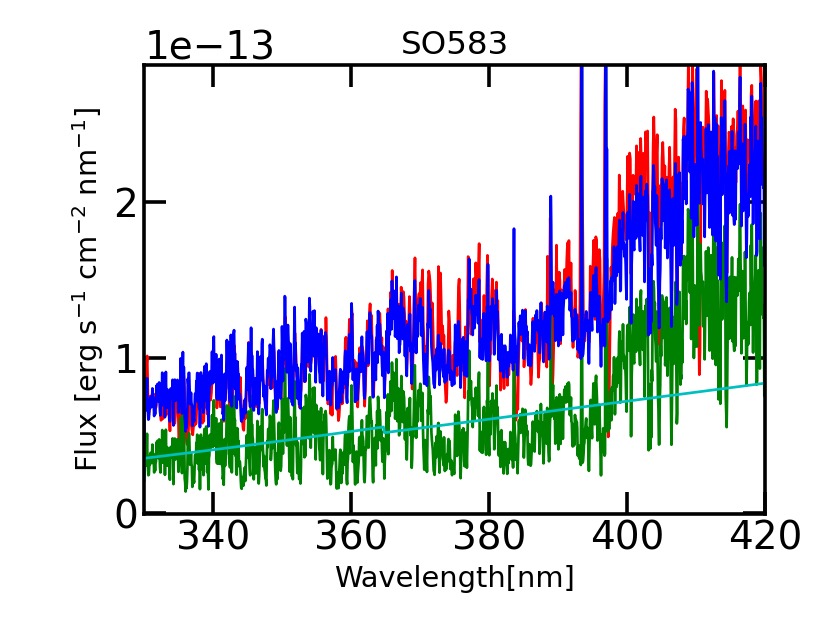}
\includegraphics[width=0.4\textwidth]{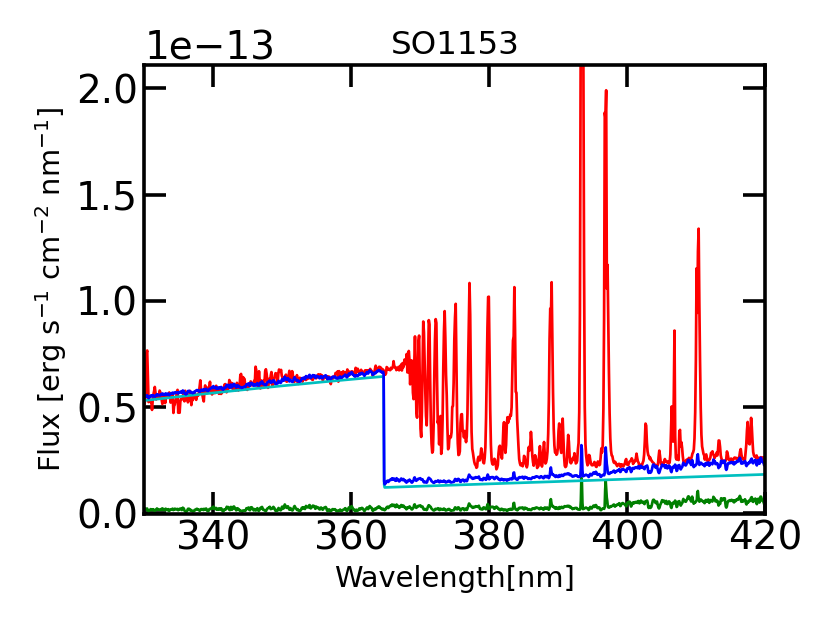}
\caption{Best fit for the Balmer continuum region for the targets in the OB1 association and $\sigma$-Orionis cluster.
     \label{fig::fit_bj_sOri}}
\end{figure*}

\begin{figure*}[]
\centering
\includegraphics[width=0.4\textwidth]{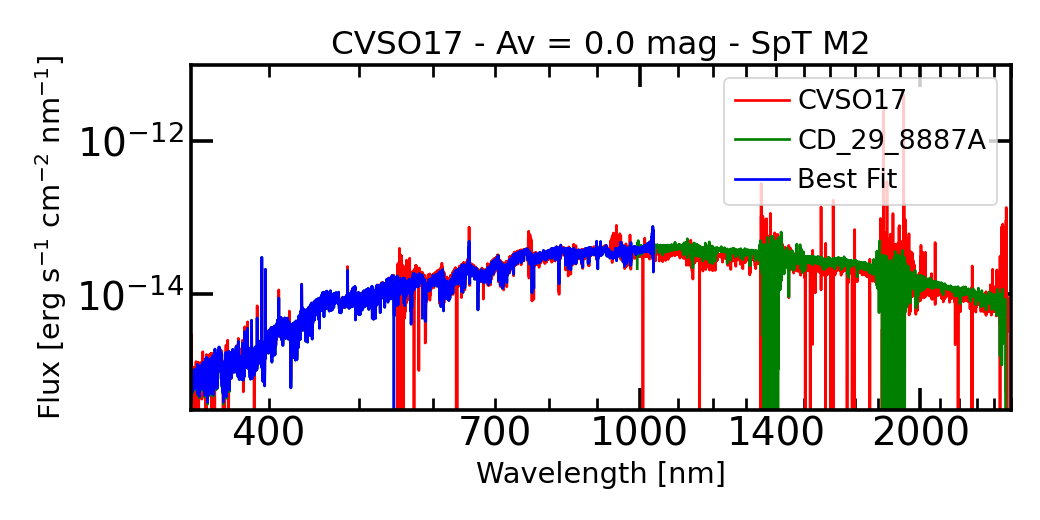}
\includegraphics[width=0.4\textwidth]{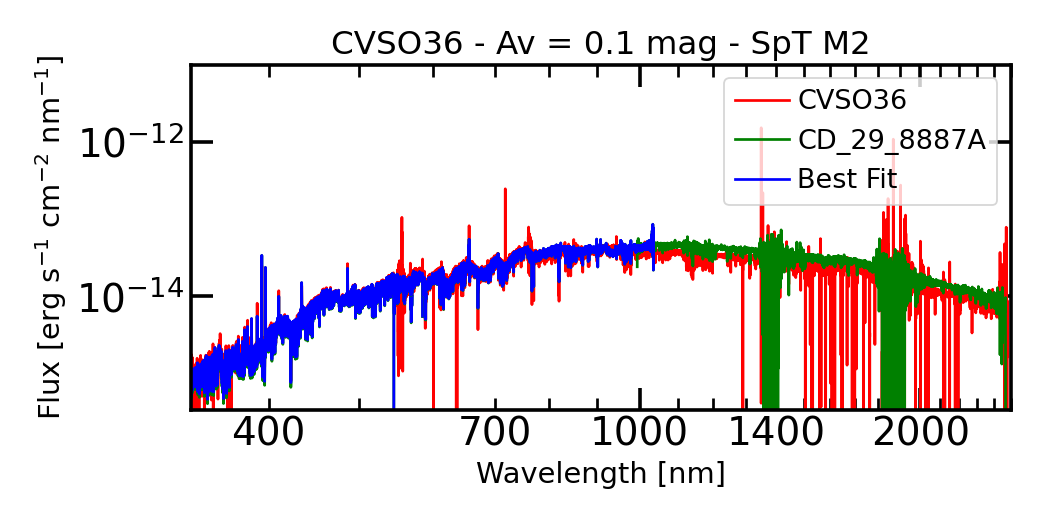}
\includegraphics[width=0.4\textwidth]{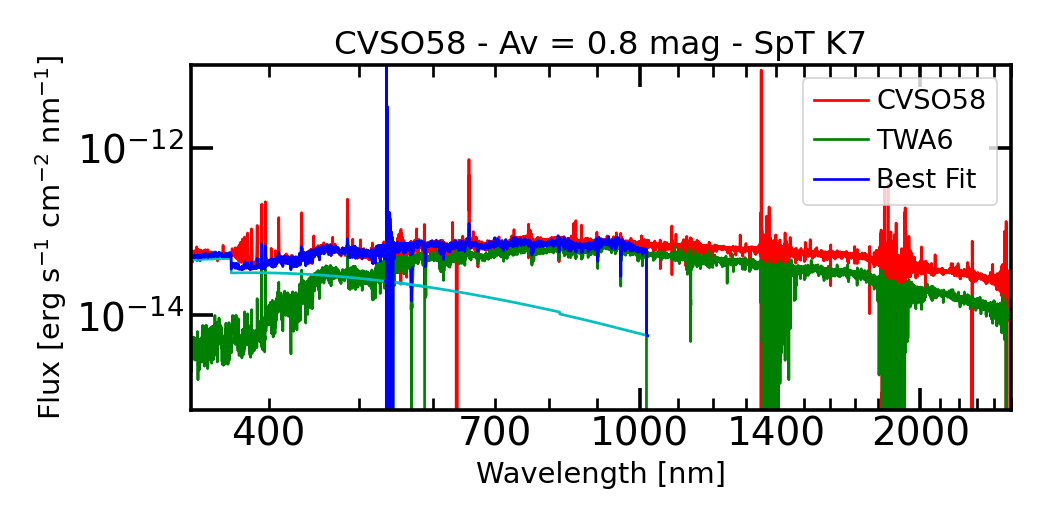}
\includegraphics[width=0.4\textwidth]{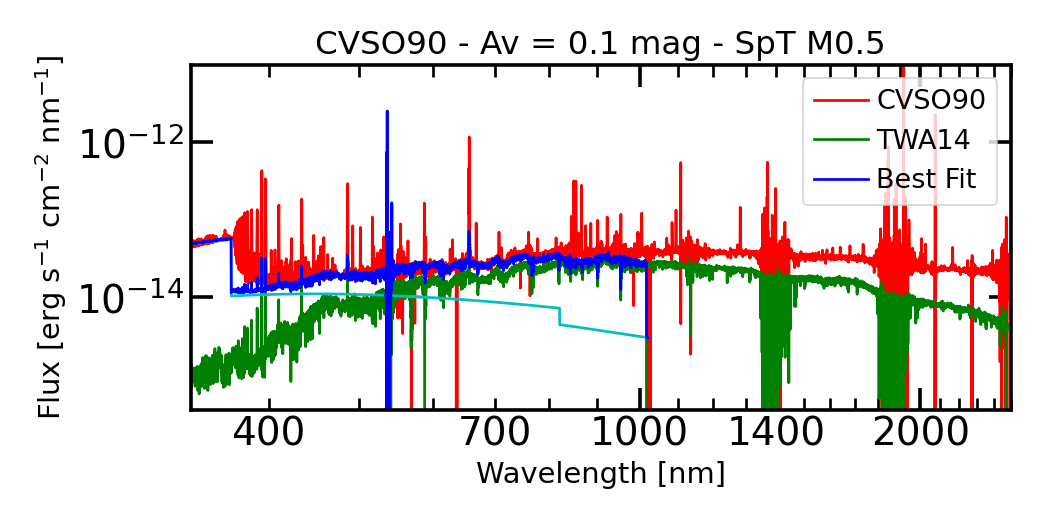}
\includegraphics[width=0.4\textwidth]{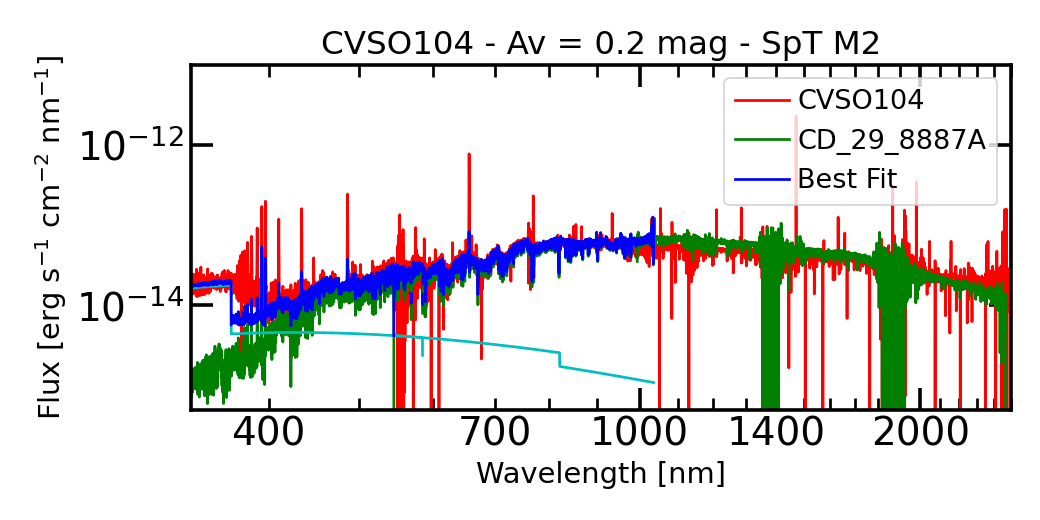}
\includegraphics[width=0.4\textwidth]{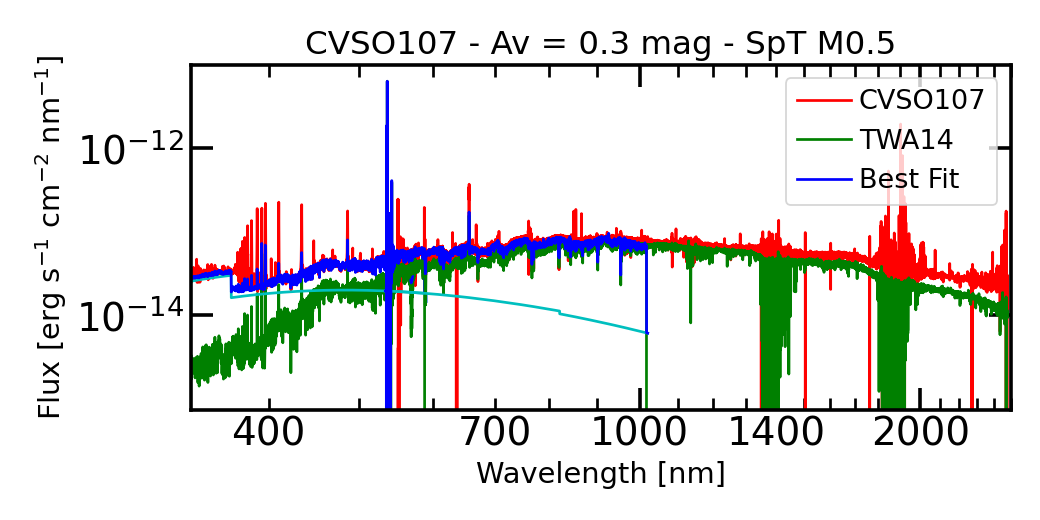}
\includegraphics[width=0.4\textwidth]{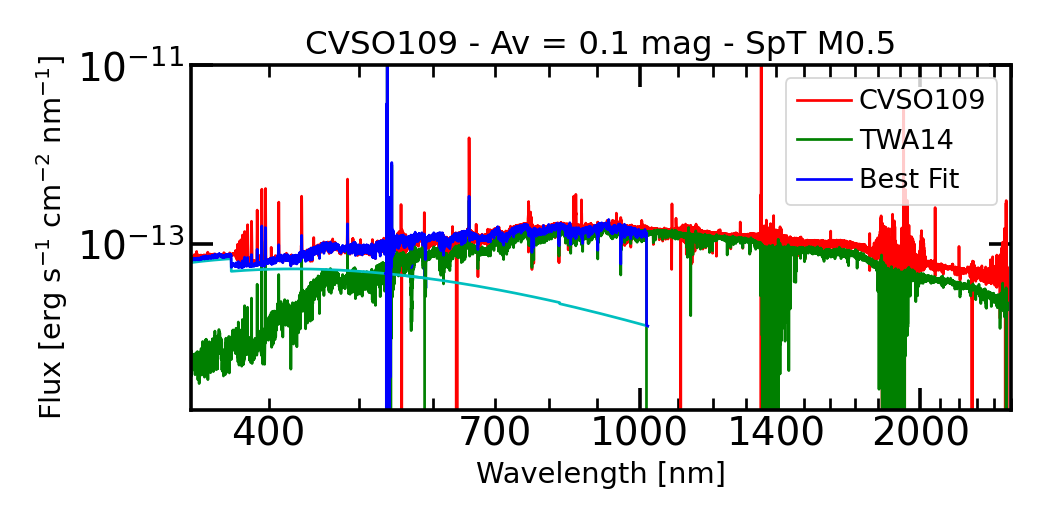}
\includegraphics[width=0.4\textwidth]{CVSO176_all.png}
\includegraphics[width=0.4\textwidth]{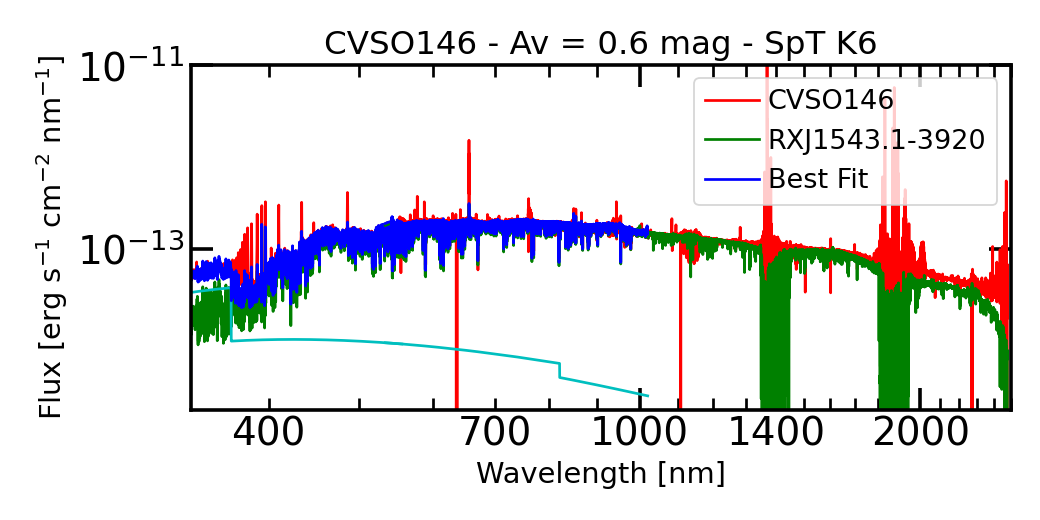}
\includegraphics[width=0.4\textwidth]{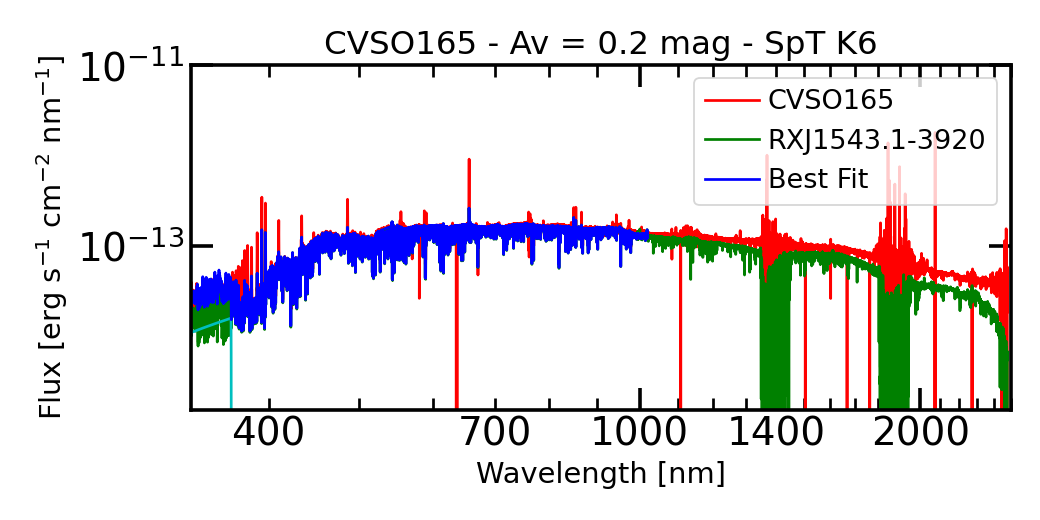}
\caption{X-Shooter spectra of Orion OB1 targets (red) with the best fit, composed of a photospheric template (green) and a slab model for the accretion spectrum (cyan). 
     \label{fig::fit_all_ob1}}
\end{figure*}

\begin{figure*}[]
\centering
\includegraphics[width=0.4\textwidth]{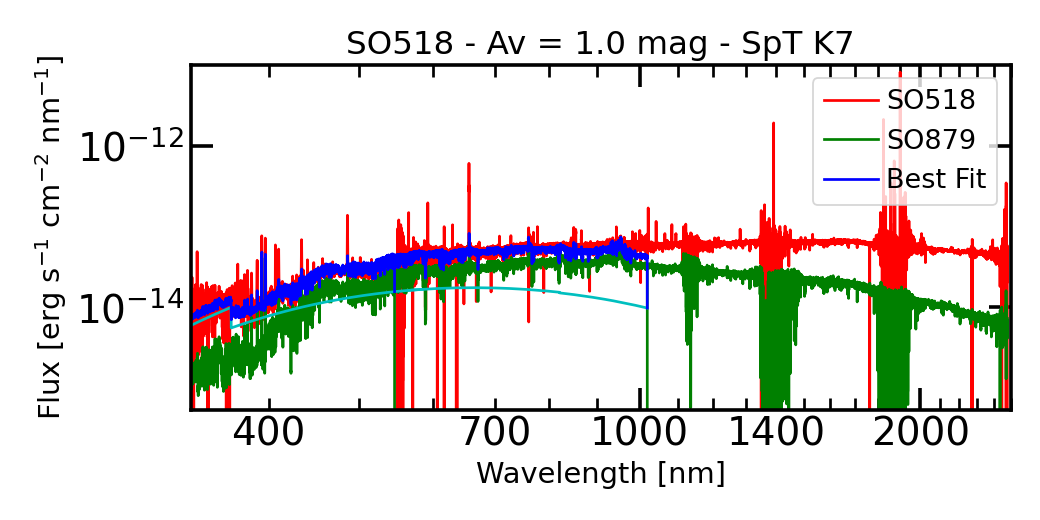}
\includegraphics[width=0.4\textwidth]{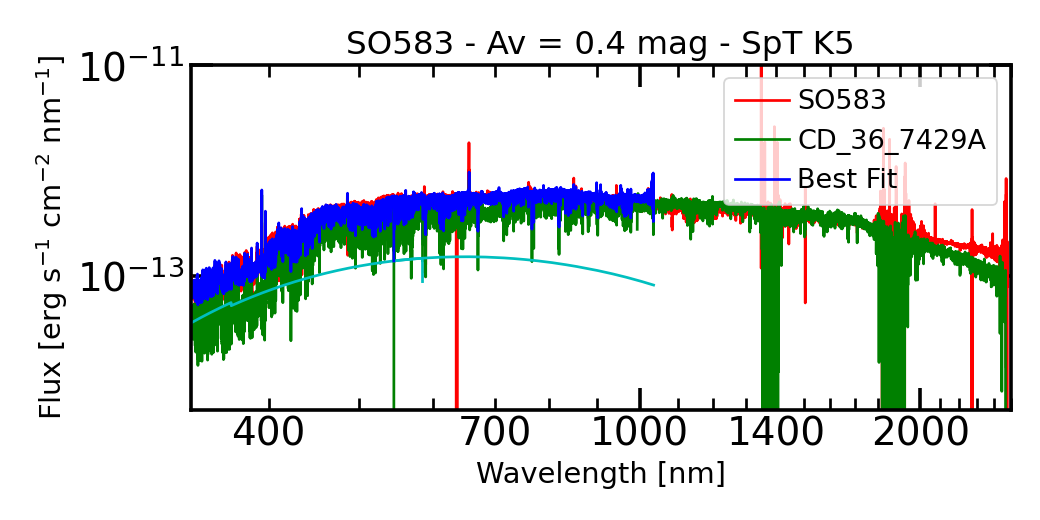}
\includegraphics[width=0.4\textwidth]{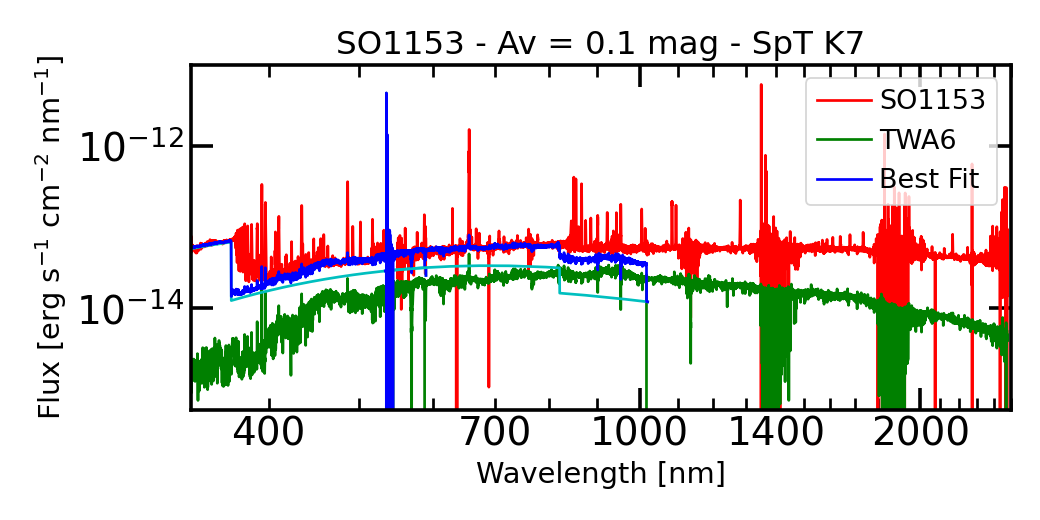}
\caption{X-Shooter spectra of $\sigma$-Orionis targets (red) with the best fit, composed of a photospheric template (green) and a slab model for the accretion spectrum (cyan). 
     \label{fig::fit_all_sOri}}
\end{figure*}

Here, we show the best fit of the X-Shooter spectra of our targets obtained with the method by \citet{manara13b} described in Sect.~\ref{sect::method_fitter}. The Balmer jump region of the spectra is shown in Figures~\ref{fig::fit_bj_ob1}-\ref{fig::fit_bj_sOri}, while the whole best fit spectra are shown in Figures~\ref{fig::fit_all_ob1}-\ref{fig::fit_all_sOri}.

\section{Comparison between HST and VLT/X-Shooter spectra}

\begin{figure*}[]
\centering
\includegraphics[width=0.4\textwidth]{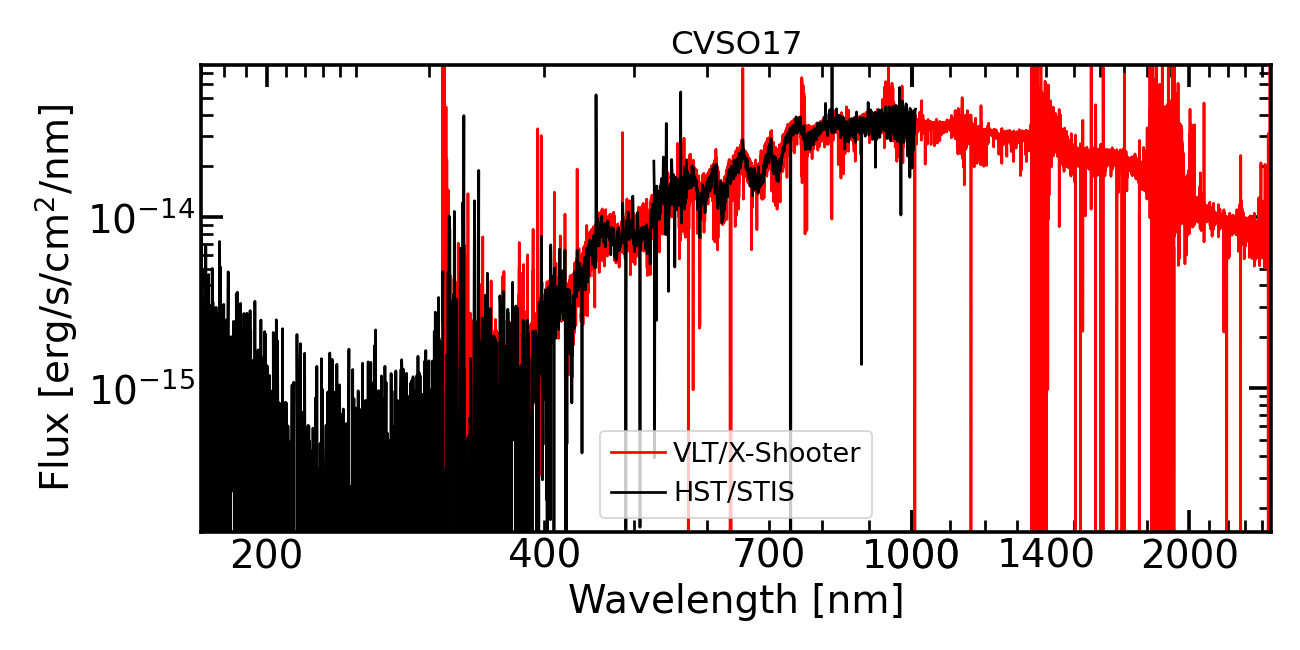}
\includegraphics[width=0.4\textwidth]{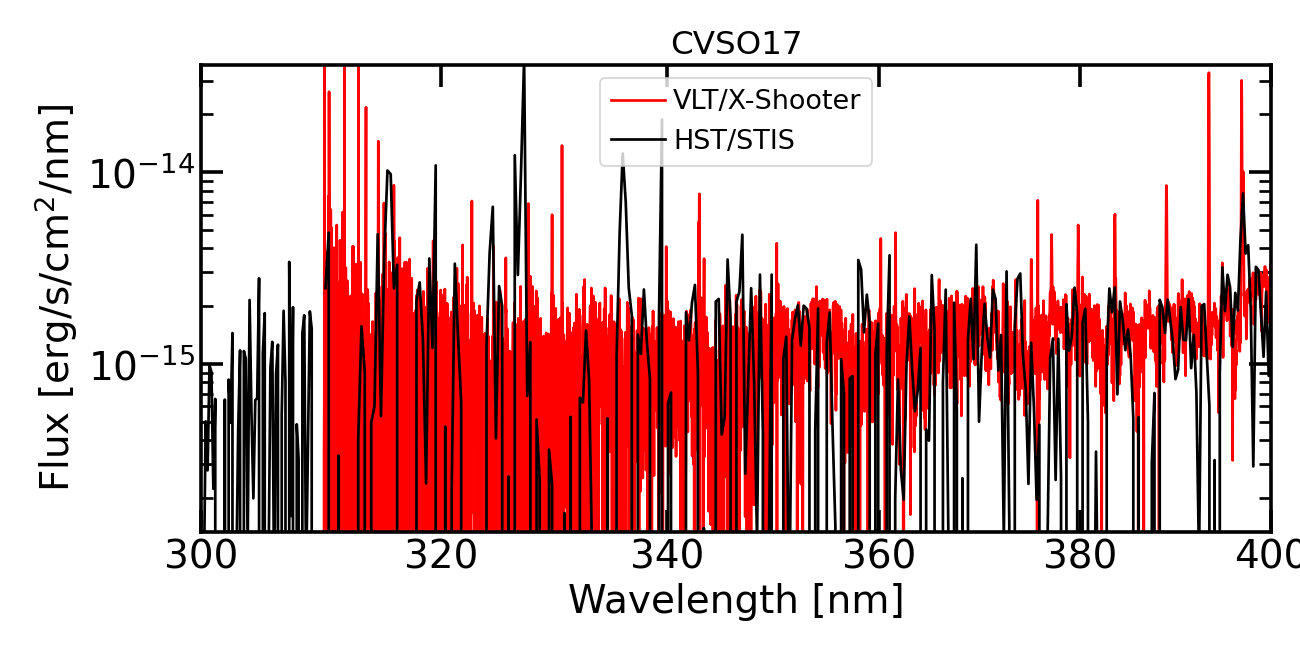}
\includegraphics[width=0.4\textwidth]{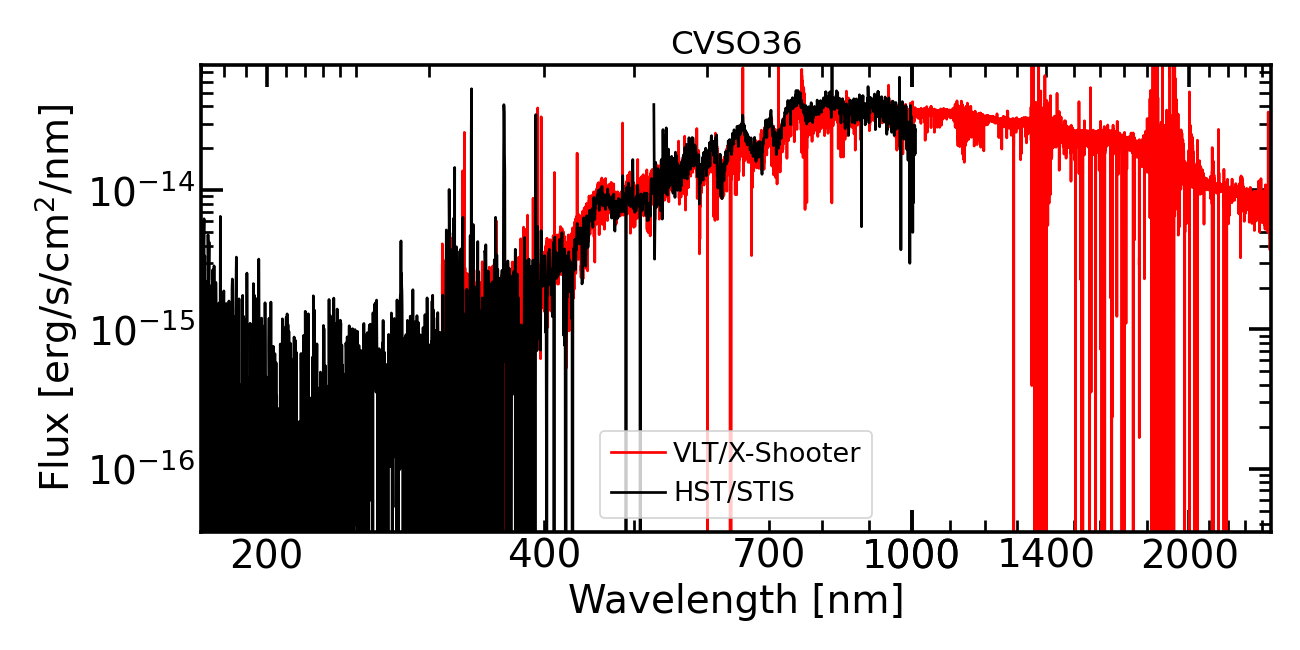}
\includegraphics[width=0.4\textwidth]{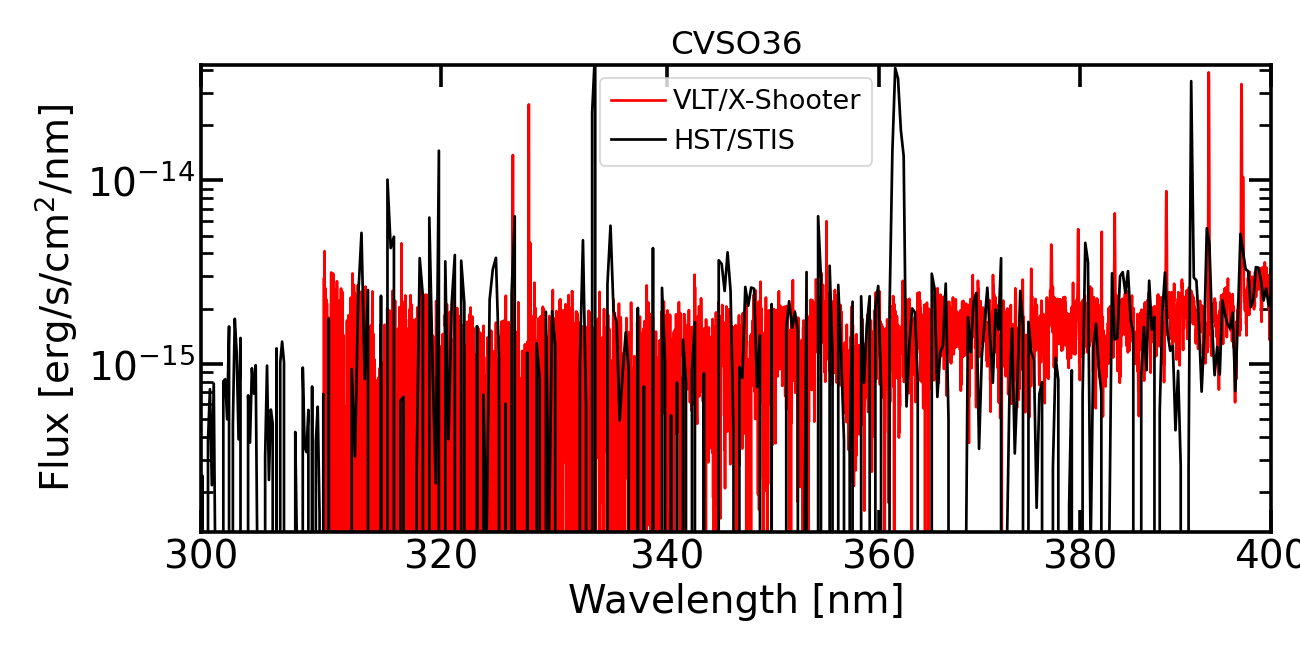}
\includegraphics[width=0.4\textwidth]{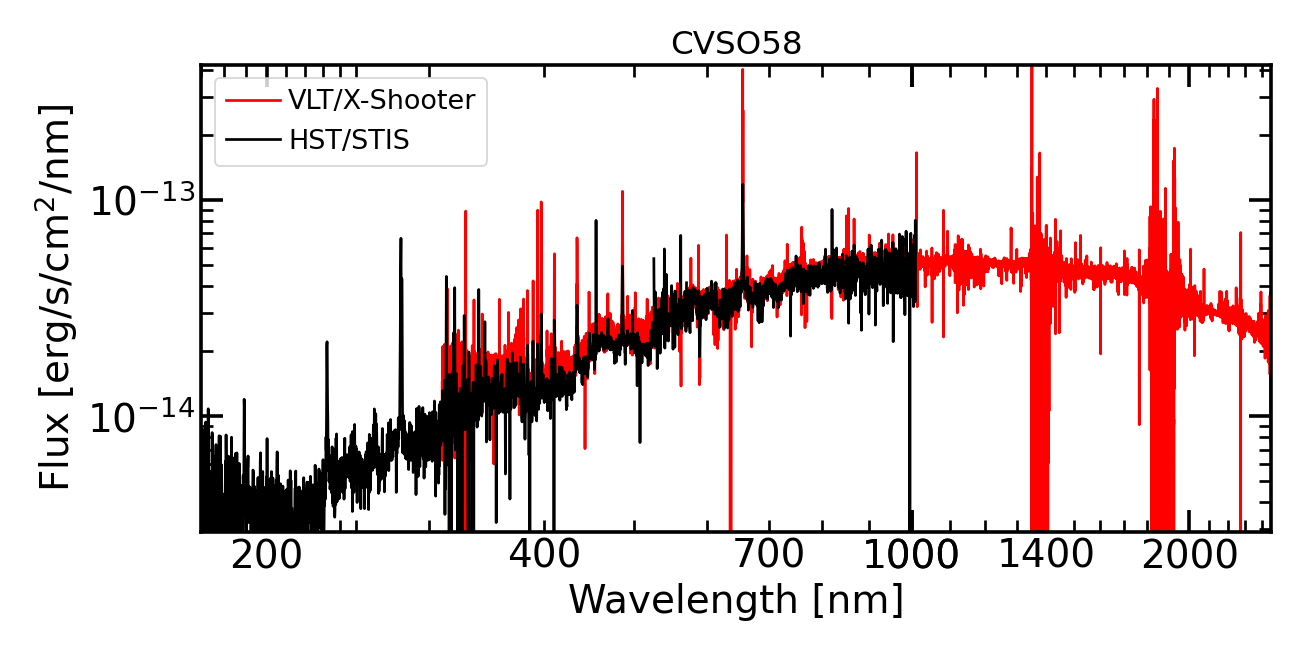}
\includegraphics[width=0.4\textwidth]{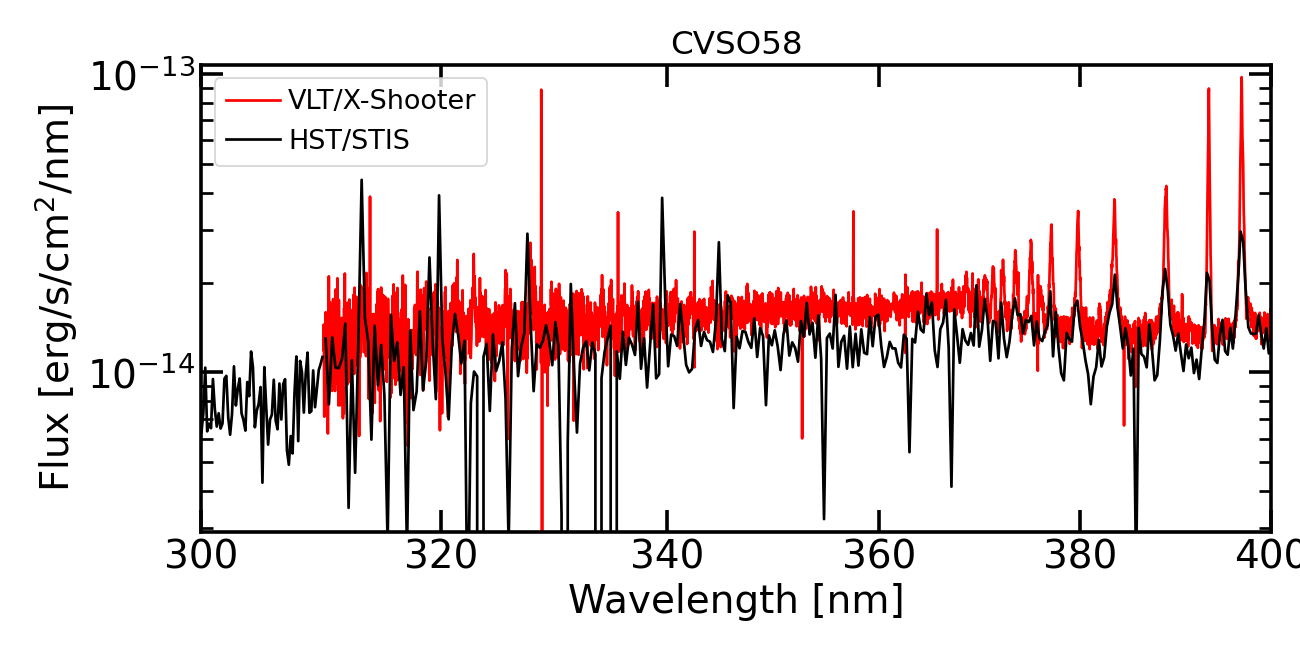}
\includegraphics[width=0.4\textwidth]{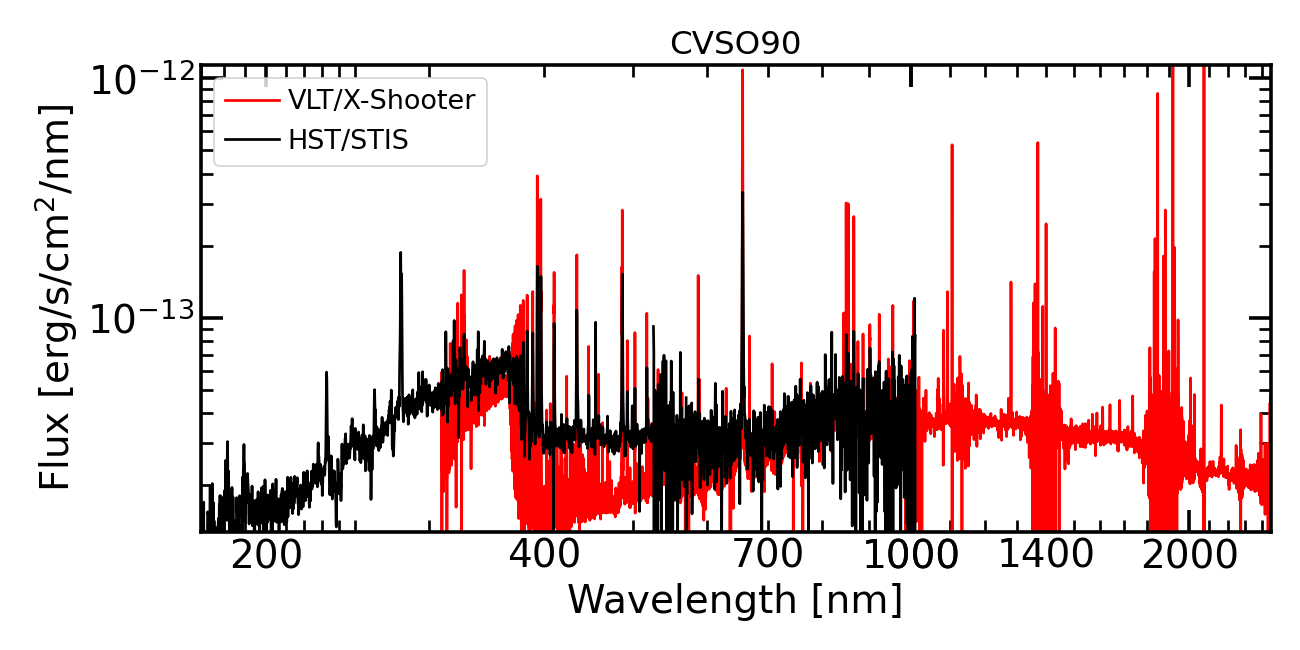}
\includegraphics[width=0.4\textwidth]{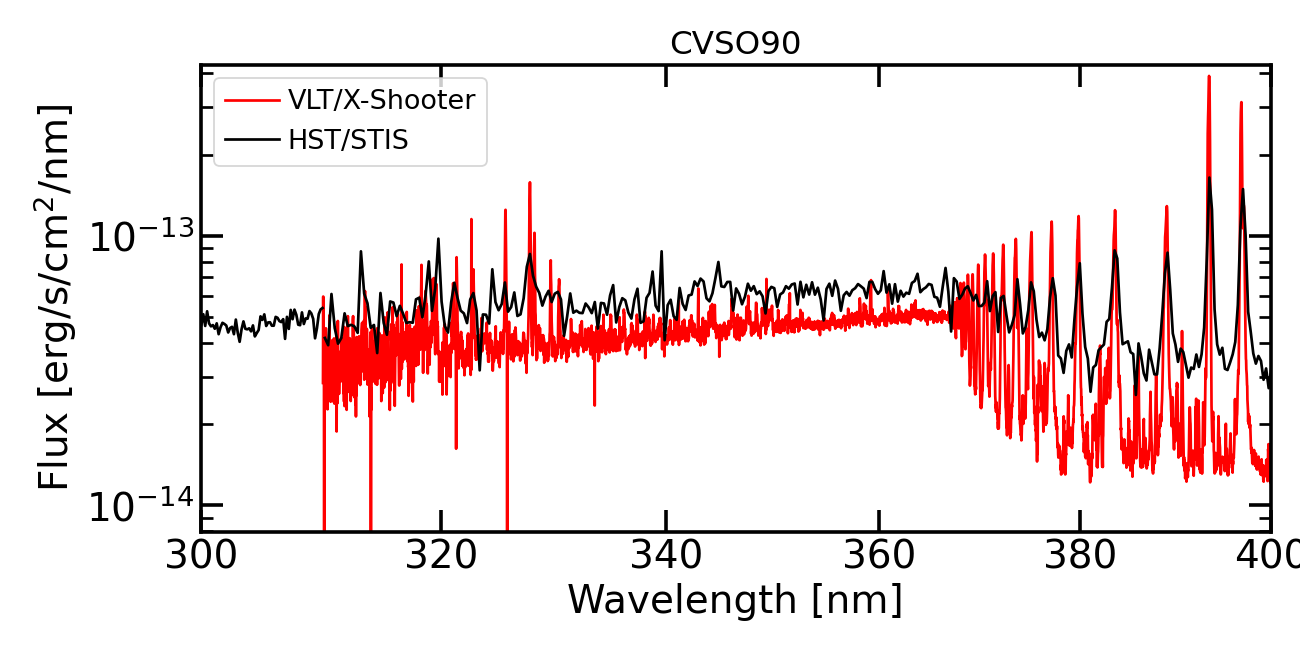}
\includegraphics[width=0.4\textwidth]{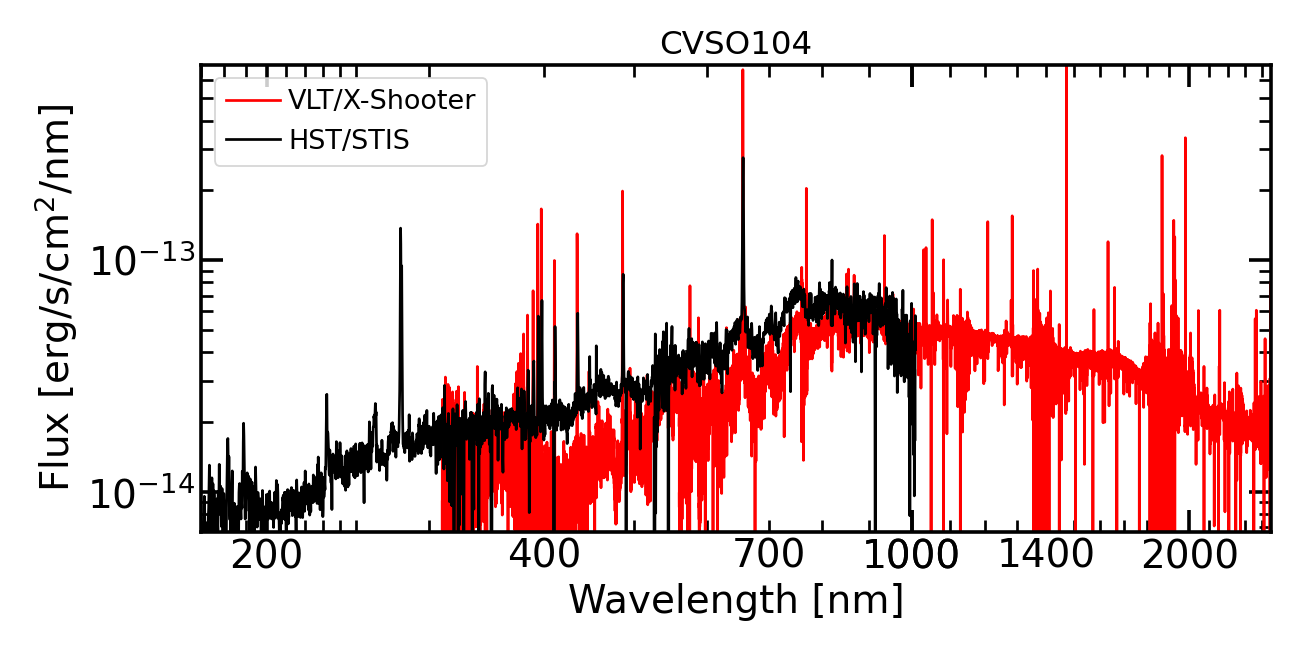}
\includegraphics[width=0.4\textwidth]{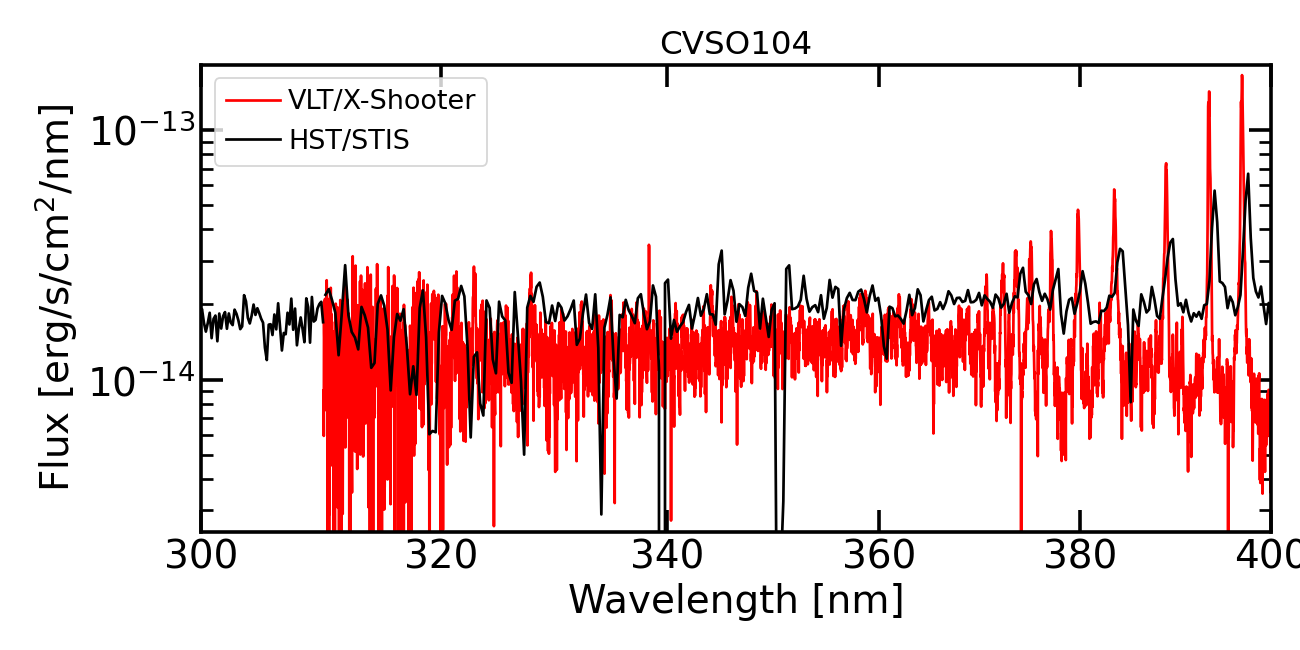}
\includegraphics[width=0.4\textwidth]{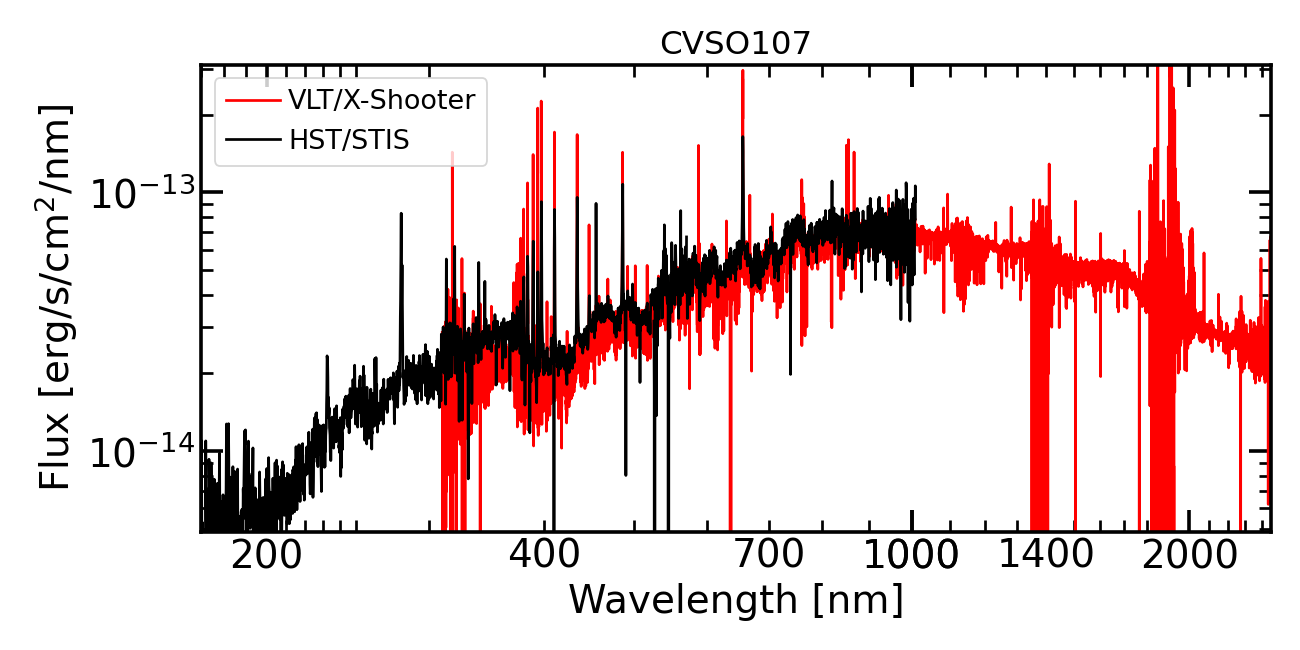}
\includegraphics[width=0.4\textwidth]{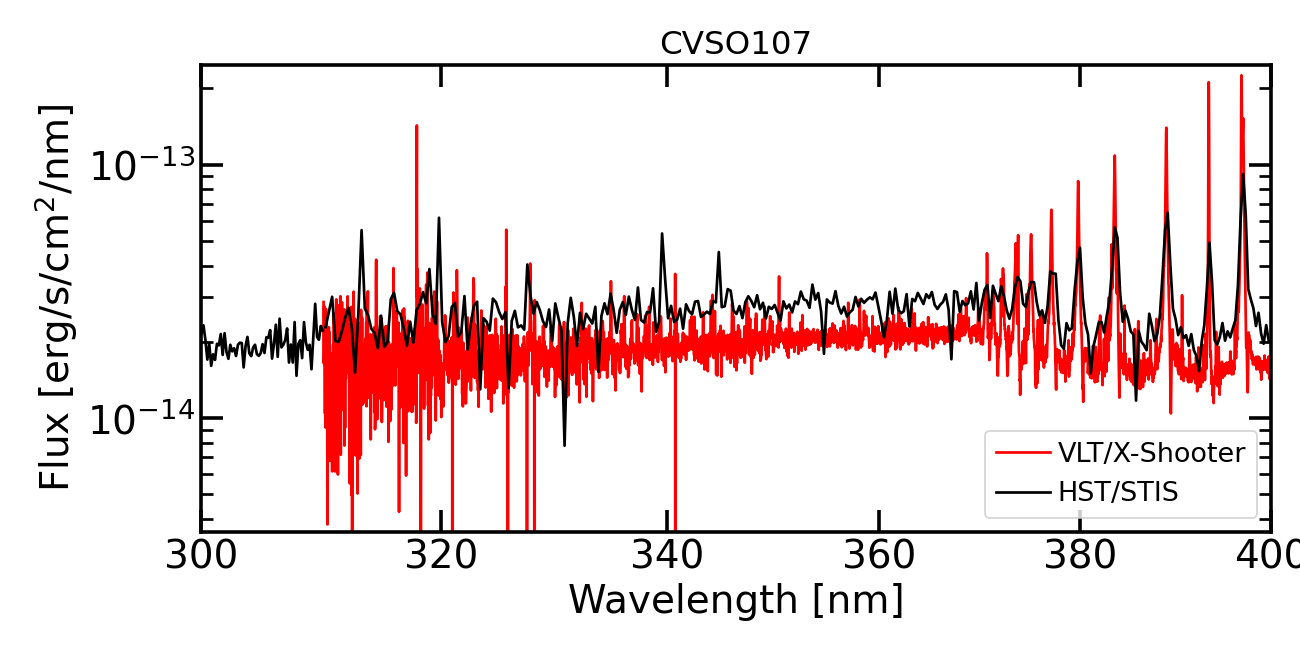}
\caption{Comparison between the observed VLT/X-Shooter (red) and HST/STIS (black) spectra.
     \label{fig::hst_vs_xs}}
\end{figure*}

\begin{figure*}[]
\centering
\includegraphics[width=0.4\textwidth]{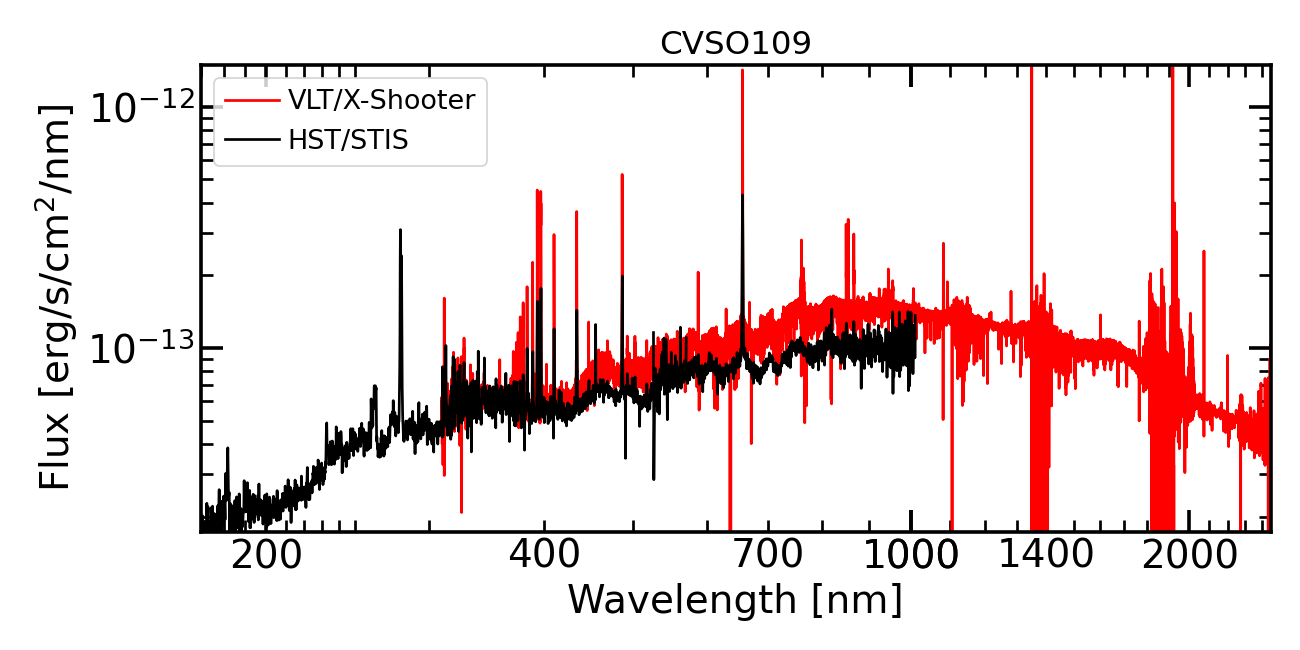}
\includegraphics[width=0.4\textwidth]{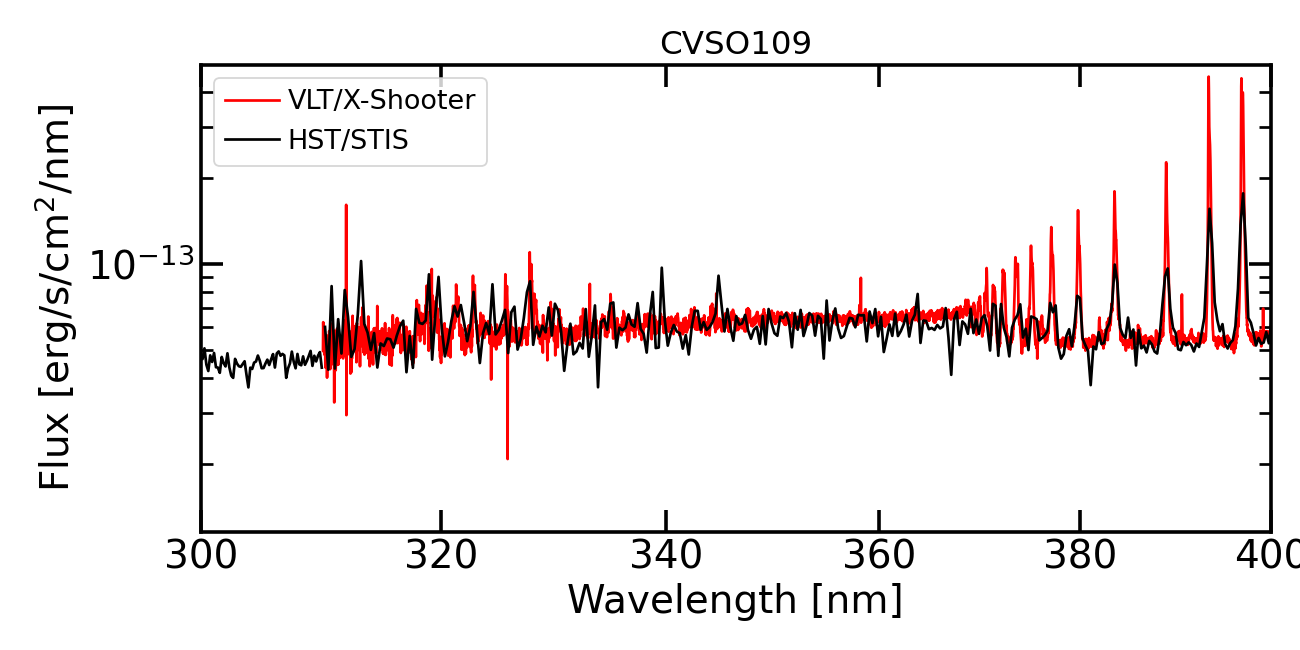}
\includegraphics[width=0.4\textwidth]{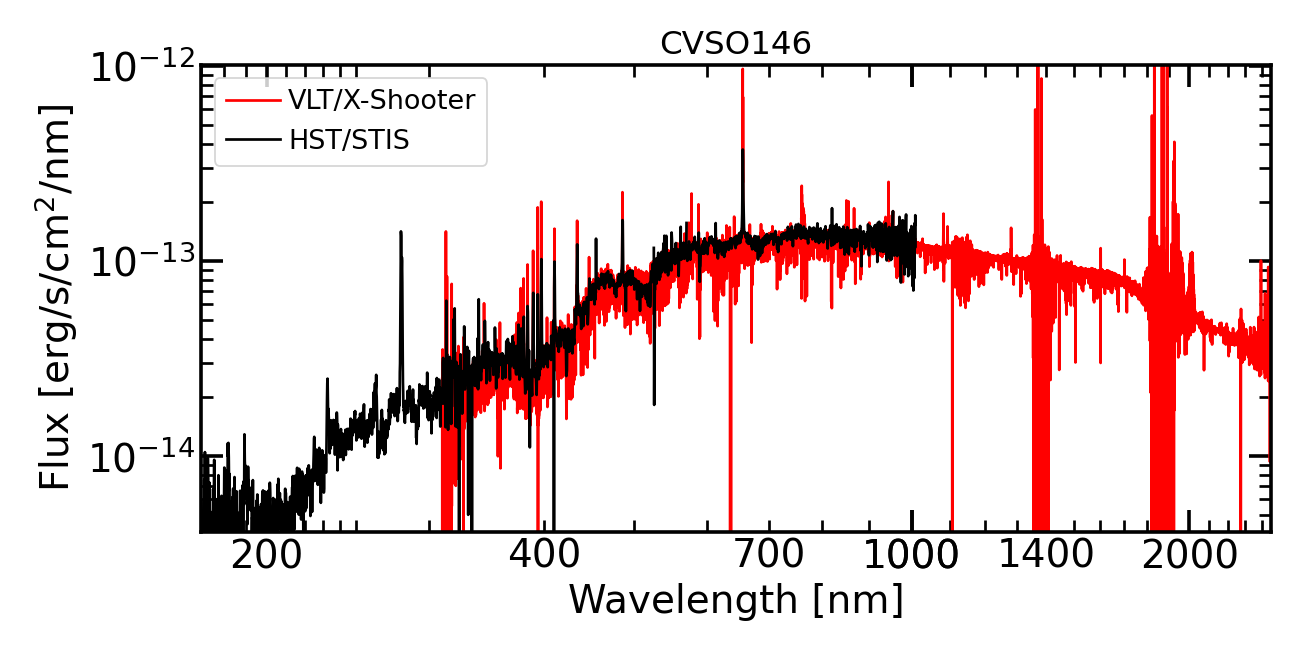}
\includegraphics[width=0.4\textwidth]{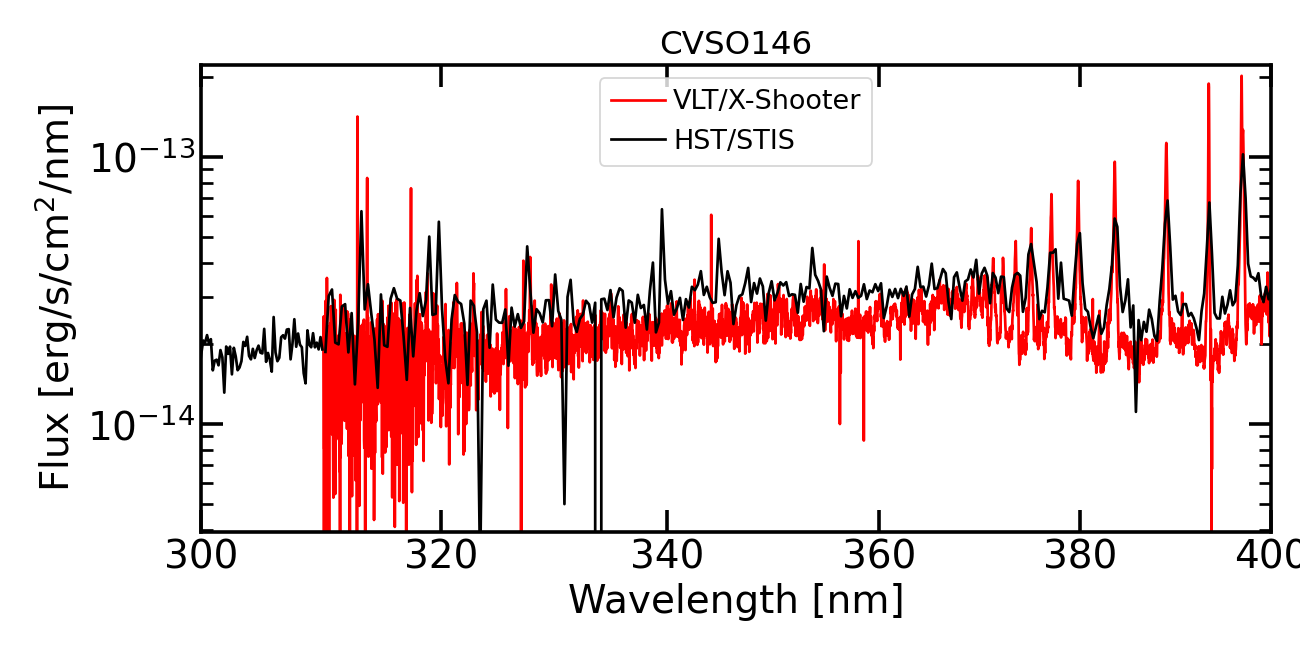}
\includegraphics[width=0.4\textwidth]{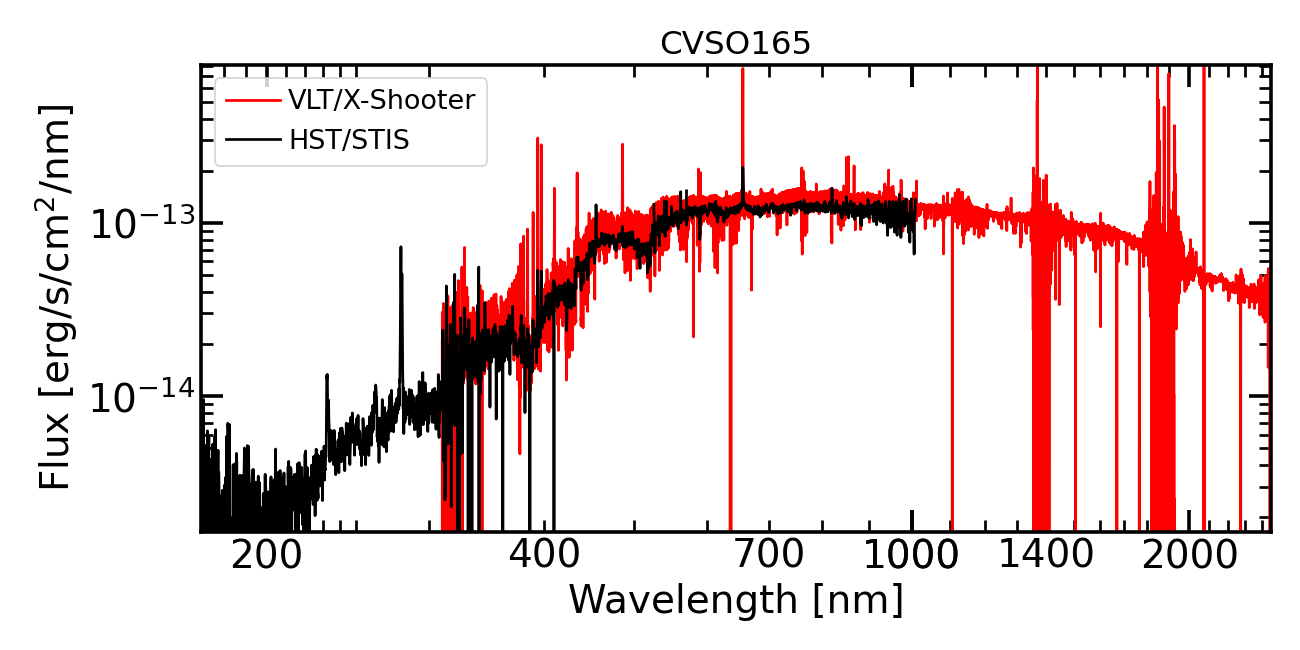}
\includegraphics[width=0.4\textwidth]{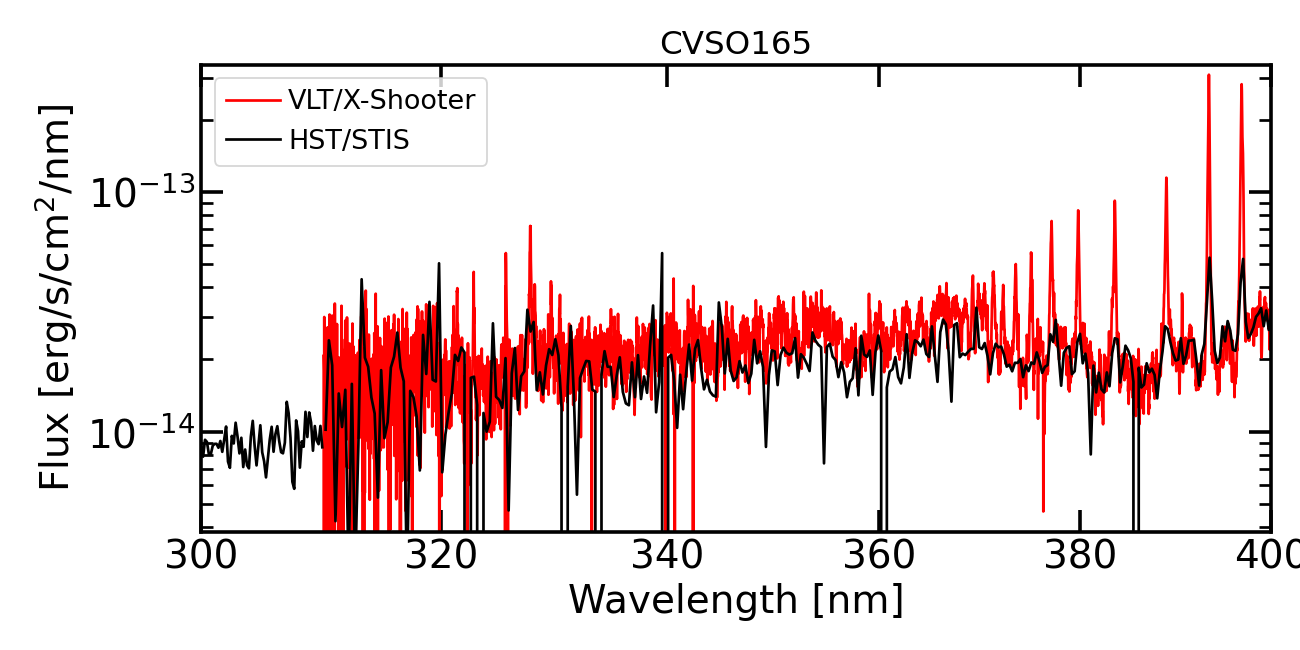}
\includegraphics[width=0.4\textwidth]{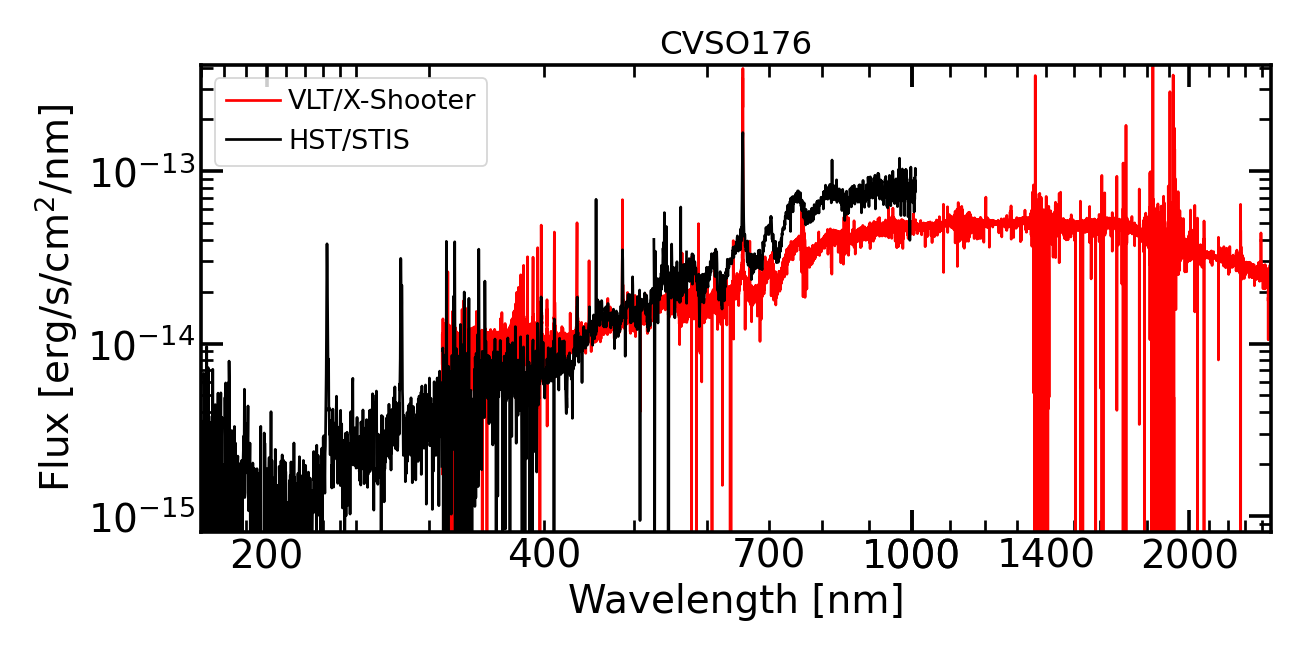}
\includegraphics[width=0.4\textwidth]{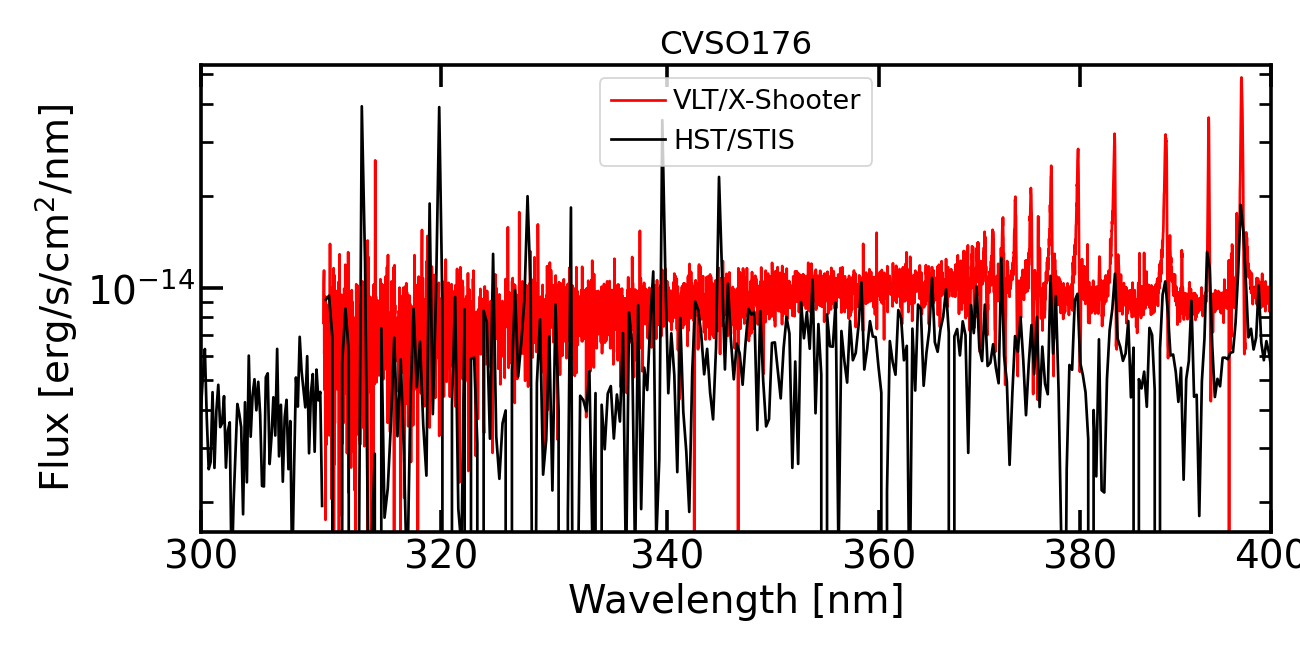}
\includegraphics[width=0.4\textwidth]{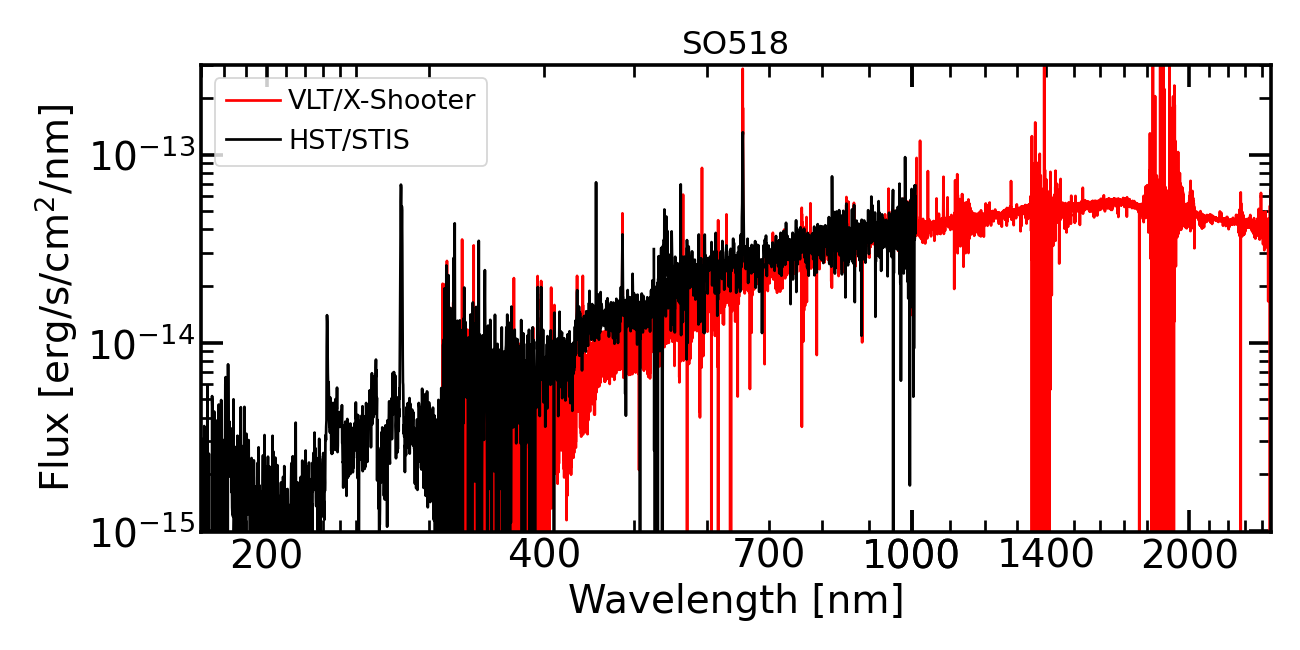}
\includegraphics[width=0.4\textwidth]{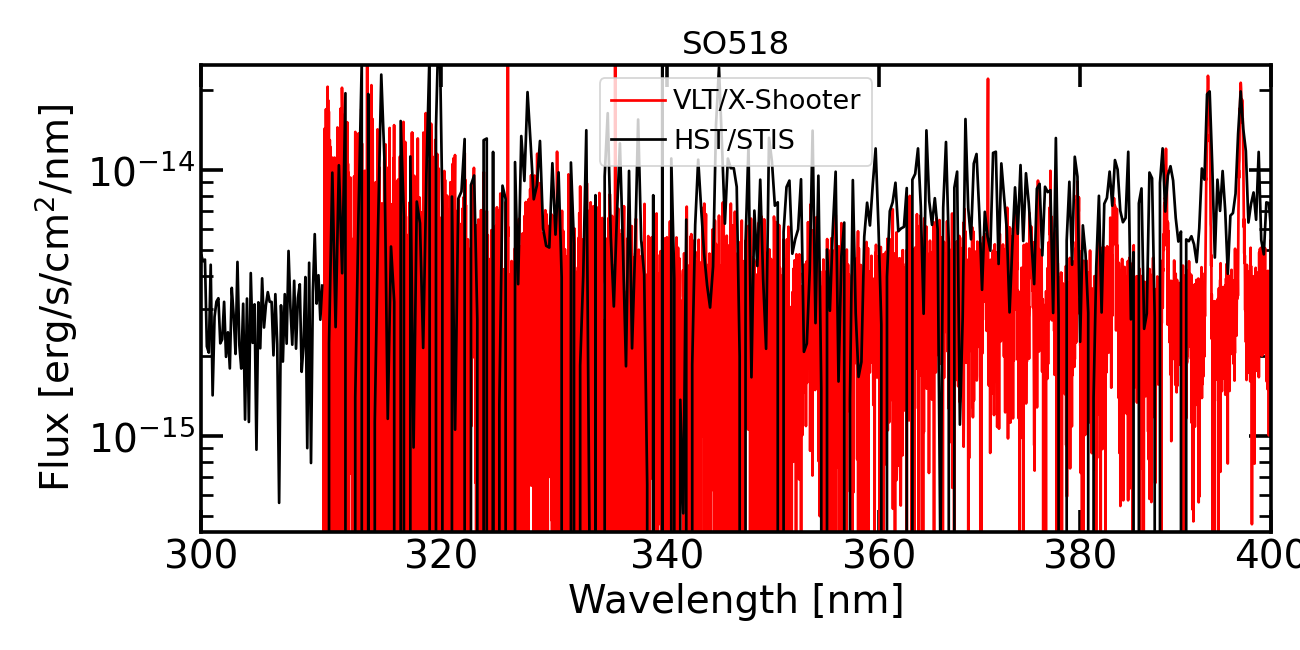}
\includegraphics[width=0.4\textwidth]{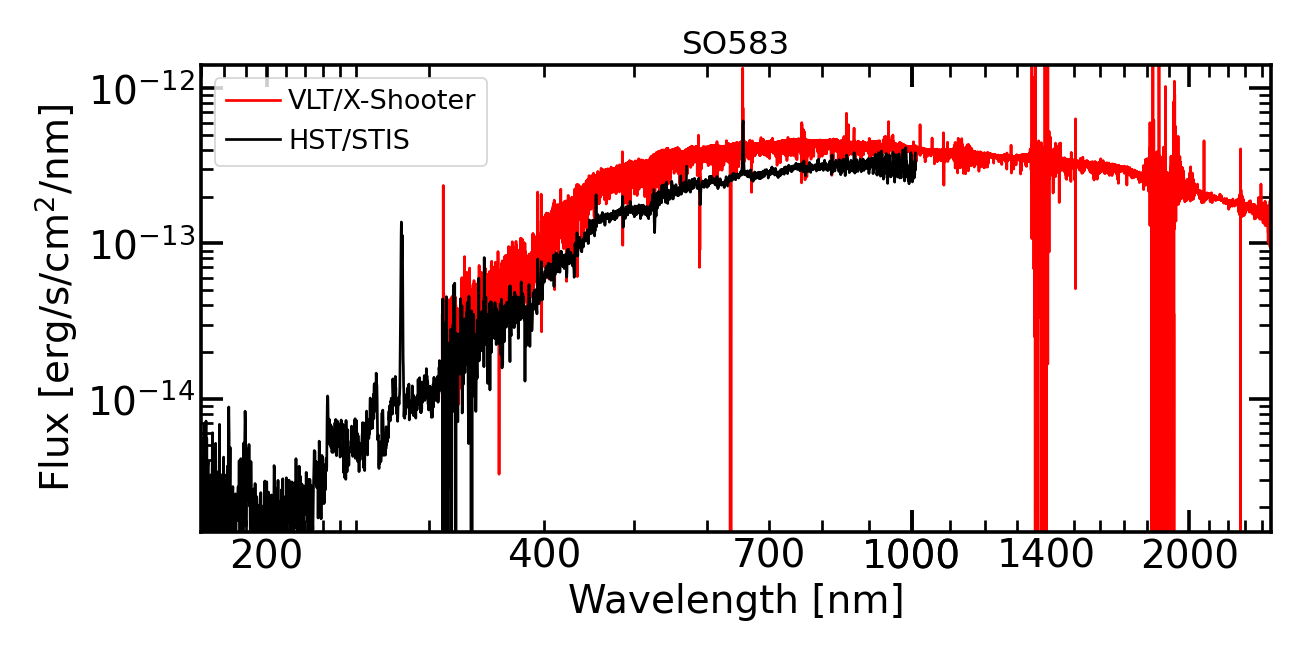}
\includegraphics[width=0.4\textwidth]{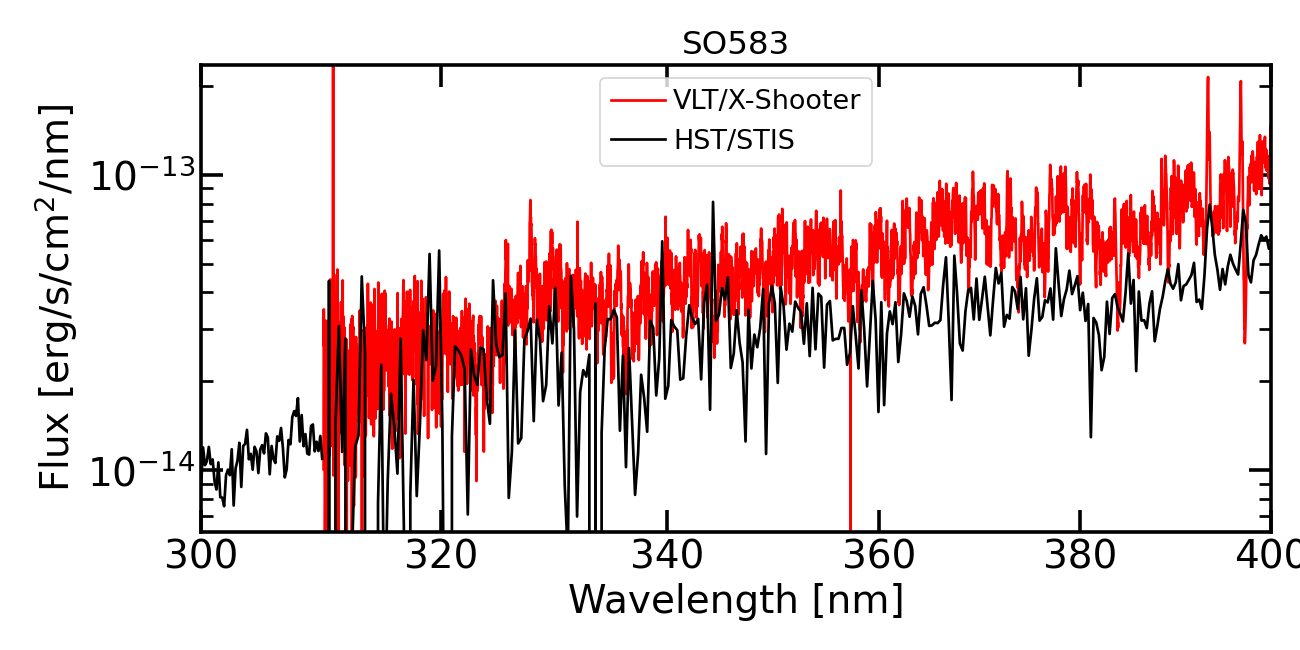}
\caption{Comparison between the observed VLT/X-Shooter (red) and HST/STIS (black) spectra.
     \label{fig::hst_vs_xs2}}
\end{figure*}

The comparison between the observed VLT/X-Shooter spectra and the HST/STIS ones is shown in Figures~\ref{fig::hst_vs_xs}-\ref{fig::hst_vs_xs2}.

\section{Line profiles for the observed Orion targets}\label{sect::profiles}
\begin{figure*}[]
\centering
\includegraphics[width=0.4\textwidth]{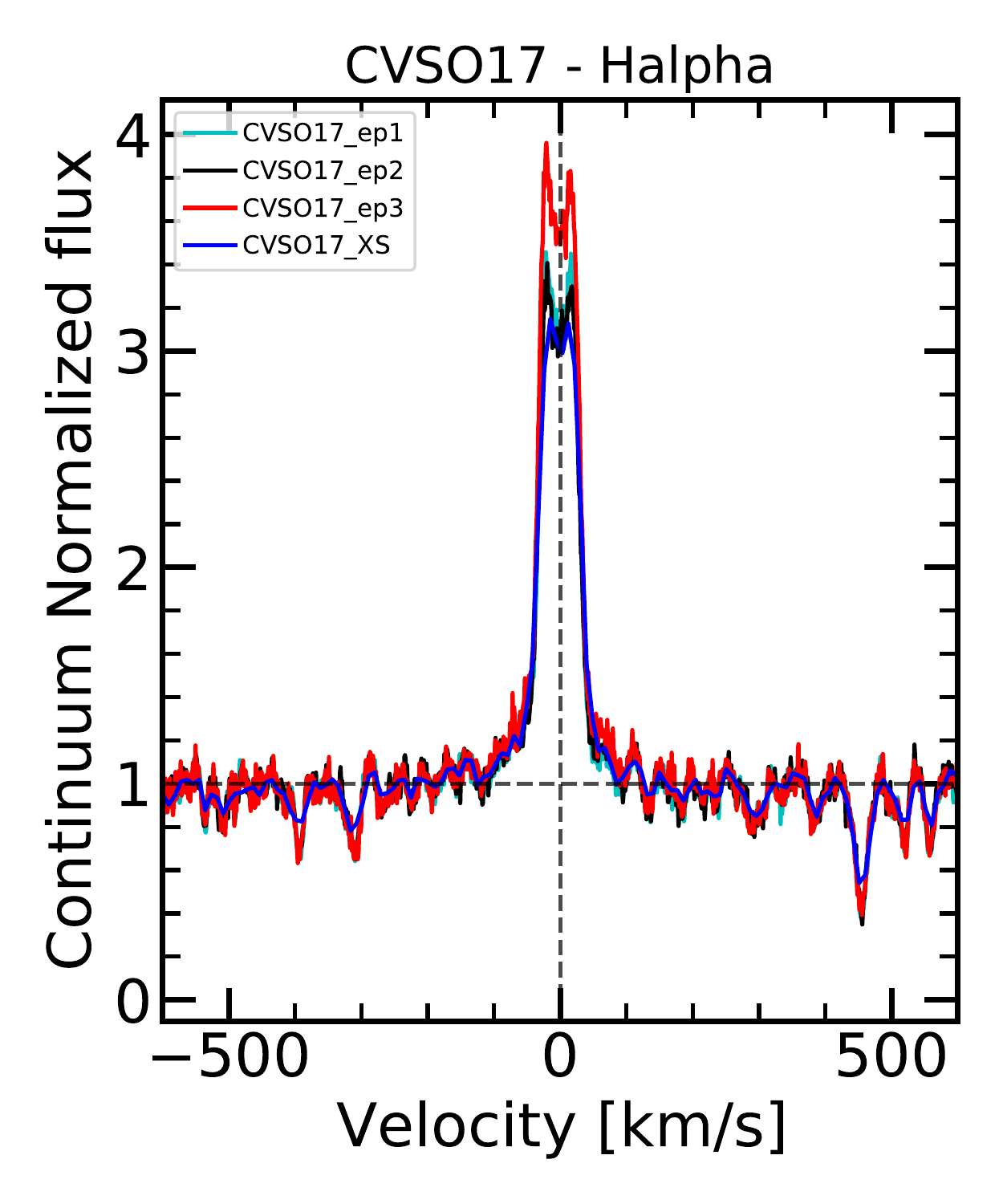}
\includegraphics[width=0.4\textwidth]{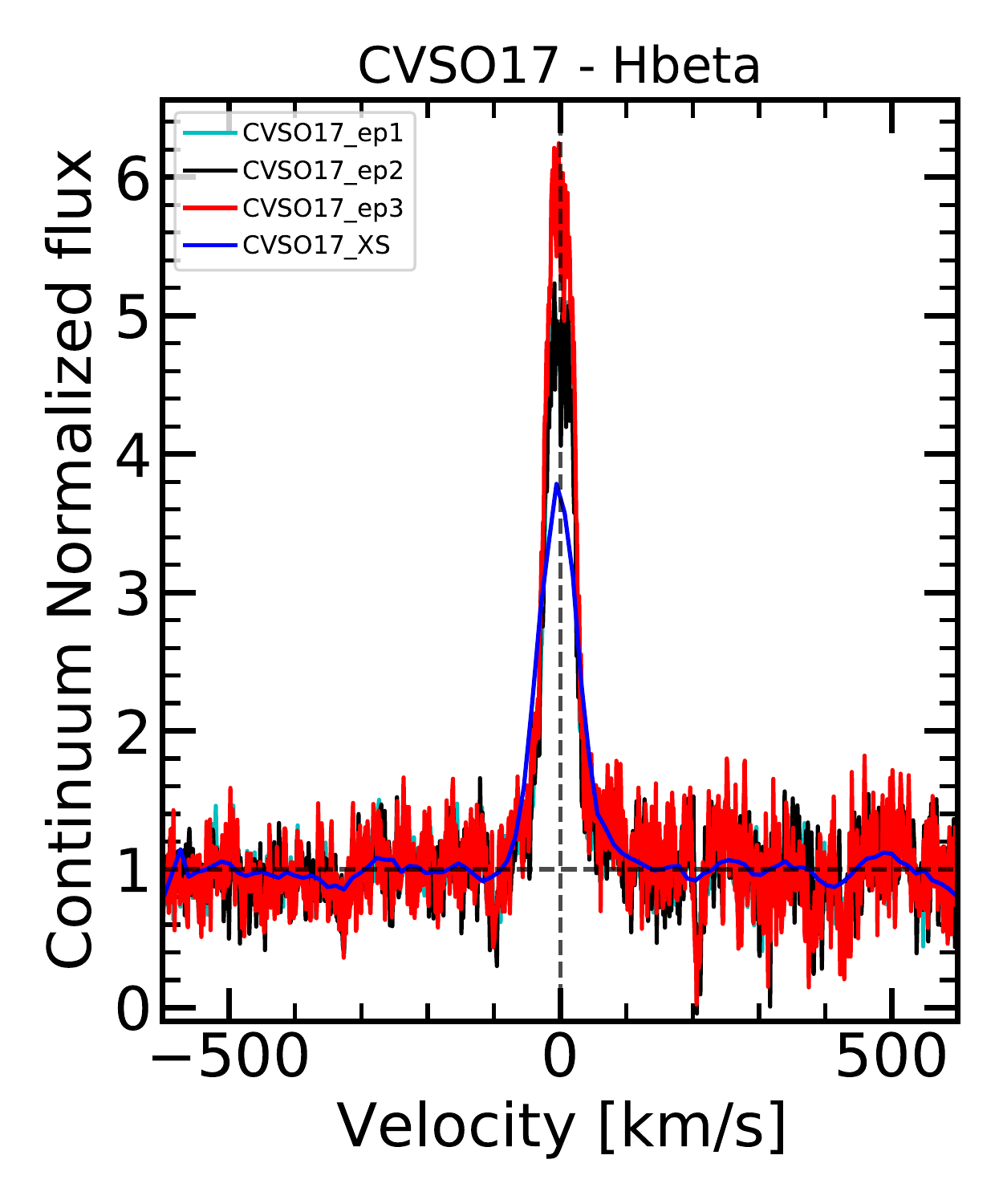}
\includegraphics[width=0.4\textwidth]{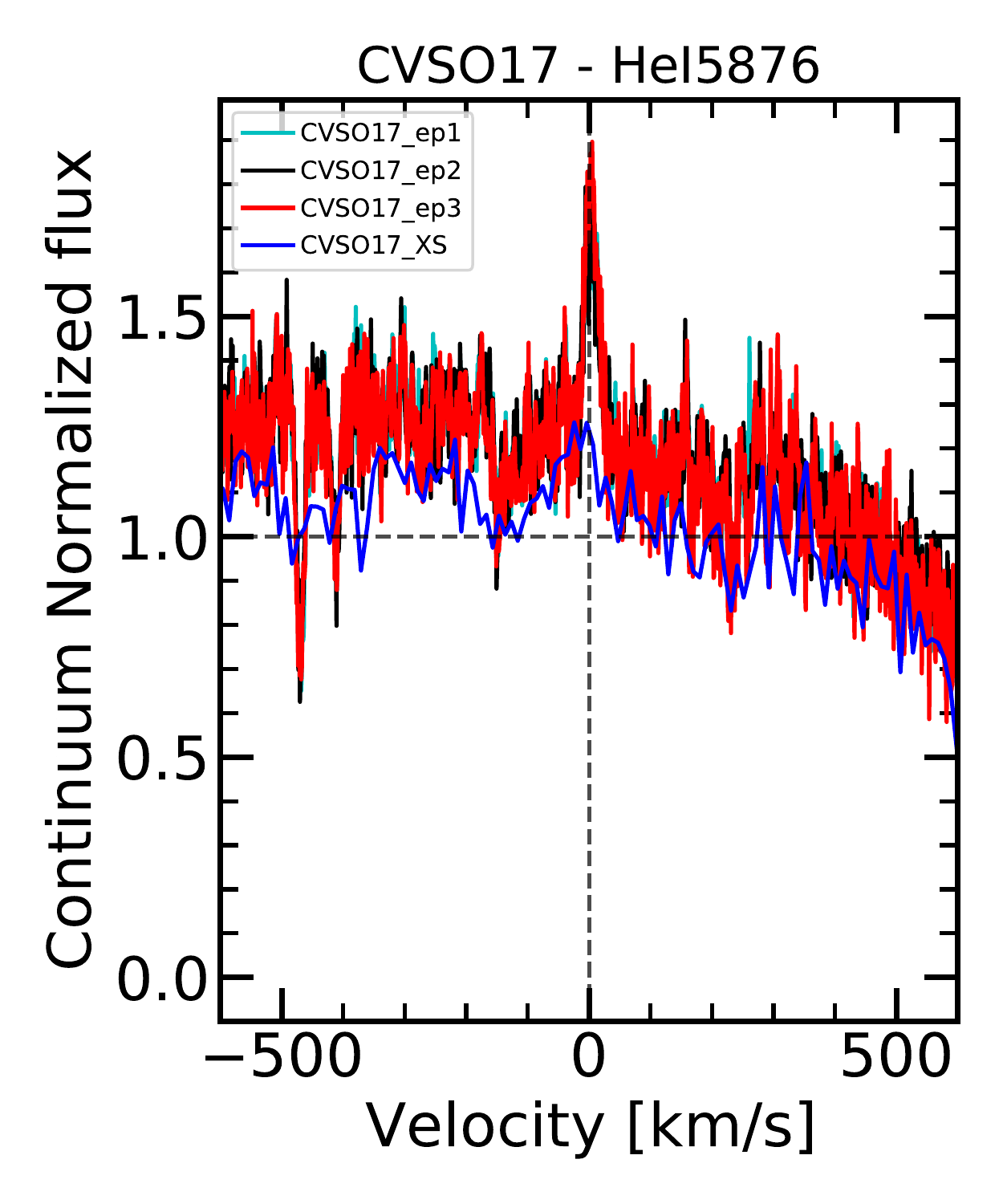}
\includegraphics[width=0.4\textwidth]{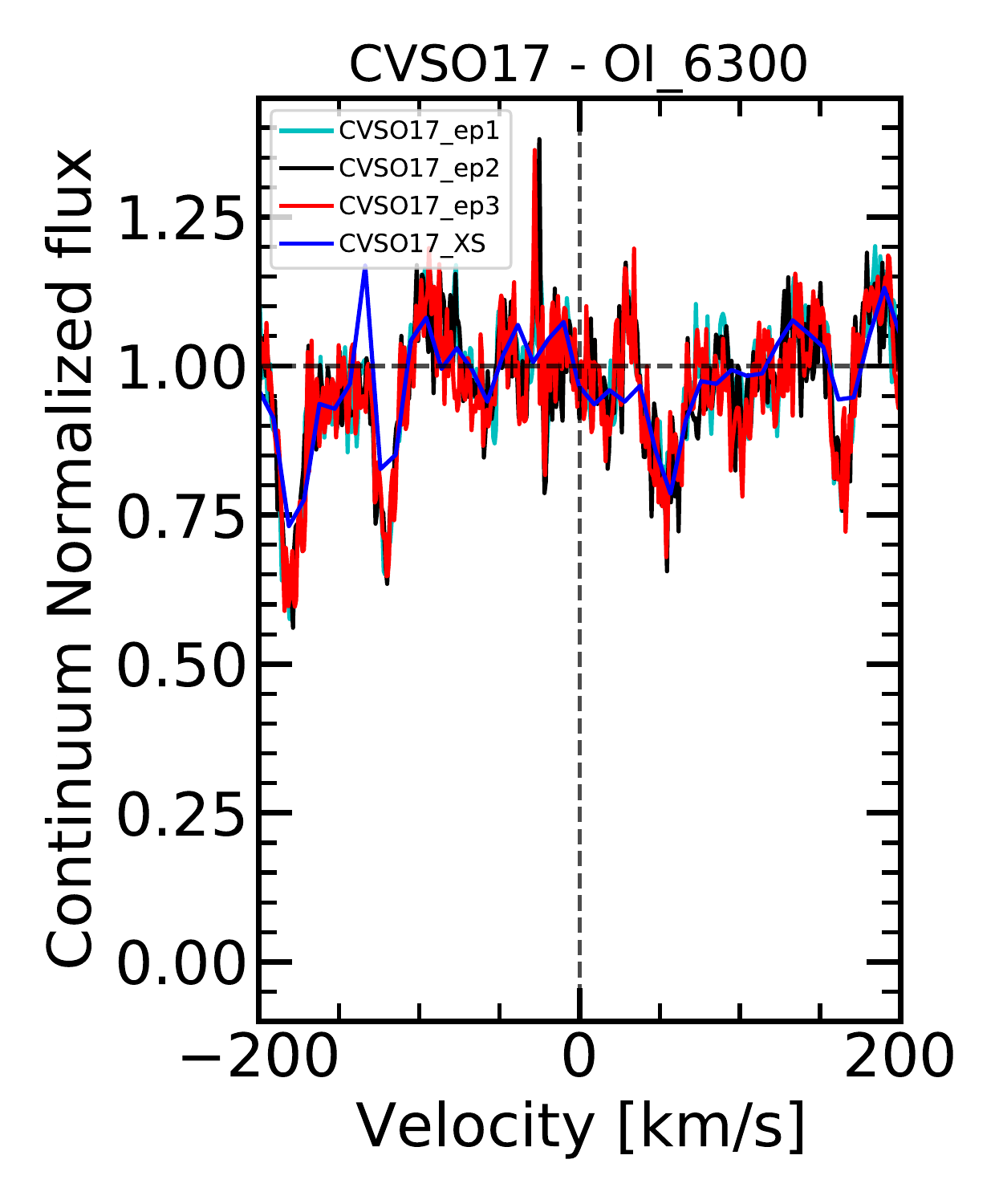}
\caption{Emission lines of the target CVSO17 observed with UVES and X-Shooter.
     \label{fig::lines_CVSO17}}
\end{figure*}

\begin{figure*}[]
\centering
\includegraphics[width=0.4\textwidth]{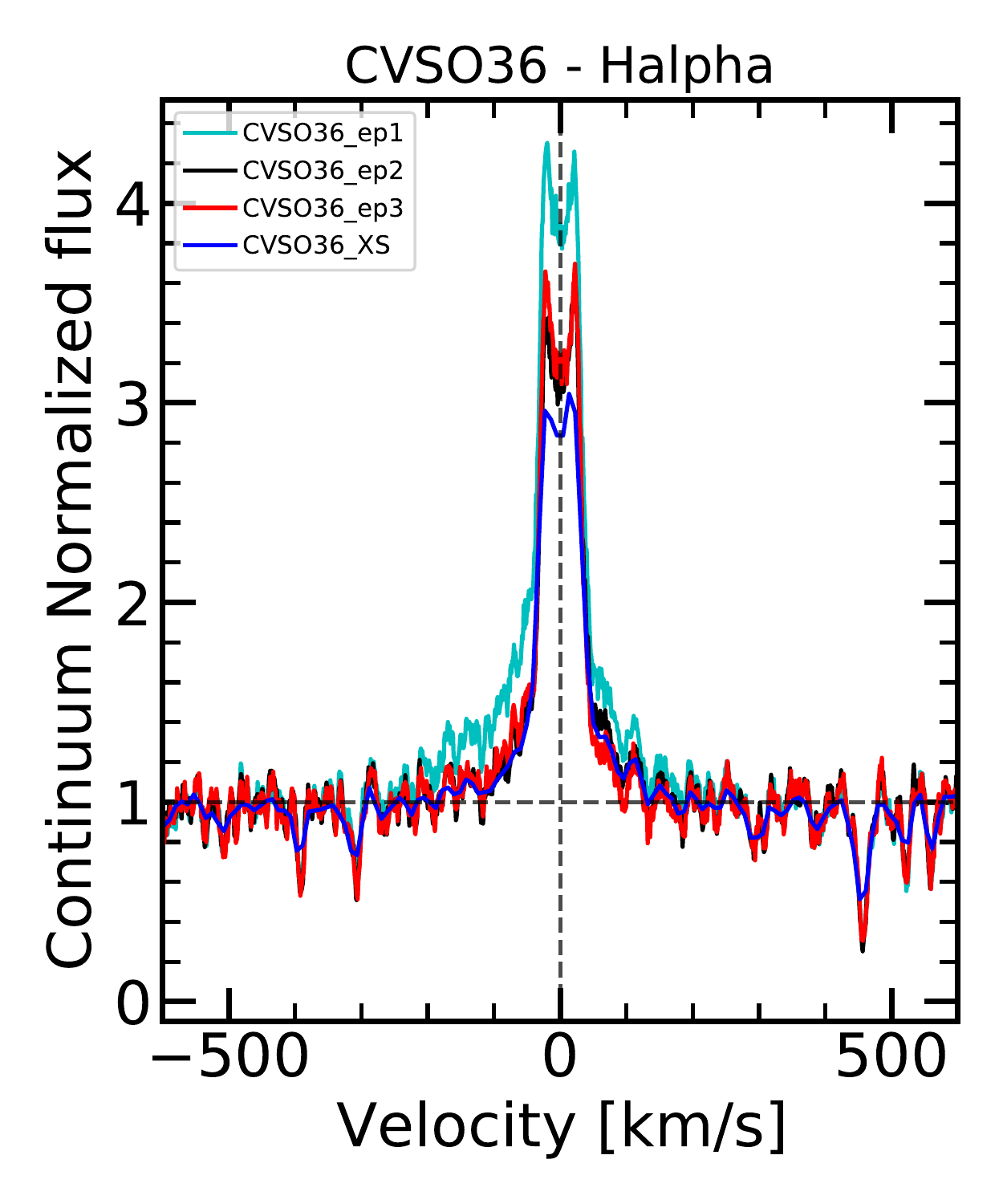}
\includegraphics[width=0.4\textwidth]{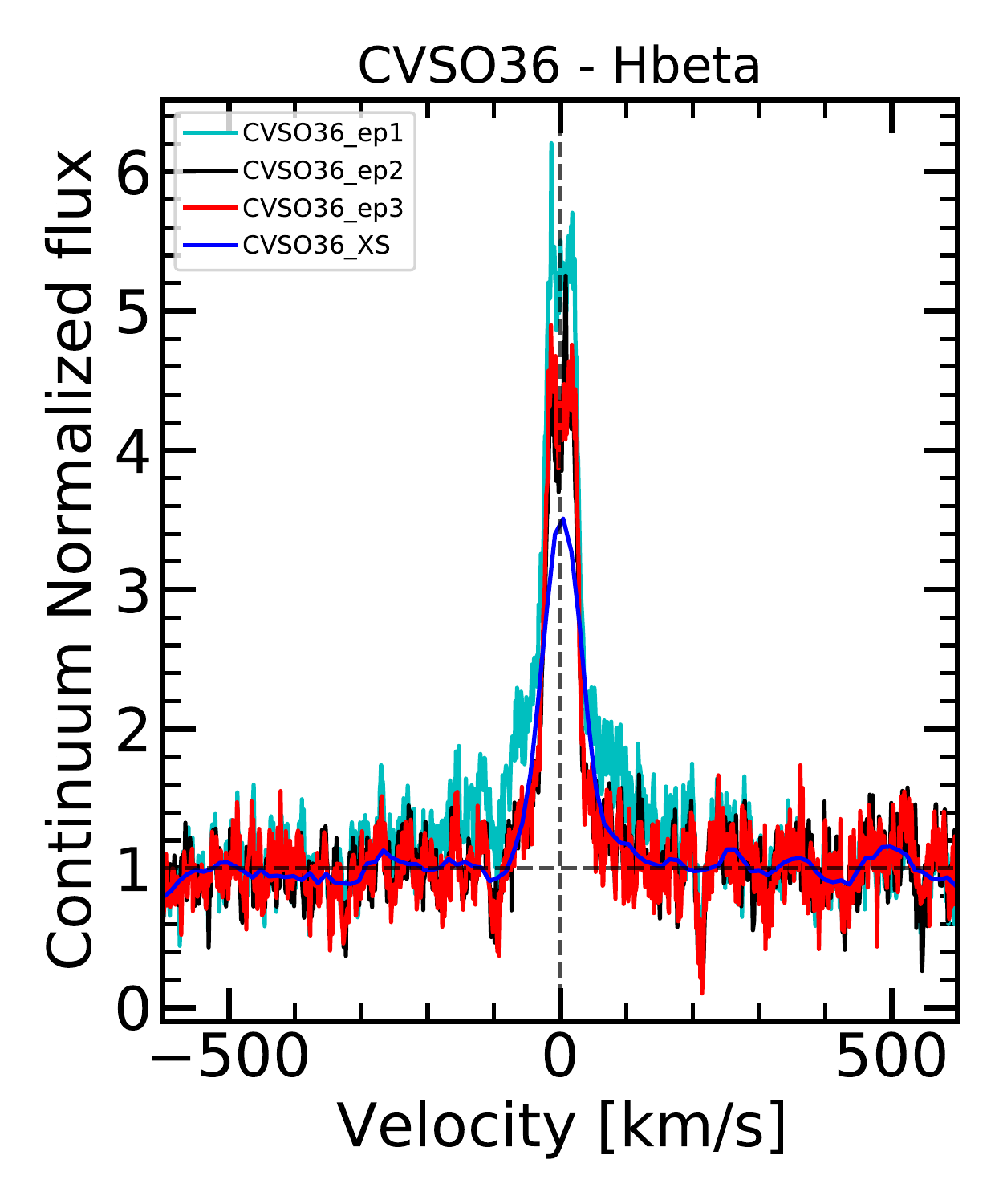}
\includegraphics[width=0.4\textwidth]{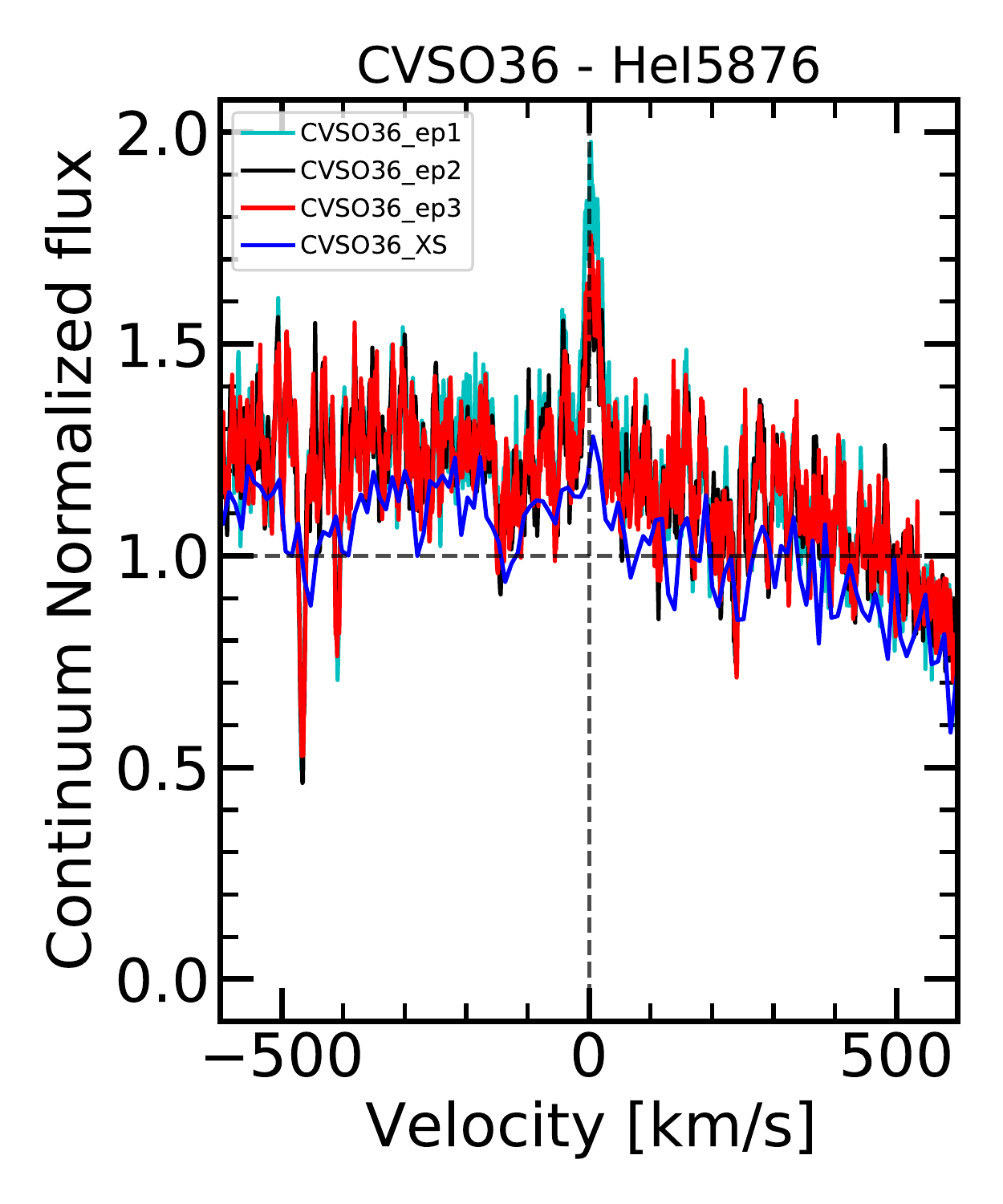}
\includegraphics[width=0.4\textwidth]{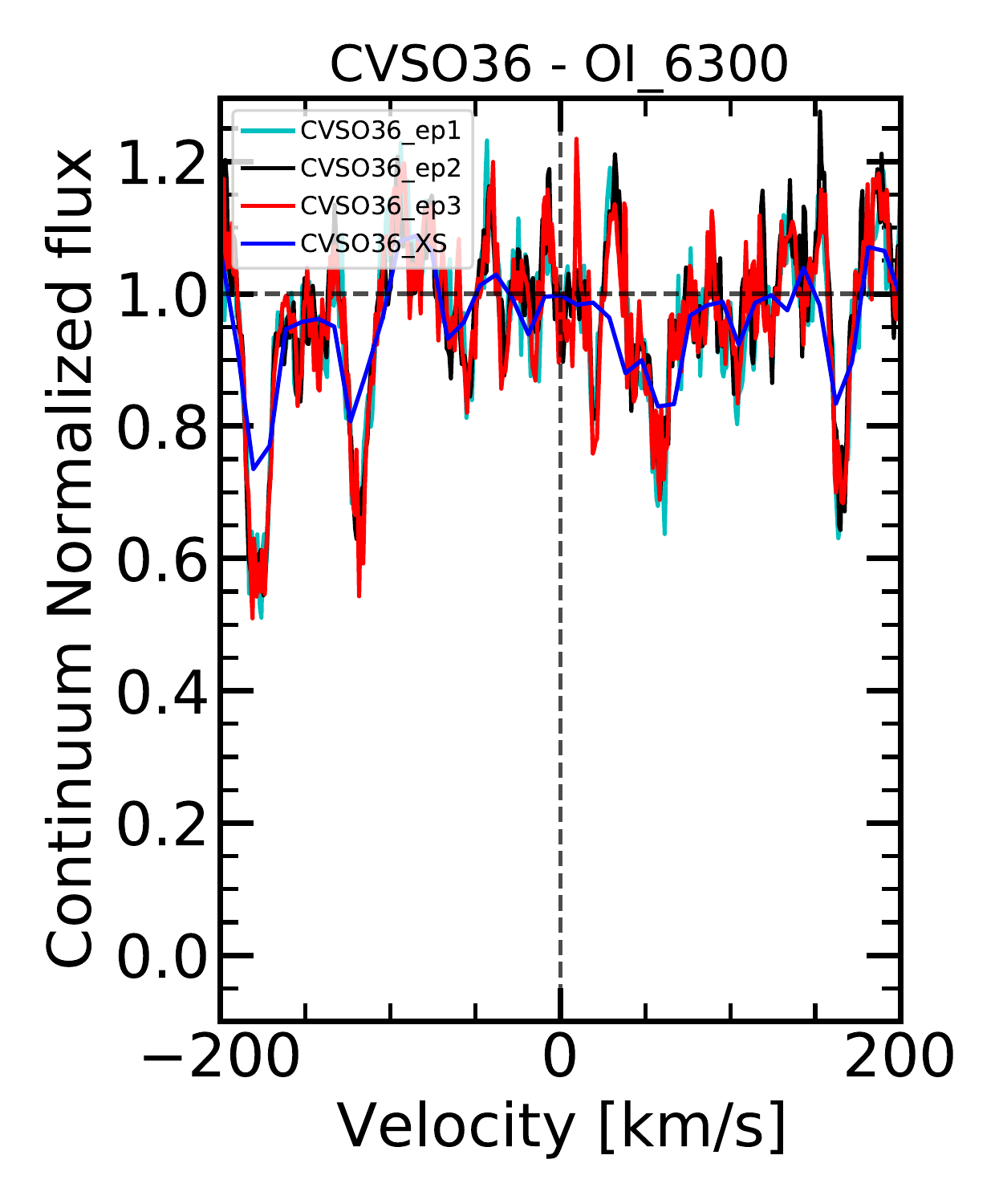}
\caption{Emission lines of the target CVSO36 observed with UVES and X-Shooter.
     \label{fig::lines_CVSO36}}
\end{figure*}

\begin{figure*}[]
\centering
\includegraphics[width=0.4\textwidth]{CVSO58_Halpha_6562_compare.pdf}
\includegraphics[width=0.4\textwidth]{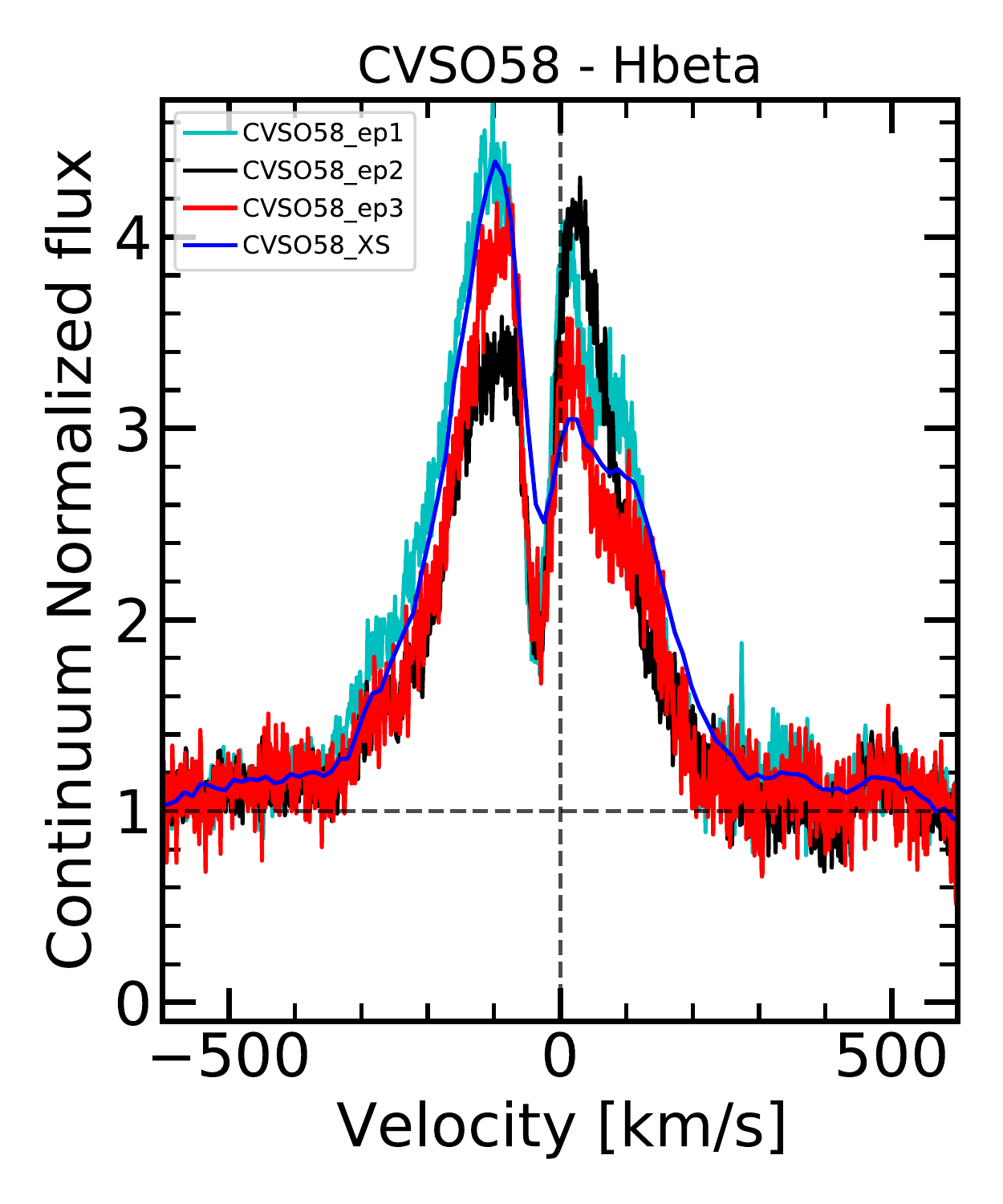}
\includegraphics[width=0.4\textwidth]{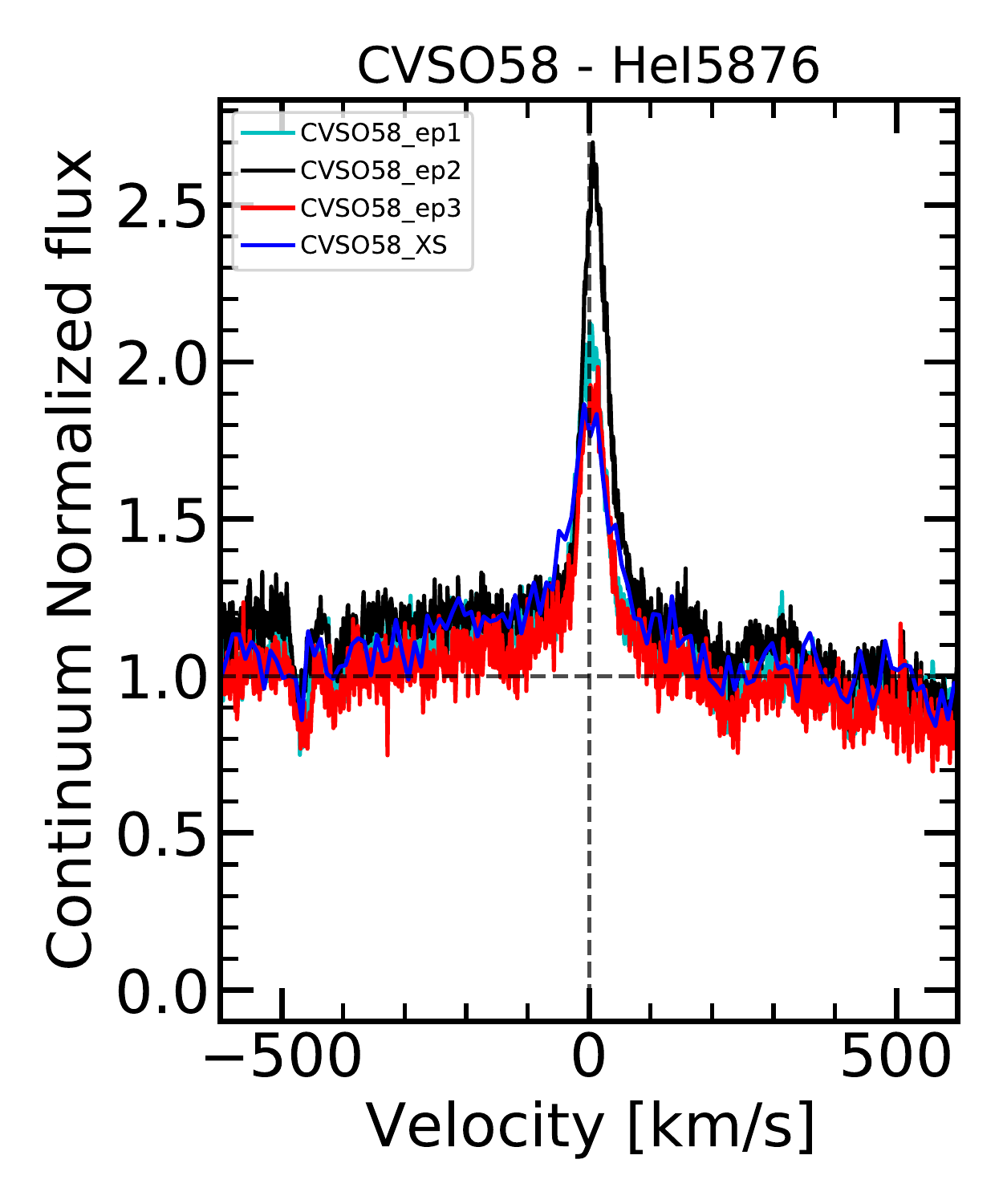}
\includegraphics[width=0.4\textwidth]{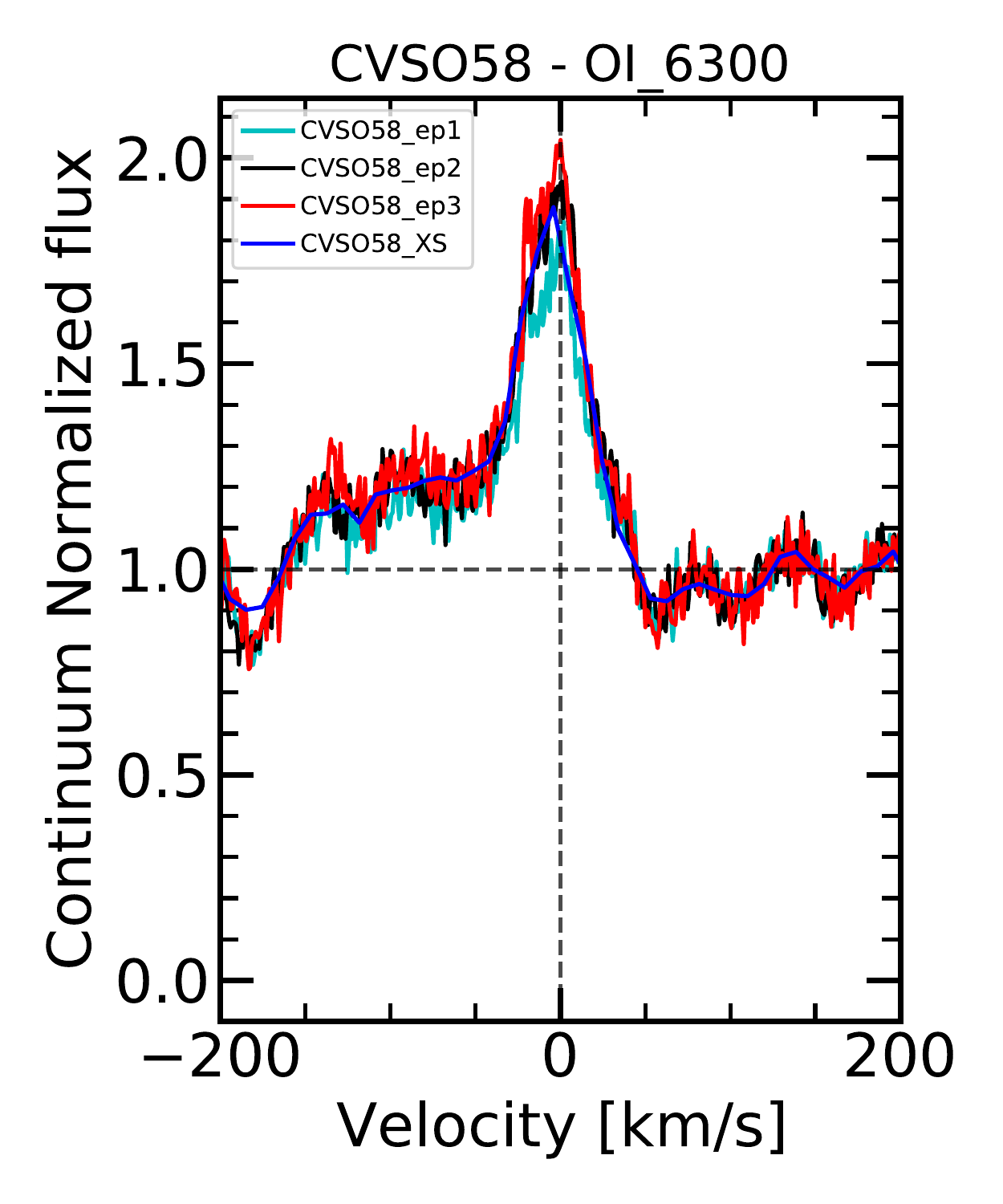}
\caption{Emission lines of the target CVSO58 observed with UVES and X-Shooter.
     \label{fig::lines_CVSO58}}
\end{figure*}

\begin{figure*}[]
\centering
\includegraphics[width=0.4\textwidth]{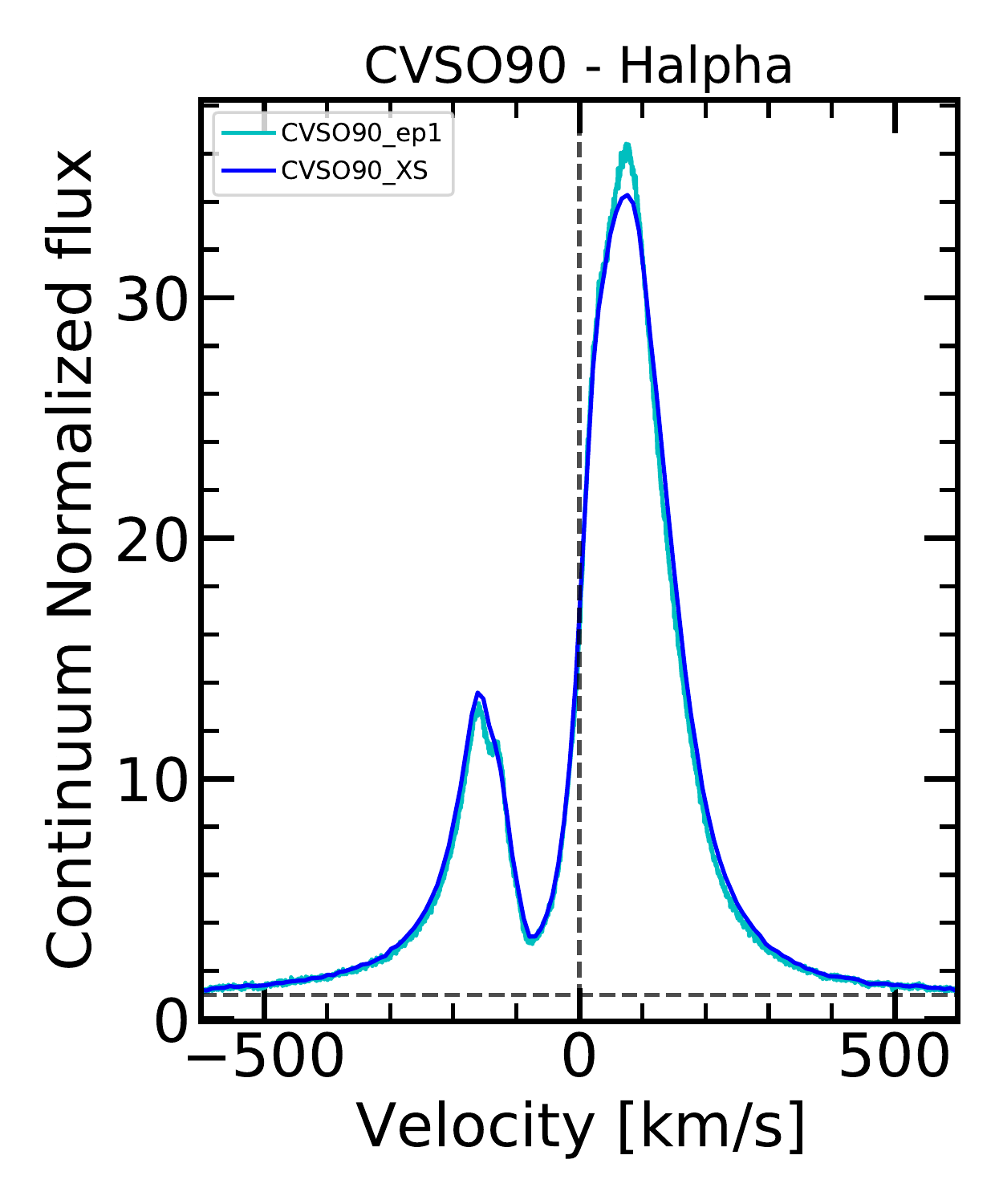}
\includegraphics[width=0.4\textwidth]{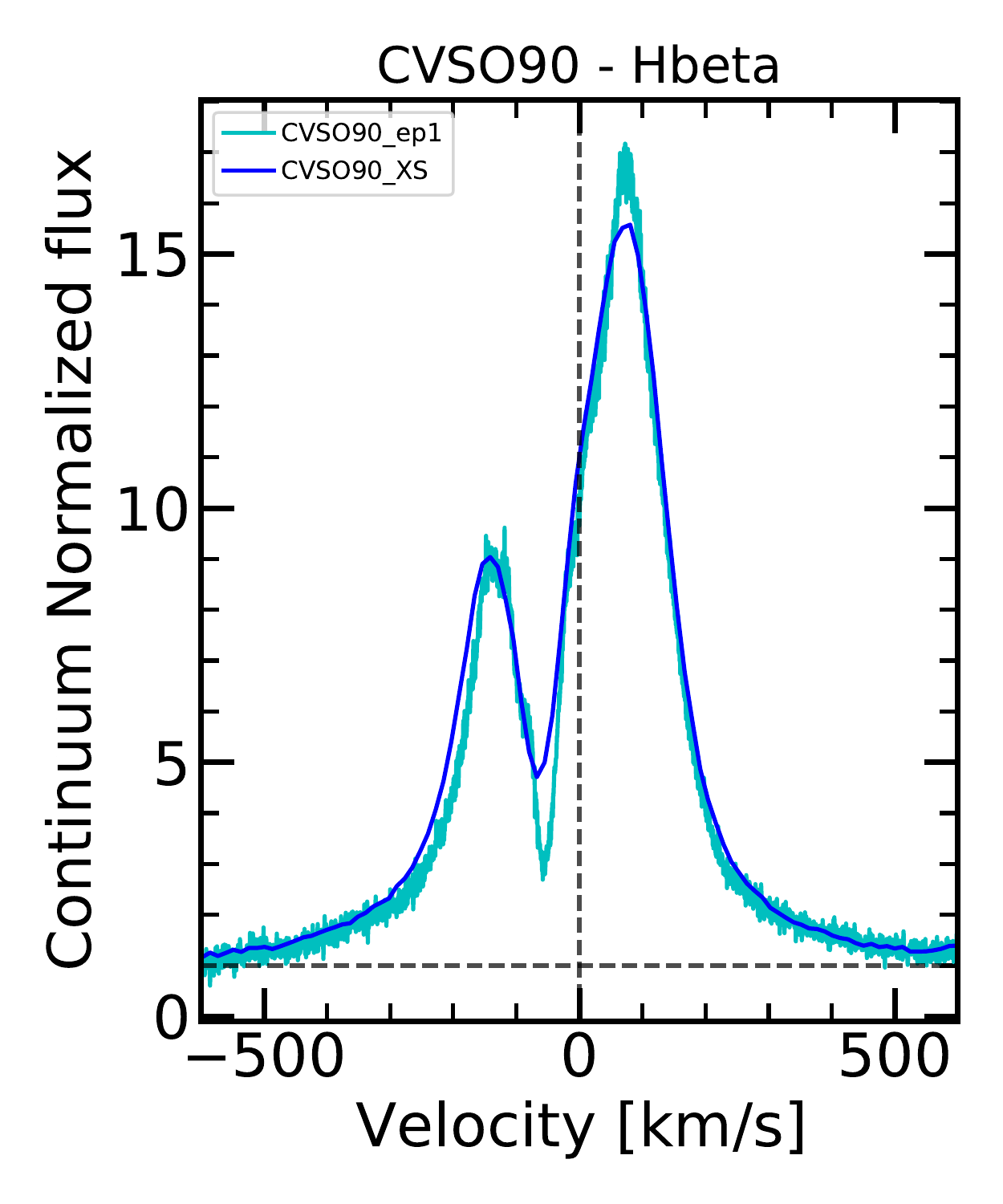}
\includegraphics[width=0.4\textwidth]{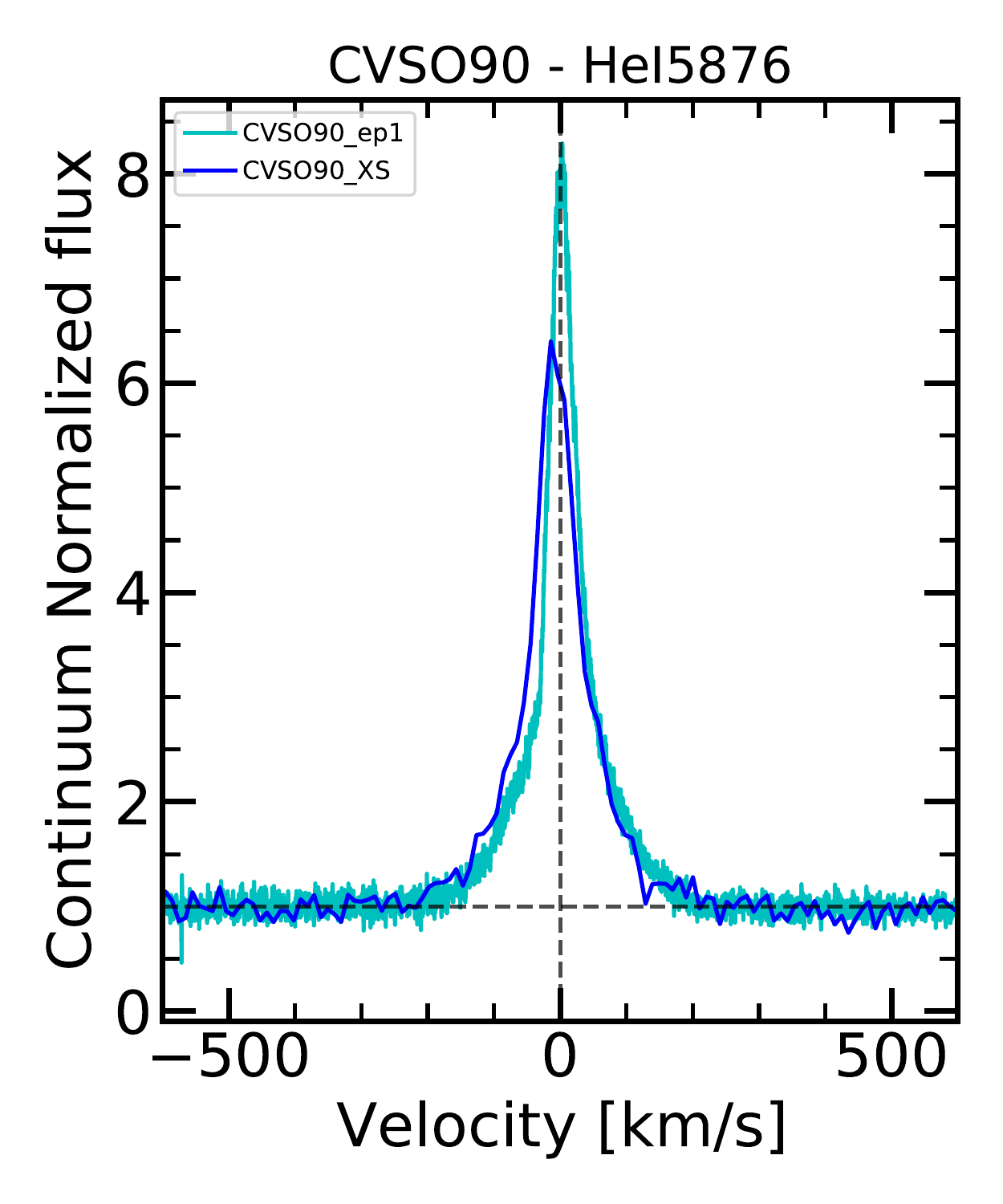}
\includegraphics[width=0.4\textwidth]{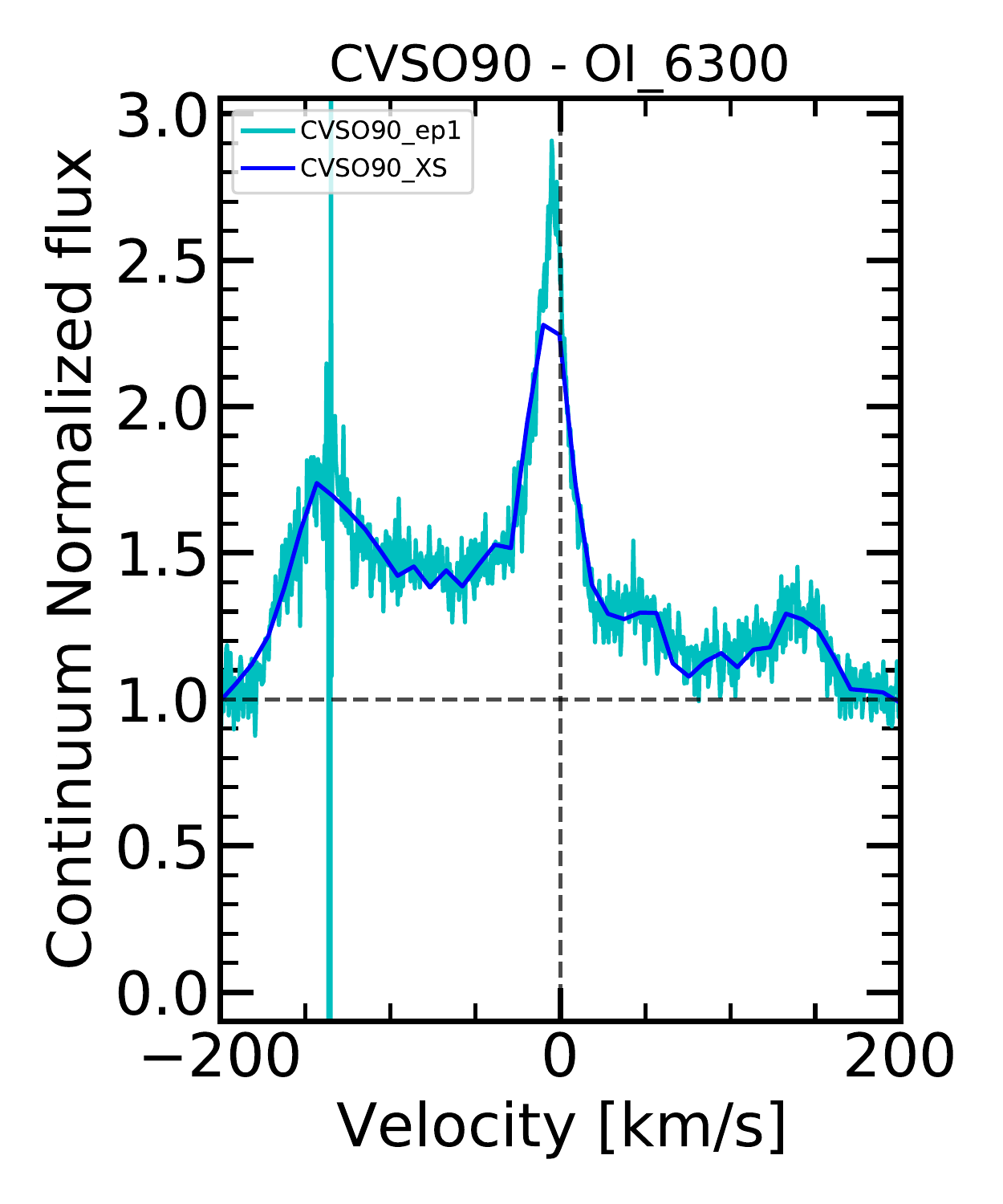}
\caption{Emission lines of the target CVSO90 observed with ESPRESSO and X-Shooter. Only one epoch of ESPRESSO data was taken.
     \label{fig::lines_CVSO90}}
\end{figure*}

\begin{figure*}[]
\centering
\includegraphics[width=0.4\textwidth]{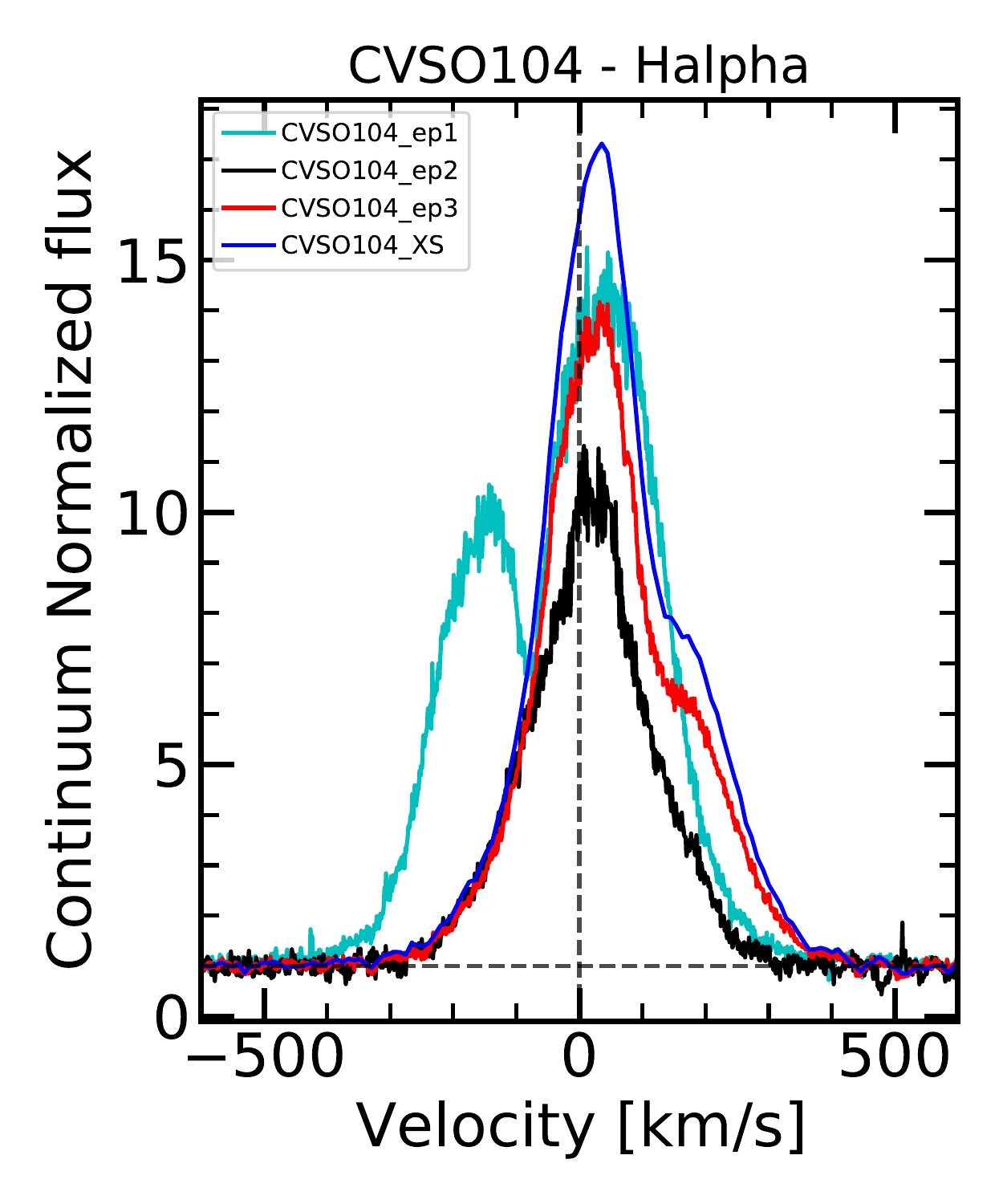}
\includegraphics[width=0.4\textwidth]{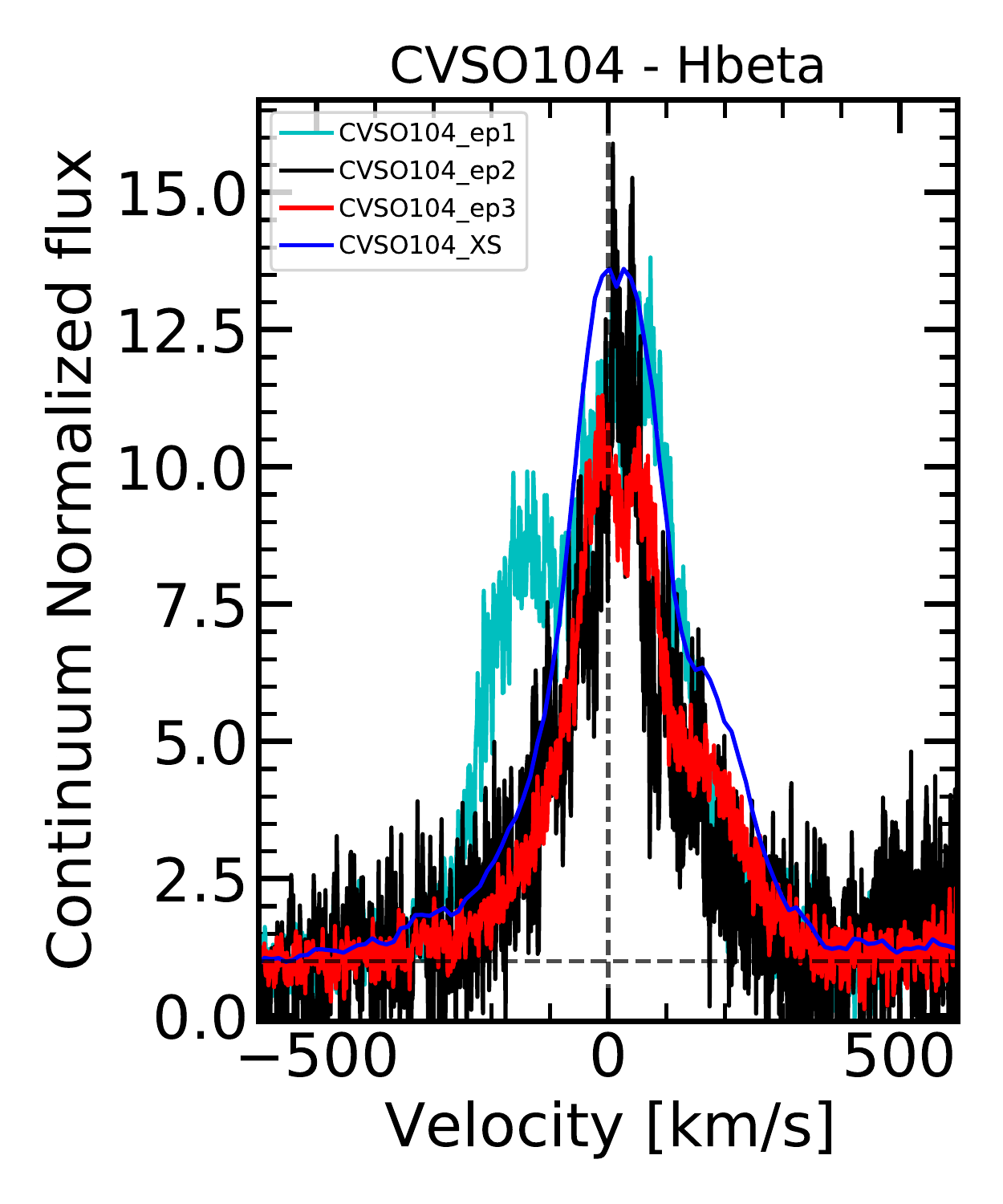}
\includegraphics[width=0.4\textwidth]{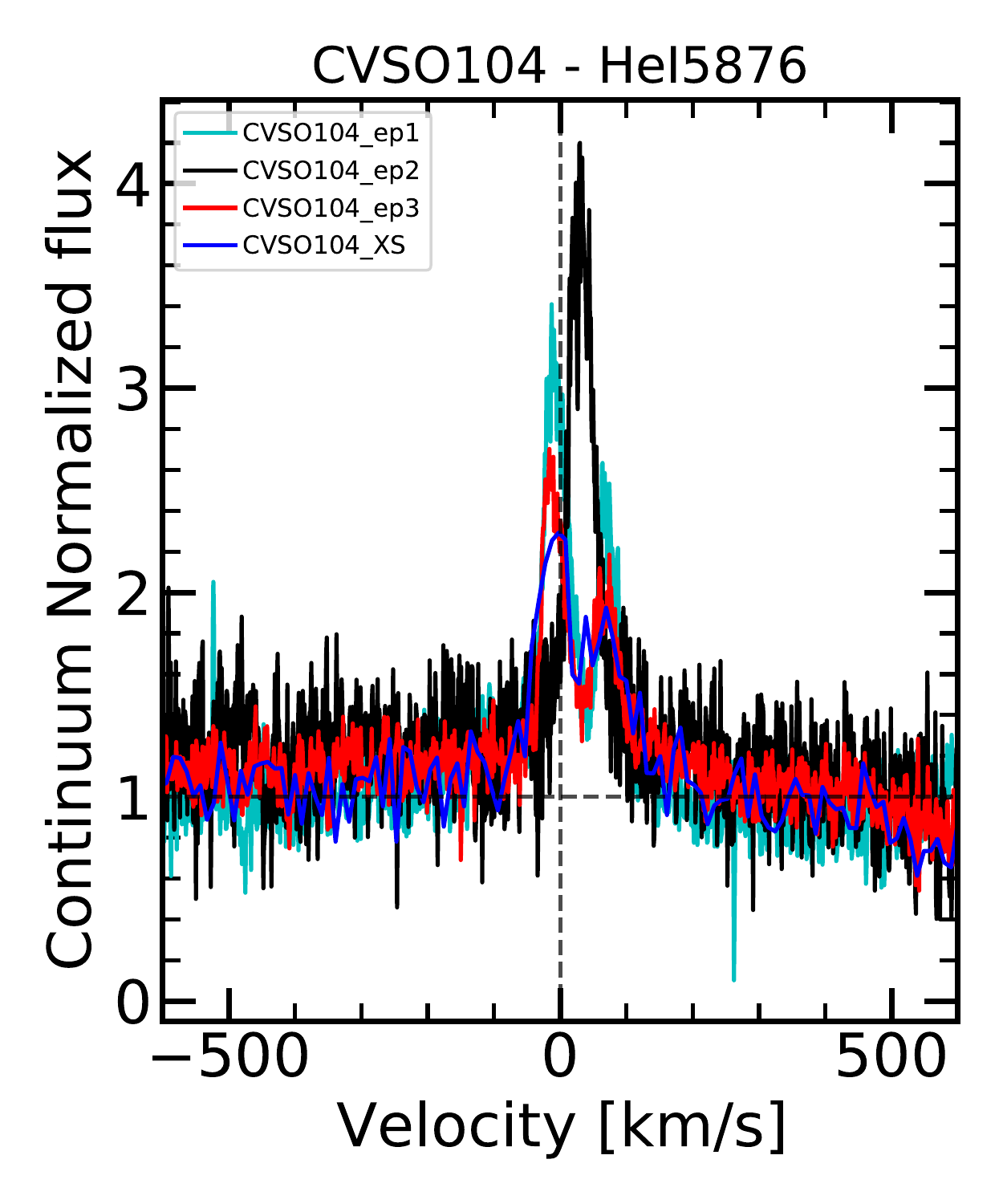}
\includegraphics[width=0.4\textwidth]{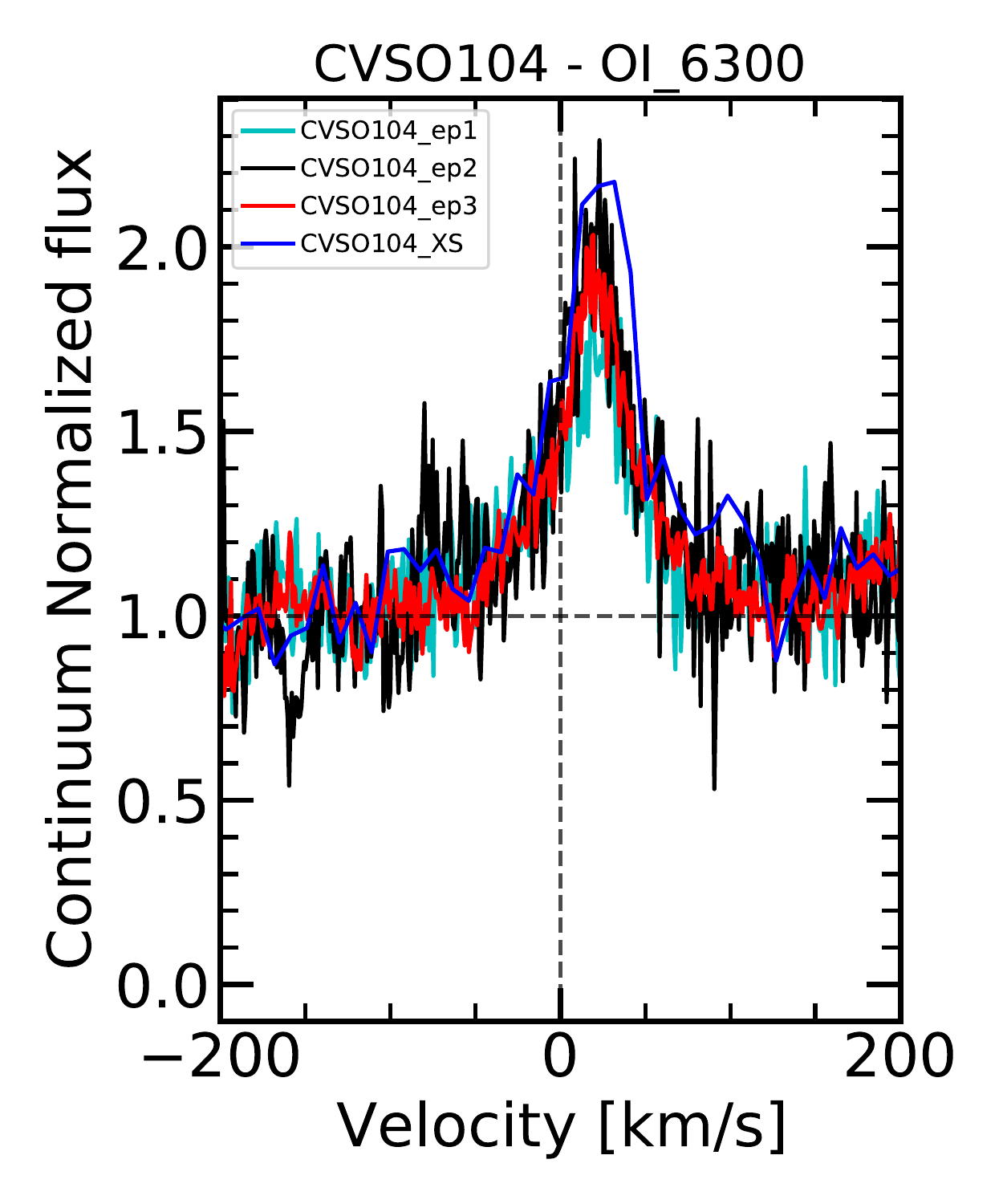}
\caption{Emission lines of the target CVSO104 observed with UVES and X-Shooter. This target is a spectroscopic binary.
     \label{fig::lines_CVSO104}}
\end{figure*}

\begin{figure*}[]
\centering
\includegraphics[width=0.4\textwidth]{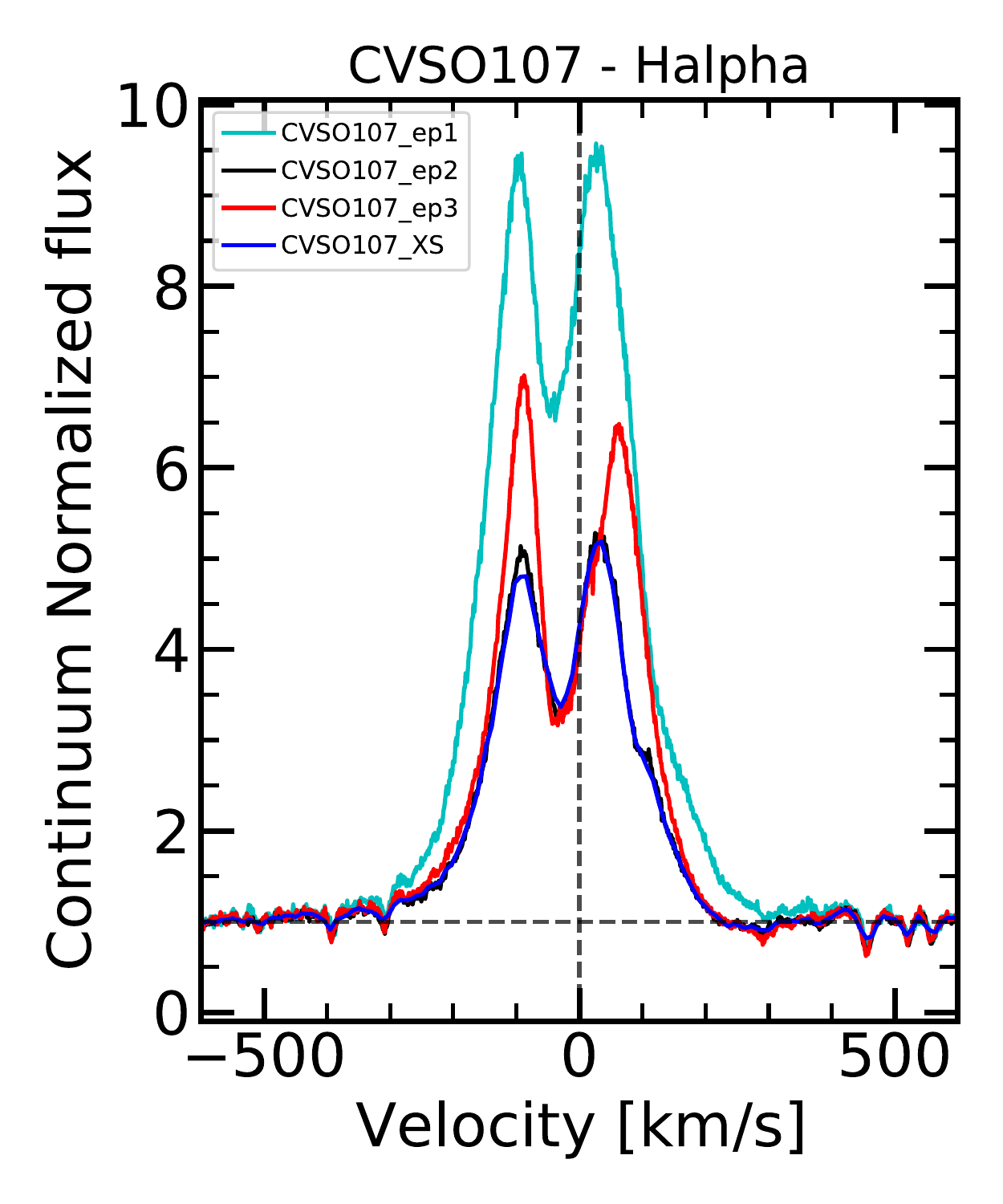}
\includegraphics[width=0.4\textwidth]{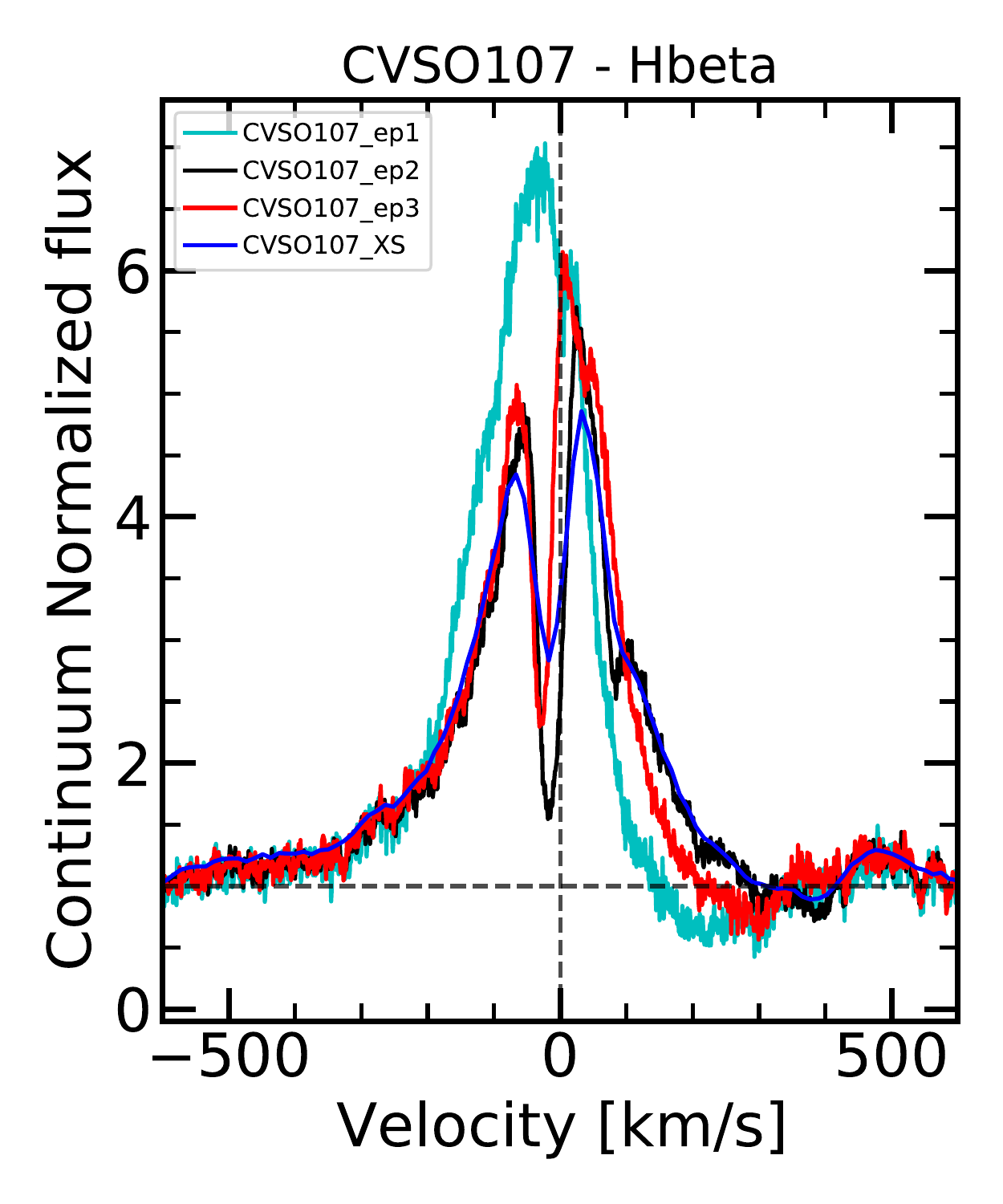}
\includegraphics[width=0.4\textwidth]{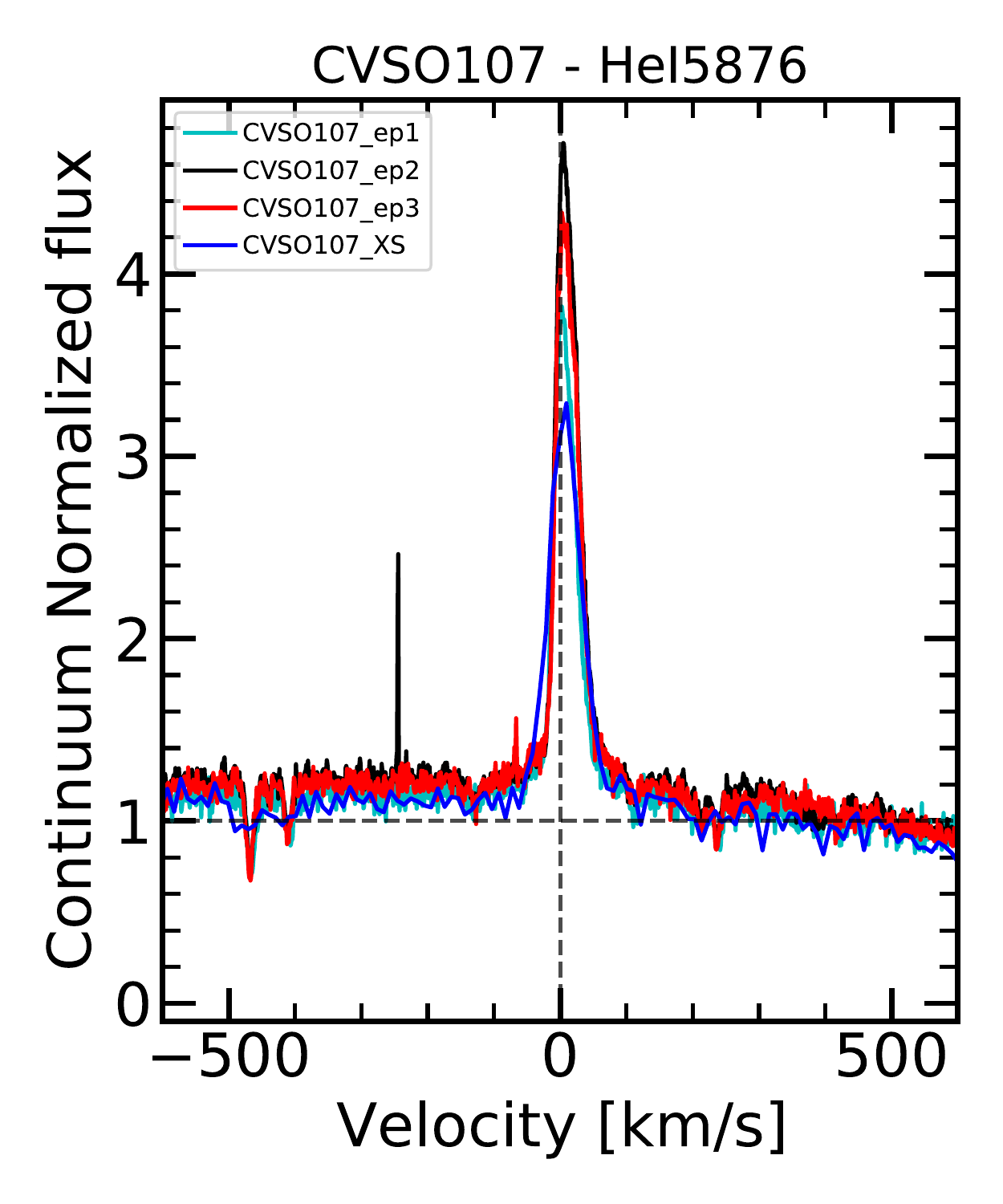}
\includegraphics[width=0.4\textwidth]{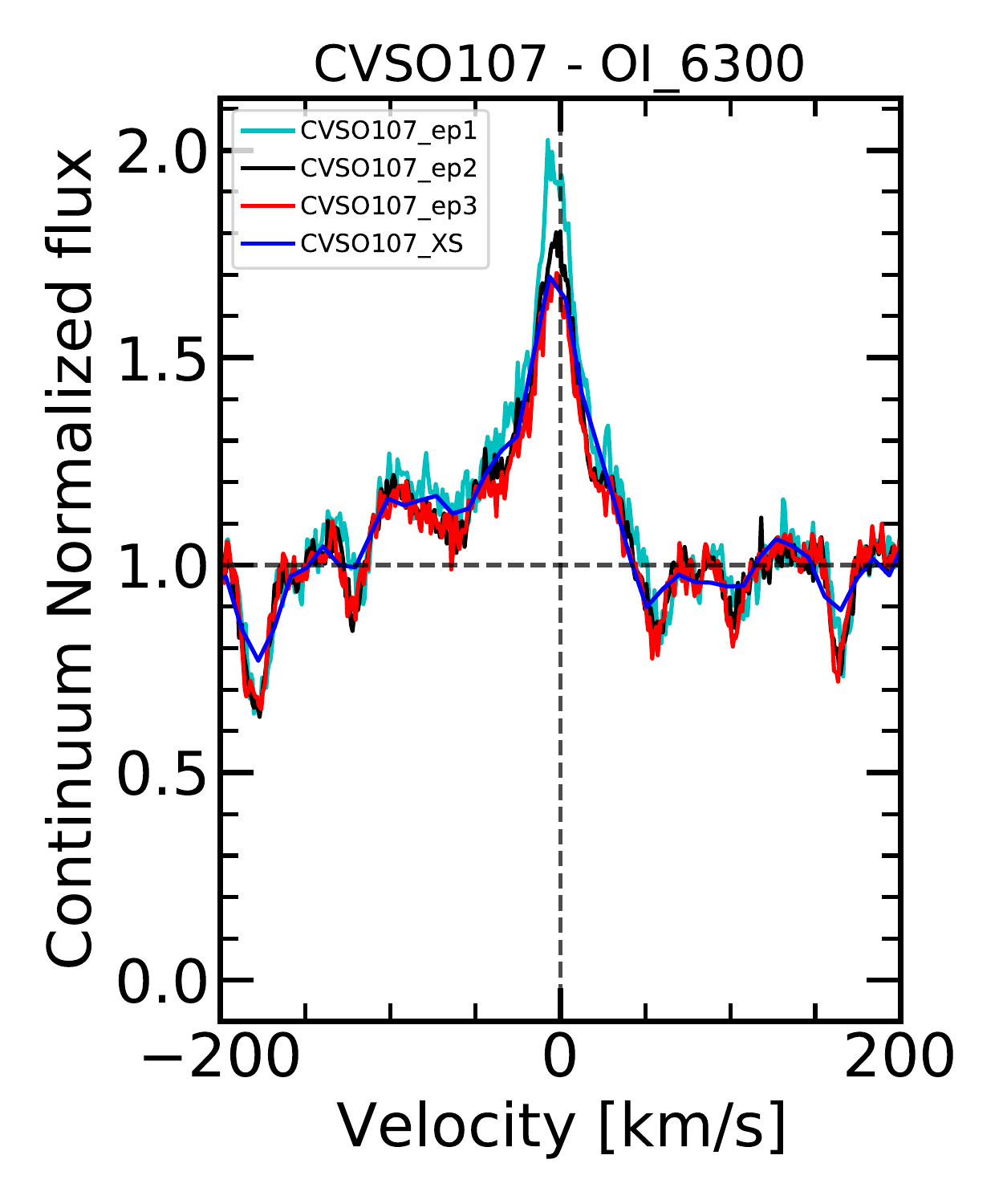}
\caption{Emission lines of the target CVSO107 observed with UVES and X-Shooter.
     \label{fig::lines_CVSO107}}
\end{figure*}

\begin{figure*}[]
\centering
\includegraphics[width=0.4\textwidth]{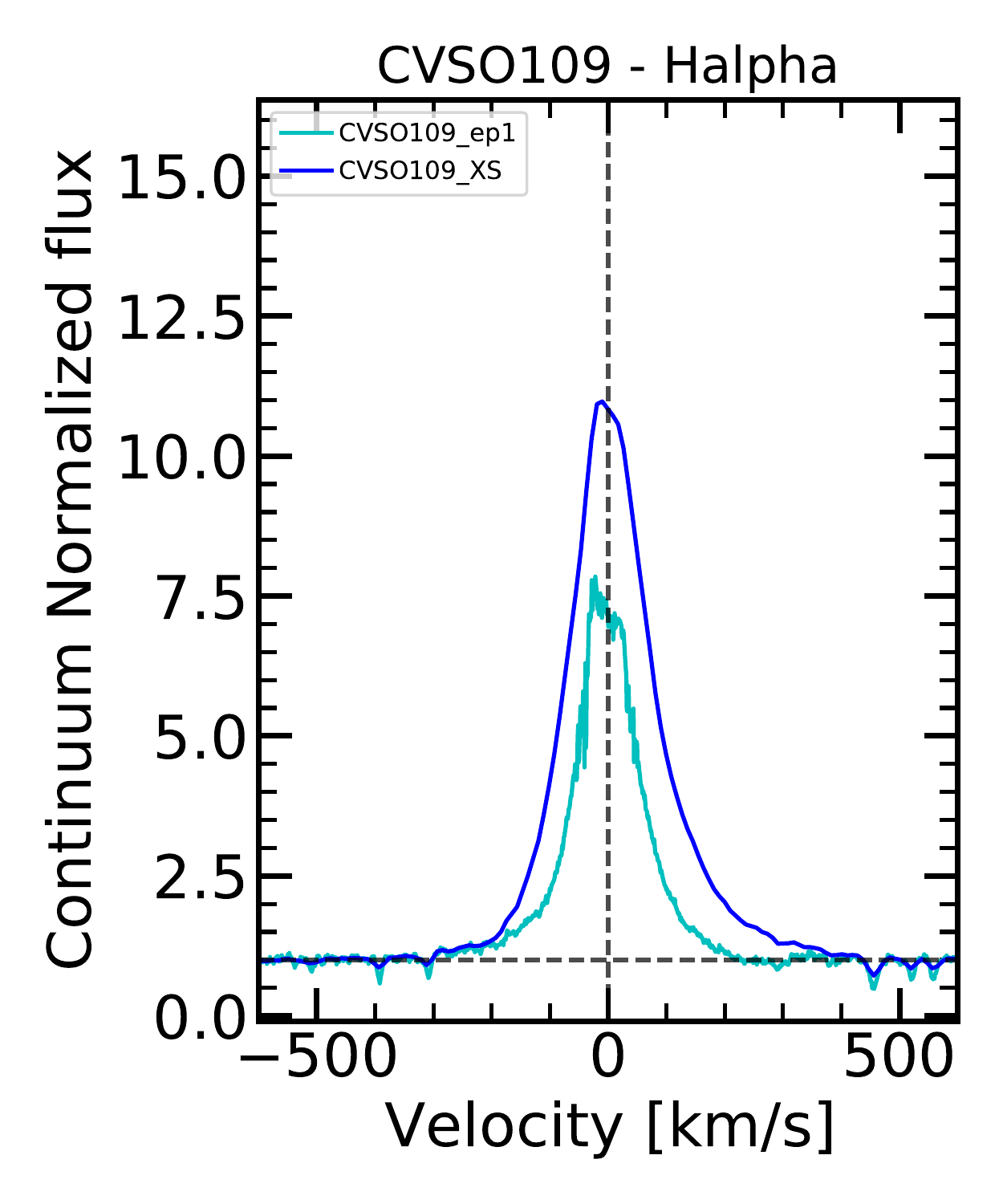}
\includegraphics[width=0.4\textwidth]{CVSO109_Hbeta_4861_compare.pdf}
\includegraphics[width=0.4\textwidth]{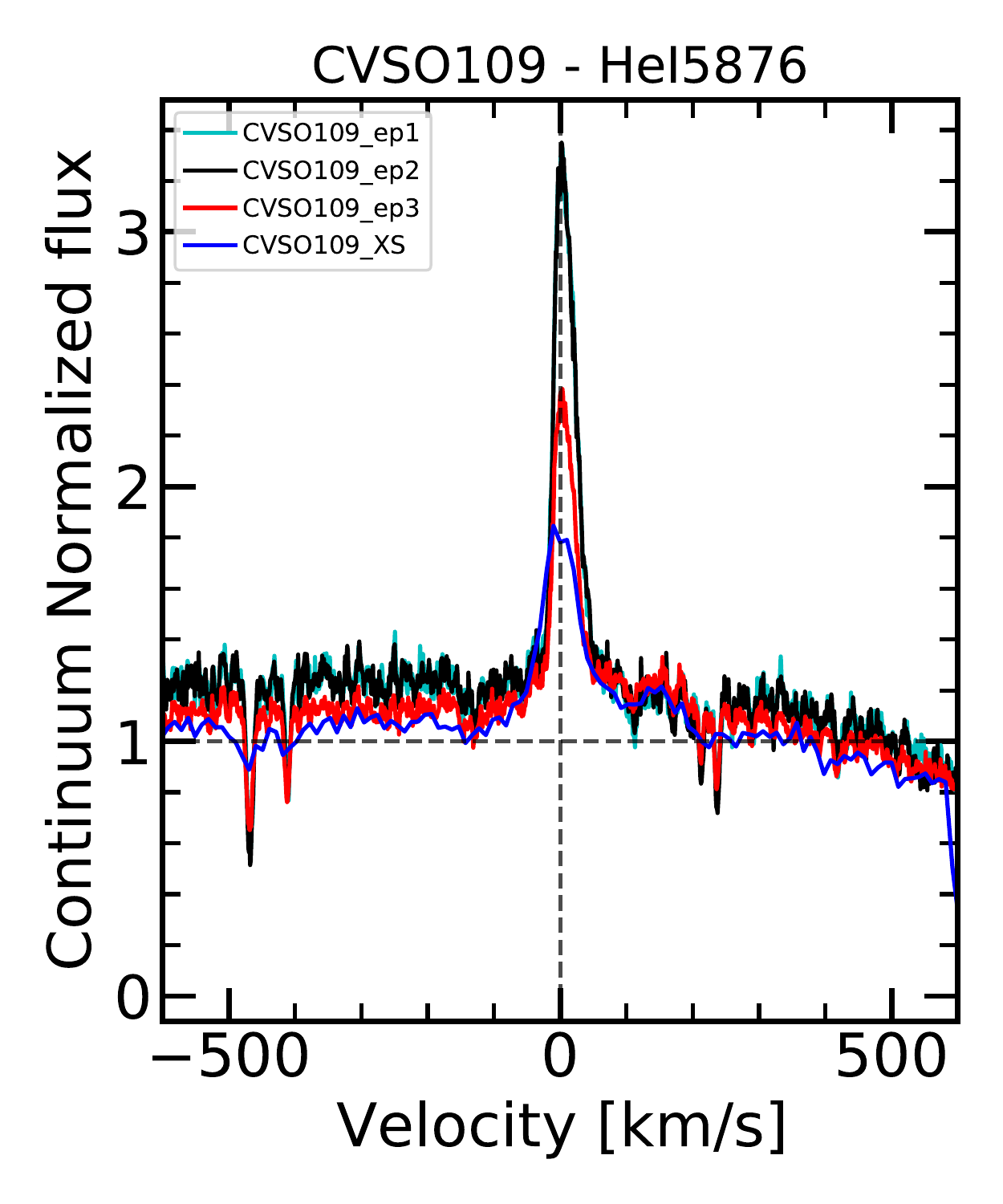}
\includegraphics[width=0.4\textwidth]{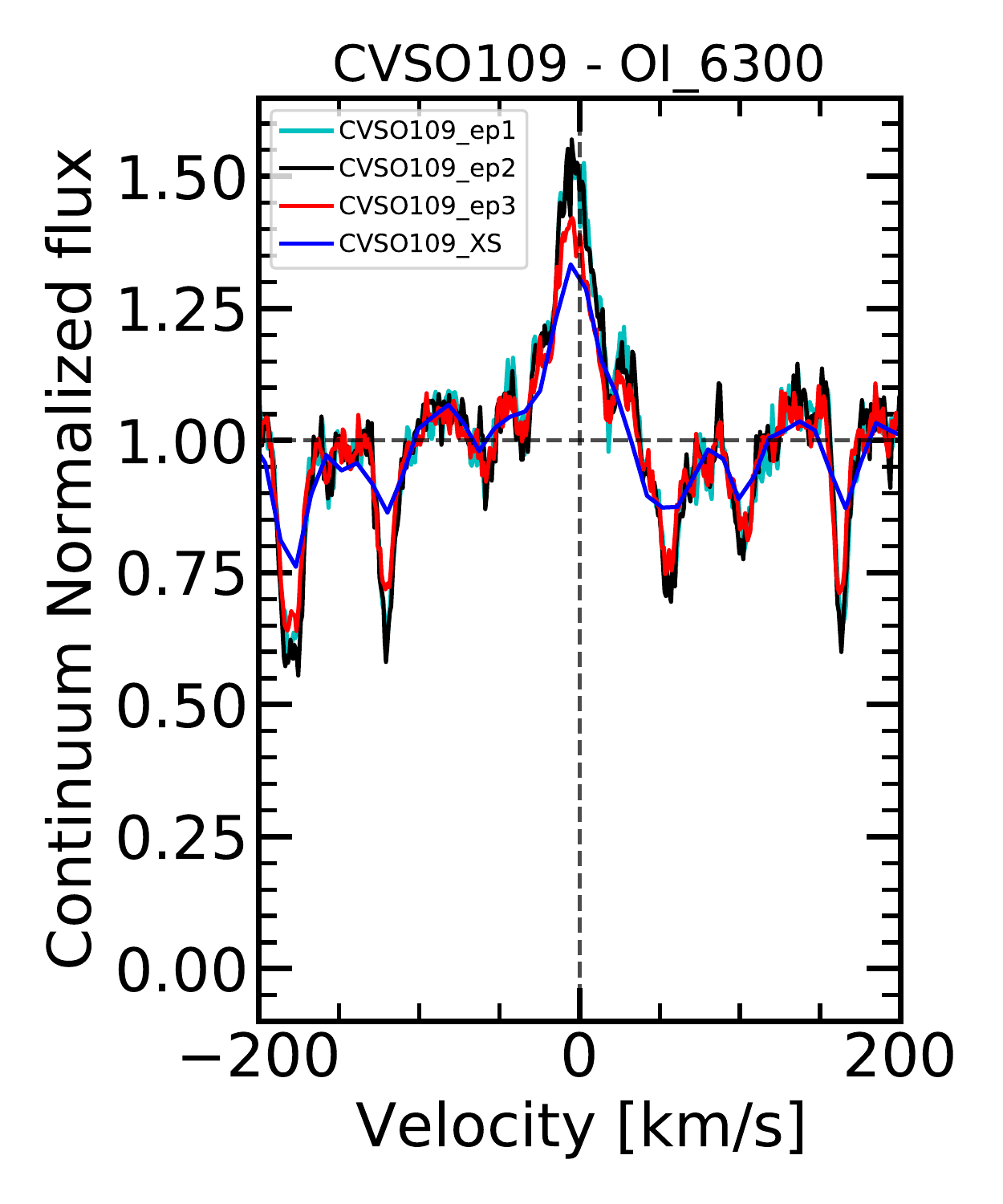}
\caption{Emission lines of the target CVSO109 observed with UVES and X-Shooter. In the second and third epoch of UVES observations of CVSO109 the H$\alpha$ line is saturated, and this is therefore not shown here. 
     \label{fig::lines_CVSO109}}
\end{figure*}

\begin{figure*}[]
\centering
\includegraphics[width=0.4\textwidth]{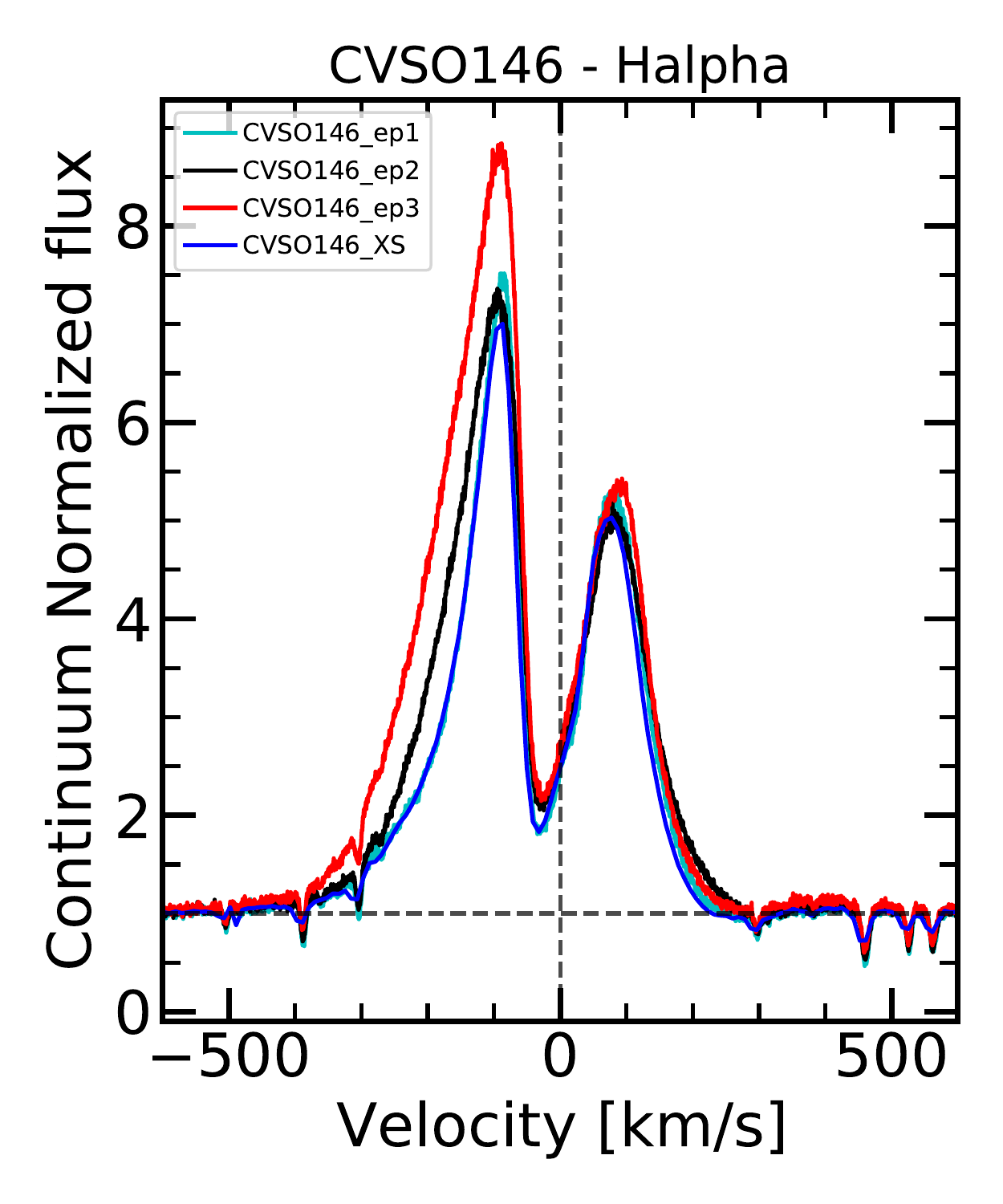}
\includegraphics[width=0.4\textwidth]{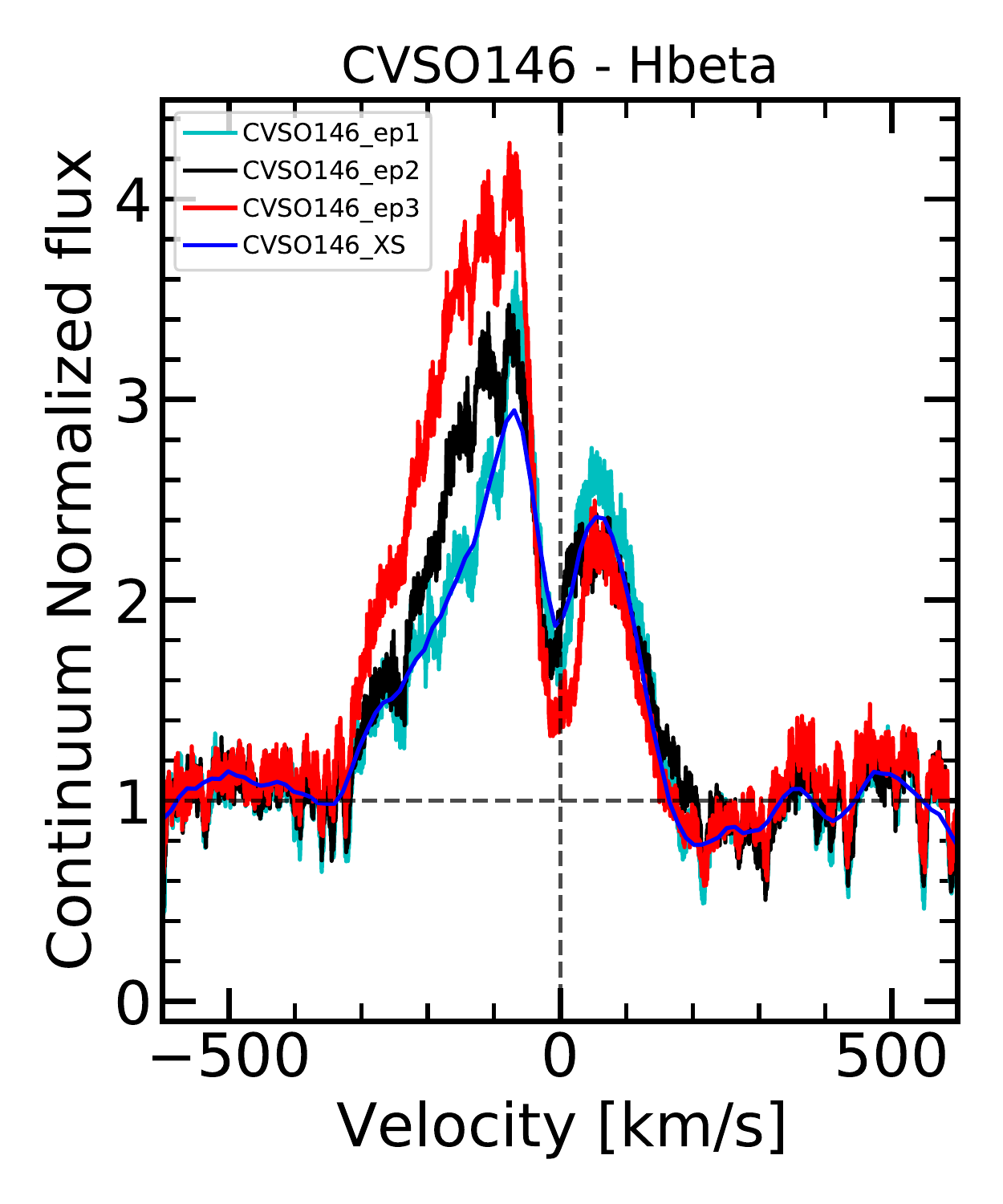}
\includegraphics[width=0.4\textwidth]{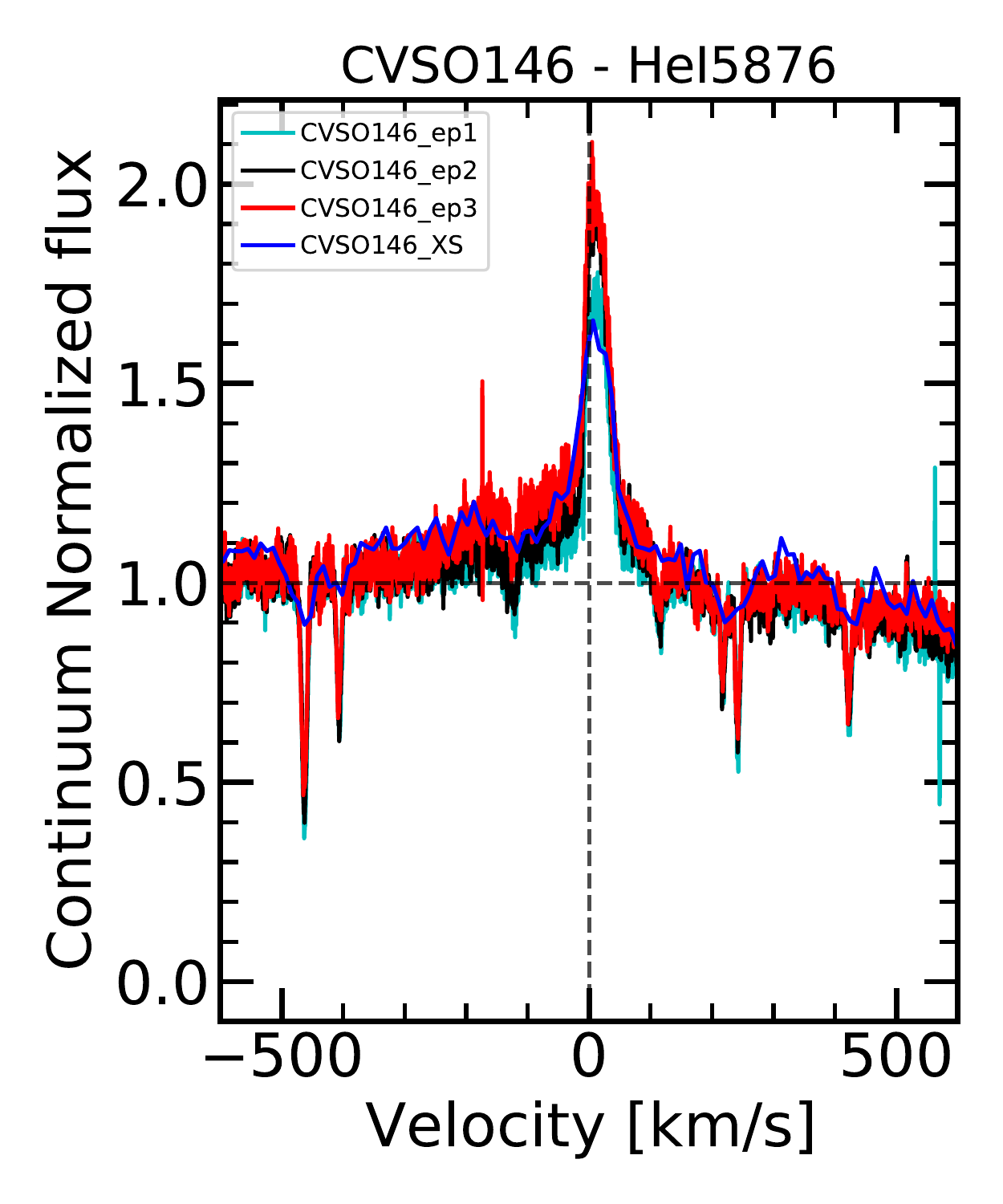}
\includegraphics[width=0.4\textwidth]{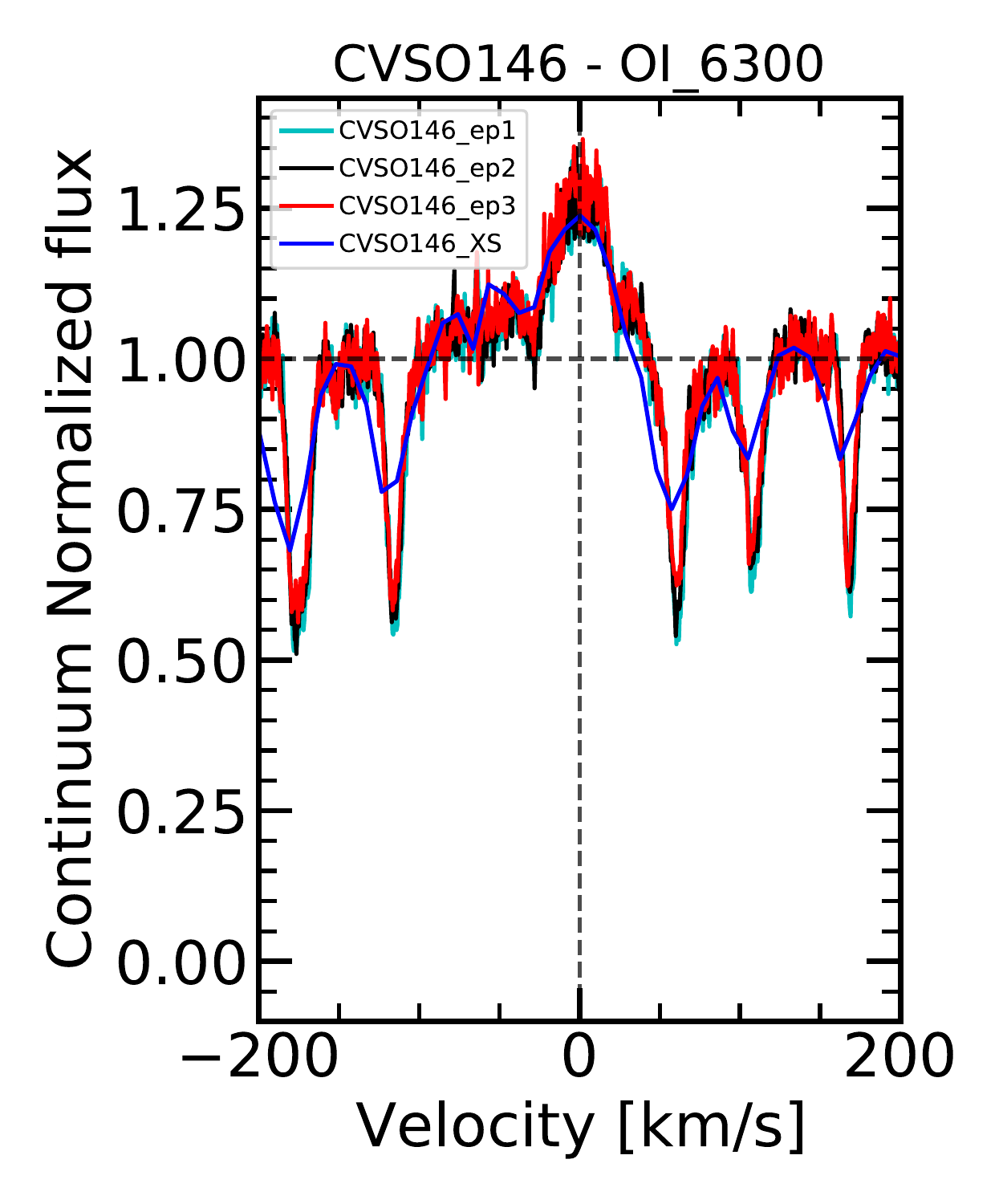}
\caption{Emission lines of the target CVSO146 observed with ESPRESSO and X-Shooter.
     \label{fig::lines_CVSO146}}
\end{figure*}

\begin{figure*}[]
\centering
\includegraphics[width=0.4\textwidth]{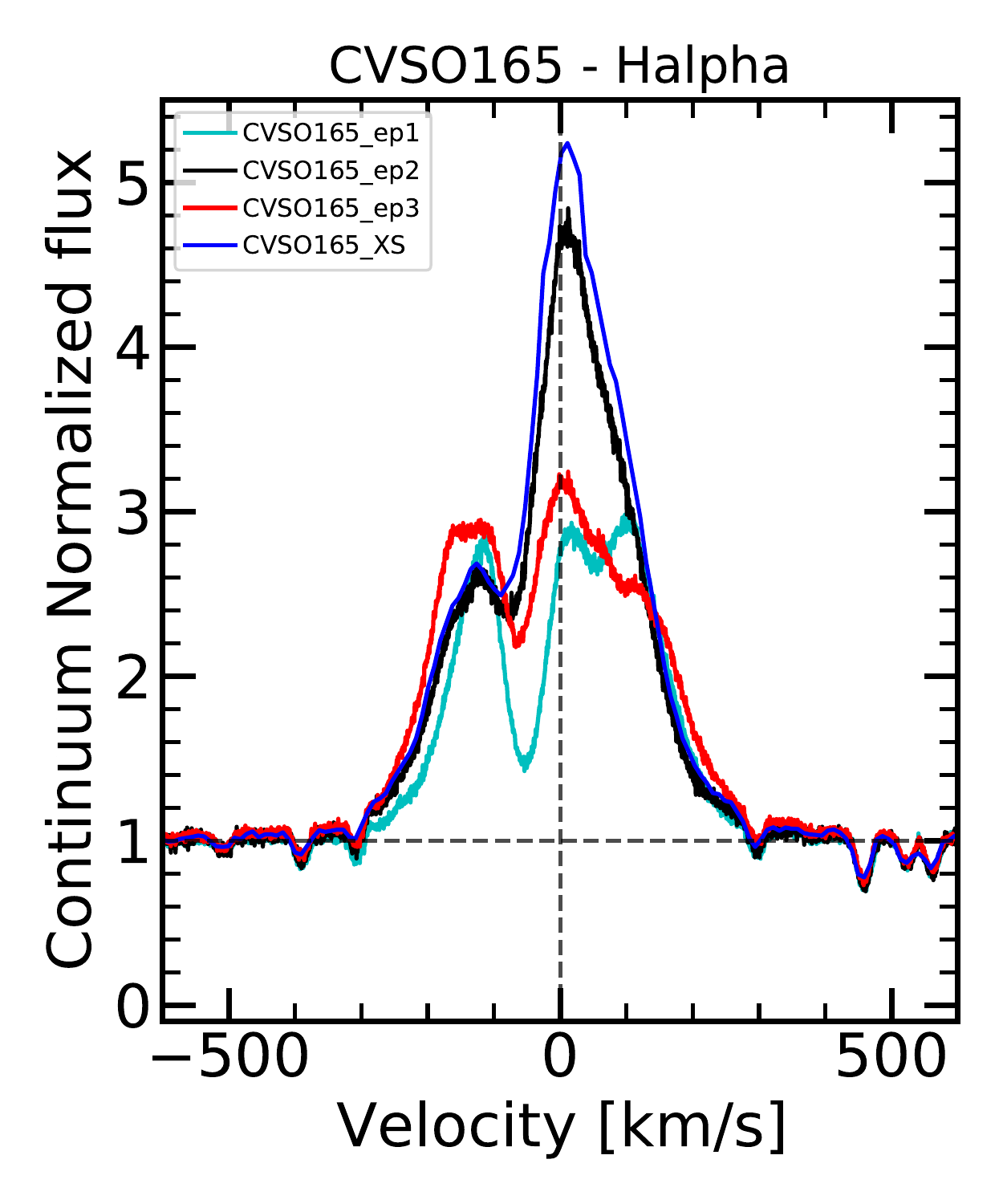}
\includegraphics[width=0.4\textwidth]{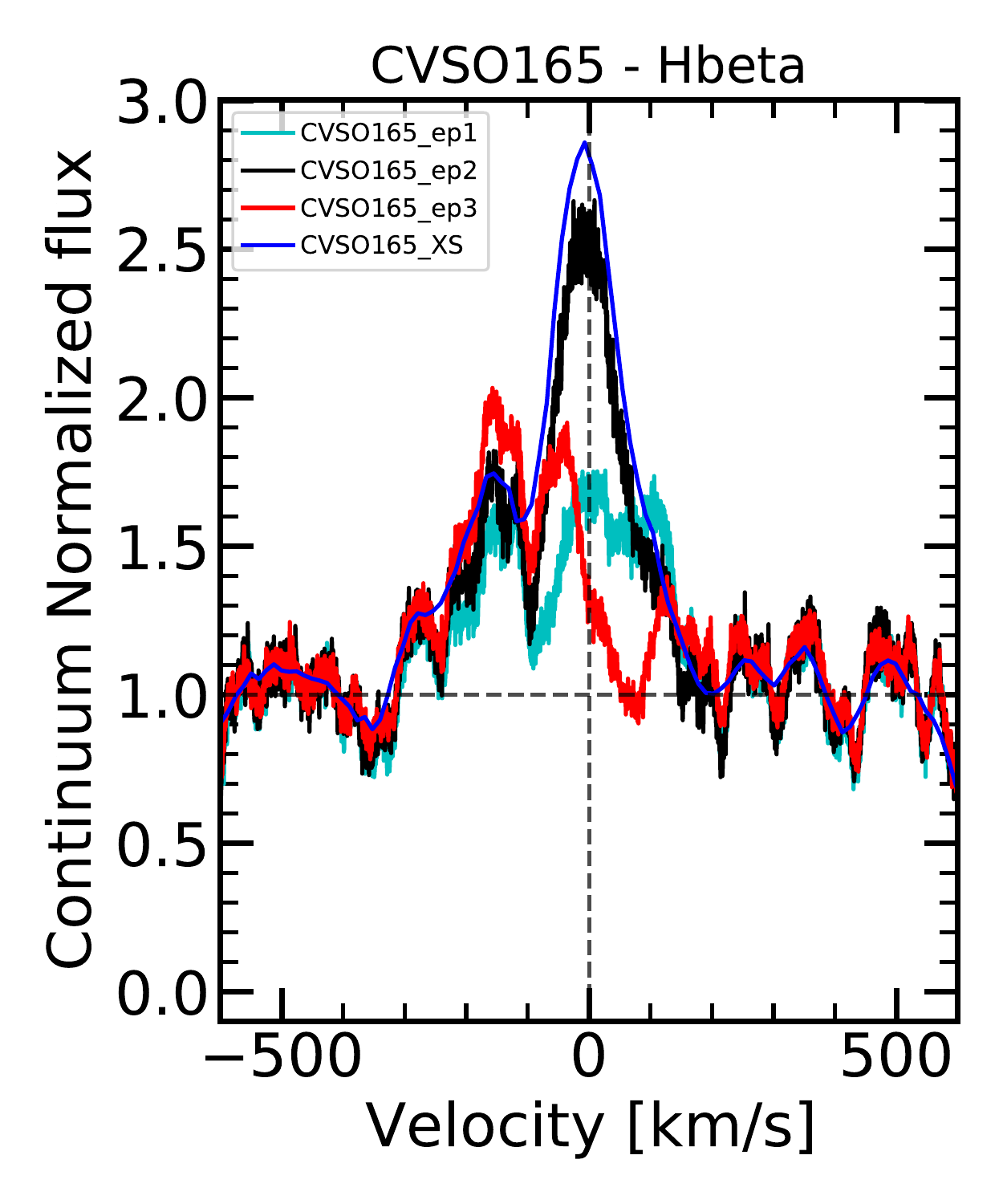}
\includegraphics[width=0.4\textwidth]{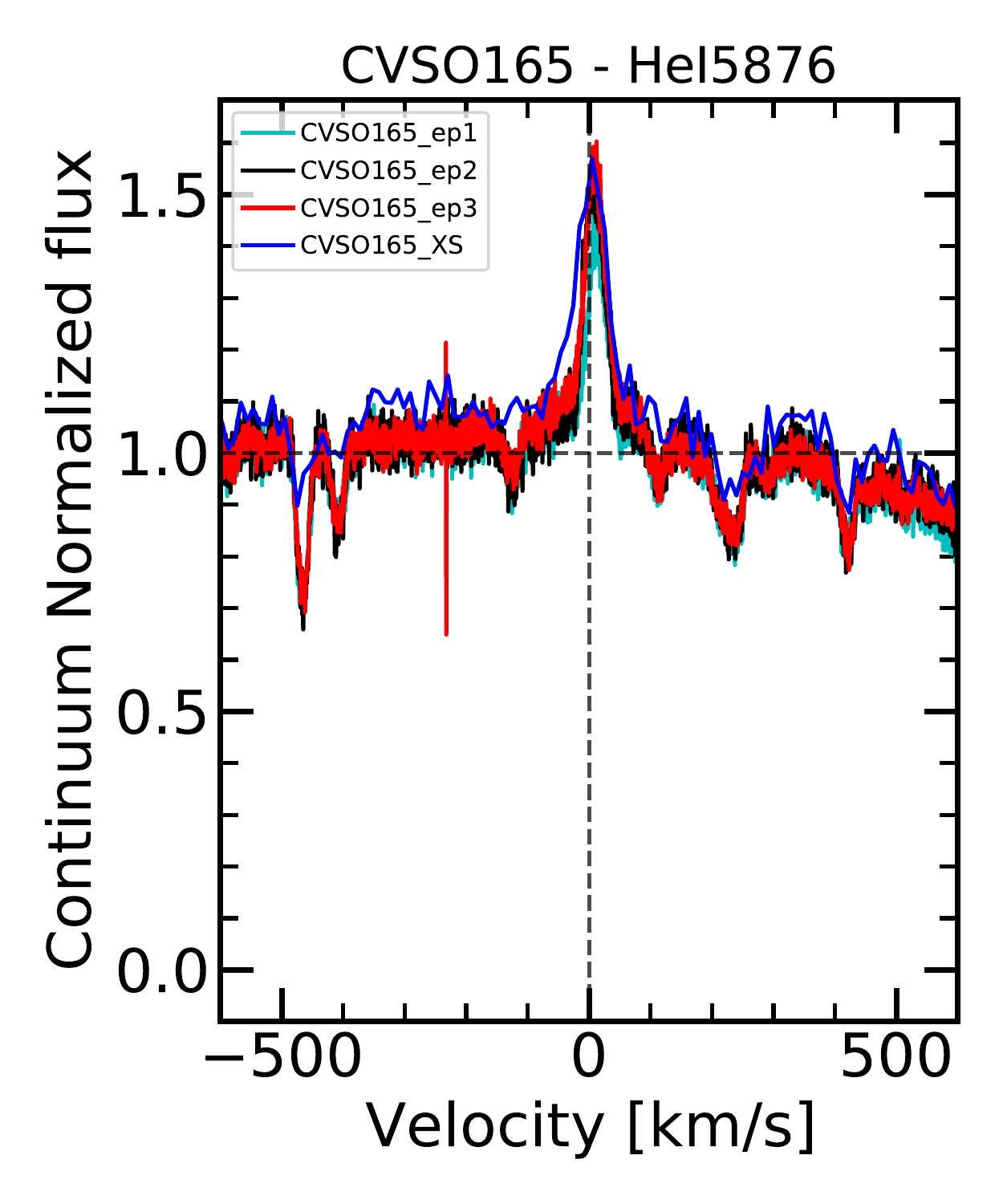}
\includegraphics[width=0.4\textwidth]{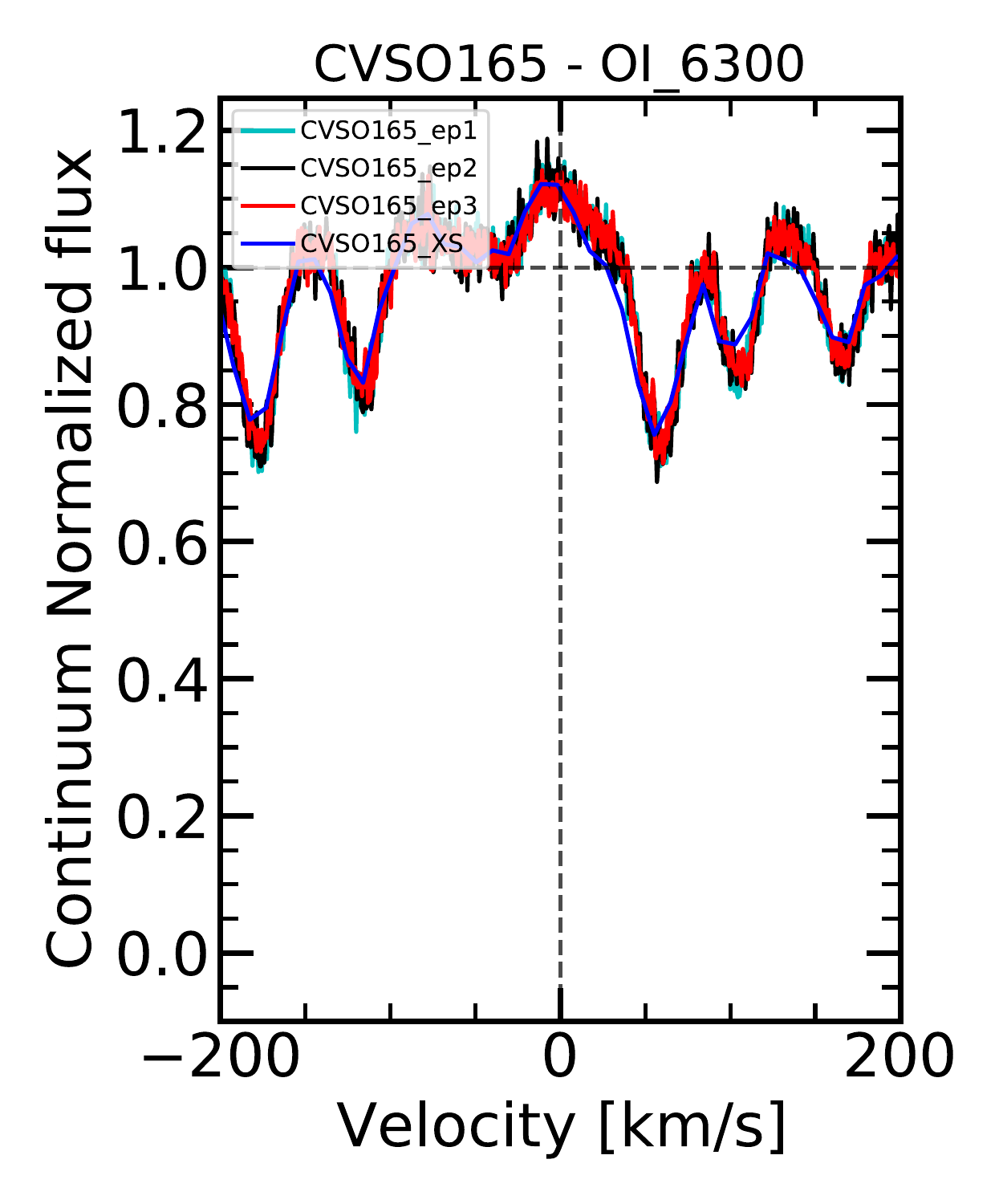}
\caption{Emission lines of the target CVSO165 observed with ESPRESSO and X-Shooter.
     \label{fig::lines_CVSO165}}
\end{figure*}

\begin{figure*}[]
\centering
\includegraphics[width=0.4\textwidth]{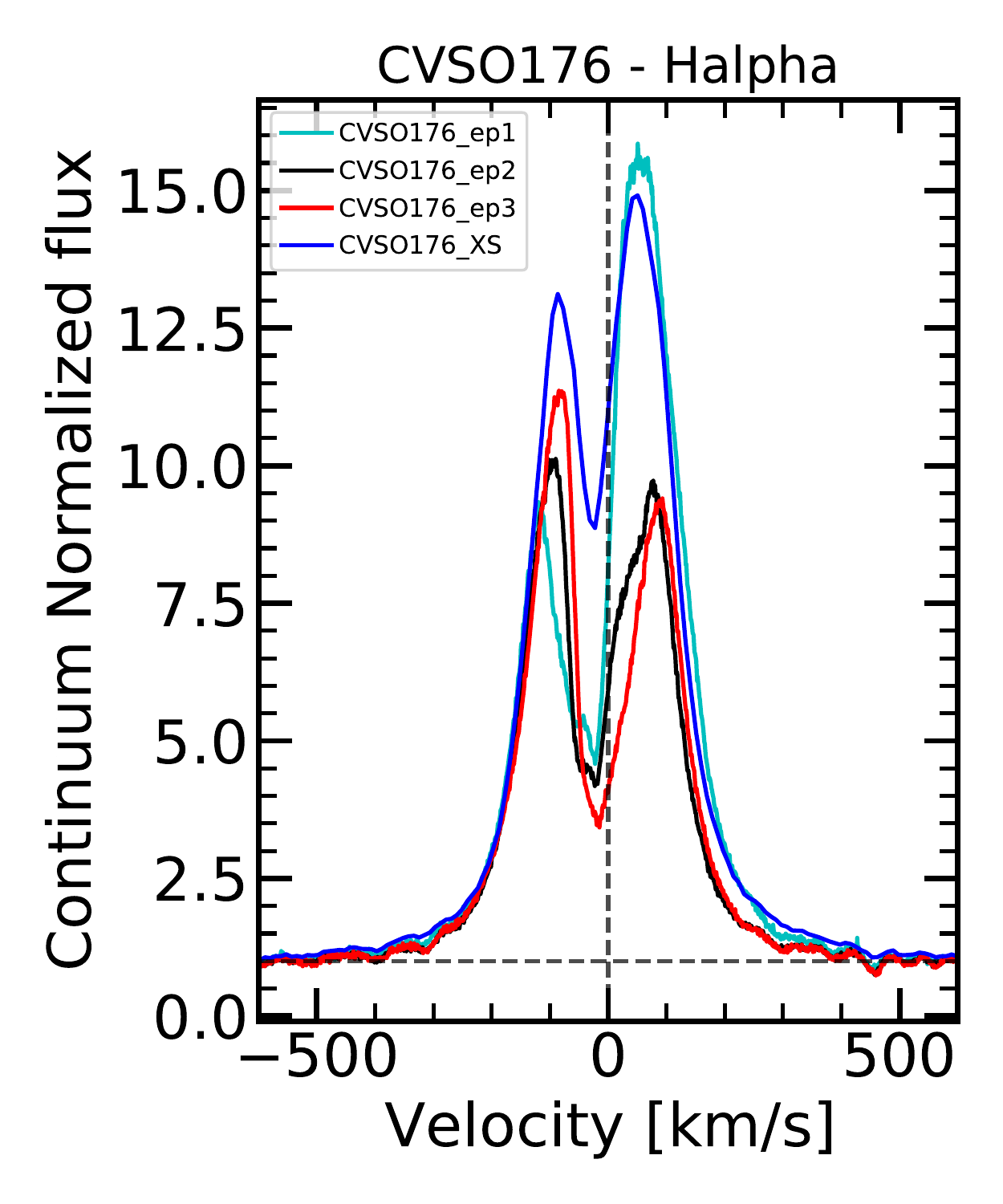}
\includegraphics[width=0.4\textwidth]{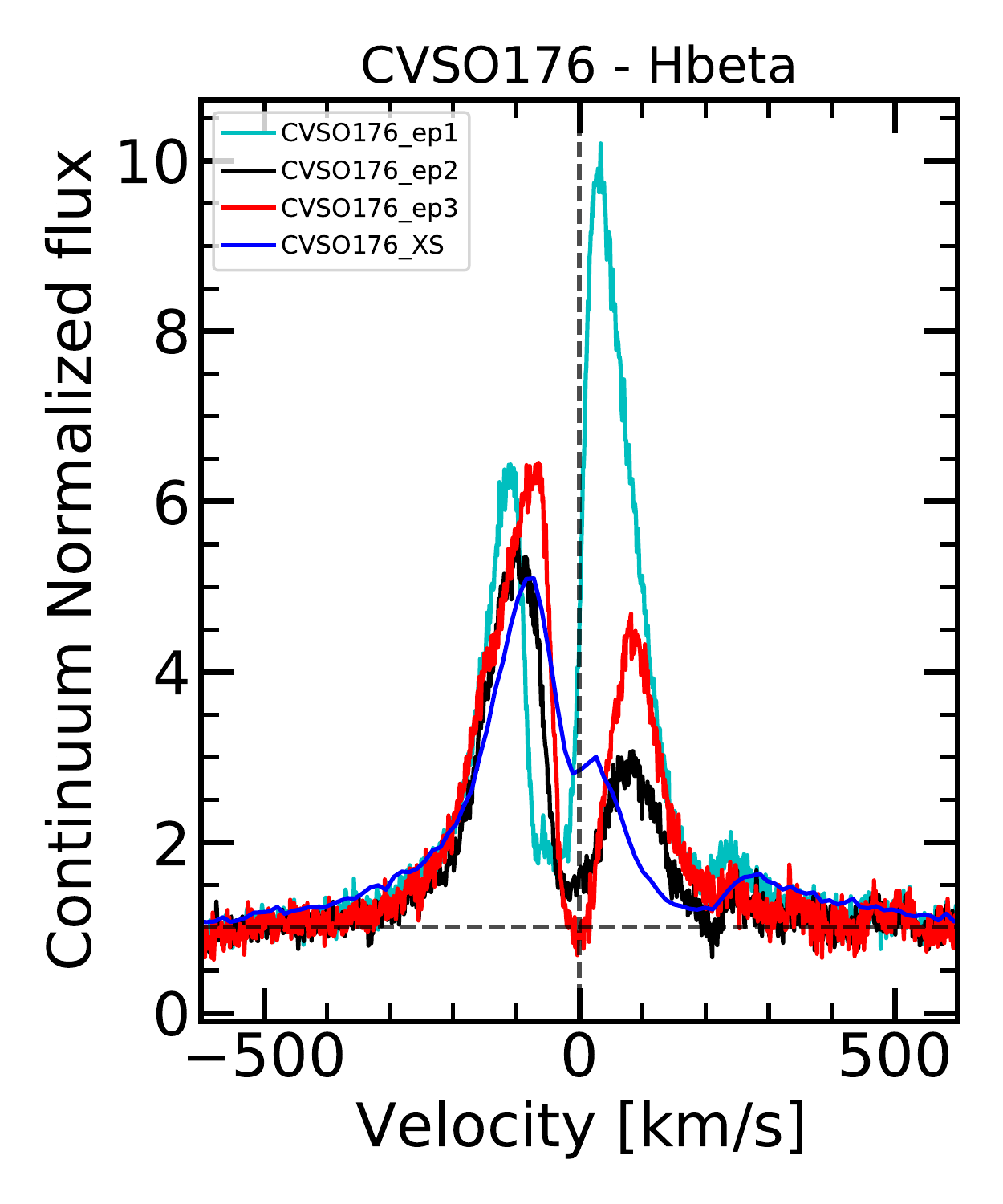}
\includegraphics[width=0.4\textwidth]{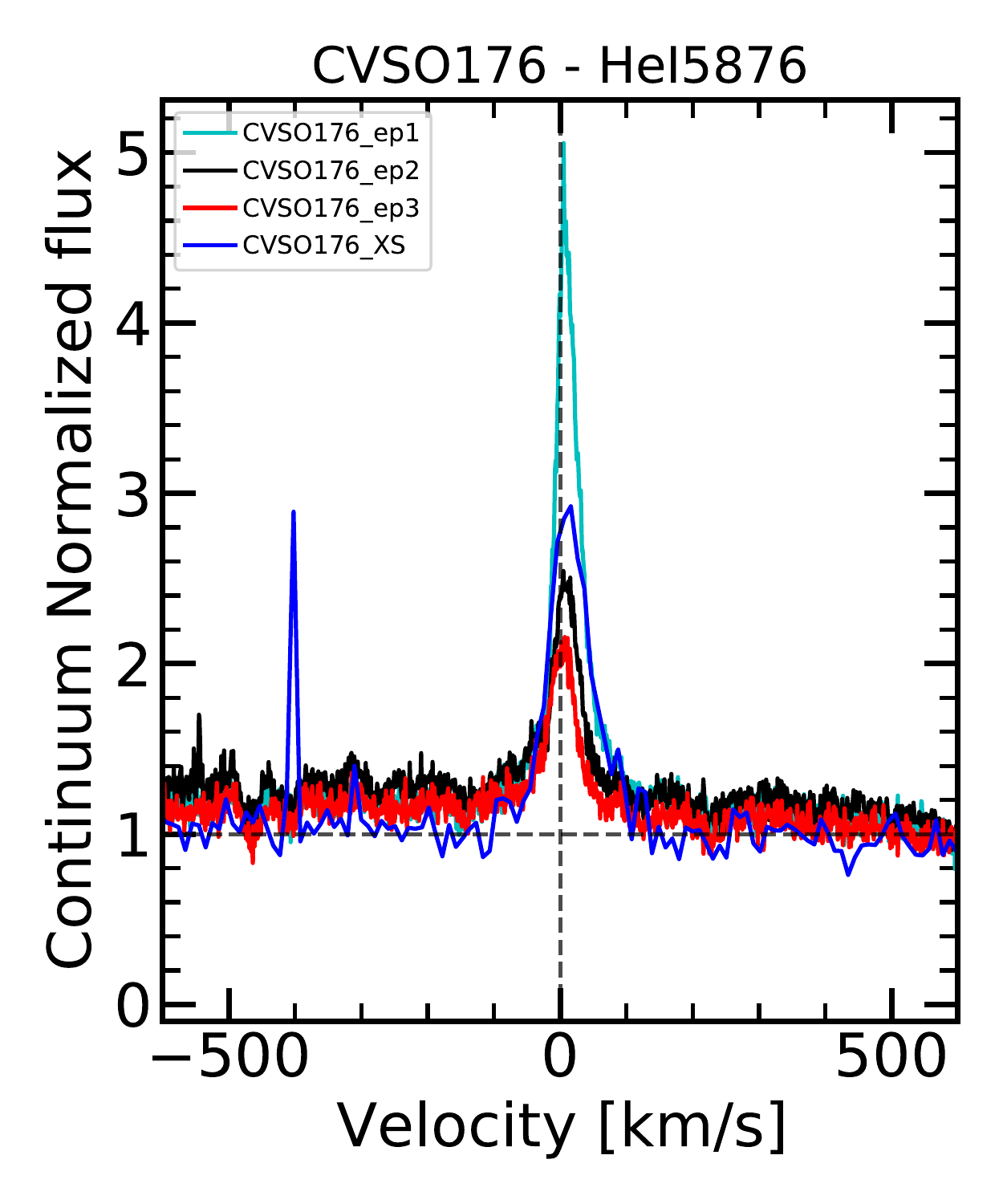}
\includegraphics[width=0.4\textwidth]{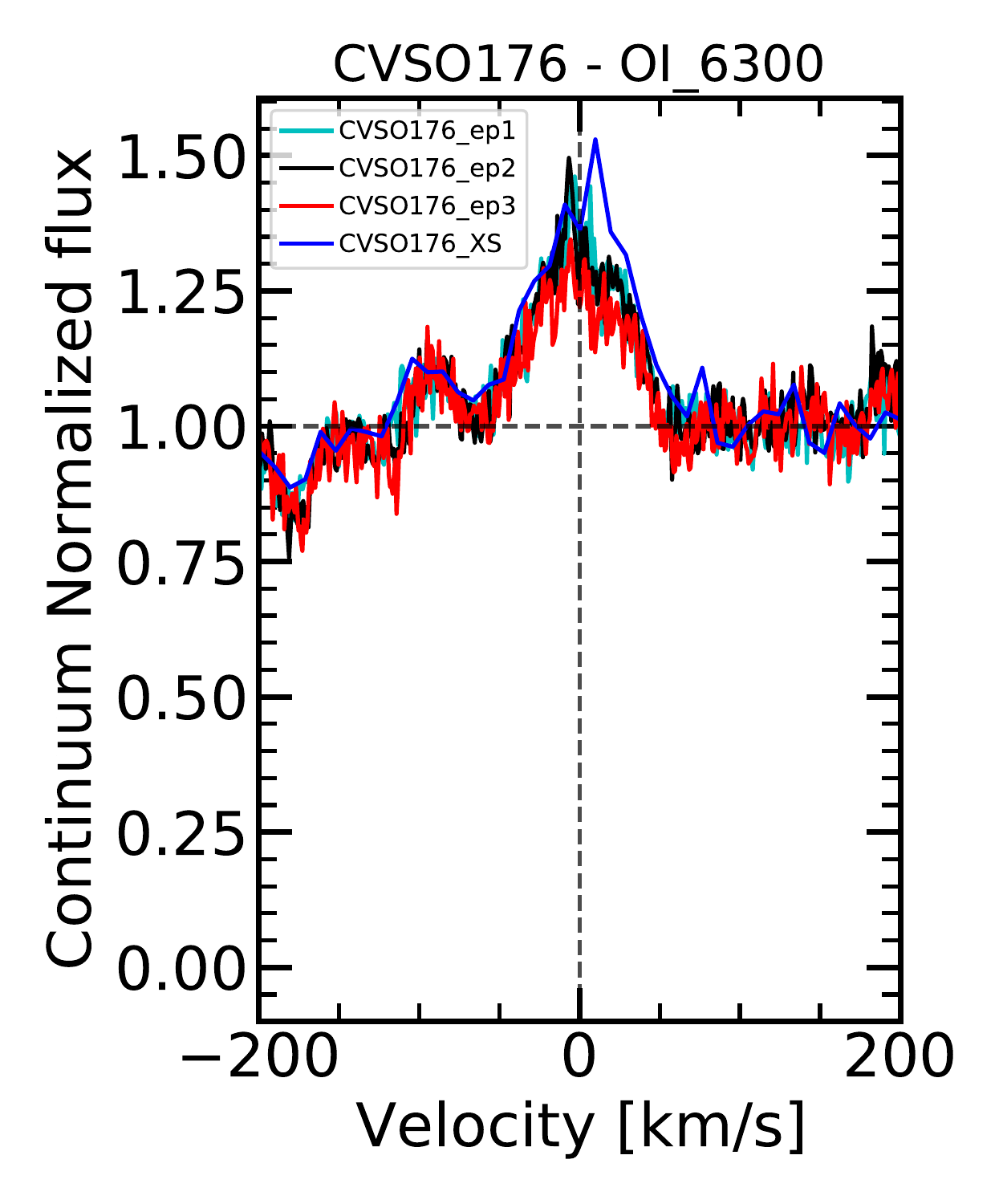}
\caption{Emission lines of the target CVSO176 observed with UVES and X-Shooter.
     \label{fig::lines_CVSO176}}
\end{figure*}

\begin{figure*}[]
\centering
\includegraphics[width=0.4\textwidth]{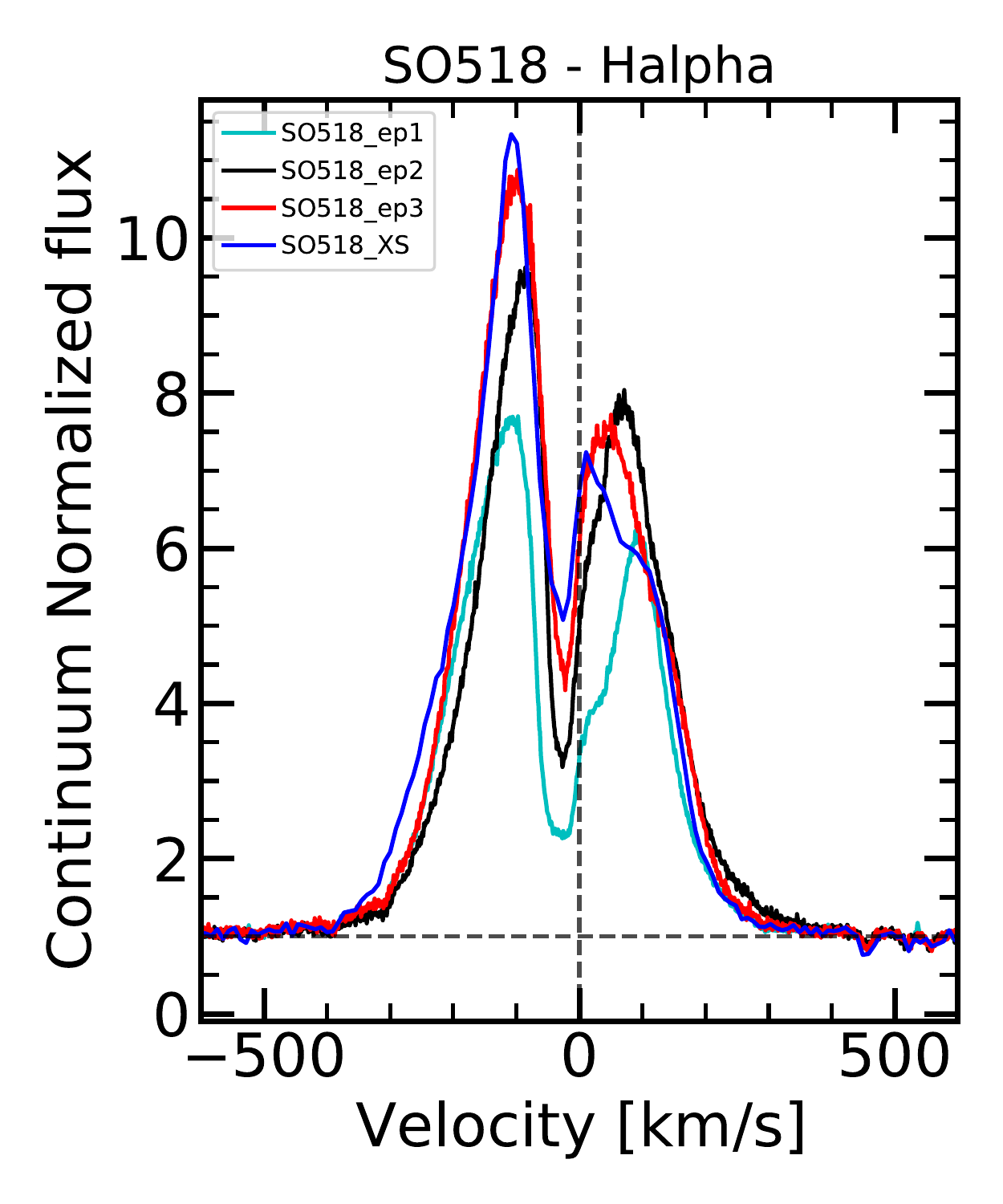}
\includegraphics[width=0.4\textwidth]{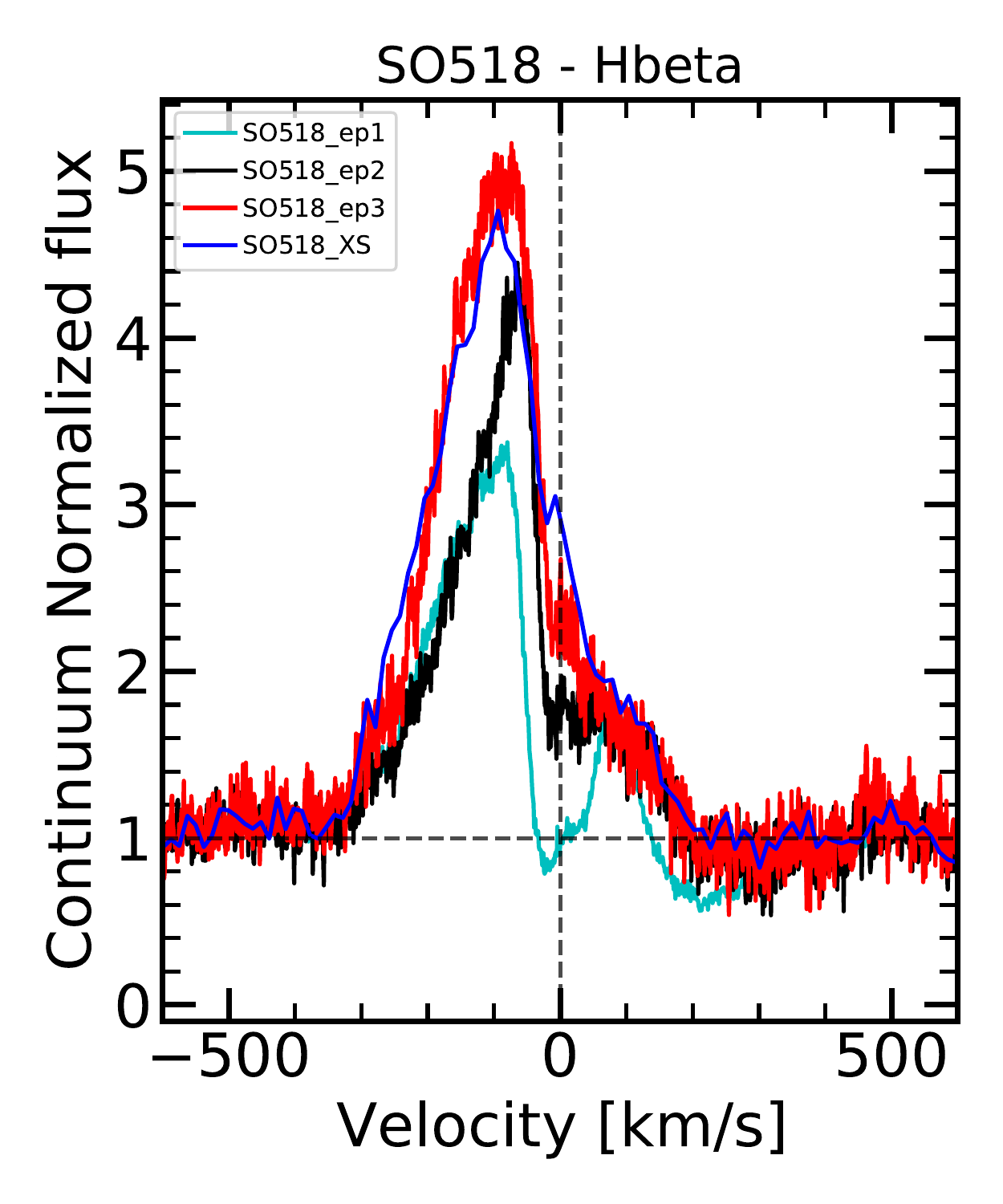}
\includegraphics[width=0.4\textwidth]{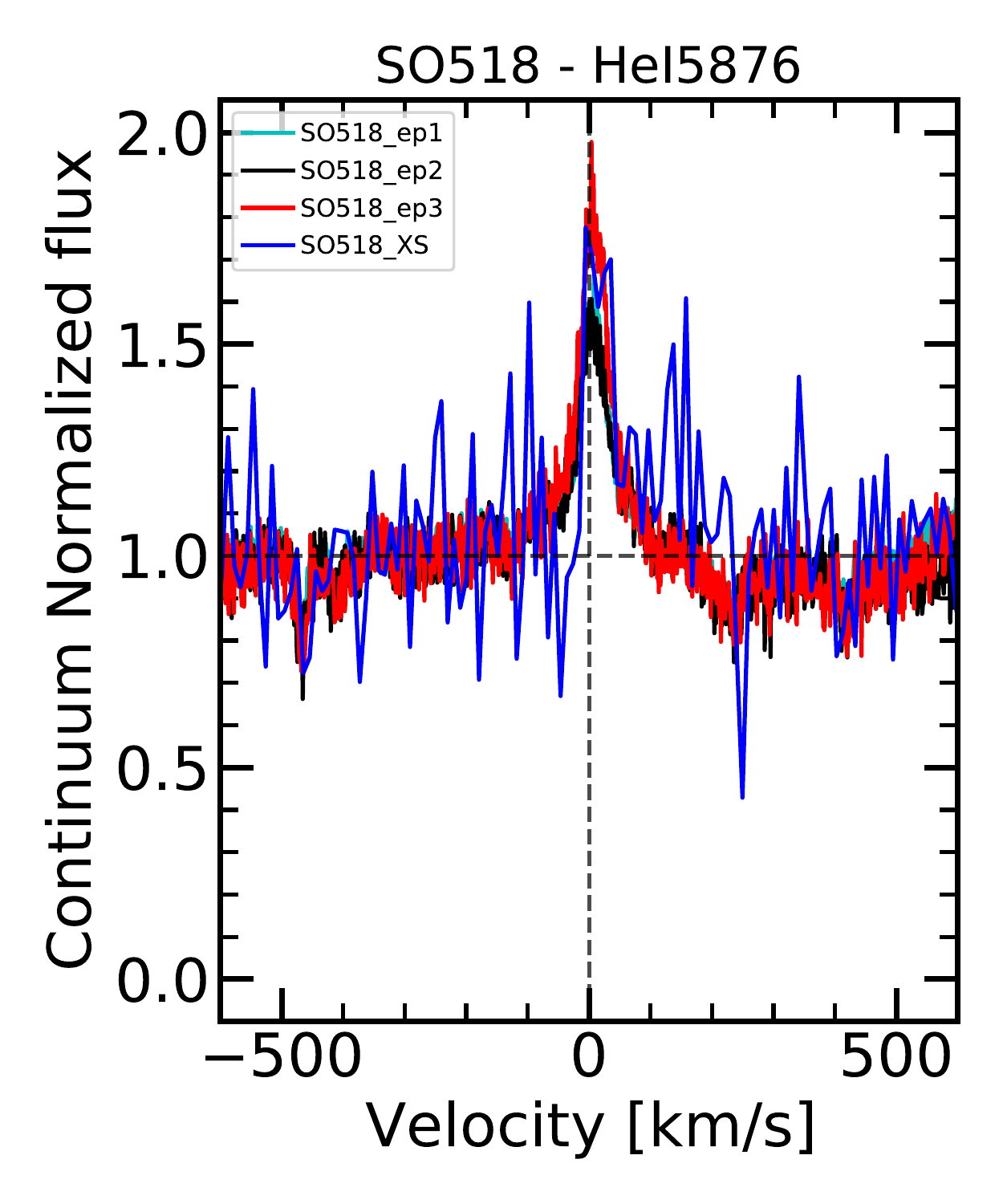}
\includegraphics[width=0.4\textwidth]{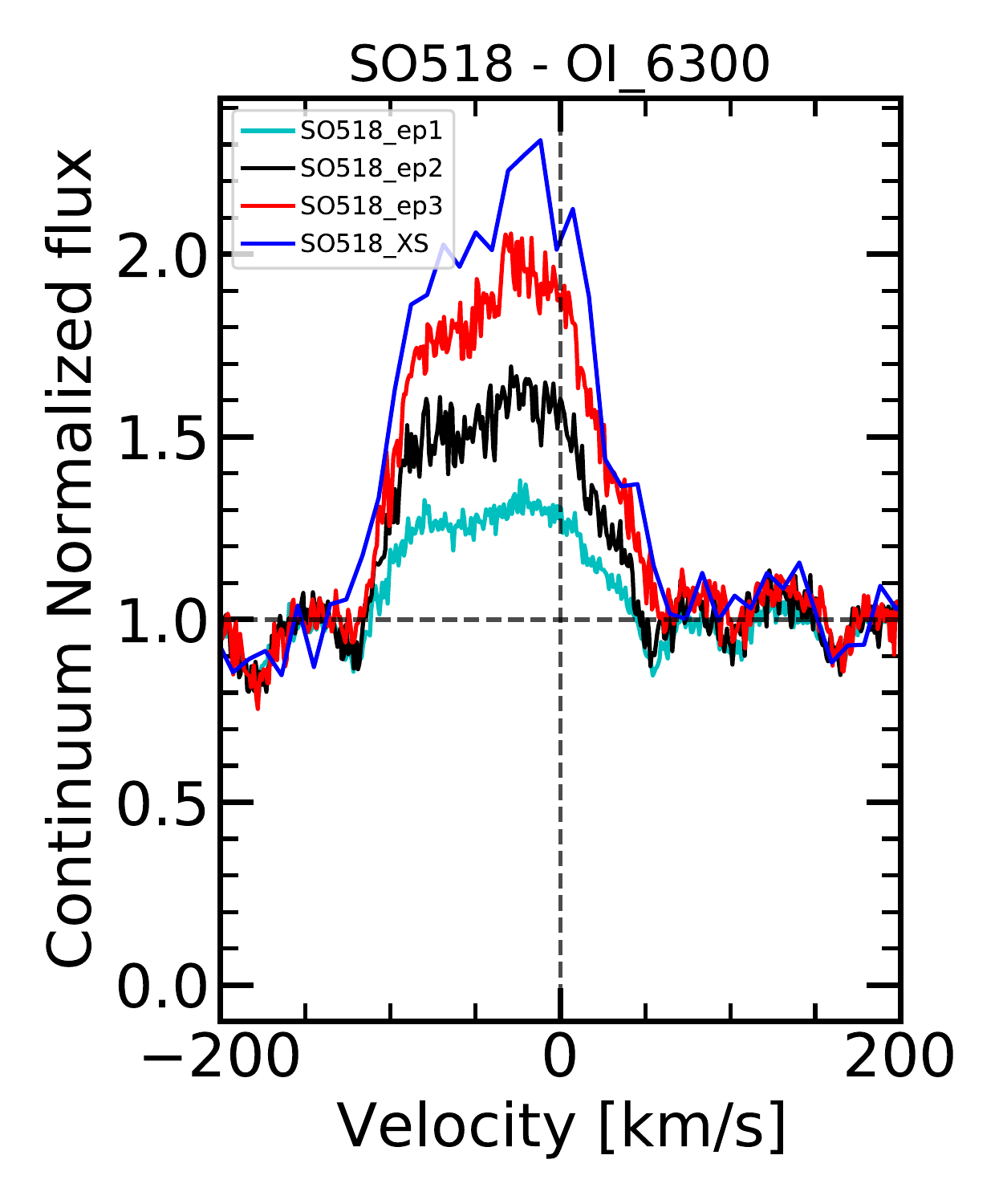}
\caption{Emission lines of the target SO\,518 observed with UVES and X-Shooter.
     \label{fig::lines_SO518}}
\end{figure*}

\begin{figure*}[]
\centering
\includegraphics[width=0.4\textwidth]{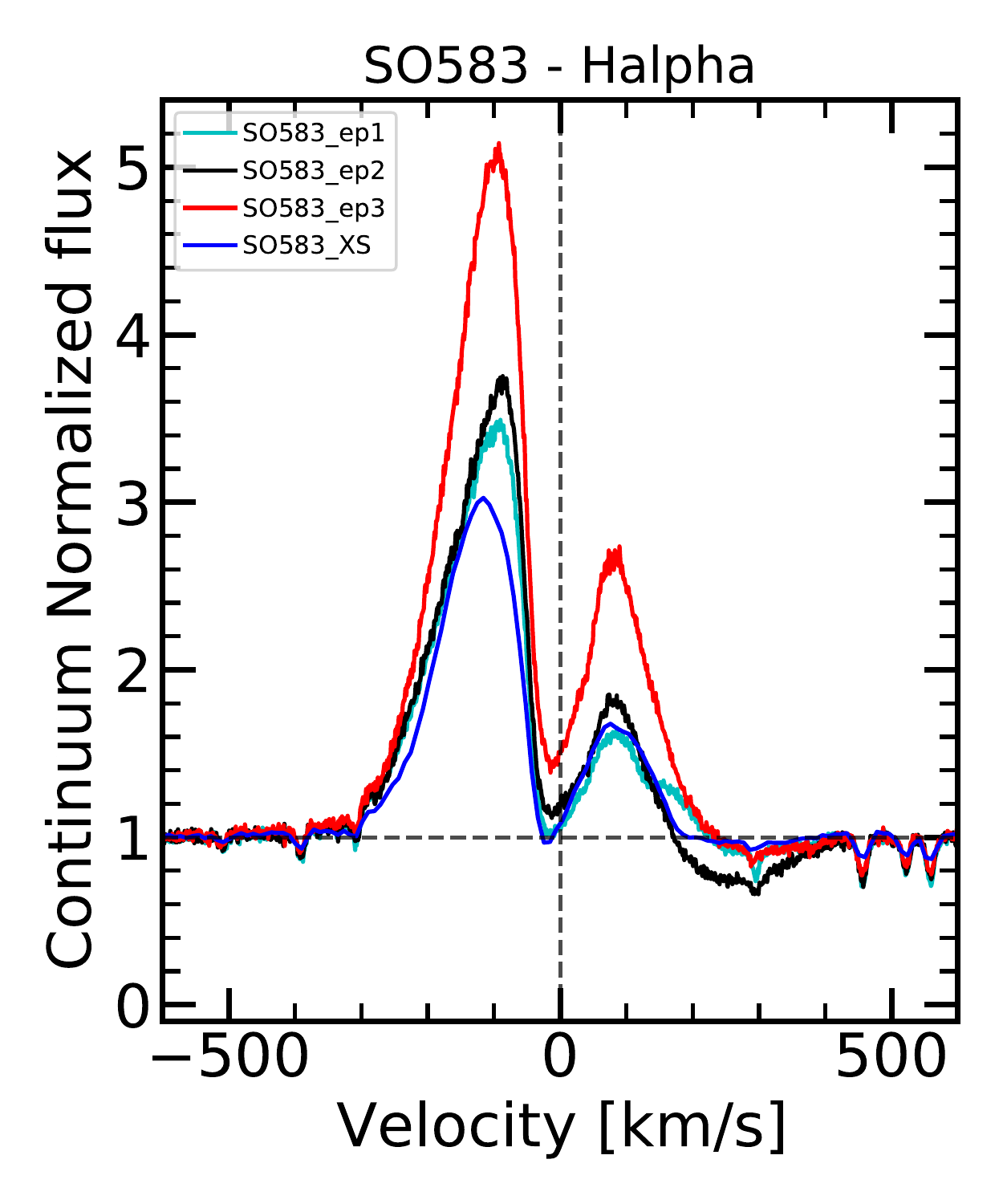}
\includegraphics[width=0.4\textwidth]{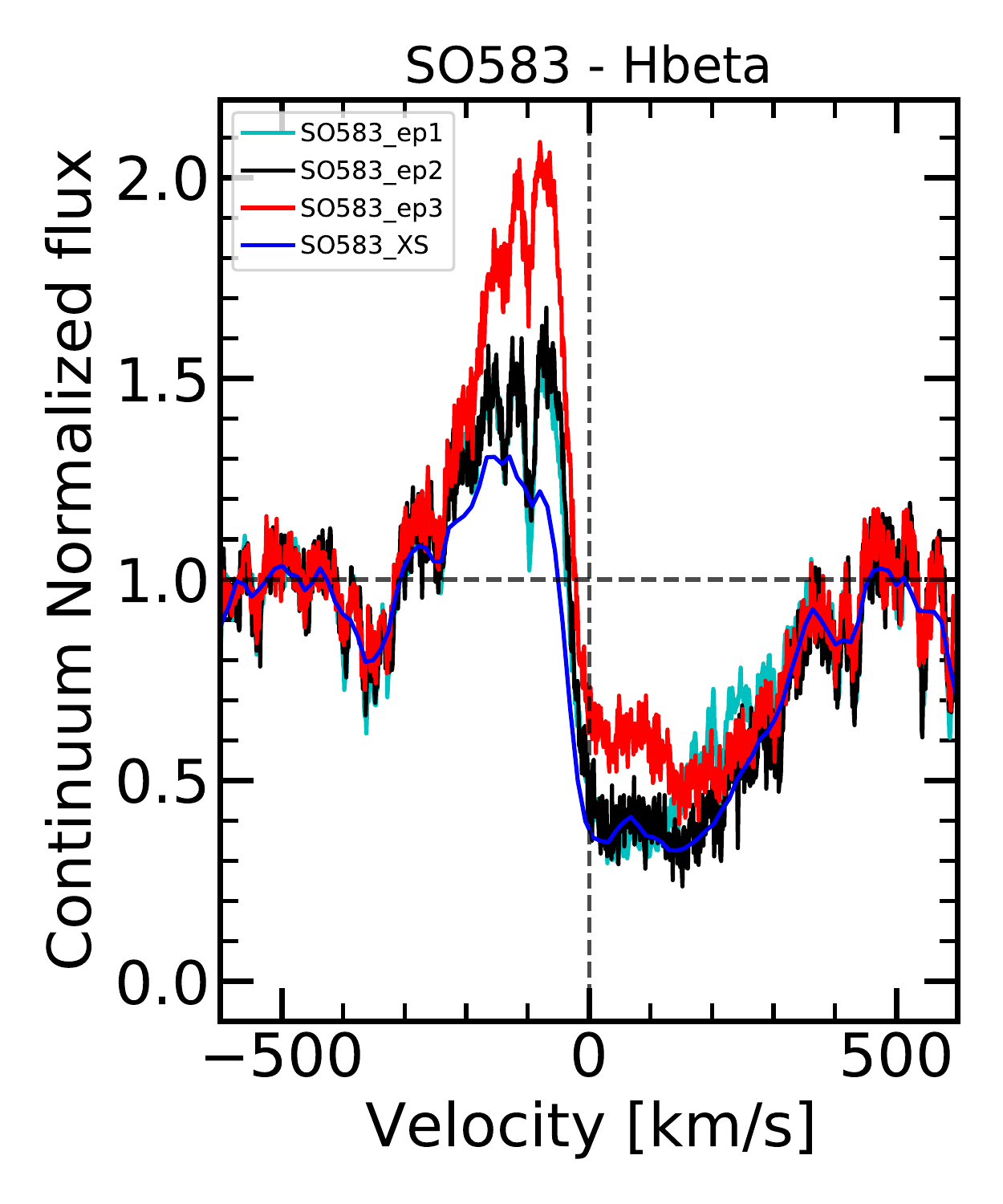}
\includegraphics[width=0.4\textwidth]{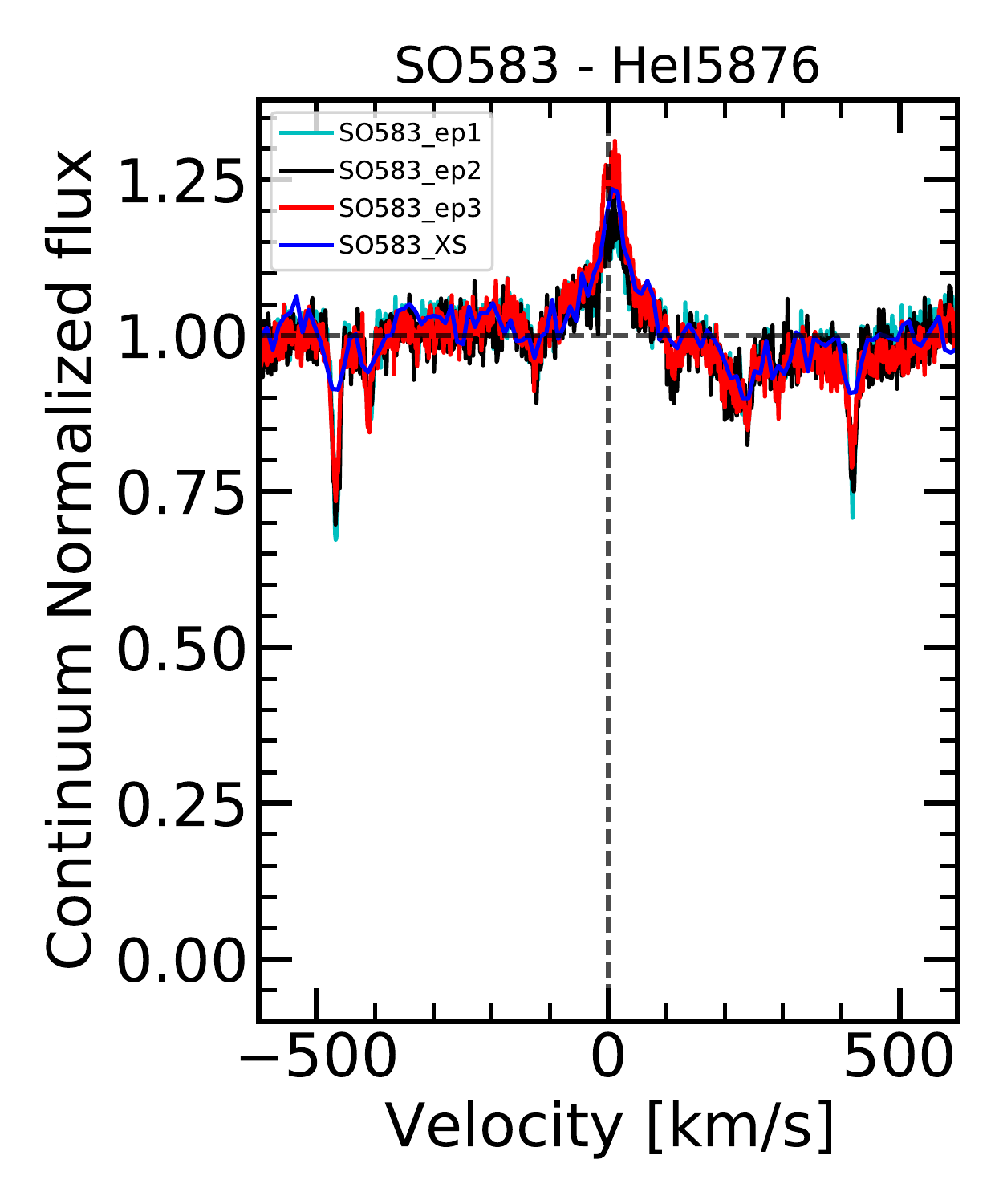}
\includegraphics[width=0.4\textwidth]{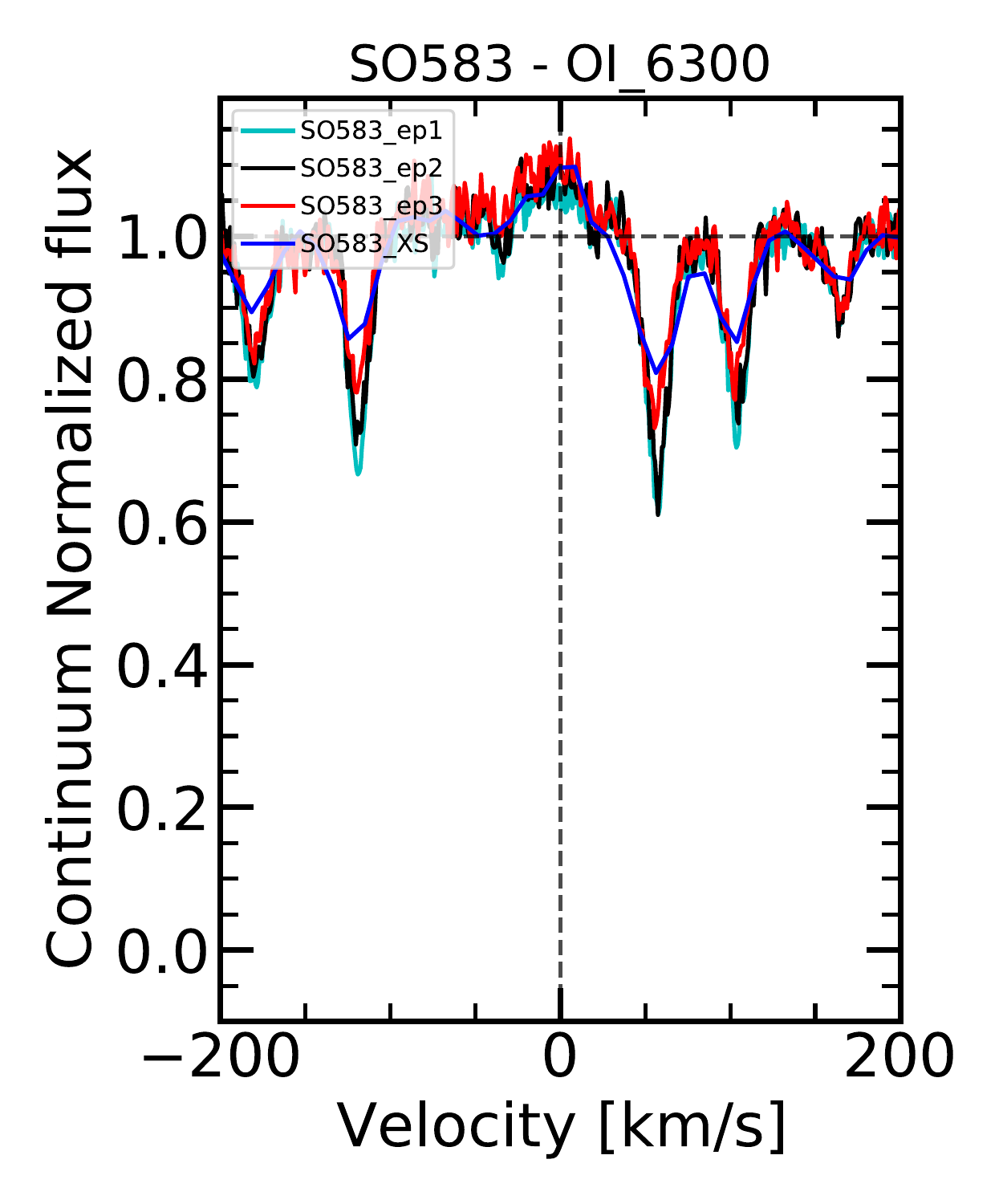}
\caption{Emission lines of the target SO\,583 observed with UVES and X-Shooter.
     \label{fig::lines_SO583}}
\end{figure*}

\begin{figure*}[]
\centering
\includegraphics[width=0.4\textwidth]{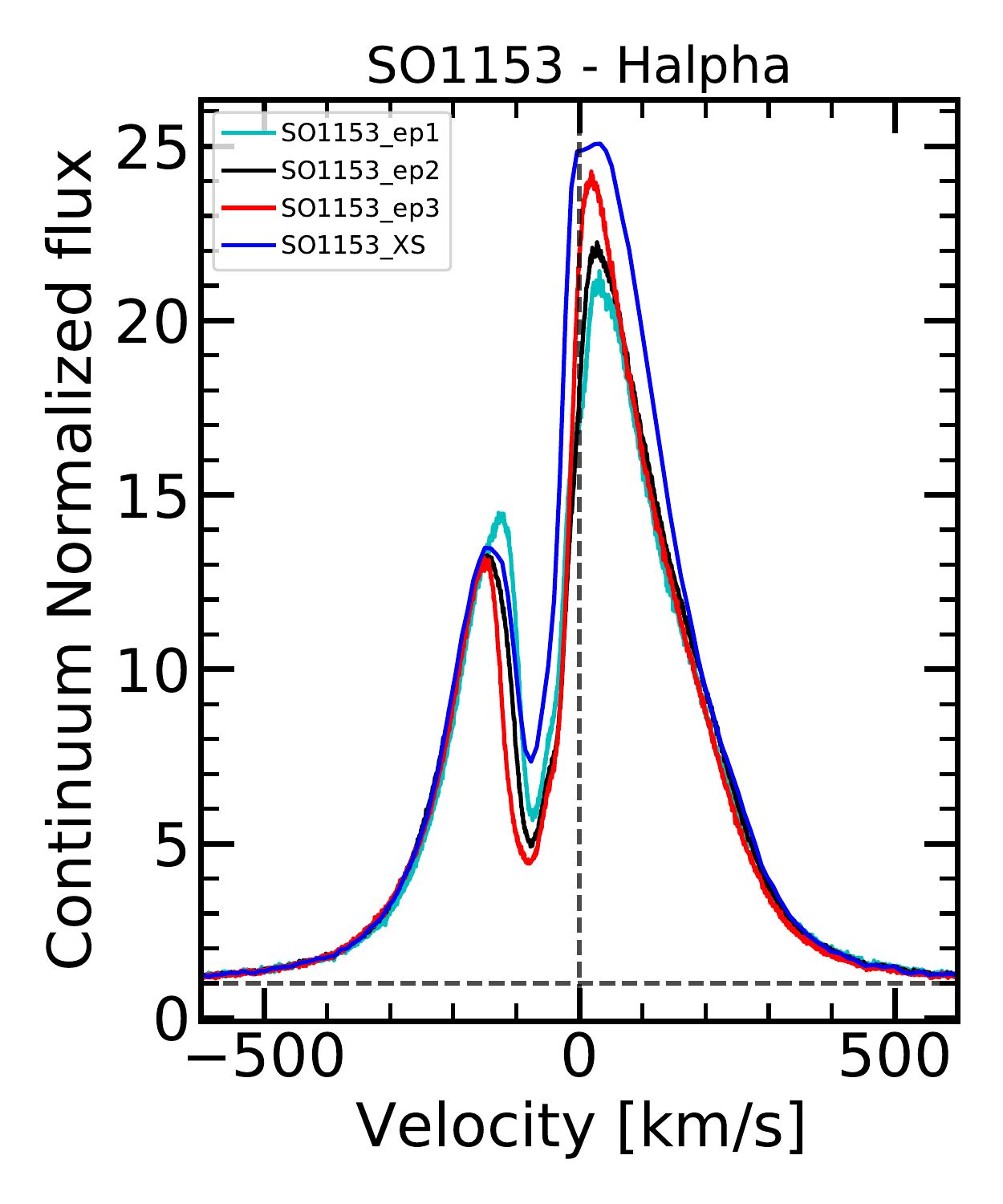}
\includegraphics[width=0.4\textwidth]{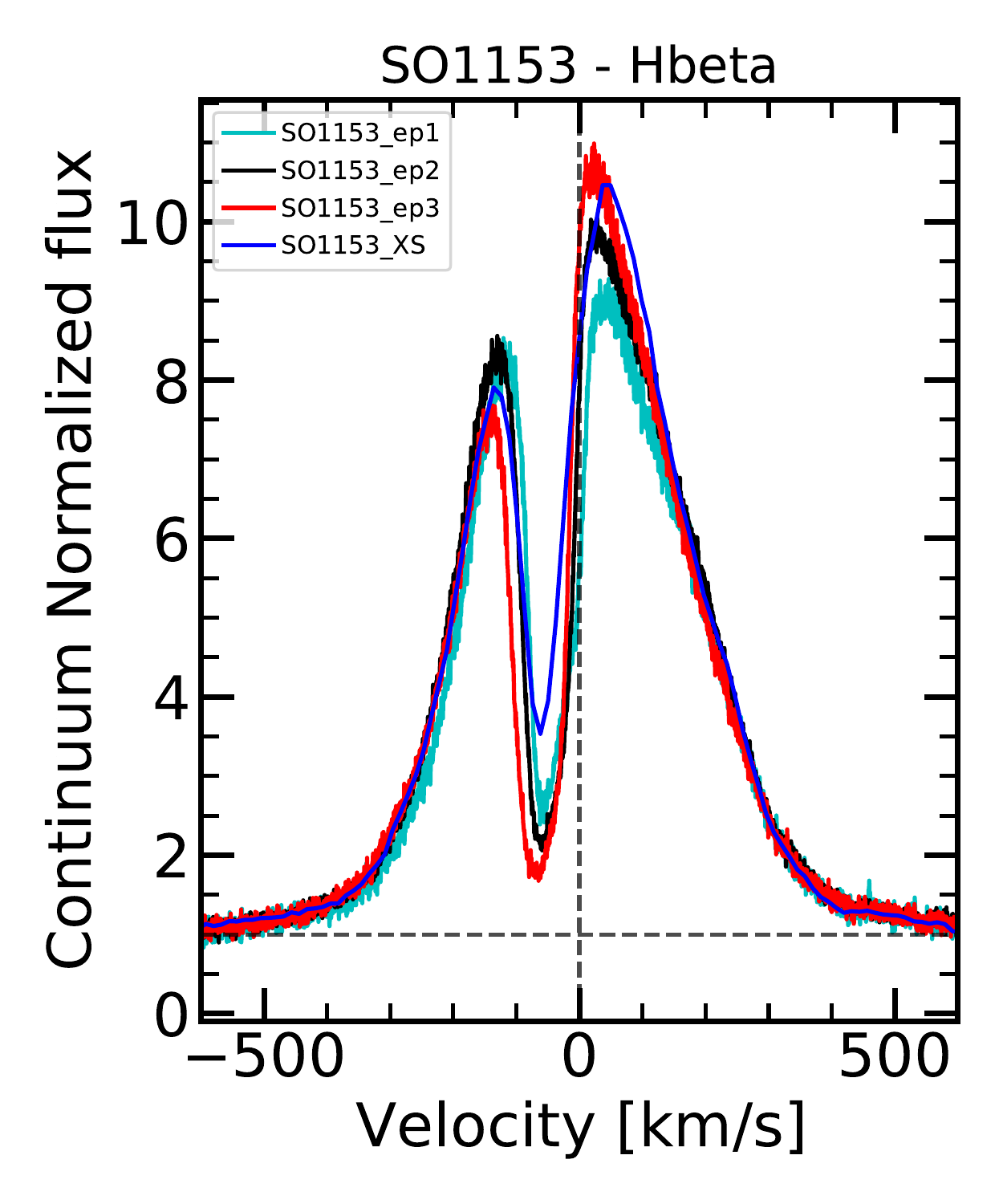}
\includegraphics[width=0.4\textwidth]{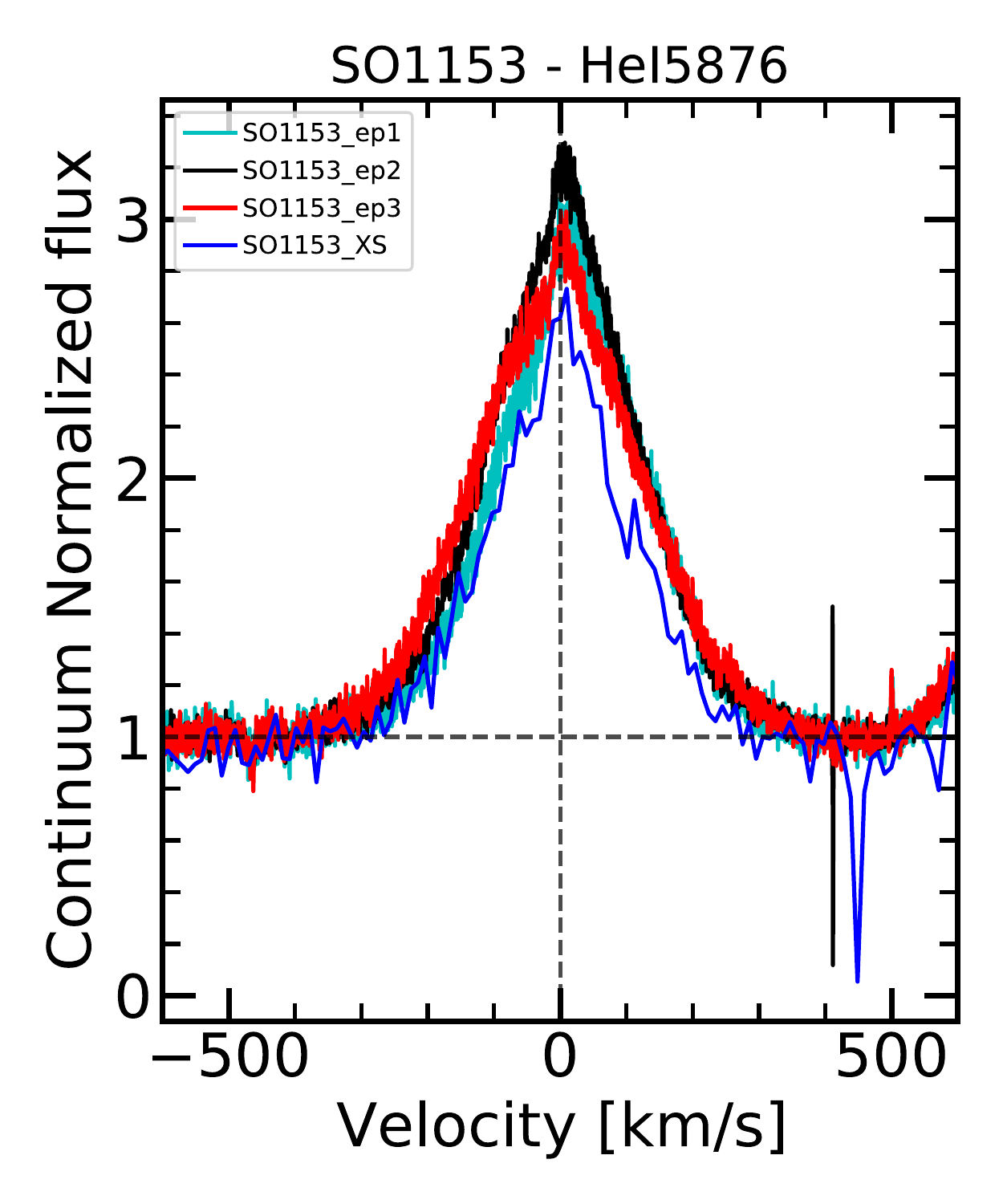}
\includegraphics[width=0.4\textwidth]{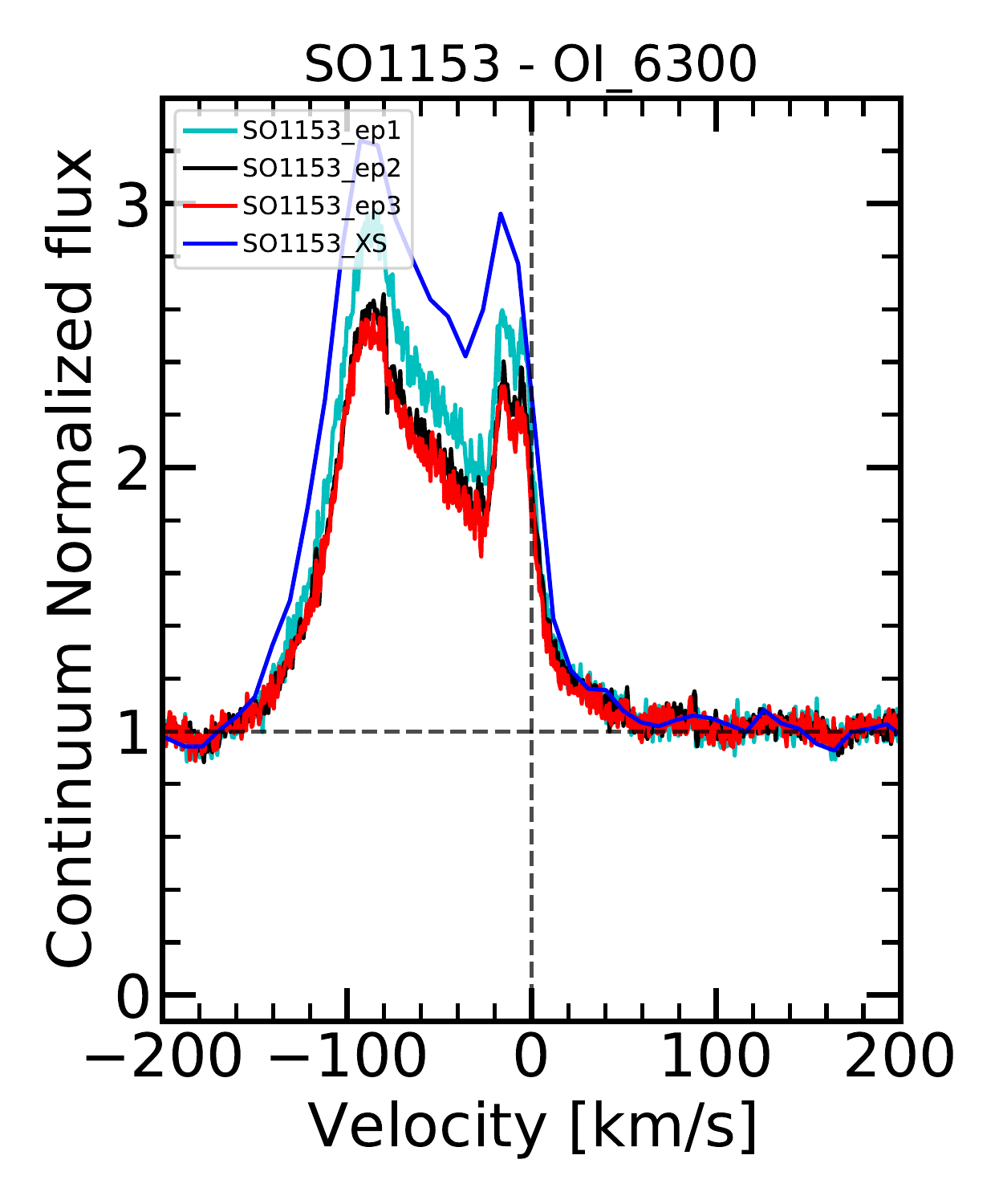}
\caption{Emission lines of the target SO\,1153 observed with UVES and X-Shooter.
     \label{fig::lines_SO1153}}
\end{figure*}

The profile of the H$\alpha$, H$\beta$, HeI 5876\AA,~ and [OI]$\lambda$6300\AA~ lines are shown in Figures~\ref{fig::lines_CVSO17}-\ref{fig::lines_SO1153}. The X-Shooter spectra, shown in blue in all figures, have lower resolution (R$\sim$10,000 - 20,000) than the UVES (R$\sim$70,000) and ESPRESSO (R$\sim$140,000) ones. 

\section{Examples of spectral subtraction with ROTFIT}\label{sect::rotfit_sub}
\begin{figure*}
\begin{center}
\hspace{-0.7cm}
\includegraphics[width=18cm]{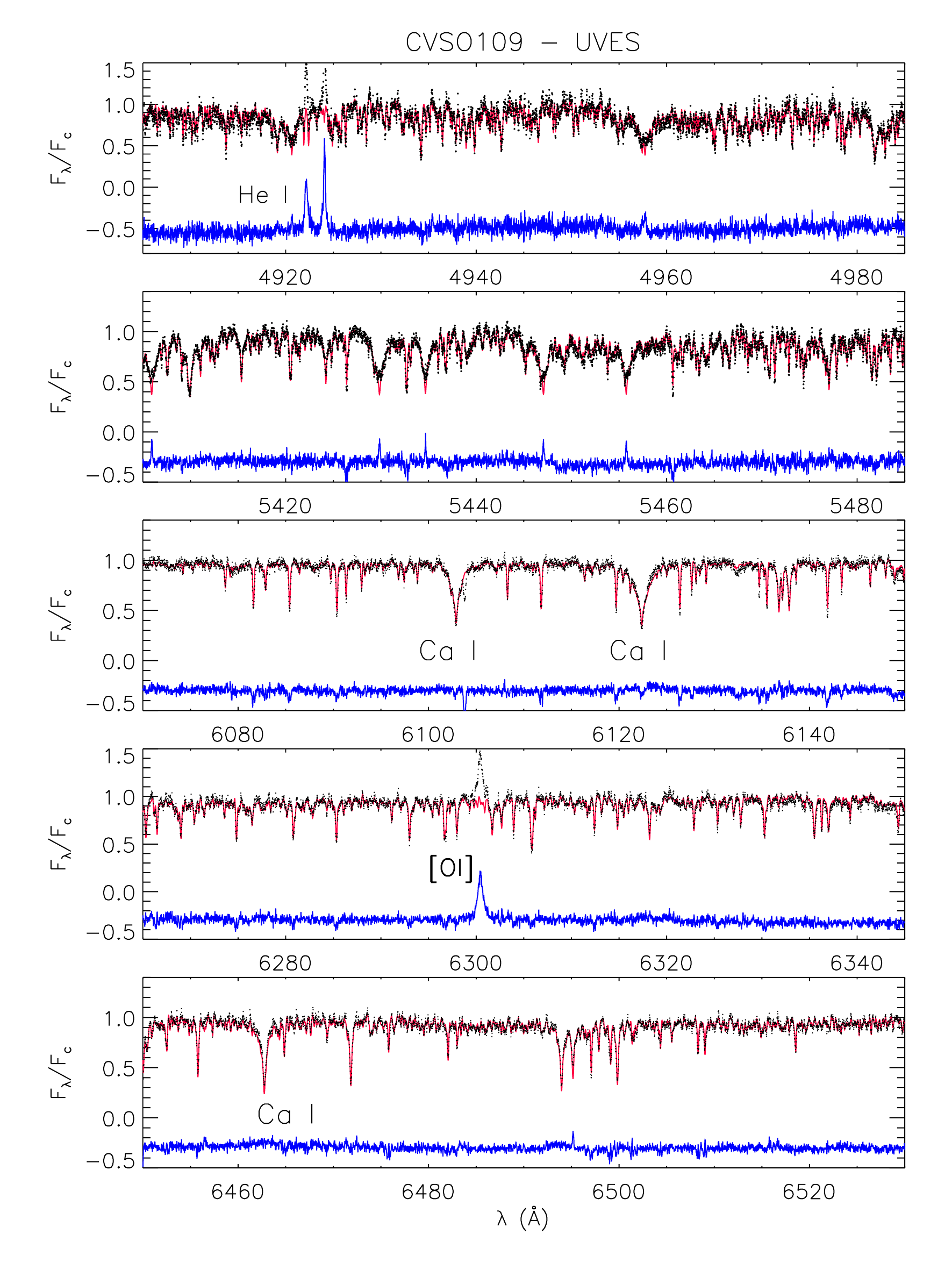} %
\caption{Subtraction of the non-active, lithium-poor template (red lines) from
the spectrum of CVSO\,109 (black dots) in the five different spectral regions. The most prominent emission and absorption lines have been indicated.
}
\label{fig:subtraction_suppl}
\end{center}
\end{figure*}

Two examples of spectral subtraction for two stars with different mass accretion rate observed with UVES and ESPRESSO are shown in the five spectral regions in Fig.~\ref{fig:subtraction_suppl}.

\begin{figure*}
\begin{center}
\hspace{-0.7cm}
\includegraphics[width=18cm]{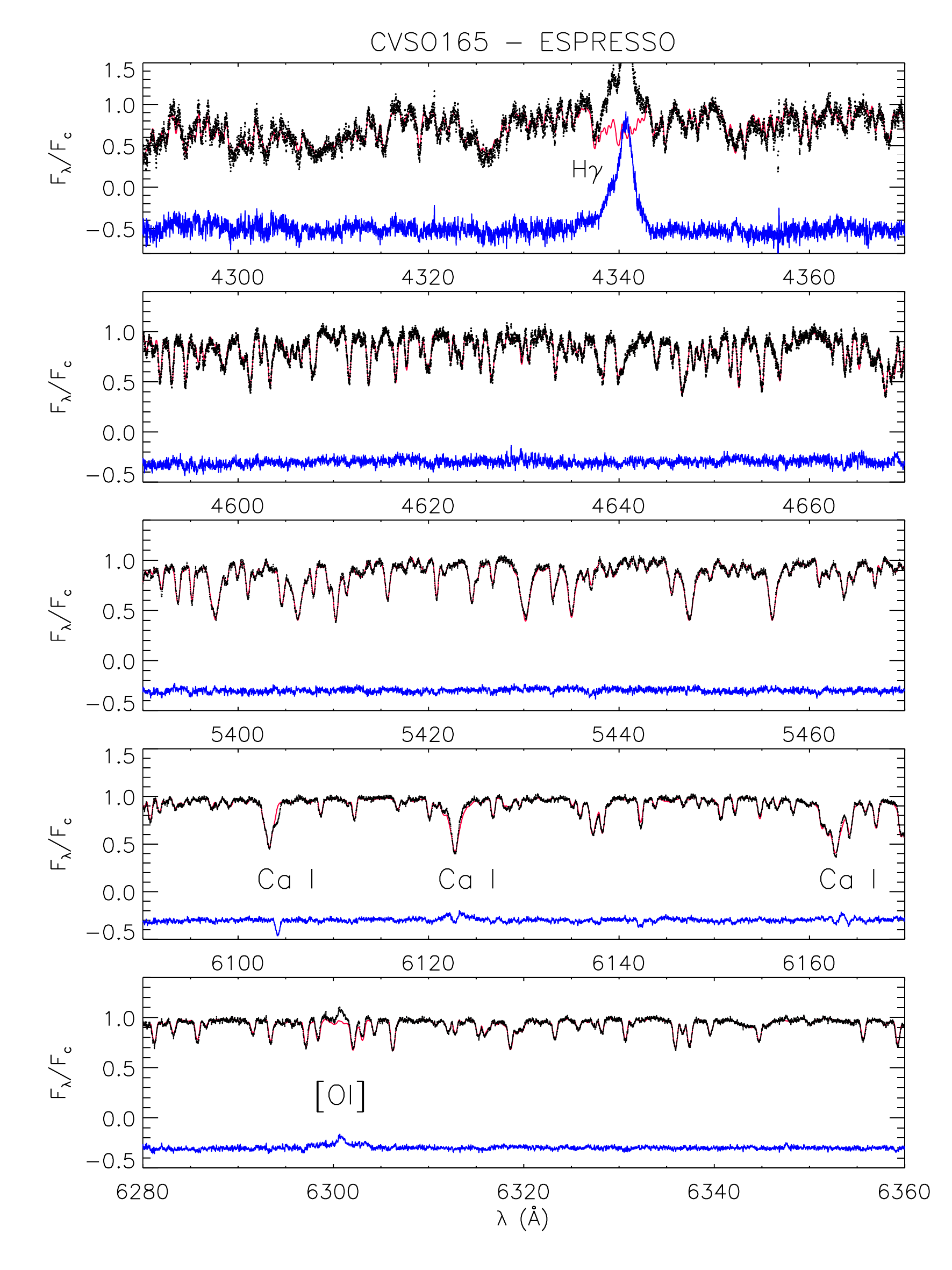} 
\caption{Subtraction of the non-active, lithium-poor template (red lines) from
the spectrum of CVSO\,165 (black dots) in five different spectral regions. The most prominent emission and absorption lines have been indicated.
}
\label{fig:subtraction_suppl2}
\end{center}
\end{figure*}

\end{document}